\documentclass[%
 reprint,
 superscriptaddress,
 nofootinbib,
 amsmath,amssymb,
 aps,prd
]{revtex4-2}

\usepackage[dvipsnames]{xcolor}
\usepackage{graphicx}
\usepackage{dcolumn}
\usepackage{bm}
\usepackage{placeins}
\usepackage{float}
\usepackage{comment}
\usepackage{graphicx}

\usepackage{hyperref}
\usepackage{makecell}

\usepackage[caption=false]{subfig}

\usepackage{hhline}

\newcommand{\apjl}{Astrophys. J. Lett.}%
\newcommand{\mnras}{Mon. Not. Roy. Astron. Soc.}%
\begin{document}

\preprint{APS/123-QED}

\title{Analytic models of the spectral properties of gravitational waves from neutron star merger remnants}

\author{Theodoros Soultanis}\email{theodoros.soultanis@h-its.org}
\affiliation{Heidelberg Institute for Theoretical Studies, Schloss-Wolfsbrunnenweg 35,
69118 Heidelberg, Germany}
\affiliation{Max-Planck-Institut f\"ur Astronomie, K\"onigstuhl 17, 69117 Heidelberg, Germany}

\author{Andreas Bauswein}
\affiliation{GSI  Helmholtzzentrum  f\"ur  Schwerionenforschung,  Planckstra{\ss}e  1,  64291  Darmstadt,  Germany}
\affiliation{Helmholtz Research Academy Hesse for FAIR (HFHF), GSI Helmholtz Center for Heavy Ion Research, Campus Darmstadt,  Germany}

\author{Nikolaos Stergioulas}
\affiliation{Department of Physics, Aristotle University of Thessaloniki, 54124 Thessaloniki, Greece}

\date{\today}

\begin{abstract}
We present a new analytic model describing gravitational wave emission in the post-merger phase of binary neutron star mergers. The model is described by a number of physical parameters that are related to various oscillation modes, quasi-linear combination tones or non-linear features that appear in the post-merger phase. The time evolution of the main post-merger frequency peak is taken into account and it is described by a two-segment linear expression. The effectiveness of the model, in terms of the fitting factor or, equivalently, the reduction in the detection rate, is evaluated along a sequence of equal-mass simulations of varying mass. We find that all parameters of the analytic model correlate with the total binary mass of the system. For high masses, we identify new spectral features originating from the non-linear coupling between the quasi-radial oscillation and the antipodal tidal deformation, the inclusion of which significantly improves the fitting factors achieved by the model. We can thus model the post-merger gravitational-wave emission with an analytic model that achieves high fitting factors for a wide range of total binary masses.  Our model can be used for the detection and parameter estimation of the post-merger phase in upcoming searches with upgraded second-generation detectors, such as aLIGO+ and aVirgo+, with future, third-generation detectors (Einstein Telescope and Cosmic Explorer) or with dedicated, high-frequency detectors.

\end{abstract}

\maketitle

\section{\label{sec:level1}Introduction  }

The two gravitational-wave (GW) events that have been identified as binary neutron star (BNS) mergers, GW170817 \citep{2017PhRvL.119p1101A} and GW190425 \citep{2020ApJ...892L...3A} offer a glimpse into the many more observations that are anticipated for the next years \citep{Aasi:2013wya}. Already, the detection of GWs from the inspiral phase of GW170817 produced new constraints on the dimensionless tidal deformability of neutron stars and thus on their equation of state (EoS)~\cite{2017PhRvL.119p1101A,LIGOScientific:2018hze}, see \citep{2020GReGr..52..109C,2021GReGr..53...27D} for recent reviews. Those detections can be  combined with information extracted from the electromagnetic counterpart of GW170817 or other measurements, e.g.~\cite{Bauswein_etal_2017,Abbott:2018exr,Radice2019Apr,2020NatAs...4..625C,2020NatAs...4..625C,2020Sci...370.1450D,2020PhRvD.101l3007L,2020Sci...370.1450D,2021arXiv210101201B,Raaijmakers2021sep,2021PhRvD.104f3003L,2021arXiv210508688P,2021arXiv210101201B,2021MNRAS.505.3016N} and references therein. Significant improvement on these EoS constraints are expected by combining a larger number of detections in the near future \cite{DelPozzo2013,2015PhRvD..92j4008C,2015PhRvD..91d3002L,2019PhRvD.100j3009H,2020PhRvD.101d4019C}. Although the sensitivity of the Advanced LIGO and Advanced Virgo detectors was not sufficient to detect the post-merger phase in GW170817  \citep{2017PhRvL.119p1101A,2017ApJ...851L..16A,LIGOScientific:2018hze}, such detections are likely to be achieved in the future, with upgraded \citep{KAGRA:2020npa}, with dedicated high-frequency \citep{2019PhRvD..99j2004M,Ackley:2020atn,2021PhRvD.103b2002G,2021CmPhy...4...27P,2021arXiv211010892S} or with third-generation \citep{LIGOScientific:2016wof,Maggiore:2019uih} detectors. Such observations of GWs in the post-merger phase of BNS mergers would offer a tremendous opportunity to probe the high-density EoS, see \citep{1992ApJ...401..226R,2005PhRvL..94t1101S,PhysRevLett.108.011101,Hotokezaka2013,2014PhRvD..90f2004C,AB2015,Bernuzzi2015,Clark2016,AB2016,PhysRevD.93.124051,Foucart2016,Lehner2016,PhysRevD.96.124035,PhysRevD.96.063011,2018PhRvD..98d3015E,2019PhRvL.122f1102B,Bauswein2019sep,PhysRevD.100.104029,Easter2019,2020PhRvL.124q1103W,PhysRevD.99.044014,PhysRevD.100.044047,2020PhRvD.102l3023B,2020EPJST.229.3595B,2020PhRvD.101h4039V,Easter2020,2020PhRvL.125z1101H,2020IJMPD..2941015F,2020PhRvD.101h4006K,2021PhRvD.104h3029P,2021arXiv210706804M,2021arXiv210101201B,2021arXiv211006957B,Liebling2021,Lioutas2021,2021arXiv211011968R,PhysRevD.104.083004} and references therein.

In order to detect the post-merger GW phase, robust and efficient data analysis techniques are needed, and currently, there are two main approaches. One employs morphology-independent signals \cite{PhysRevD.96.124035,PhysRevD.99.044014}, while the other is based on matched-filtering techniques, which require accurate GW post-merger template banks. We focus our discussion on the latter method. In \cite{AB2016} we introduced a time-domain analytic model, which utilizes a combination of three exponentially decaying sinusoids. The model incorporated the dominant post-merger peak ($f_{\rm peak}$) and the two most significant secondary frequency components ($f_\mathrm{spiral}$, $f_\mathrm{2-0}$) that also correlate with the binary's properties. Informed by numerical relativity simulations, frequency-domain models were introduced in \cite{Messenger2014,Clark2016,Easter2019,PhysRevD.100.044047} and time-domain models were introduced in \cite{Hotokezaka2013,Bose2018,Yang2018,PhysRevD.100.104029,Easter2020,2022arXiv220106461W}. In \cite{Clark2016} a frequency-domain model (which can be inverted to the time domain) was constructed for the amplitude and phase of the spectra, using a principal component analysis (PCA).  \cite{Easter2019} introduced a frequency domain hierarchical model, which generates amplitude spectra. In \cite{PhysRevD.100.044047} a frequency-domain amplitude model, which utilizes Lorentzian functions, in combination with relations connecting binary properties to the post-merger characteristic features was introduced.  \cite{PhysRevD.100.104029} developed a time-domain analytical model based on the morphology of the post-merger waveforms (employing numerical relativity informed relations), which can be combined with an inspiral waveform. In \cite{Easter2020} a time-domain analytic model which uses exponentially damped sinusoids (as in \cite{AB2016}) and includes three frequency components was developed. In this model, for all the frequency components, a constant linear frequency-drift term was introduced. 

In order to construct faithful post-merger GW templates, it is important to understand the underlying physical mechanisms, which dictate the different features of the GW spectrum. The latter is complex and even though some of its properties are well studied and understood others are not. 

In this work we compare the spectral properties of the post-merger phase along a sequence of equal-mass BNS merger simulations with increasing total mass ($M_\mathrm{tot}$). We find a smooth transition of spectral features along this sequence as anticipated in \cite{AB2015}. In particular, we study the time evolution of the dominant frequency component ($f_\mathrm{peak}$) in the GW spectra (see also \cite{Easter2020}) and using spectrograms we introduce a time-dependent 2-segment piecewise analytic function, which models such a frequency drift.

In addition, we identify a new coupling mechanism between tidal antipodal bulges  ($f_\mathrm{spiral}$, see \cite{AB2015}) and the quasi-radial mode ($f_0$), which results in two frequency peaks in the GW spectra of high mass models. The inclusion of this new feature significantly improves the fitting factors achieved for systems with binary masses near the threshold mass to prompt collapse.

We develop a time-domain analytic model (based on \cite{AB2016}) for the post-merger GW emission, which incorporates the four frequency components ($f_\mathrm{peak}$, $f_\mathrm{spiral}$, $f_{2\pm0}$) and allows a time-dependent description of the $f_\mathrm{peak}$ component. We introduce a hierarchical procedure to determine the analytic model's parameters. We evaluate the performance of our analytic model using the noise-weighted fitting factor ($FF$) and show that this remains higher than $\sim 0.96$ along the whole sequence of binary models considered.

Our new analytic model is described by physical parameters only and we find that all parameters correlate with the total binary mass of the system. It can be used for the detection and parameter estimation of the post-merger phase in upcoming searches with upgraded second-generation detectors, such as aLIGO+ and aVirgo+ (see \citep{KAGRA:2020npa}), with future, third-generation (Einstein Telescope \citep{2010CQGra..27s4002P} and Cosmic Explorer \citep{LIGOScientific:2016wof}) or with dedicated high-frequency detectors \citep{2019PhRvD..99j2004M,Ackley:2020atn,2021PhRvD.103b2002G,2021CmPhy...4...27P,2021arXiv211010892S}. Because the model is based on physical parameters, it elucidates the mechanisms shaping the spectra and how those depend on the binary masss.

This paper is structured as follows: In Sec.~\ref{Sec:Methods} we describe the physical systems we simulate and our numerical setup. In Sec.~\ref{Sec:Spectral analysis of the post-merger GW emission} we discuss particular features of the GW signal in the post-merger phase for a reference simulation. In Sec.~\ref{Sec. mass sequence} we consider a sequence of models with increasing total binary mass $M_\mathrm{tot}$ and describe how the spectral properties depend on $M_\mathrm{tot}$. In Sec.~\ref{Analytic post-merger model} we introduce an analytic time-domain model for the post-merger phase. In Sec.~\ref{Sec:Performance of analytic model} we discuss the fits of the analytic model to the simulation data and evaluate its performance using the noise-weighted fitting factor $FF$. Sec.~\ref{Parameters of the analytic model} addresses the parameters of the analytic model and their dependencies on the total binary mass $M_\mathrm{tot}$. In Sec.~\ref{Clode to prompt collapse} we focus on specific configurations with a total mass $M_\mathrm{tot}$ close to the threshold mass for prompt collapse $M_\mathrm{thres}$. In different appendices, we include more detailed information on various aspects described in the main text. 

Unless otherwise note, we employ a dimensionless system of units for which $c=G=M_\odot=1$. In Appendix~\ref{Empirical relations} we summarize the notation and the units for all the parameters of our analytic model.

\section{\label{Sec:Methods} Methods  }
We perform three-dimensional fully general relativistic simulations of binary neutron star mergers and discuss the spectral features of the post-merger gravitational wave (GW) emission. We use the MPA1 \cite{MUTHER1987469} EoS. This EoS model is compatible with constraints from GW170817 \cite{GW170817} and with the mass measurement of $2.01\pm0.04~M_\odot$  for pulsar PSR J0348+0432~\cite{Antoniadis:2013pzd}. We simulate a sequence of symmetric binaries (with mass ratio $\mathrm{q}=1$) varying the total binary mass $M_\mathrm{tot}$. We consider 8 models with ${M_{\mathrm{tot}}}=2.4, 2.5, 2.6, 2.7, 2.8, 2.9, 3.0$ and $3.1~{M_{\mathrm{\odot}}}$. None of the models collapses to a black hole (BH) within the simulation time of up to 25 milliseconds after merging, although the total mass of the most massive binary is close to the threshold binary mass for prompt BH formation $M_\mathrm{thres}$~\cite{AB2021june}. In the following sections we will introduce our results by discussing the model with $M_\mathrm{tot}=2.5~M_\odot$ as a reference simulation and then extend the analysis by including models with other binary masses.

We construct initial data (ID) of circular quasi-equilibrium solutions with the LORENE code \cite{Lorene:web,Gourgoulhon2001}. The initial separation between the centers of the NSs is~40 km, which results in a few revolutions before merging. We assess the impact of residual eccentricity $e$ in the simulations in Appendix~\ref{Effect of the eccentricity in the ID}. We show that the GW spectral features are hardly affected by eccentricity $e<0.01$.

For the evolution we employ the Einstein Toolkit code \cite{EinsteinToolkit:2021_05}. The hydrodynamics are solved by the \texttt{GRHydro} module \cite{Baiotti:2004wn,Moesta:2013dna} adopting the Valencia formulation \cite{Banyuls:1997,Font:2007zz}. We use the HLLE Riemann solver \cite{Harten:1983on} and WENO reconstruction \cite{WENO:1994,WENO:1996}. The spacetime evolution  is carried out in the Z4c formulation \cite{PhysRevD.81.084003,PhysRevD.88.084057} as implemented in the \texttt{CTGamma} module \cite{PhysRevD.83.044045,PhysRevLett.111.151101}. The computational domain consists of 7 refinement levels where the inner one has the finest resolution ($dx=277$ m) while the grid spacing is doubled at each successive level. The box size corresponds to $x_{\mathrm{max}}=2126.276$ km. In Appendix~\ref{Resolution study}) we describe an additional simulation with better resolution ($dx=185$ m) and find an only weak influence on the GW spectral features. In the following sections we will refer to this calculation as HR simulation and will indicate the respective results in various figures.

To reduce the computational costs we impose reflection symmetry with respect to the orbital plane and pi-symmetry with respect to the axis normal to this plane. We have also performed additional simulations without the pi-symmetry and find that the impact on the spectral properties is negligible (see Appendix~\ref{Effect of pi-symmetry}).

The EoS is implemented as a 7-segment piecewise polytrope \cite{PhysRevD.79.124032} and is supplemented with an ideal-gas pressure component to approximate thermal effects, where we set $\Gamma_\mathrm{th}=1.75$ (see e.g.~\cite{AB2010} justifying this value as a reasonable choice to model the post-merger GW emission).

We extract GWs employing the $\Psi_4$-formalism. The Weyl scalar $\Psi_4$ is decomposed in spin-weighted spherical harmonics at a finite coordinate radius $R$, where the radially averaged component is denoted by $\Psi_4^\mathrm{l,m}(t,R)$. We focus on the dominant mode $(l,m)=(2,2)$. We use an extraction coordinate radius of $R\simeq443~$km (but we tested also a larger extraction coordinate radius of $R\simeq1033~$km and find that the GW spectra are essentially unaffected). Computing the strain requires a double integration of $\Psi_4^\mathrm{l,m}(t,R)$ with respect to coordinate time $t$, which leads to non-linear drifts in the strain. To avoid this problem, we perform the integration in the frequency domain using a fixed frequency integration (FFI) scheme \cite{FFI-Reisswig_2011}.

We define the merging time $t_\mathrm{merge}$ as the time at which $|h(t)|=\sqrt{h_+^2(t)+h_\times^2(t)}$ reaches the maximum. We perform a time shift ($t\rightarrow t-t_\mathrm{merge}$) so that $t=0$ corresponds to the merging time. We split the GW signals into two phases accordingly: a) the inspiral phase ($t<0$) b) the post-merger phase ($t\ge0$). All Figures associated with GW quantities (such as GW spectrograms) use the aforementioned convention meaning that $t=0$ corresponds to the merging time obtained from $\max{|h(t)|}$. 
Figures and measures associated with the lapse function (such as spectrograms of the minimum lapse function $\alpha_\mathrm{min}$) define the merging time using the maximum of strain obtained from the quadrupole formula $|h^\mathrm{QF}(t)|$. We note that (prior to our time shifting) the two times of the $\max{|h(t)|}$ and $\max{|h^\mathrm{QF}(t)|}$ should differ by approximately $\Delta t \simeq \frac{1}{c}\ R\simeq 1.5$~ms, which thus can be removed as appropriate.

\section{Spectral analysis of the post-merger GW emission for a reference simulation}\label{Sec:Spectral analysis of the post-merger GW emission}

First, we consider the model with $M_\mathrm{tot}=2.5~M_\odot$ as a {\it reference simulation} and discuss the different features of the post-merger GW signal. We describe how we extract those features from the simulation data and how we  include them in an analytic model for the post-merger phase. Apart from considering the features present in the GW spectrum, we also also extract the time evolution of certain features by computing spectrograms. Throughout this work and for the sake of simplicity, we will often use a frequency, for example $f_\mathrm{peak}$, to refer to a specific peak in the spectrum or mode of the GW signal.

\subsection{Evolution of $\mathbf{f_{\mathrm{peak}}}$ and analytic fit}\label{fpeakevol}

The strongest feature in the post-merger gravitational wave signal is attributed to the fundamental quadrupolar oscillation mode (see \cite{Zhuge1996,Shibata2005a,Shibata2005b,Oechslin2007,Stergioulas2011,Bauswein2012Jan,Bauswein2012Sep,Hotokezaka2013,Takami2015,Bernuzzi2015,Clark2016,AB2016,Bauswein2019sep,Baiotti2019,Friedman2020,Bernuzzi2020,Diedrich2021mar,Sarin2021jun}). Its frequency, usually denoted as  $f_{\mathrm{peak}}$ (or $f_2$), dominantly depends on the equation of state (EoS) and the total binary mass. This is expected since the high-density regime of the EoS dictates the size of the remnant. As the remnant undergoes further evolution, $f_{\mathrm{peak}}$ shifts to higher or lower frequencies. The interplay of cooling and angular momentum redistribution as well as losses leads to a change of the stellar structure and thus to a change in the dominant oscillation frequency. 

\begin{figure}[h]
	\includegraphics[scale=0.34]{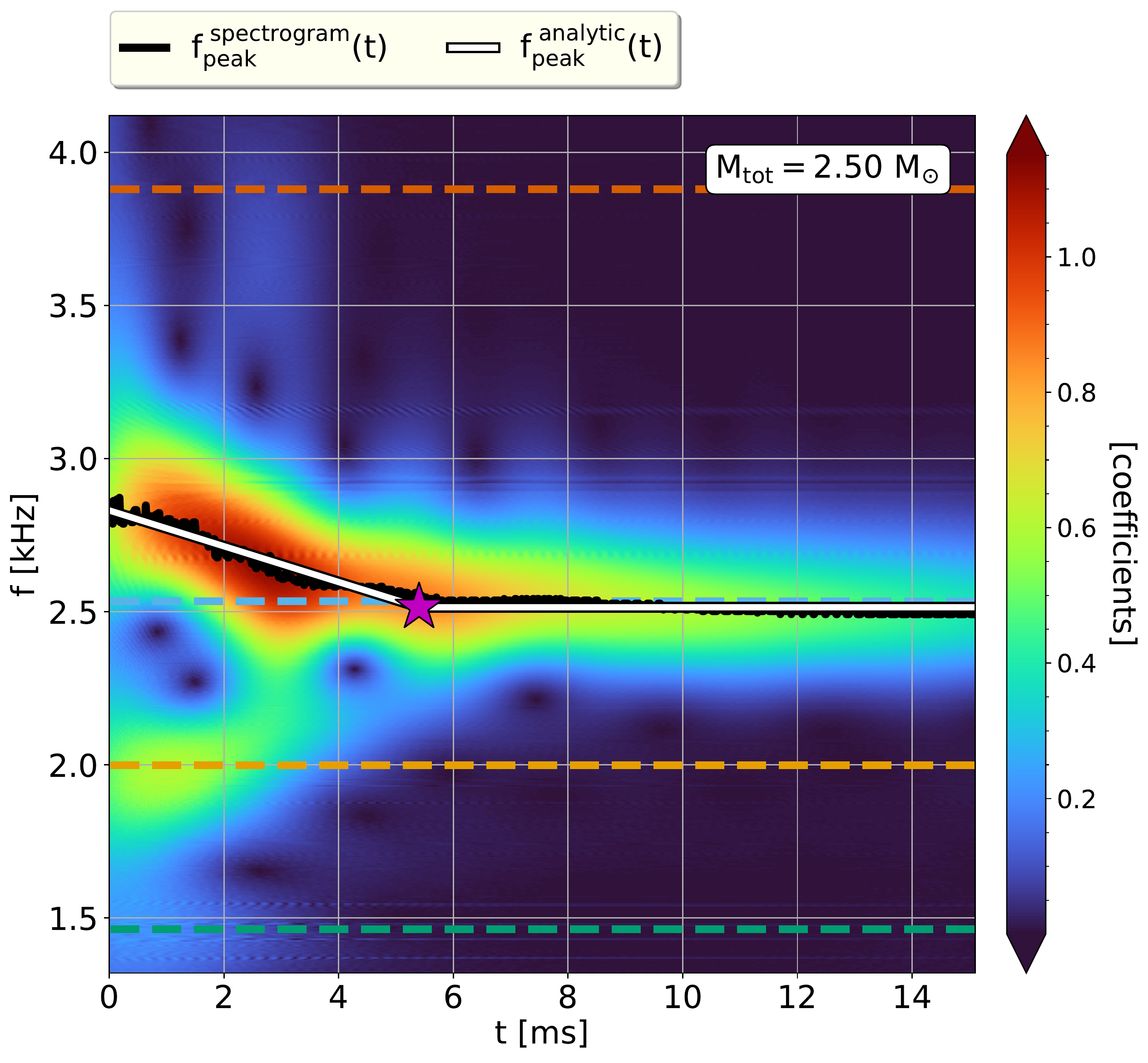} 
	\caption{\label{fig:dumpSPEC_mpa1-m1.25} Spectrogram of strain $h_{+}(t)$ for the reference simulation. The black curve illustrates $f_{\mathrm{peak}}(t)$ determined by the maximum wavelet coefficient at given time $t$. The white curve shows the 2-segment piecewise analytic fit $f_\mathrm{peak}^\mathrm{{analytic}(t)}$ of Eq.~(\ref{fpeak_piecewise}). The purple star indicates $t=t_*$, after which the frequency remains constant. The cyan, yellow, green, and orange dashed horizontal lines indicate $f_\mathrm{peak}, f_\mathrm{spiral}, f_{2-0}$, $f_{2+0}$, respectively, as extracted from the spectrum shown in Fig. \ref{fig:dumpFFT_evolution_mpa1-m1.25.pdf}. }
\end{figure}

In order to understand the frequency evolution of particular components in the GW signal, we compute spectrograms that employ a wavelet-based scheme \cite{Lee2019}. The spectrogram in Fig.~\ref{fig:dumpSPEC_mpa1-m1.25} displays the time evolution of the dominant component $f_\mathrm{peak}$ of our reference simulation. In the first few milliseconds, $f_{\mathrm{peak}}$ undergoes a rapid evolution, and the signal can be split in two phases: a) for $t\lesssim6$ ms, $f_{\mathrm{peak}}$ follows a decreasing trend approximately from about 2.8kHz to 2.5kHz, b) for $t\gtrsim 6$ ms, $f_{\mathrm{peak}}$ is approximately constant with $f_{\mathrm{peak}}=2.5$ kHz. We quantify the drift by extracting the evolution of $f_\mathrm{peak}(t)$ from the spectrogram (black curve) as the frequency which corresponds to the maximum wavelet coefficient at time $t$. We model $f_{\mathrm{peak}}(t)$ with a simple 2-segment piecewise {\it analytic fit} with respect to the time coordinate $t$. The first segment describes the initial drift as a linear function in the frequency-time plane, while the second segment assumes a constant $f_{\mathrm{peak}}$, imposing continuity as
\begin{eqnarray}\label{fpeak_piecewise}
f_{\mathrm{peak}}^\mathrm{{analytic}}(t) =    \left\{
\begin{array}{ll}
\zeta_\mathrm{drift}\cdot t + f_\mathrm{peak,0}  & \mbox{for } t\leq t_* \\
f_\mathrm{peak}(t_*) &\mbox{for } t> t_* \\
\end{array} .
\right. 
\end{eqnarray}
The analytic fit is shown as two white line segments in Fig.~\ref{fig:dumpSPEC_mpa1-m1.25} and is in good agreement with the numerically extracted $f_{\mathrm{peak}}(t)$. Notice that Fig.~\ref{fig:dumpSPEC_mpa1-m1.25} shows a particular example, with an initial negative drift. In the parameter space of different EoSs and masses, the initial drift can be positive or nearly zero and there are also cases which can be better described by a constant drift up to the delayed collapse to a black hole (these cases can also be covered by the above analytic description).
 
\begin{figure}[h!]
	\includegraphics[scale=0.33]{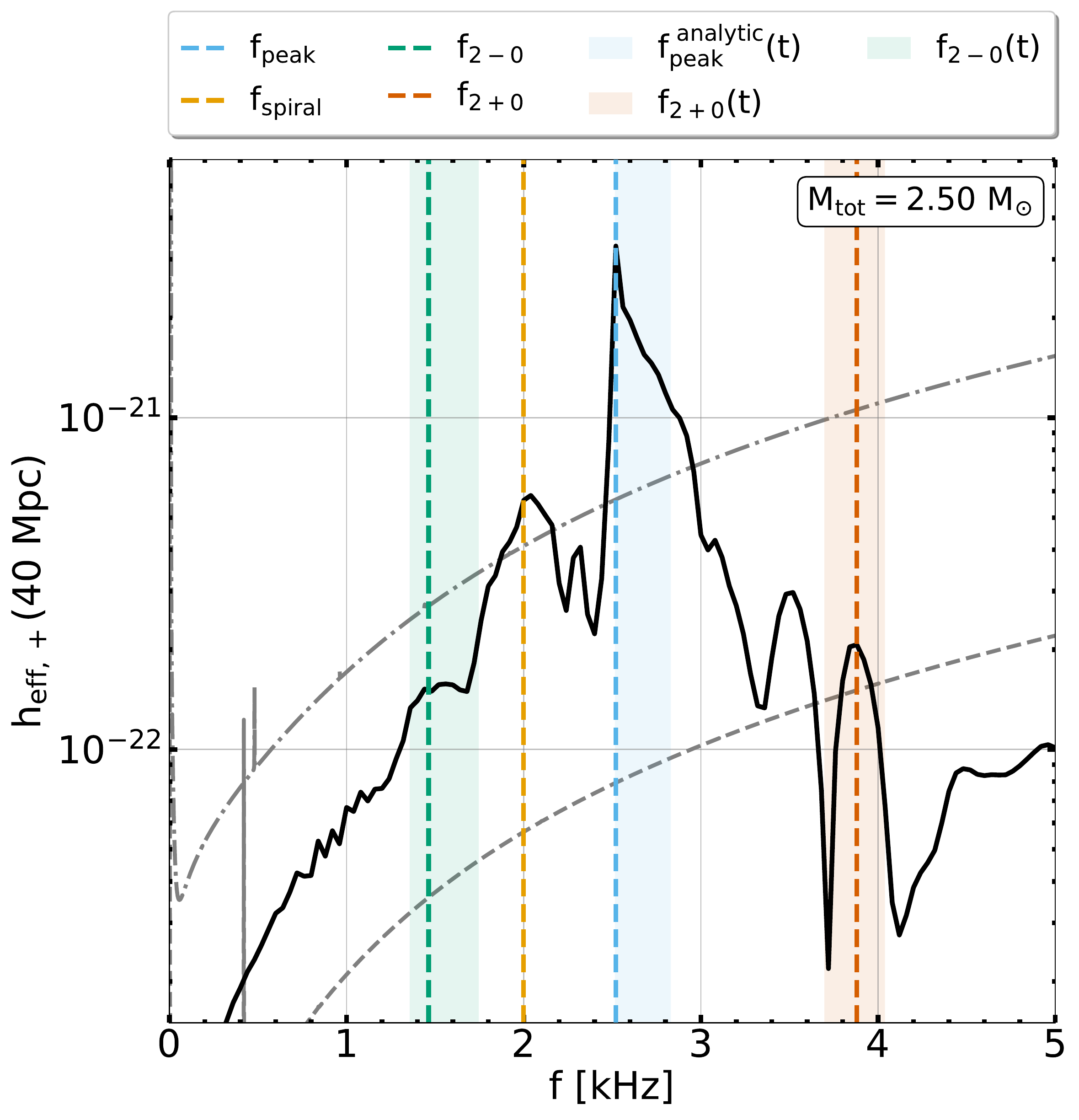} 
	\caption{\label{fig:dumpFFT_evolution_mpa1-m1.25.pdf} Effective GW spectrum $h_\mathrm{eff,+}(f)$ for the post-merger phase of the reference simulation. Colored dashed vertical lines indicate the frequency peaks $f_\mathrm{peak},f_\mathrm{spiral}, f_{2-0}, f_{2+0}$. Shaded areas correspond to the frequency range of $f_\mathrm{peak}, f_{2-0}, f_{2+0}$ (see text for details). The dash dotted curves denote the design sensitivity Advanced LIGO \cite{LIGOScientific:2014pky} and of the Einstein Telescope \cite{EinsteinTelescope2010}, respectively.}
\end{figure}

Throughout this work, $f_{\mathrm{peak}}$, i.e. without explicit time argument, denotes the frequency corresponding to the maximum amplitude of $h_\mathrm{eff,+}(f)=f\cdot \widetilde{h}_+(f)$, where $\widetilde{h}_+(f)$ is the Fourier transform of ${h}_+(t)$ (in agreement with the definition currently used in literature), $f_{\mathrm{peak}}(t)$ refers to dominant frequency as a dynamical quantity, which is extracted from the spectrogram. Below we use this notation for other components of the signal as well. As shown in Fig.~\ref{fig:dumpFFT_evolution_mpa1-m1.25.pdf}, the frequency peak may not be symmetric, but it can have a broader, one-sided distribution towards higher frequencies. This feature is explained by an evolving $f_{\mathrm{peak}}(t)$ which covers the corresponding frequency range, see also Section \ref{GW fits}. The cyan-shaded area in Fig.~\ref{fig:dumpFFT_evolution_mpa1-m1.25.pdf} shows the frequency range as covered by our analytic piecewise function $f_{\mathrm{peak}}^\mathrm{{analytic}}(t)$, in agreement with the one-sided peak of the dominant mode.  

From the spectrograms we also extract a mean value of $f_{\mathrm{peak}}(t)$ averaged over the initial interval from 0 to $t_*$. This mean value $\langle f_{\mathrm{peak}}^{t\in [0, t*]}(t) \rangle$ does not necessarily coincide very well with the maximum in the power spectrum, but it provides a measure for $f_{\mathrm{peak}}(t)$ at early times.

\subsection{Secondary GW peaks $\mathbf{f_\mathrm{2\pm0}}, \mathbf{f_{\mathrm{spiral}}}$}\label{SubSec. secondary peaks}
As it is apparent from the spectrogram (see Fig.~\ref{fig:dumpSPEC_mpa1-m1.25}) and the spectrum (see Fig.~\ref{fig:dumpFFT_evolution_mpa1-m1.25.pdf}) the post-merger GW signal contains several additional secondary features apart from the dominant oscillation mode. Two of those subdominant features originate from a non-linear coupling between the quadrupolar mode and the quasi-radial oscillation mode $f_0$. This coupling is expected to produce side peaks (combination tones) of the dominant peak at frequencies $f_{2\pm0}\approx f_\mathrm{peak} \pm f_0$. Inspecting the GW spectrum in  Fig.~\ref{fig:dumpFFT_evolution_mpa1-m1.25.pdf}, we indeed identify secondary peaks at approximately $f_\mathrm{peak}\pm f_0$, where we estimate $f_0$ from a Fourier transform of the evolution of the minimum lapse function, since $f_0$ (being a quasi-radial oscillation) does not occur prominently in the GW spectrum. 

In our analysis we extract and define $f_{2\pm0}$ as the corresponding local maxima in the effective power spectrum $h_\mathrm{eff,+}(f)$ (employing the full signal including the inspiral), where we note that the relation $f_{2\pm0}=f_\mathrm{peak}\pm f_0$ holds only approximately. This slight inequality is due to the fact that the frequencies of the $f_{2\pm0}$ peaks are determined during the early, very dynamical evolution of the remnant, when the radial oscillation is still strongly excited. In this very early post-merger period, the main frequency peaks, in particular $f_\mathrm{peak}(t)$, can evolve rapidly.

\begin{figure}[h]
	\includegraphics[scale=0.34]{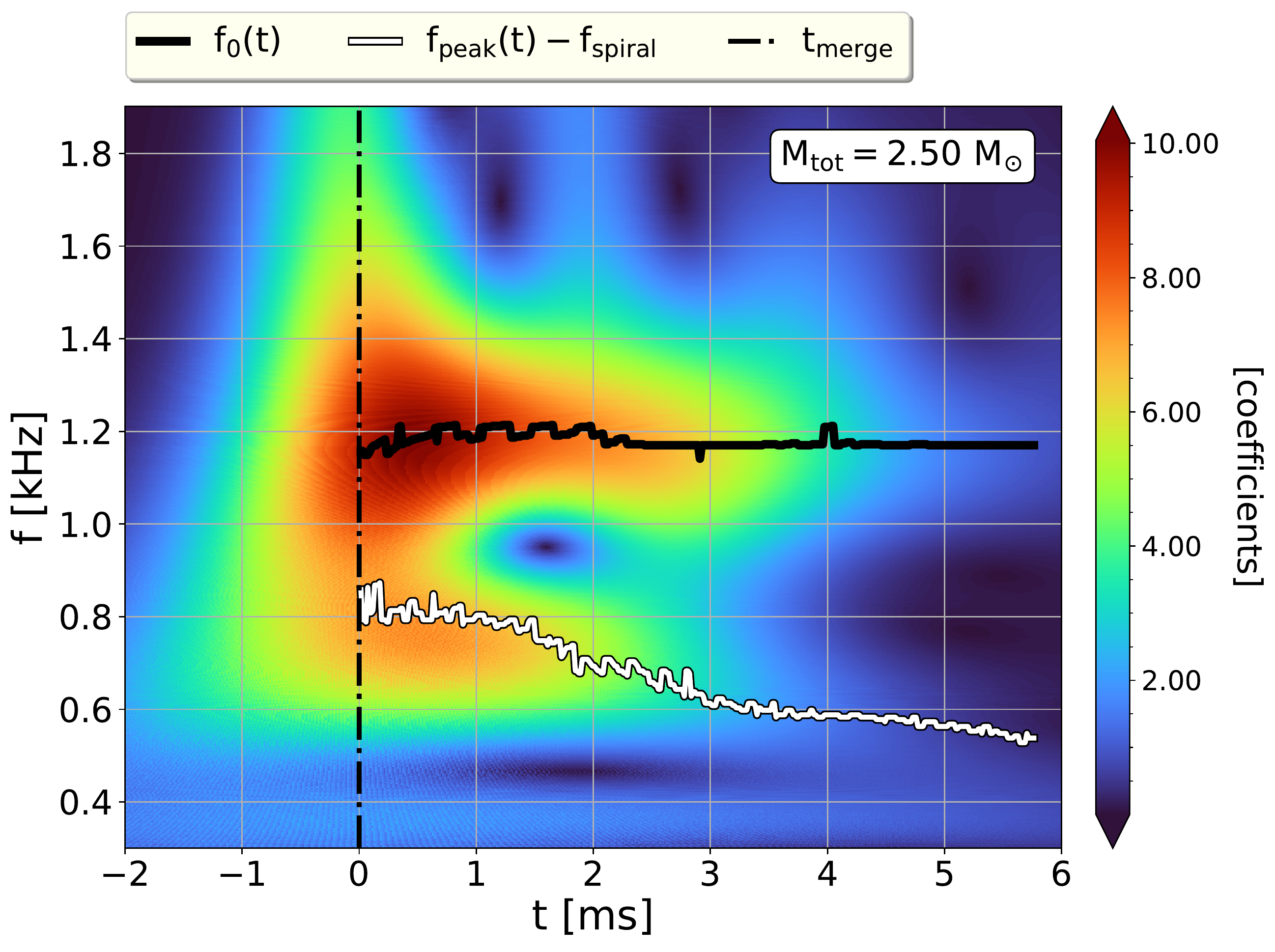} 
	\caption{\label{fig:dumpSPEC_f0_from_ac_mpa1-m1.25.pdf} 
	Spectrogram of minimum lapse function $\alpha_{\mathrm{min}}(t)$ for the reference simulation. The black curve shows $f_0(t)$, as determined by the maximum wavelet coefficient at time $t$. The white line shows $f_{\mathrm{peak}}(t)-f_{\mathrm{spiral}}$. The vertical dash-dotted line indicates the merging time $t_\mathrm{merge}$.}
\end{figure}

In this regard, we further investigate the time evolution of $f_0(t)$. We extract the quasi-radial oscillation from the time evolution of minimum lapse function $\alpha_\mathrm{min}$ (or from the maximum rest mass density $\rho_c$). Figure \ref{fig:dumpSPEC_f0_from_ac_mpa1-m1.25.pdf} shows the spectrogram of the minimum lapse function $\alpha_\mathrm{min}(t)$. For our reference simulation, the frequency change of $f_0(t)$ is small and comparable to the noise associated to the spectrogram scheme (for high-mass configurations the frequency drift of $f_0(t)$ is slightly more pronounced, see Sec. \ref{Sec. mass sequence}). The roughly constant frequency $f_0(t)$ is in good agreement with the maximum in the power spectrum of $\alpha_{\mathrm{min}}$, as shown in Fig.~\ref{sequence-lapse-FFT.pdf}. 

Finally, we consider the time evolution of both $f_{\mathrm{peak}}(t)$ and $f_{0}(t)$ to obtain $f_{2\pm0}(t)=f_{\mathrm{peak}}(t)\pm f_0(t)$. In Fig.~\ref{fig:dumpFFT_evolution_mpa1-m1.25.pdf} the green and orange bands indicate the ranges in which $f_{2\pm0}(t)$ varies and which coincide well with the secondary peaks. 

As already noted, the exact values of the $f_{2\pm0}$ peaks deviate by some per cent from $f_\mathrm{peak}\pm f_0$, i.e. the frequencies extracted from the full signal, which is a consequence of the initial evolution of the main peak frequency $f_\mathrm{peak}(t)$. Employing the average $\langle f_{\mathrm{peak}}^{t\in[0,t_*]}(t)\rangle$ being more representative for the initial phase, we find an excellent agreement between $f_{2\pm0}$ and $\langle f_{\mathrm{peak}}^{t\in[0,t_*]}(t)\rangle\pm f_0$. This is understandable, because $f_0$ decays in comparison to $f_{\mathrm{peak}}$ relatively fast, which is why one may expect that the coupling between both modes is shaped by the early $f_{\mathrm{peak}}(t)$.

Another secondary peak, $f_{\mathrm{spiral}}$, originates from the orbital motion of tidal antipodal bulges \cite{AB2015} formed at the merging phase. Their angular frequency is lower than the one of the inner remnant, and this component is present only for a few cycles \cite{AB2015}. We consider $f_{\mathrm{spiral}}$ to be constant in time and define $f_\mathrm{spiral}$ as the maximum of the corresponding peak at the GW spectrum. There may be a slight evolution of the frequency of $f_\mathrm{spiral}$ as the central remnant evolves in time and thus affects the motion of the bulges generating $f_\mathrm{spiral}$. At approximately $t=3$~ms (and $f=2.20$~kHz) in Fig.~\ref{fig:dumpSPEC_mpa1-m1.25} a frequency increase of $f_{\mathrm{spiral}}$ is observed. This drift can be seen more clearly in spectrograms with different wavelet parameters, which enhance the frequency resolution. However, since the amplitude of the $f_\mathrm{spiral}$ feature decays rapidly, we expect the impact of the frequency evolution to be small. In our reference simulation $f_{\mathrm{spiral}}$ is the strongest secondary frequency peak and therefore an additional low frequency modulation $f_{\mathrm{peak}}(t)-f_{\mathrm{spiral}}$ is expected to affect the remnant's compactness and thus the evolution of $\alpha_{\mathrm{min}}(t)$ \cite{AB2015}. This modulation is indeed seen in our reference simulation's spectrogram of $\alpha_{\mathrm{min}}(t)$ (see Fig.~\ref{fig:dumpSPEC_f0_from_ac_mpa1-m1.25.pdf}), where we overplot the extracted $f_{\mathrm{peak}}(t)-f_{\mathrm{spiral}}$.

\subsection{$\mathbf{f_{\mathrm{spiral-0}}}$ coupling}\label{SubSec. spiral-0 coupling}

\begin{figure}[h!]
	\includegraphics[scale=0.33]{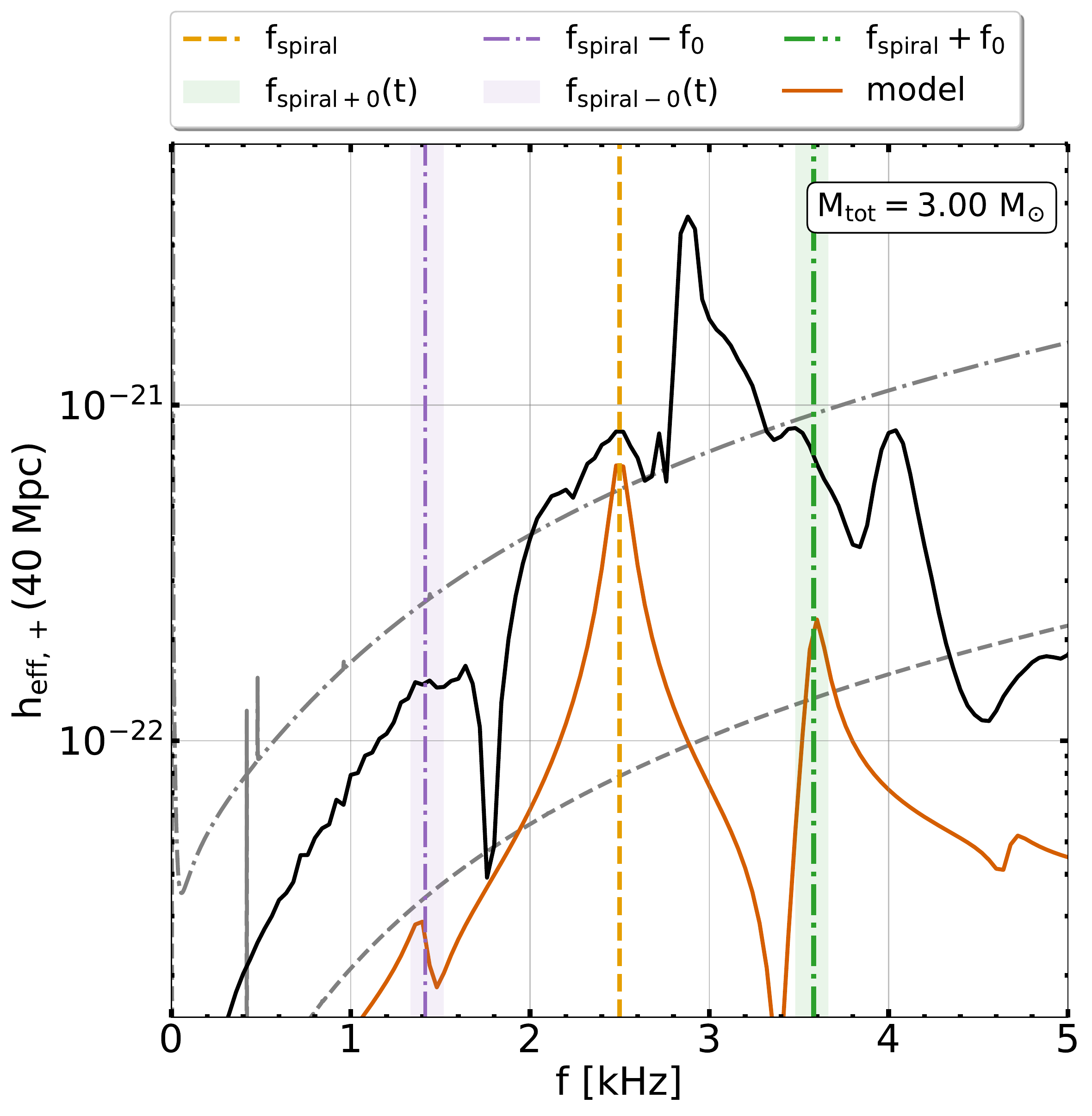} 
	\caption{\label{fig:dumpFFT_fsp0_mpa1-m1.50.pdf} 
	Effective GW spectrum $h_\mathrm{eff,+}(f)$ for the ${M_\mathrm{tot}=3.0M_\odot}$ model at post-merger phase. Colored vertical lines indicate $f_{\mathrm{spiral}},f_{\mathrm{spiral}-0}, f_{\mathrm{spiral}+0}$. Shaded areas correspond to their frequency range visualized by the same colors respectively. Orange curve shows the effective GW spectrum of a simple analytic toy model discussed in Section \ref{SubSec. spiral-0 coupling}.}
\end{figure}

In this subsection we present our findings about a new mechanism which explains additional frequency peaks in the GW power spectrum, specifically, a coupling between $f_{\mathrm{spiral}}$ and the quasi-radial oscillation mode $f_0$. To illustrate this, we discuss the model with a total binary mass $M_\mathrm{tot}=3.0\ M_{\odot}$ (see Fig.~\ref{fig:dumpFFT_fsp0_mpa1-m1.50.pdf}) where this feature is more pronounced. In this {configuration}, the total binary mass $M_\mathrm{tot}$ is close to $M_\mathrm{thres}$ and therefore the quasi-radial mode is strongly excited \cite{AB2015}. In this model, an $f_{\mathrm{spiral}}$ component is clearly present. We thus conjecture that the strong radial oscillation affects the motion of the bulges and leads to a coupling between $f_0$ and $f_{\mathrm{spiral}}$. And indeed, we find additional frequency peaks at approximately $f_{\mathrm{spiral}\pm0} = f_{\mathrm{spiral}}\pm f_0$. 

Figure \ref{fig:dumpFFT_fsp0_mpa1-m1.50.pdf} illustrates the post-merger power spectrum for this simulation. As before, we extract the quasi-radial oscillation frequency $f_0$ from the maximum in the Fourier transform $\tilde{\alpha}_\mathrm{min}(f)$ of the minimum lapse function  (see Fig.~\ref{sequence-lapse-FFT.pdf}) and obtain the estimates for $f_\mathrm{spiral\pm 0}$ using $f_{\mathrm{spiral}}$ from the GW spectrum. In Fig.~\ref{fig:dumpFFT_fsp0_mpa1-m1.50.pdf} the estimates $f_\mathrm{spiral\pm 0}$ match very well with additional frequency peaks in the power spectrum. The frequency $f_{\mathrm{spiral}}-f_0$ is in better agreement with the corresponding frequency peak while the high frequency $f_{\mathrm{spiral}}+f_0$ deviates by roughly 200 Hz. To further assess our conjecture, we extract the time evolution of $f_0(t)$ from the spectrogram of $\alpha_\mathrm{min}(t)$ (see Fig.~\ref{fig:dumpSPEC_f0_from_ac_mpa1-m1.25.pdf}) to estimate the frequency ranges of $f_{\mathrm{spiral}\pm0}$. These ranges are in good agreement with the additional peaks {in the GW spectrum}. In particular, the frequency peak in the vicinity of $f_\mathrm{spiral}+f_0$ lies in the corresponding range. Since in reality $f_{\mathrm{spiral}}$ is not exactly constant, a 200 Hz deviation may be understandable. We emphasize that our finding is not unique for this EoS model but it is a general feature in merger simulations. We observe it in additional GW spectra of additional simulations carried out with an SPH code \cite{SPH1,SPH2} with varying EoSs and total mass $M_\mathrm{tot}$.  

We further corroborate our finding by considering a simple analytic toy model: We adopt two point-particles with individual masses $m_1=m_2=0.2\ M_\odot$ on an orbit with an orbital frequency $f_\mathrm{orb}=\frac{f_{\mathrm{spiral}}}{2}$ at a radius $R=9$~km. We add a radial oscillation with frequency $f_0$ superimposed on the circular orbit with amplitude $A=1.0$ km. These values may be representative of typical simulations. To mimic the fact that the bulges disappear after a few milliseconds we assume an exponential decay of the point-particle masses with a timescale $\tau_\mathrm{m}=5.0$~ms. Finally, we compute the corresponding GW radiation employing the quadrupole formula and derive the Fourier transform, which we overplot in Fig \ref{fig:dumpFFT_fsp0_mpa1-m1.50.pdf}. Interestingly, this simple model produces a strong peak at $f_{\mathrm{spiral}}$ (as expected) and two secondary peaks which coincide with  $f_{\mathrm{spiral}}\pm f_0$. Note the same pattern of the relative amplitudes of $f_{\mathrm{spiral}}\pm f_0$ in the simulation and the analytic toy model; $f_\mathrm{spiral}+f_0$ is significantly enhanced. 

Finally, we note that the coupling to the quasi-radial mode $f_0$ may result in frequency peaks at approximately $f_{\mathrm{peak}}(t) \pm 2\cdot f_0$ and $f_{\mathrm{spiral}} \pm 2\cdot f_0$. These components are expected to be weak, however for high total mass models, where the $f_0$ mode is strongly excited, they may become more significant. Our simple analytic toy model generates a peak (in its spectrum) at approximately $f=4.7$~kHz which coincides with $f_{\mathrm{spiral}}+2\cdot f_0$. A weak bump in the GW spectrum can be seen in the vicinity of $f_{\mathrm{spiral}}+2\cdot f_0$. In Sec.~\ref{Clode to prompt collapse} we identify more features in the GW spectrum which can be associated to such couplings.

\section{Sequence of merger simulations with different total binary mass}\label{Sec. mass sequence}
\begin{figure*}[t!]
	\includegraphics[ scale=0.36]{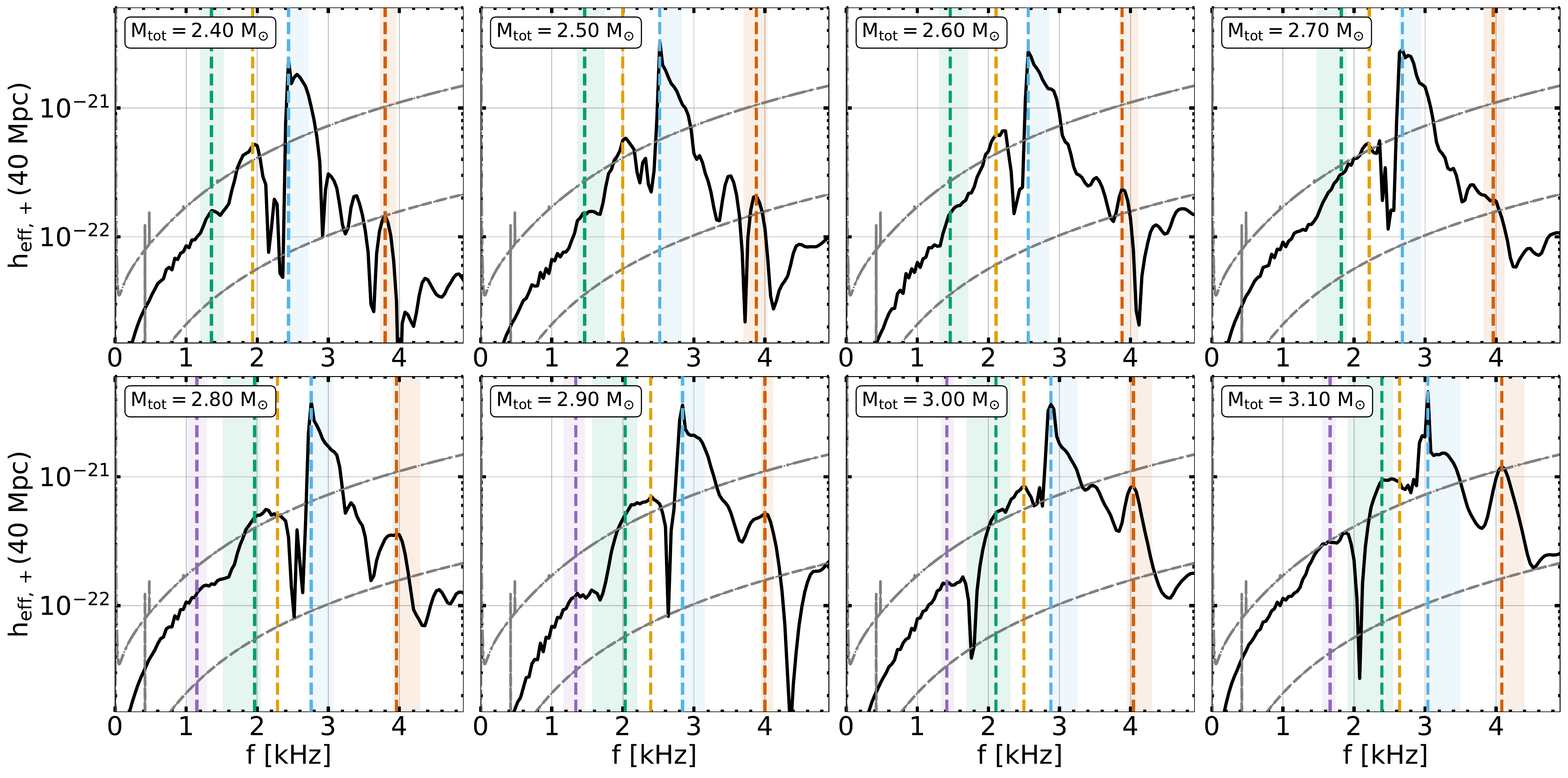} 
	\caption{\label{fig:sequence-FFT.pdf} Effective GW spectra  $h_\mathrm{eff,+}(f)$ for the mass sequence. Purple dashed lines indicate $f_{\mathrm{spiral}-0}$. Purple shaded areas correspond to frequency ranges. The other colors follow the notation of Fig.~\ref{fig:dumpFFT_evolution_mpa1-m1.25.pdf}.}
\end{figure*}

In this section we discuss a sequence of merger simulations with different total binary masses, in the range of $2.4-3.0M_\odot$, with a step size of $0.1 M_\odot$ (which includes the reference simulation) and we describe how the different components of the post-merger GW signal depend on $M_\mathrm{tot}$. We find a smooth transition between the GW spectra along the sequence and observe that the presence and strength of the different spectral features continuously change with total binary mass $M_\mathrm{tot}$ as it approaches the binary mass for prompt BH formation $M_\mathrm{thres}$. Figure \ref{fig:sequence-FFT.pdf} shows the effective GW spectra $h_\mathrm{eff,+}(f)$ for different $M_\mathrm{tot}$, where the $f_\mathrm{peak}$, $f_\mathrm{spiral}$, $f_{2\pm0}$ and $f_{\mathrm{spiral-0}}$ peaks are indicated. As in the previous sections, we assume a distance of 40~Mpc, and overplot the sensitivity curves of Advanced LIGO and the Einstein Telescope for reference.

\subsection{Secondary GW peaks}

 In Figure \ref{fig:sequence-FFT.pdf}, the main as well as the secondary peaks show a clear dependence on the total binary mass. The morphology of the GW spectra broadly follows the classification of the post-merger GW signals as in \cite{AB2015}, which is based on the presence and relative strength of the secondary peaks. In \textit{low-mass} configurations ($M_\mathrm{tot}\leq 2.6 M_\odot$) $f_{2-0}$ and $f_{\mathrm{spiral}}$ are well separated and $f_{2-0}$ is relatively weak because the quasi-radial mode is not strongly excited. For \textit{high-mass} configurations ($M_\mathrm{tot}\geq 2.8 M_\odot$), $f_{2-0}$ becomes more pronounced and there is a noticeable overlap between $f_{2-0}$ and $f_{\mathrm{spiral}}$. The absolute height of the $f_{\mathrm{spiral}}$ peak is roughly constant in all models, whereas the $f_{2-0}$ feature becomes stronger with higher total binary mass (by nearly one order of magnitude in $h_\mathrm{eff,+}(f)$). 

The secondary frequency peak $f_{2+0}$ is, in most models, observationally less interesting because of its lower amplitude, when compared to the other secondary peaks and because of the lower sensitivity of current detectors at higher frequencies. However, for the two models with the highest mass within our sequence, the amplitude of $f_{2+0}$ becomes comparable to the amplitude of the other secondary peaks and so it becomes observationally relevant.   Interestingly, the frequency $f_{2+0}$ only mildly depends on $M_\mathrm{tot}$ and ranges between 3.8-4.0 kHz for the whole mass sequence. The latter is due to $f_{\mathrm{peak}}$ being an increasing function of $M_\mathrm{tot}$ while $f_0$ decreases.

Only the high mass configurations (and especially the ones with total mass $M_\mathrm{tot}\geq2.9M_\odot$) exhibit a significant frequency peak $f_{\mathrm{spiral-0}}$ (see purple dashed line in Fig.~\ref{fig:sequence-FFT.pdf}). Since $f_{\mathrm{spiral}}$ grows with $M_\mathrm{tot}$ while $f_0$ slightly decreases, the frequency $f_{\mathrm{spiral-0}}$ increases as the total binary mass approaches the threshold binary mass for prompt collapse $M_\mathrm{thres}$. The strength of the $f_{\mathrm{spiral-0}}$ peak increases with the total binary mass $M_\mathrm{tot}$. Its absolute height is always smaller than that of the other secondary features, but relative to the projected detector sensitivity curves, the signal to noise ratio of the $f_{\mathrm{spiral-0}}$ coupling is roughly comparable to that of $f_{2+0}$. The $f_{\mathrm{spiral-0}}$ feature is thus important for configurations with binary masses close to $M_\mathrm{thres}$, where the quasi-radial mode is strongly excited, which enhances both $f_{\mathrm{spiral-0}}$ and $f_{2\pm 0}$ (see lower right panel in Fig.~\ref{fig:sequence-FFT.pdf}).

\subsection{Minimum of the lapse function}

\begin{figure}[h!]
	\includegraphics[scale=0.33]{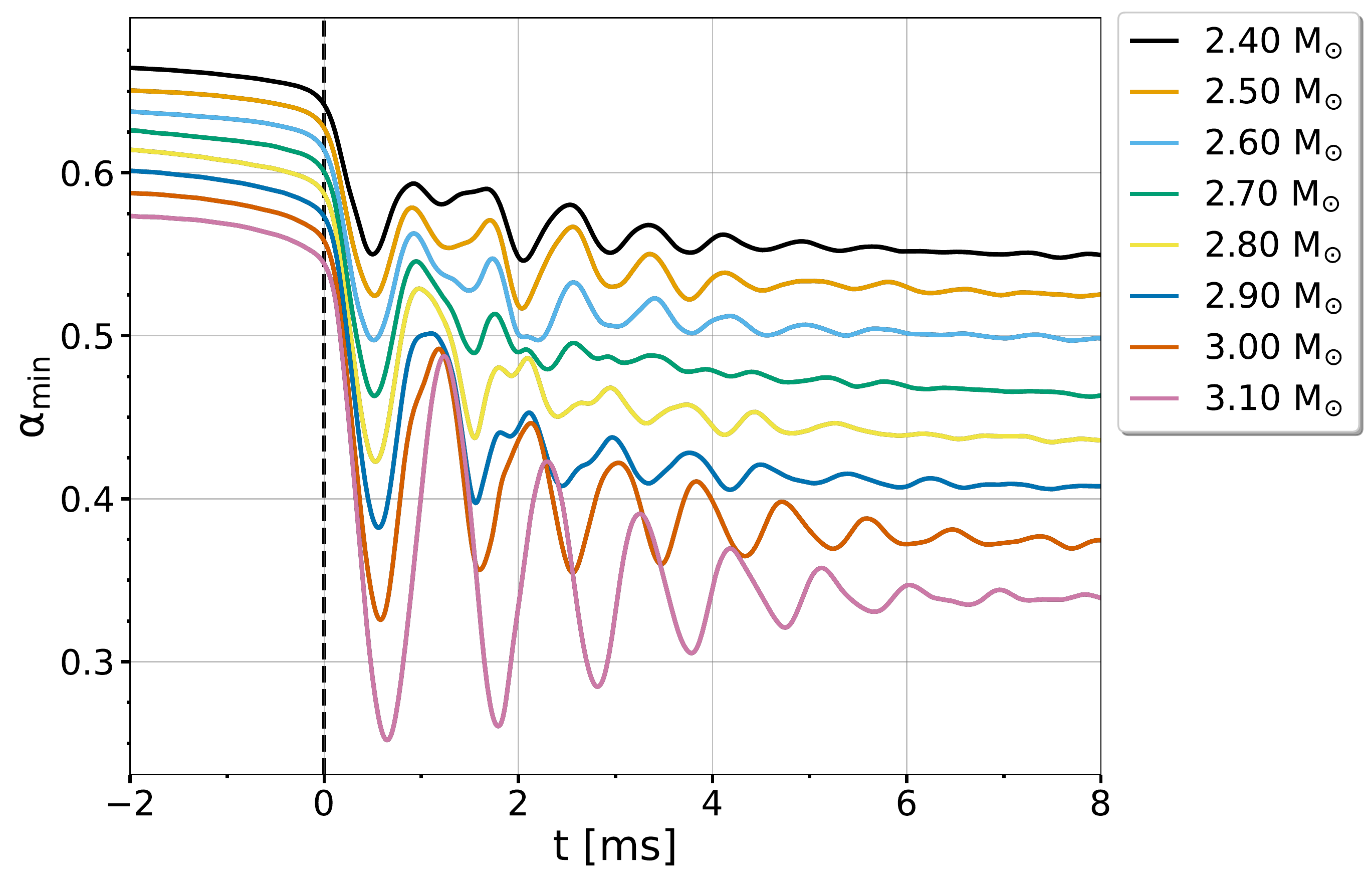} 
	\caption{\label{fig:lapses_all} Time evolution for minimum lapse function $\alpha_{\mathrm{min}}(t)$ normalized to merging time $t_\mathrm{merge}$ along the sequence of models with varying $M_{\rm tot}$. Black vertical dashed line shows the merging time $t_\mathrm{merge}$. }
\end{figure}

As in \cite{AB2015}, we investigate the time evolution of the minimum lapse function  $\alpha_\mathrm{min}$. Figure~\ref{fig:lapses_all} shows that the behaviour of  $\alpha_\mathrm{min}(t)$ for all models along the sequence is consistent with the respective GW spectra and shows a clear dependence on $M_\mathrm{tot}$. As already noted for the models in \cite{AB2015}, the quasi-radial mode is stronger excited with increasing total binary mass $M_\mathrm{tot}$, which explains the enhancement of those GW features that involve a coupling to this particular oscillation mode. For lower-mass and intermediate-mass models the quasi-radial mode is only weakly excited. Instead, $\alpha_\mathrm{min}(t)$ features an additional oscillation with lower frequency $f_{\mathrm{peak}}(t)-f_{\mathrm{spiral}}$, which dominates during the early phase of the remnant evolution. This modulation is explained by the impact of the massive orbiting bulges generating $f_{\mathrm{spiral}}$ on the remnant compactness (see \cite{AB2015} for details). 

In general, the behavior in Fig.~\ref{fig:lapses_all} can be understood from the merger dynamics and remnant properties. High-mass models lead to a collision with a higher impact velocity and thus the quasi-radial oscillation is strongly excited. 

\begin{figure}[h!]
	\includegraphics[scale=0.33]{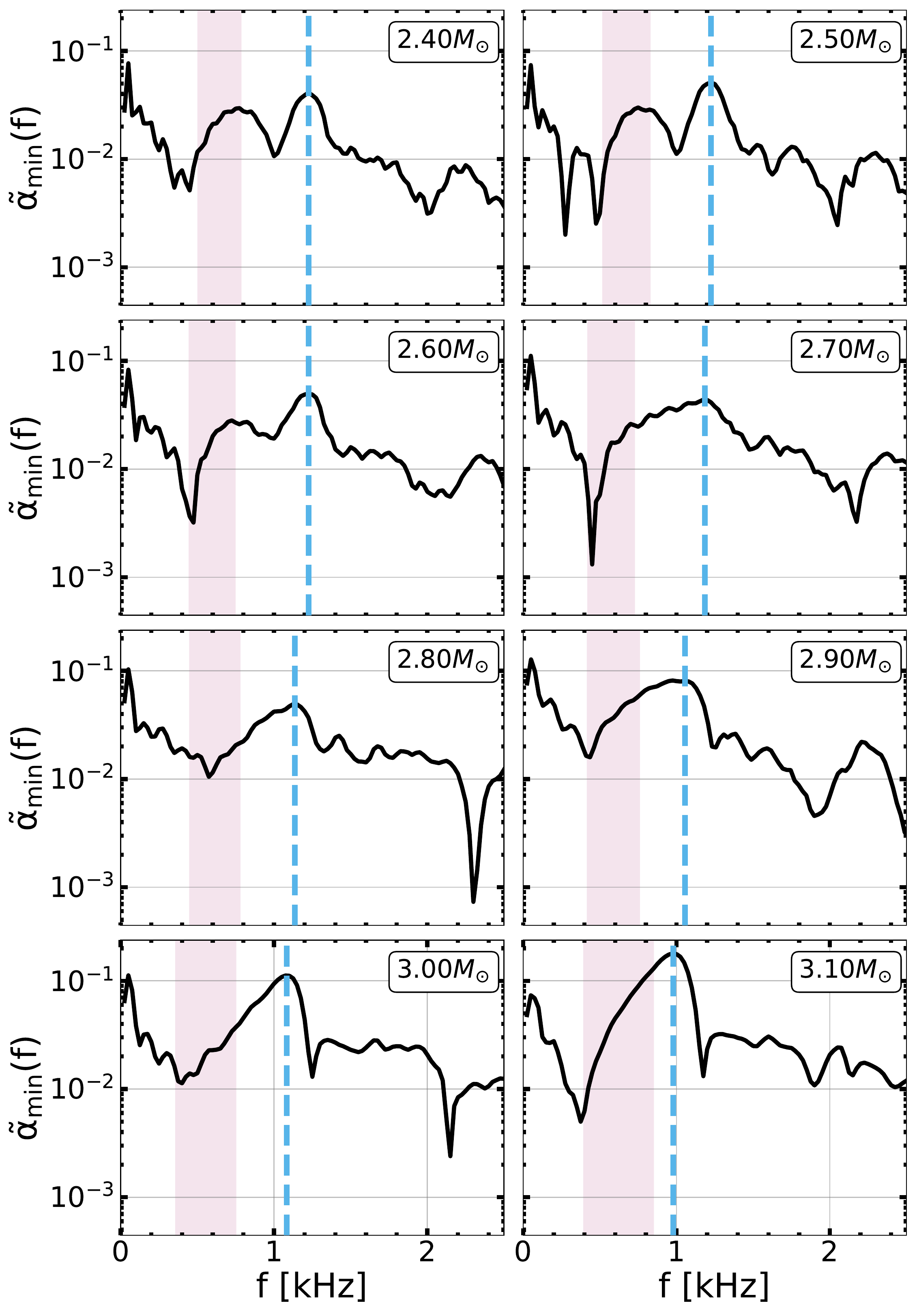} 
	\caption{\label{sequence-lapse-FFT.pdf} Fourier transform of the minimum lapse function $\alpha_{\rm min}$ along the sequence models with varying $M_{\rm tot}$.  The vertical dashed line indicates the quasi-radial frequency $f_0$. The red band indicates the frequency range of $f_{\mathrm{peak}}(t)-f_{\mathrm{spiral}}$.}
\end{figure}

The aforementioned features in the minimum lapse function can be identified in the power spectra of the Fourier transform $\tilde{\alpha}_\mathrm{min}(f)$, as shown in Fig.~\ref{sequence-lapse-FFT.pdf} (see also \cite{PhysRevD.96.063011}). We compute $\tilde{\alpha}_\mathrm{min}(f)$ using an appropriate window function to select only the post-merger phase. 

A common feature in all models is the pronounced quasi-radial oscillation frequency peak $f_0$ in the vicinity of 1 kHz. In low-mass configurations we observe an additional peak at lower frequencies, which corresponds to the $f_{\mathrm{peak}}(t)-f_{\mathrm{spiral}}$ modulation with a strength comparable to that of the quasi-radial mode. Even in cases where the  $f_{\mathrm{peak}}(t)-f_{\mathrm{spiral}}$ modulation appears dominant in the initial phase in the time domain (see Fig.~\ref{fig:lapses_all}), the $f_0$ peak in the post-merger spectrum is stronger, because the quasi-radial mode oscillates longer.

High-mass configurations show a very dominant frequency peak $f_0$. The strength of the peak increases with $M_\mathrm{tot}$, and the peak becomes broader and one-sided. This suggests that the quasi-radial frequency undergoes an evolution, which can be verified by spectrograms of $\alpha_\mathrm{min}$ (see Appendix \ref{A:Spectral properties of the mass sequence models}). Intermediate mass models show that the $f_{\mathrm{peak}}(t)-f_{\mathrm{spiral}}$ peak overlaps and merges with the $f_0$ peak, as $M_\mathrm{tot}$ increases. We remark in particular for models with higher masses, that $f_\mathrm{peak}(t)$ initially evolves rapidly towards lower values and is initially higher than the $f_\mathrm{peak}$ identified in the GW spectrum (see Appendix \ref{A:Spectral properties of the mass sequence models}). Hence, at early times, when the low-frequency modulation is present, the difference $f_{\mathrm{peak}}(t)-f_{\mathrm{spiral}}$ is in fact larger than one would infer from an inspection of the GW spectrum alone and in high-mass models it is roughly consistent with the left side of the main peaks in  Fig.~\ref{sequence-lapse-FFT.pdf}. In Fig.~\ref{sequence-lapse-FFT.pdf} we show the estimates for the frequency ranges for $f_{\mathrm{peak}}(t)-f_{\mathrm{spiral}}$ (red band).

\subsection{Evolution of frequencies}

As in the case of the reference simulation of Section \ref{Sec:Spectral analysis of the post-merger GW emission}, all post-merger GW spectra along the sequence of models (see Fig.~\ref{fig:sequence-FFT.pdf})
exhibit an asymmetric (one-sided) peak, due to the time evolution of $f_{\mathrm{peak}}(t)$. The exact morphology somewhat varies as the total mass increases.  We quantify the respective drifts in $f_{\mathrm{peak}}(t)$ and fit a 2-segment piecewise linear/constant function $f_{\mathrm{peak}}^{\mathrm{analytic}}(t)$, as shown for the reference simulation in Section
\ref{fpeakevol} (see Appendix \ref{A:Spectral properties of the mass sequence models} for the spectrograms used in extracting the time evolution of $f_{\mathrm{peak}}(t)$ 
and Section \ref{fpeak parameters} for empirical fits as a function of $M_{\rm tot}$ of the parameters of $f_{\mathrm{peak}}^{\mathrm{analytic}}(t)$ along our sequence of models).
For each model along the sequence, the analytic fit $f_{\mathrm{peak}}^{\mathrm{analytic}}(t)$ provides a frequency range, which we indicate by cyan bands in Fig.~\ref{fig:sequence-FFT.pdf} and which coincides well with the full structure of the main peak in the different spectra.

In a similar way, we proceed with the non-linear couplings between the quadrupolar and the quasi-radial mode to estimate a frequency range of these secondary peaks. We extract $f_0(t)$ from the spectrogram of the minimum lapse function $\alpha_{\mathrm{min}}(t)$ and employ $f_{\mathrm{peak}}^\mathrm{analytic}(t)$ to obtain a time evolution of $f_{2\pm 0}(t)=f_{\mathrm{peak}}^\mathrm{analytic}(t)\pm f_0(t)$. Considering the evolution of $f_{2\pm 0}(t)$ during the first milliseconds provides frequency ranges, which we overplot in Fig.~\ref{fig:sequence-FFT.pdf} (green and orange bands) and which very well agree with the  $f_{2\pm0}$ peaks. 

We also estimate the frequency range of the coupling between the $f_\mathrm{spiral}$ and the quasi-radial mode (for high-mass configurations) using the time-dependent $f_\mathrm{spiral-0}(t)=f_\mathrm{spiral} - f_0(t)$. We overplot them (see Fig.~\ref{fig:sequence-FFT.pdf}) and find a very good agreement with the $f_\mathrm{spiral-0}$ peak.   

As in the case of the reference simulation, it is seen that also for other binary masses the main frequency $f_{\rm peak}$ in Fig.~\ref{fig:sequence-FFT.pdf} does not exactly occur in the middle between $f_{2-0}$ and $f_{2+0}$ as one would expect for $f_{2\pm0}$. Instead, since the combination tones are rapidly evolving features, we find that $\langle f_{\mathrm{peak}}^{t\in[0,t_*]}\rangle$ (which is higher than $f_{\mathrm{peak}}$ in all models and more representative for $f_{\mathrm{peak}}(t)$ at early times) does indeed agree very well with $\frac{1}{2}(f_{2-0}+f_{2+0})$, i.e. it lies as expected in the middle between the two secondary peaks.

\section{Analytic and semi-analytic post-merger models} \label{Analytic post-merger model}

In this section we use our analysis of the post-merger GW signal to build analytic waveform models for the post-merger phase. We construct an {\it accurate analytic model} of the post-merger GW signal, as an extension of \cite{AB2016}, which included fixed $f_\mathrm{peak}$, $f_\mathrm{spiral}$ and $f_{2-0}$ frequencies with exponential damping, and of \cite{Easter2020}, which introduced a linear time-dependence of $f_\mathrm{peak}(t)$ with a constant slope throughout the time evolution.
In comparison to these previous works, we also include the higher-frequency combination tone $f_\mathrm{2+0}$ and use the two-segment piecewise linear model of Eq. (\ref{fpeak_piecewise}) to describe the time evolution of $f_\mathrm{peak}(t)$. The model can easily be extended to include additional features, such as $f_\mathrm{spiral-0}$ for high masses and we do so in Section \ref{Clode to prompt collapse}.

In addition to the fully analytic model, we also consider a {\it semi-analytic model} which incorporates directly a numerical representation of $f_\mathrm{peak}(t)$ extracted from the spectrograms. This is extended in Section \ref{Clode to prompt collapse} to also include time-dependent secondary components $f_{2\pm0}(t)$.

Since the models include a relatively large  number of parameters, we employ several successive steps in order to determine the model's parameters. We describe these steps in the following subsections. 

\subsection{Analytic Model}\label{Model description}

The analytic model employs exponentially decaying sinusoids and except for the main frequency $f_\mathrm{peak}(t)$ we assume all other frequencies of the model to be constant in time. The model reads
\begin{eqnarray} 
\label{analytic model equations}
h_\mathrm{+}(t) &=& A_\mathrm{peak}\  e^{(-t/\tau_\mathrm{peak})}\cdot\sin(\phi_\mathrm{peak}(t))\nonumber\\ 
&+& A_\mathrm{spiral} \  e^{(-t/\tau_\mathrm{spiral})}\cdot\sin(2\pi f_\mathrm{spiral}\cdot t+\phi_\mathrm{spiral})\nonumber\\
&+& A_{2-0}\  e^{(-t/\tau_{2-0})}\cdot \sin(2\pi f_{2-0}\cdot t+\phi_{2-0})\nonumber\\
&+& A_{2+0}\  e^{(-t/\tau_{2+0})}\cdot \sin(2\pi f_{2+0}\cdot t+\phi_{2+0}), 
\end{eqnarray}
where the $f_\mathrm{peak}$ component's phase, $\phi_\mathrm{peak}(t)$, is
\begin{eqnarray}\label{phipeak(t)}
\phi_\mathrm{peak}(t) =    \left\{
\begin{array}{ll}
2\pi \left(f_\mathrm{peak,0}+\frac{\zeta_\mathrm{drift}}{2}t \right) t+\phi_\mathrm{peak}, & \mbox{for } t\leq t_* \\
2\pi \ f_\mathrm{peak}(t_*)\Big(t-t_*\Big)+ \phi_\mathrm{peak}(t_*), &\mbox{for } t> t_*
\end{array} 
\right.\quad 
\end{eqnarray}
Using the above expression, the phase $\phi_\mathrm{peak}(t)$ is continuous and the frequency $f_\mathrm{peak}(t)=\frac{1}{2\pi}\frac{d\phi_\mathrm{peak}(t)}{dt}$ features a time-dependence as in Eq.~\eqref{fpeak_piecewise}. 

We employ several steps to determine the analytic model's parameters and the model contains several frequency components, which is why it is not straightforward to find the optimal values describing the data. We find that by introducing a {\it normalization factor} $\mathcal{N}$ we obtain better fits. We thus define
\begin{eqnarray}
\label{Fplusfit}
h_{+}^\mathrm{Fit}(t) &=& \mathcal{N} \cdot {h_\mathrm{+}}(t),
\end{eqnarray}
with ${h_\mathrm{+}}(t)$ given as in  Eq.~\eqref{analytic model equations}. We note that when simpler (under-performing) analytic models are employed (consisting of only one or two frequency components) the normalization factor $\mathcal{N}$ is dropped. We stress that the normalization factor is only introduced as part of our procedure for determining the best fit - with other fitting procedures it may not be required.

Lastly, in order to improve the fits for this particular mass sequence and EoS we introduce a {\it phenomenological modification} to the analytic model in the description of the $f_\mathrm{peak}(t)$ component. Our quasi-linear model of Eq.~(\ref{analytic model equations}) does not accurately capture the very early evolution presumably because of the nonlinearities that are present immediately after merger. We observe a mild delay in the starting times of the exponentially decaying sinusoids between the $f_\mathrm{peak}$ component and the secondary components $f_\mathrm{spiral}$ and $f_{2\pm0}$ during the first $\approx1.0$ ms (see the spectograms in Appendix \ref{A:Spectral properties of the mass sequence models}). This delay is more pronounced in high-mass configurations. We mimic this delay by multiplying the first line of Eq.~(\ref{analytic model equations}), which corresponds to the $f_\mathrm{peak}$ component, by a Tukey window function, denoted here by $\mathcal{W}(t; s)$, where $s$ is the roll-off parameter. We use a roll-off parameter $s=0.075$ for models with $M_\mathrm{tot}\leq 2.9M_\odot$ and $s=0.1$ for models with $M_\mathrm{tot}> 2.9M_\odot$. 

The above phenomenological introduction of non-linear effects leads to more accurate fits of the initial phases of the secondary components. We note after $\approx1.0$ ms from the onset of the post-merger phase that the evolution is close to quasi-linear (linear plus quasi-linear combination tones) and the analytic model  of  Eq.~(\ref{analytic model equations}) is sufficient for its description. 

To summarize, the complete analytic model of the $+$ polarization of the signal amplitude reads
\begin{eqnarray}
\label{analytic model equations-2}
h_{+}^\mathrm{Fit}(t) &=& \mathcal{N}\cdot\Big(h_{+}^\mathrm{peak}(t)\cdot \mathcal{W}(t; s)  +\sum_\mathrm{i} h_{+}^\mathrm{i}(t)\Big),\\
&\mbox{for}&\ i=\mbox{spiral, }2\pm0, \nonumber
\end{eqnarray}
where $h_{+}^\mathrm{i}(t)=A_\mathrm{i}\ e^{(-t/\tau_\mathrm{i})}\cdot \sin(\phi_\mathrm{i}(t))$ .

To obtain the cross polarization $h_{\times}^\mathrm{Fit}(t)$, we adopt the parameters for the amplitudes, damping time scales and frequencies from $h_+^\mathrm{Fit}(t)$ and assume a phase shift of $90^o$~degrees to the individual initial phases $\phi_\mathrm{i}$ (for i=peak, spiral, $2\pm0$). 

\subsection{Semi-analytic model}\label{semi-analytic model}
The semi-analytic model differs from the analytic model by the substitution of $\phi_\mathrm{peak}(t)$ with the numerical phase $\phi_\mathrm{peak}^\mathrm{numerical}(t)$. The latter is obtained by first extracting the instantaneous frequency  $f_\mathrm{peak}^\mathrm{spectrogram}(t)$ from the spectrograms, which is then integrated in time to obtain the phase at a particular time step $t_i$ as $\phi_\mathrm{peak}^\mathrm{numerical}(t)$ using the iterative formula
\begin{eqnarray}
\label{numerical phase(t)}
\phi_\mathrm{i+1} &=& \phi_\mathrm{i} + 2\pi f_\mathrm{peak,i} \cdot \left(t_\mathrm{i+1} - t_\mathrm{i} \right)
\end{eqnarray}
where $\phi_\mathrm{i} \equiv \phi(t=t_\mathrm{i})$ and $f_\mathrm{peak,i} \equiv f_\mathrm{peak}(t=t_\mathrm{i})$. The initial phase $\phi_\mathrm{peak,0}\equiv  \phi(t=0)$ is a parameter (like $\phi_\mathrm{peak}$ in the analytic model). 

In Sec.~\ref{Clode to prompt collapse} we consider an extended semi-analytic model, which includes time-dependent secondary components $f_{2\pm0}(t)$ where the phases $\phi_{2\pm0}(t)$ are extracted from the spectrograms in a similar way.

\subsection{Parameter extraction procedure} \label{Parameter extraction}
This subsection is structured as follows: In Section \ref{fpeak(t) parameters} we discuss the analytic description of $f_\mathrm{peak}(t)$. In Section \ref{Secondary frequencies} we describe how we obtain the secondary frequencies $f_\mathrm{spiral},f_{2\pm0},f_\mathrm{spiral-0}$ from GW spectra. In Section \ref{ATs for secondary components} we describe the method for the extraction of model parameters ($A_\mathrm{i},\tau_\mathrm{i}$ for $i=\mathrm{spiral,2\pm0}$) of the secondary components. In Section \ref{Fit to sim data} we discuss the determination of the remaining parameters $A_\mathrm{peak}$, $\tau_\mathrm{peak}$, $\mathcal{N}$, $\phi_\mathrm{i}$ for $i=\mathrm{peak, spiral},2\pm0$ and the fit to the simulation data. For all models of our mass sequence we proceed as follows.

\subsubsection{Analytic description of $f_\mathrm{peak}(t)$}\label{fpeak(t) parameters}

As mentioned before, we extract the evolving $f_\mathrm{peak}(t)$ from spectrograms as the frequency of the maximum wavelet coefficient at time $t$. We parametrize $f_\mathrm{peak}(t)$ as 2-segment piecewise function Eq.~\eqref{fpeak_piecewise}. We obtain the parameters $[\zeta_\mathrm{drift},f_\mathrm{peak,0},t_*]$ from a fit to the extracted $f_\mathrm{peak}(t)$. The fit is done in one step using the analytic function of Eq.~\eqref{fpeak_piecewise}. The extracted parameters are finally inserted to the analytic model via $\phi_\mathrm{peak}(t)$ as in Eq.~\eqref{phipeak(t)}. 

\subsubsection{Secondary frequency peaks}\label{Secondary frequencies}
We compute the secondary frequencies $f_\mathrm{spiral}$, $f_{2\pm0}$ in two steps. First we obtain a rough estimate of the ranges of the different components, which is necessary to correctly identify the different features. Then we pick the frequency at the maximum in the GW spectrum within the estimated frequency ranges of the different components. For the estimate of $f_\mathrm{spiral}$ we use the rest-mass density profiles on the equatorial plane (as done in \cite{AB2015}). We estimate of $f_{2\pm0}$ using the relation $f_{2\pm0}\approx f_\mathrm{peak} \pm f_0$. We replace $f_\mathrm{peak}$ by the mean value of $f_\mathrm{peak}(t)$ during the first milliseconds, while $f_0$ is the dominant frequency peak in the Fourier transform of the minimum lapse function $\alpha_\mathrm{min}$ (see Fig.~\ref{sequence-lapse-FFT.pdf}). We note that our choices are in agreement with the empirical relations in \cite{Vretinaris2020}.

\subsubsection{Amplitudes $A_i$ and decay timescales $\tau_i$ for secondary components}\label{ATs for secondary components}
We describe the technique to estimate the amplitudes $A_\mathrm{spiral}$, $A_\mathrm{2\pm0}$ and timescales $\tau_\mathrm{spiral}$, $\tau_\mathrm{2\pm0}$ using the spectrograms. We find that the following procedure leads to results which better reproduce the secondary frequency peaks in GW spectrum.

First we employ spectrograms and extract the wavelet coefficients as functions of time $t$ for the frequency components $f_\mathrm{spiral}$, $f_\mathrm{2\pm0}$ and obtain $\mathcal{A}_\mathrm{spiral}(t)$, $\mathcal{A}_\mathrm{2\pm0}(t)$. We then assume a signal of the form $A_i\ e^{-t/\tau_i}\cos(2\pi\ f_i \cdot t)$ and consider each component separately. Within a curve fitting procedure we compute the coefficients at $f_\mathrm{i}$ of this model's signal and determine $A_\mathrm{i}$ and $\tau_\mathrm{i}$ such that the coefficient function matches the extracted $\mathcal{A}_\mathrm{i}(t)$. The curve fitting procedure adopts a trust-region-reflective algorithm \cite{TRF1,TRF2,2020SciPy-NMeth}.

By this method the various components are treated independently, thus in the case of overlapping frequencies the method loses accuracy, since each component is amplified by its neighboring component. The scheme may thus overestimate the amplitudes $A_i$. To compensate this, we introduced the aforementioned normalization factor $\mathcal{N}$, which we determine in the next step. 

\subsubsection{Fit to simulation data}\label{Fit to sim data}
In the final step we determine the remaining parameters $A_\mathrm{peak}$, $\tau_\mathrm{peak}$, $\phi_\mathrm{i}$ for $\mathrm{i=spiral,2\pm0}$, $\mathcal{N}$. We perform a fit of the analytic model to the simulation data, using the aforementioned curve fitting routine, employing a trust-region-reflective algorithm. The previously determined parameters are inserted in the analytic model. 

We found that the secondary features of the signal are better reproduced from the spectrogarms as described in \ref{Amplitudes and timescales}, whereas the $f_\mathrm{peak}$ feature as the dominant component is well determined by the fitting routine.

\section{Performance of the analytic and semi-analytic models}\label{Sec:Performance of analytic model}

In this Section, we discuss fits of the analytic and semi-analytic models to the GW signals extracted from simulations and quantify their performances. We compare the fits to the actual numerical waveform in the time and frequency domains and examine how well certain GW features are reproduced. 
 
We evaluate the performance of the models with the (noise-weighted) fitting factor ($FF$) defined by 
\begin{eqnarray}\label{Fitting factor definition}
FF &\equiv& \frac{\left( h_1, h_2 \right)}{\sqrt{\left( h_1, h_1 \right)\left( h_2, h_2 \right)}},
\end{eqnarray}
using the noise-weighted inner product $\left( h_1, h_2 \right)$ between two waveforms given by
\begin{eqnarray}\label{noise-weighted product definition}
\left( h_1, h_2 \right) &\equiv& 4 \mathrm{Re} \int_0^\infty df \frac{\tilde{h}_1(f)\cdot \tilde{h}_2^*(f)}{S_h(f)},
\end{eqnarray}
where $S_h(f)$ is the detector's noise spectral density, and $\tilde{h}_\mathrm{i}(f)$ is the Fourier transform of the waveform $h_\mathrm{i}(t)$ (for $i=1,2$).

Moreover, we consider simpler versions of our analytic model, where we include only a subset of GW features. By this we assess the significance of the individual components of the GW signal.

\subsection{GW fits} \label{GW fits}
First we focus on the analysis of the reference simulation ($M_\mathrm{tot}=2.5M_\odot$), and later extend the discussion to the whole sequence of models.

\subsubsection{Reference simulation}
\begin{figure}[h!]
	\includegraphics[scale=0.31]{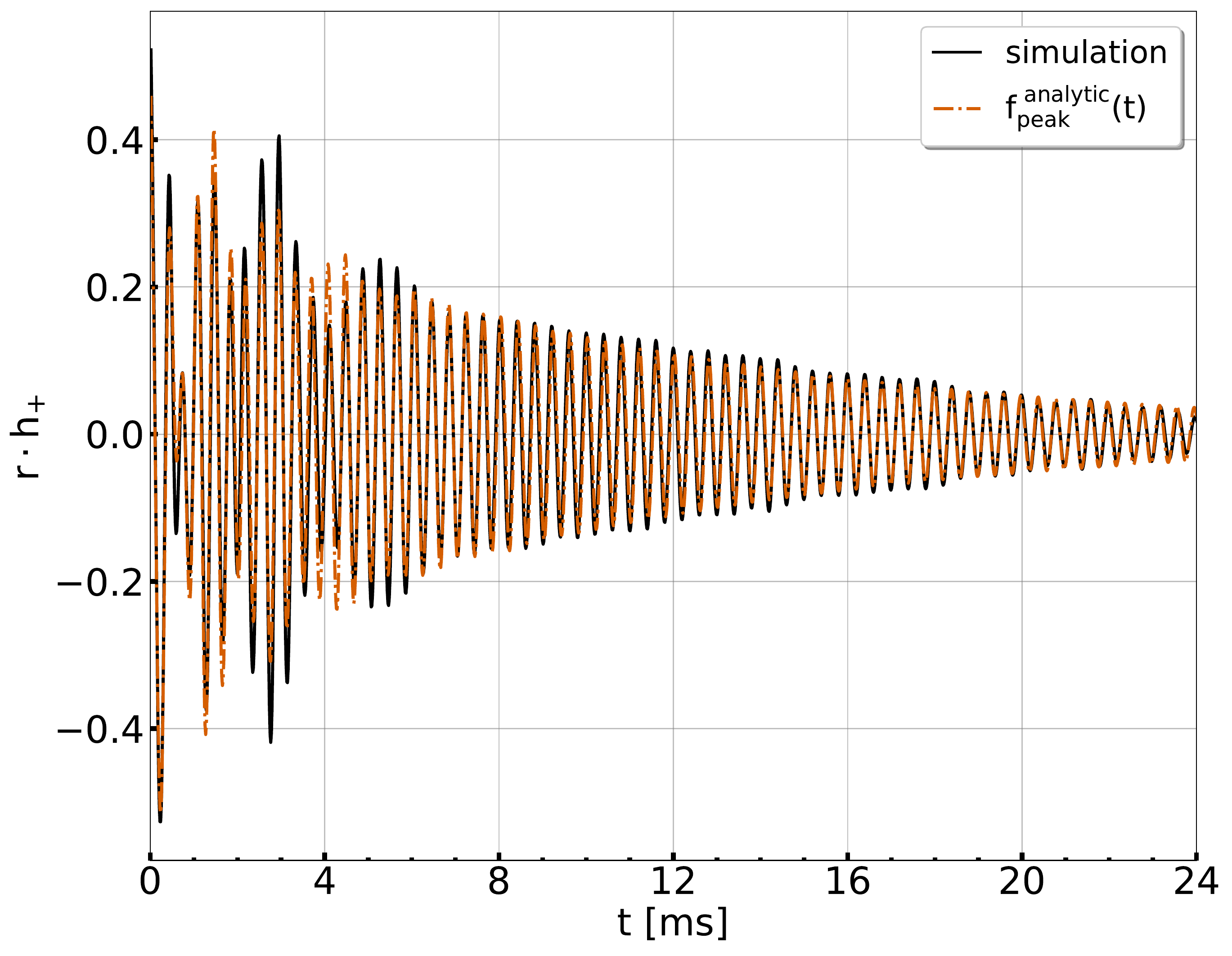} 
	\caption{\label{fig:dumpGW_model_mpa1-m1.25.pdf}GW strain $r\cdot h_{+}(t)$ for the reference simulation and for the analytic model $h_{\mathrm{+}}^\mathrm{{Fit}}(t)$ of Eq.~\eqref{analytic model equations-2}. }
\end{figure}

\begin{figure}[h!]
	\includegraphics[scale=0.33]{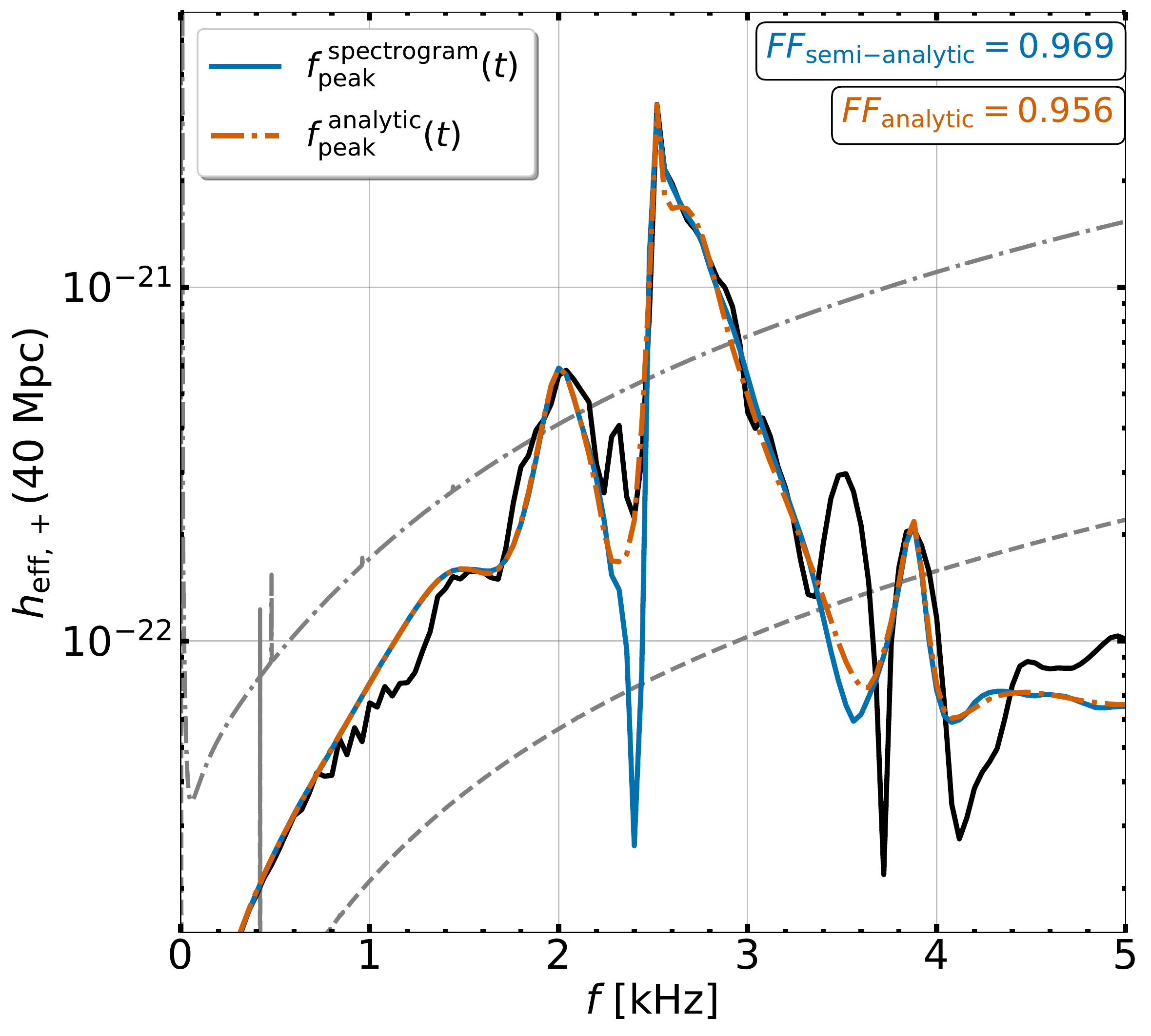} 
	\caption{\label{fig:dumpFFT_model_mpa1-m1.25.pdf} Post-merger effective GW spectra $h_\mathrm{eff,+}(f)$ for the numerical simulation (black line), for the analytic model $h_{+}^{\tiny \mbox{Fit}}(t)$ (orange dashed line) and for the  semi-analytic model (cyan line, see text), for the reference simulation. Colored boxes indicate the respective fitting  factors $FFs$. }
\end{figure}

We extract the parameters of the analytic model, Eq.~\eqref{analytic model equations-2} for our reference simulation, as described in Section~\ref{Parameter extraction}. We compare the simulation data to the analytic model in the time domain in Fig.~\ref{fig:dumpGW_model_mpa1-m1.25.pdf}. The two signals agree very well throughout the whole post-merger evolution of 24~ms. In the early phase, the dominant and the secondary components are significant, whereas during the later evolution only the $f_\mathrm{peak}$ component is present. We remark the importance of a time-dependent $f_\mathrm{peak}(t)$, which simultaneously yields a proper description of the early and the late phase. Note that the model captures the phase evolution very well at late times.

The success of the analytic model is also seen in the GW spectrum $h_\mathrm{eff}(f)$ (see Fig.~\ref{fig:dumpFFT_model_mpa1-m1.25.pdf}). The analytic model reproduces remarkably well the one-sided $f_\mathrm{peak}$ structure.

We further assess the time evolution of $f_\mathrm{peak}(t)$ and its analytic model of a 2-segment piecewise linear function $f_\mathrm{peak}^\mathrm{analytic}(t)$,  Eq.~\eqref{fpeak_piecewise}. To this end we generate the semi-analytic model, as described in  Section~\ref{semi-analytic model}. That is, we extract $f_\mathrm{peak}^\mathrm{spectrogram}(t)$ from the spectrogram and insert the numerical phase $\phi_\mathrm{peak}(t)$ using Eq.~\eqref{numerical phase(t)} in the analytic function Eq.~\eqref{analytic model equations-2}, whereas we obtain all other parameters as described in Section~\ref{Parameter extraction}. The resulting GW spectra are displayed in Fig.~\ref{fig:dumpFFT_model_mpa1-m1.25.pdf} and are compared to the numerical waveform from the simulation. Both models yield spectra that are very close to the spectrum of the numerical simulation. 
We quantify the accuracy of the models by calculating their fitting factors (with respect to the numerical simulation) assuming the projected ET sensitivity curve \cite{EinsteinTelescope2010}. We find fitting factors of  $FF=0.969$ for the semi-analytic model and $FF=0.956$ for the analytic model. The semi-analytic model yields a slightly higher $FF$ than the analytic model, which is expected since the former
contains more precise information about the $f_\mathrm{peak}$ component. However, the small difference of only $1.34\%$ between the fitting factors of the two models demonstrates that using the 
analytic model $f^{\rm analytic}_{\rm peak}$ instead of the numerically extracted $f^{\rm spectrogram}_{\rm peak}(t)$ is sufficient for the description of the time evolution of the $f_\mathrm{peak}(t)$ component.

\begin{figure}[h!]
	\includegraphics[scale=0.34]{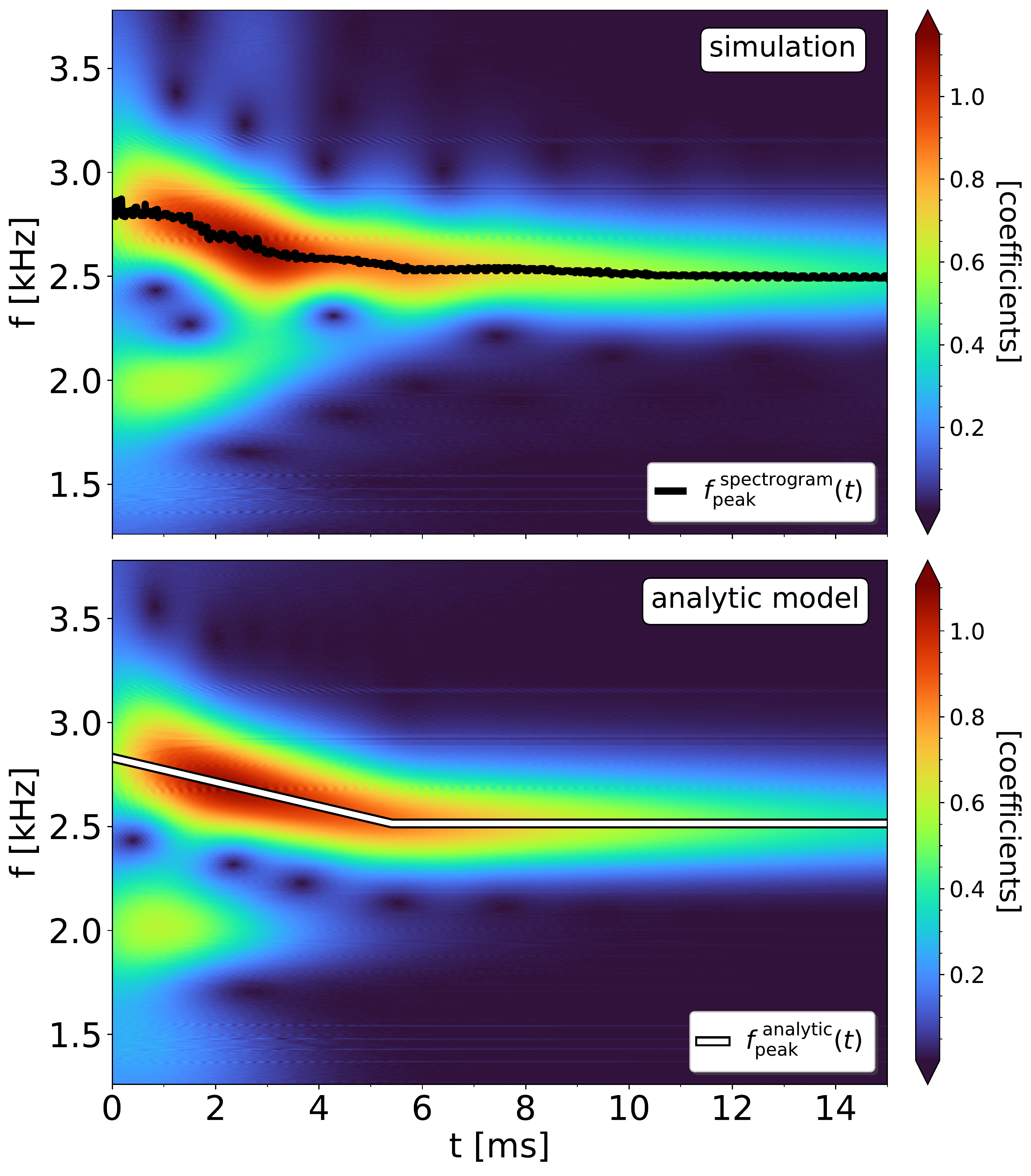} 
	\caption{\label{fig:dump_specComp_m1.25.pdf} {\it Top panel}: spectrogram of $h_+(t)$ for the reference simulation. The black line corresponds to the numerically extracted $f_{\mathrm{peak}}^{\rm spectrogram}(t)$ as described in Fig.~\ref{fig:dumpSPEC_mpa1-m1.25}. {\it Bottom panel}: spectrogram of  $h_+^{\rm Fit}(t)$ for the reference simulation. The white line illustrates $f_{\mathrm{peak}}^\mathrm{{analytic}}(t)$. }
\end{figure}

Fig.~\ref{fig:dumpFFT_model_mpa1-m1.25.pdf} also demonstrates that both the analytic and semi-analytic models successfully reproduce the triplet of secondary frequencies $f_{\mathrm{spiral}},f_{2\pm0}$, which implies that our fitting procedure yields reasonable estimates of the corresponding parameters $A_i$ and $\tau_i$. We note that for the purpose of detectability, the secondary peak $f_{2-0}$ is more important than $f_{2+0}$. Nevertheless, the inclusion of $f_{2+0}$ makes the analytic model more complete and increases the quality of the fit, since the absence of a frequency component in the early phase may spoil the determination of the other parameters. For similar reasons, we found the inclusion of the phenomenological Tukey window function $\mathcal{W}(t; s)$ for the $f_\mathrm{peak}$ component to be useful.

We note that our model does not include and hence does not reproduce the additional frequency peak at 3.5 kHz in Fig. \ref{fig:dumpFFT_model_mpa1-m1.25.pdf}, which remains to be explained and modeled.

Figure \ref{fig:dump_specComp_m1.25.pdf} directly compares the spectrogram of the simulation (upper panel) and of the analytic model (lower panel). We observe a very good agreement considering the simplicity of the analytic model. 
\begin{figure*}[t!]
	\includegraphics[scale=0.36]{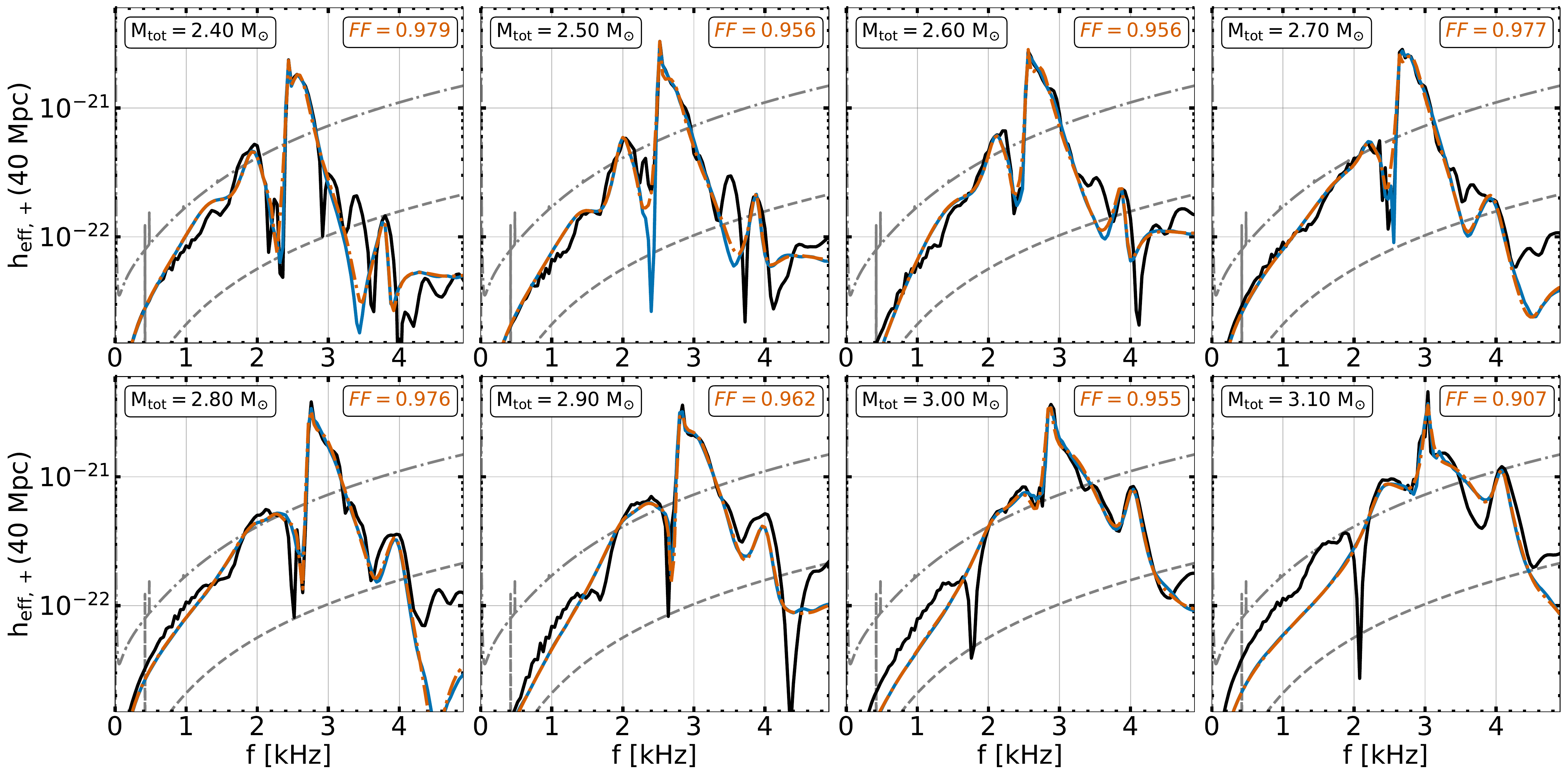} 
	\caption{\label{fig:sequence-FFT-model.pdf} Post-merger effective GW spectra $h_\mathrm{eff,+}(f)$ for the  simulations (black lines), for the analytic model (orange dashed line), and the semi-analytic model (cyan line) along the whole sequence of models. The fitting factors $FFs$ for the analytic model are reported in each case. Note that for the highest-mass model with $M_{\rm tot}=3.1 M_\odot$ an extended analytic model is introduced in Section \ref{Clode to prompt collapse}, where a higher FF is achieved.}
\end{figure*}

\subsubsection{Fitting factors along the whole sequence of merger simulations}

We test the performance of the analytic model along the sequence of models with different $M_{\rm tot}$ (as defined in Section \ref{Sec:Methods}) and display the spectra for the analytic fits (in comparison to the numerical spectra) in Fig.~\ref{fig:sequence-FFT-model.pdf}. We find that the analytic model performs well for all configurations and achieves fitting factors $FFs$ (assuming the sensitivity curve of ET \cite{EinsteinTelescope2010}) in the range $[0.955,0.979]$ for all but the most massive model of this sequence\footnote{As for the reference simulation, we obtain only slightly better FFs for the semi-analytic model - even for the most massive model - and hence we only report the FFs for the analytic model along the whole sequence.}. For the latter model (which is close to the threshold for prompt collapse) we introduce an extended analytic model in Section \ref{Clode to prompt collapse} and achieve a comparable fitting factor of 0.962.  

In the spectra of Fig.~\ref{fig:sequence-FFT-model.pdf}, the secondary frequency components are well reproduced by the analytic model and the shape of the frequency peaks agrees with that obtained from the simulations. Our fitting procedure, as described in Section \ref{Parameter extraction}, yields parameter values that capture well the secondary peaks, except for the amplitude of the $f_{2+0}$ combination tone, $A_{2+0}$. The latter would need to be individually amplified at the end of the above fitting procedure for models with $M_\mathrm{tot}\geq 2.8 M_\odot$, in order to obtain better agreement with the simulations.

\subsection{Simplified analytic models}\label{Simplified analytic models }
\begin{figure}[h]
	\includegraphics[scale=0.33]{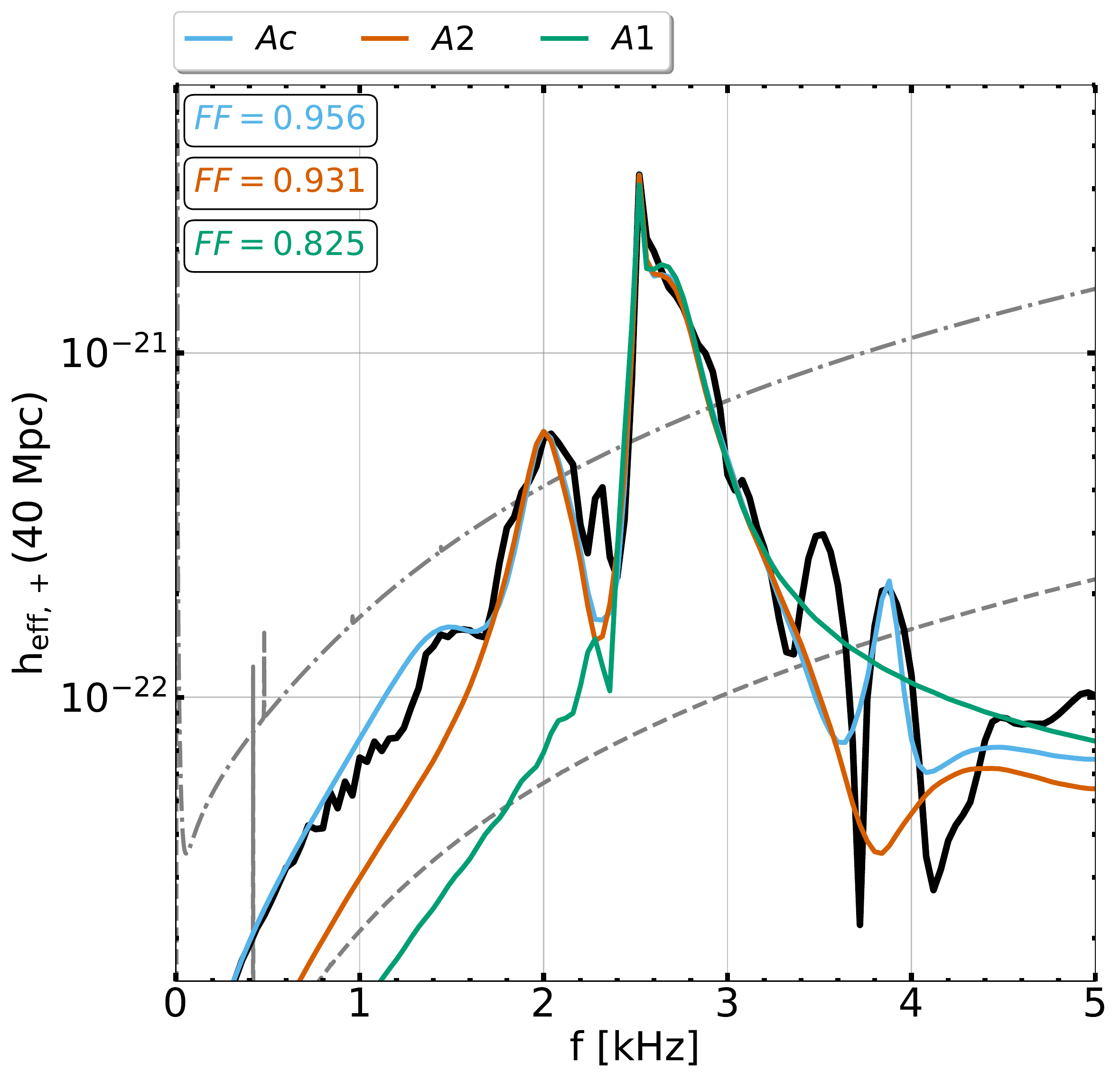} 
	\caption{\label{fig:dumpFFT_toymodel_components_mpa1-m1.25.pdf} Post-merger effective GW spectra $h_\mathrm{eff,+}(f)$ for the  simulation (black lines) and for three analytic models, Ac (cyan line), A2 (orange line) and A1 (green line) for the reference simulation. In each case, the corresponding fitting factor $FF$ is shown.}
\end{figure}

To further assess our analytic model we consider simplified analytic models and quantify their performance using $FFs$. We first discuss the reference simulation and then extend the considerations to the whole sequence of merger simulations.

\subsubsection{Definitions of the simplified analytic models}
We consider three simplified analytic models. The first one includes only the time-dependent $f_{\mathrm{peak}}(t)$ component and the second includes the $f_{\mathrm{peak}}(t)$ component plus one secondary component. As before, the  $f_\mathrm{peak}(t)$ component is modeled by the analytic 2-segment function $f_\mathrm{peak}^\mathrm{analytic}(t)$. For low-mass models, including the reference simulation, the dominant secondary component is the $f_\mathrm{spiral}$, while for higher mass configurations $f_{2-0}$ becomes the most prominent feature. In the third model, $f_{\mathrm{peak}}(t)$ is kept constant and equal to $f_{\mathrm{peak}}=\langle f_{\mathrm{peak}}^{t\in[0, t_*]}\rangle$, while all the secondary frequency components are included. 

We note that for the 2-component model (one secondary component) we do not employ the normalization factor $\mathcal{N}$. For the 1-component model we discard the phenomenological window $\mathcal{W}(t;s)$, since this leads to a slightly higher fitting factor in this case. 

Table~\ref{Table with definitions} summarizes information on the various analytic,  semi-analytic and simplified models and their assigned names. 

\begin{table}
	\begin{ruledtabular}
		\begin{tabular}
	{p{3.cm}|p{1.cm}| p{3.cm}}
		\textrm{Model description}& \textrm{Name} &
            \textrm{Included components}\\
			\colrule
\texttt{Complete analytic model} & Ac & $f_\mathrm{peak}^\mathrm{analytic}(t)$, $f_\mathrm{spiral}$, $f_{2-0}$, $f_{2+0}$ \\\hline
\texttt{Complete semi-analytic model}& Sc & $f_\mathrm{peak}^\mathrm{spectrogram}(t)$, $f_\mathrm{spiral}$, $f_{2-0}$, $f_{2+0}$\\\hline
\texttt{Simplified (2-component) analytic model} & A2 &  $f_\mathrm{peak}^\mathrm{analytic}(t)$, $f_\mathrm{spiral}$ or $f_{2-0}$\\\hline
\texttt{Simplified (1-component) analytic model} & A1 & $f_\mathrm{peak}^\mathrm{analytic}(t)$\\\hline
\texttt{Simplified (const. frequencies) complete analytic model} &  sAc & $\langle f_{\mathrm{peak}}^{t\in[0, t_*]}\rangle$, $f_\mathrm{spiral}$, $f_{2-0}$, $f_{2+0}$\\

		\end{tabular}
	\end{ruledtabular}
	\caption{Definitions for the various analytic, semi-analytic and simplified models that we consider. When the time-dependence is explicitly written, a time-dependent description is employed for that particular component.}
	\label{Table with definitions} 
\end{table}

\subsubsection{Fitting factors for the reference simulation}
We perform the fits using the aforementioned procedure for the complete analytic model (Ac), the 2-component analytic model (A2), and the 1-component analytic model (A1) and display the corresponding post-merger GW spectra in Fig.~\ref{fig:dumpFFT_toymodel_components_mpa1-m1.25.pdf} for the reference simulation. All three models reproduce well the shape of the $f_\mathrm{peak}$ peak, since they include the time-dependent description for $f_{\rm peak}^{\rm analytic}(t)$. However, there are significant differences in the $FFs$. The complete analytic model achieves  $FF=0.956$. As one would expect, the fewer components are included in the model, the worse is the value of the fitting factor. The 2-component model achieves $FF=0.931$, whereas for the 1-component model the performance  deteriorates to $FF=0.825$.  

To further understand the impact of differences in the achieved fitting factors we convert them to the reduction in detection rates, which is considered to scale as $(1-{FF}^3)\cdot 100$ \cite{Apostolatos1995}. For the reference simulation discussed in  Fig.~\ref{fig:dumpFFT_toymodel_components_mpa1-m1.25.pdf}, the complete analytic model achieves a reduction of the detection rate of only $12.63\%$, whereas the simpler, 2-component and 1-component analytic models suffer from larger reductions of $19.30\%$ and  $43.85\%$ respectively.

The above comparison quantifies the importance of including at least one secondary component to the analytic description of the post-merger phase, as this significantly increases the detectability of the signal with matched-filtered techniques, otherwise more than half of the candidate events would go undetected.

\subsubsection{Phase evolution}

\begin{figure}[h]
	\includegraphics[scale=0.33]{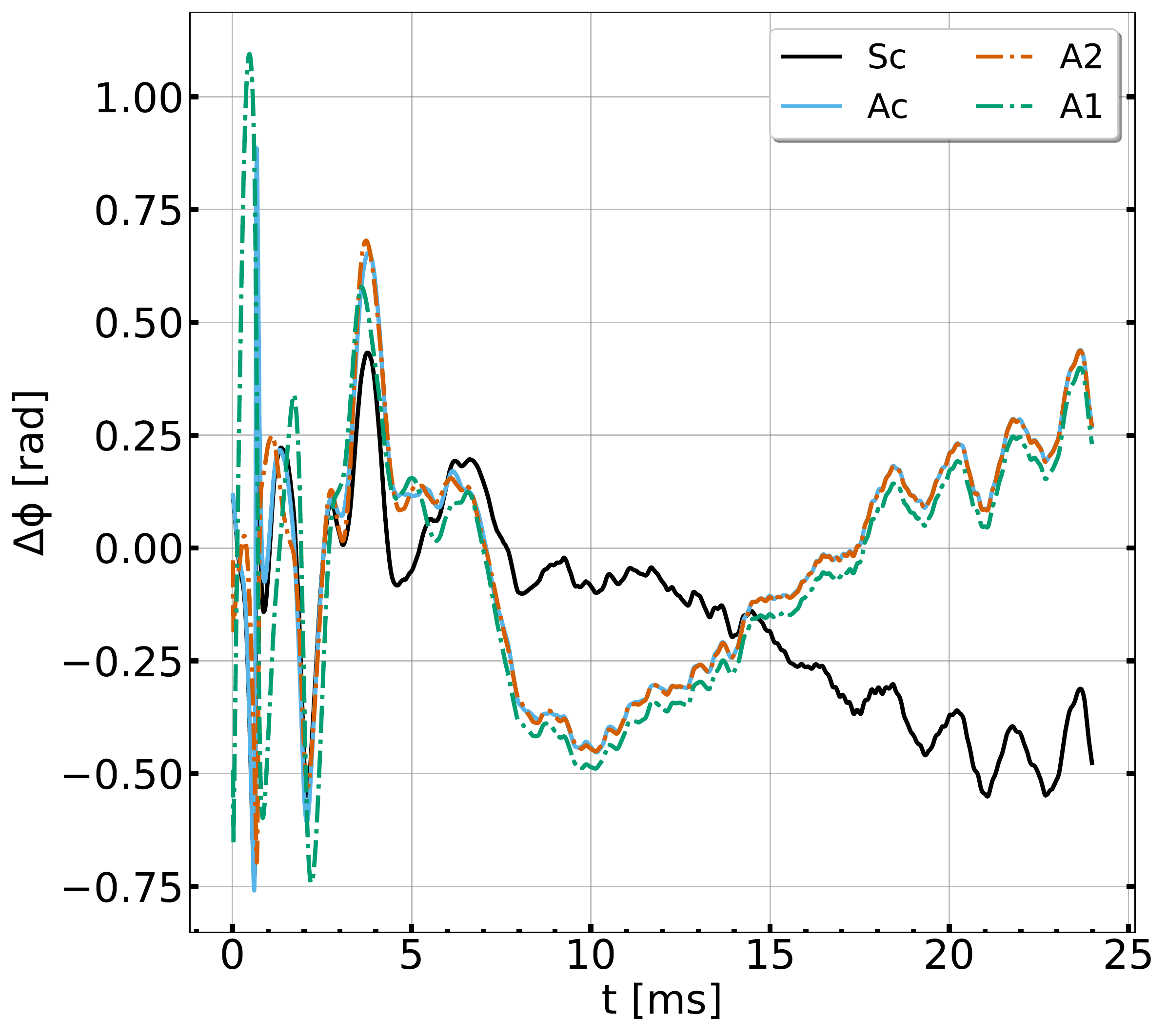} 
	\caption{\label{fig:dump_phase_model_components_mpa1-m1.25.pdf}Gravitational phase difference $\Delta \phi(t)$ between simulation and analytic or semi-analytic model fits for the reference simulation in post-merger phase. }
\end{figure}

We compare the analytic model fits with respect to the gravitational phase $\phi(t)$ defined by
\begin{eqnarray} \label{gravitaional phase}
\phi(t) &=& -\arctan\left(\frac{h_\times(t)}{h_+(t)}\right).
\end{eqnarray}
We compute the phase difference $\Delta \phi(t)=\phi^\mathrm{fit}(t)-\phi^\mathrm{simulation}(t)$ between the analytic models and the GW signal from the simulation (see Fig.~\ref{fig:dump_phase_model_components_mpa1-m1.25.pdf}). In the following analysis, we also consider the complete semi-analytic model (Sc) where the $f_\mathrm{peak}(t)$ component is modelled by $f_\mathrm{peak}^\mathrm{spectrogram}(t)$. We split the post-merger signal in two phases: the initial phase, which lasts approximately 8 milliseconds and the late phase referring to the rest of signal. 

In the early phase, the phase differences $\Delta\phi(t)$ are characterized by low amplitude spikes. These spikes are present in all of the analytic models. The semi-analytic model follows the same trends, although with slightly lower amplitudes. In the late post-merger phase, the phase difference for the analytic models is dominated by $f_{\mathrm{peak}}^\mathrm{analytic}(t)$, since by that time the secondary peaks have practically diminished. The semi-analytic model has a notably different phase evolution than the analytic models, although the absolute value $|\Delta \phi(t)|$ is comparable.

\begin{table} 
	\begin{ruledtabular}
		\begin{tabular}{c|cccc}
		\multicolumn{5}{l}{{Reduction in detection rates ($\%$)}}\vspace{0.2cm}  \\ \hline 
			\textrm{$ M_\mathrm{tot}[M_\odot]$}&
			\textrm{Sc}&
			\multicolumn{1}{c}{Ac}&
			\textrm{A2}&
			\textrm{A1}\\
			\colrule
			
            2.4& 5.01& 6.17& 9.86& 43.03\\
            2.5& 9.01& 12.63& 19.30& 43.85\\
            2.6& 7.88& 12.63& 13.45& 37.93\\
            2.7& 3.56& 6.74& 21.88& 40.52\\
            2.8& 5.30& 7.03& 9.01& 42.82\\
            2.9& 8.45& 10.97& 17.73& 41.36\\
            3.0& 11.53& 12.90& 22.38& 55.75\\
            3.1& 24.39& 25.39& 33.01& 72.28\\			
			
		\end{tabular}
	\end{ruledtabular}
	\caption{Reduction in detection rates for various analytic and semi-analytic models. The definition of each model is given in Tab.~\ref{Table with definitions}.}
	\label{reduction rates}
\end{table}

\begin{figure}[h]
	\includegraphics[scale=0.33]{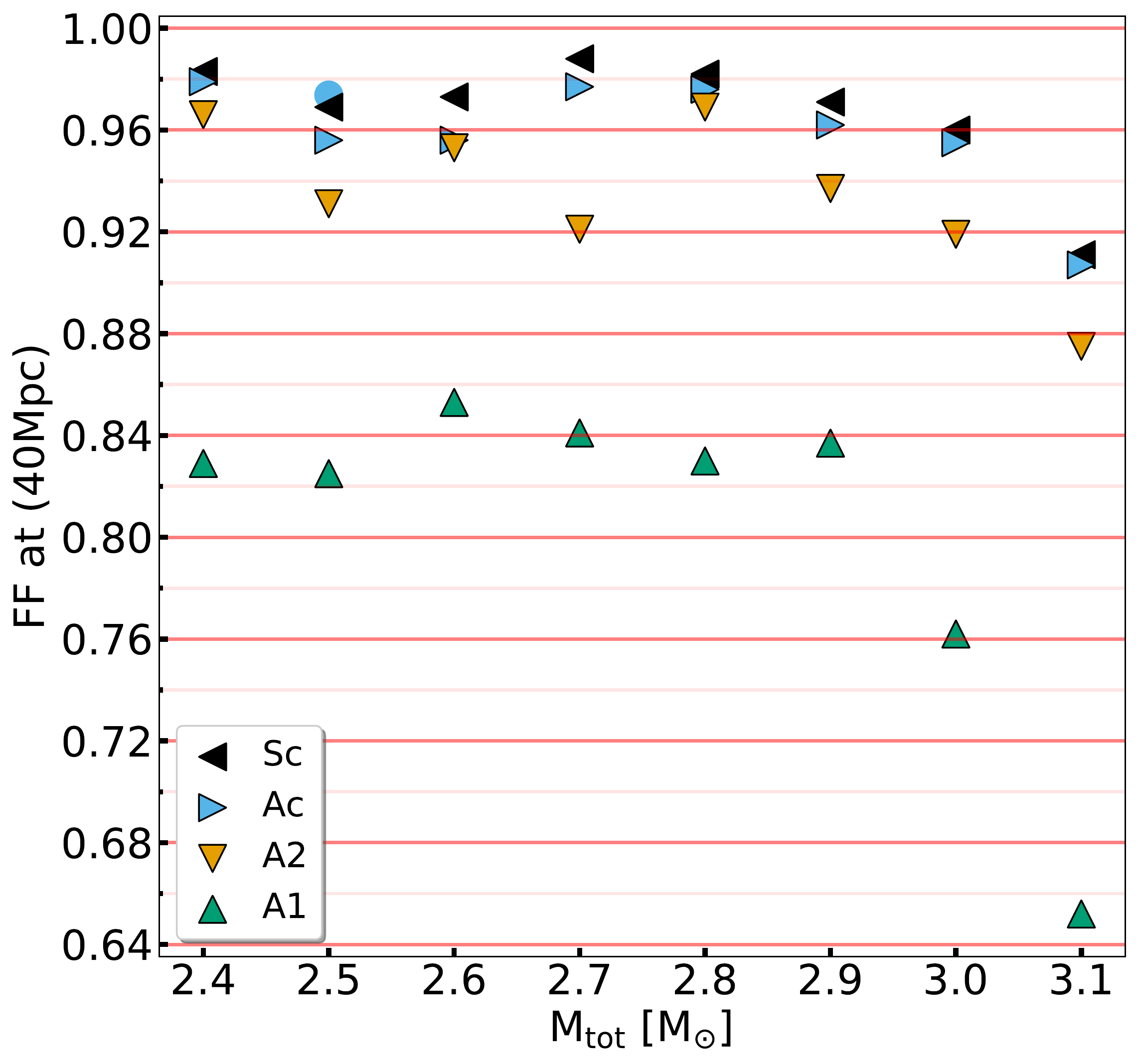} 
	\caption{\label{fig:sequence-match-ET.pdf}Fitting factors $FFs$ for the analytic, and semi-analytic model fits for a source at polar distance of 40~Mpc using the Einstein Telescope sensitivity curve \cite{EinsteinTelescope2010}. The blue circle displays the $FF$ for the Ac model fit for the HR simulation (see Appendix~\ref{Resolution study}).}
\end{figure}

\subsubsection{Fitting factors along the whole sequence of merger simulations}

We compare the fitting factors achieved by the complete  analytic (Ac) and semi-analytic (Sc) models and by the simplified analytic models (A2, A1) along the whole sequence of merger simulations in Fig.~\ref{fig:sequence-match-ET.pdf} and report the corresponding reduction in detection rates in Tab.~\ref{reduction rates}. The general trend is consistent with the findings for the reference simulation. The complete analytic and semi-analytic models perform best leading to the highest fitting factors. The fitting factors for the simple, 1-component analytic model are between 0.82 and 0.86 for most simulations, but deteriorate drastically for the two highest-mass simulations, leading to a reduction of the detection rate of up to 72.28\%. The 2-component model performs significantly better than the 1-component model, but it is still insufficient, when compared to the complete analytic or semi-analytic models.  
We thus conclude that  post-merger GW templates should include several secondary components such as $f_{\mathrm{spiral}}$ and $f_{2\pm0}$, if a small reduction of the detection is to be achieved.

\subsubsection{Importance of the 2-segment description of $f_{\rm peak}(t)$}
Furthermore, we assess the significance of the time-dependent description of $f_{\mathrm{peak}}(t)$ in the analytic model in comparison to the constant frequency description. We consider the simplified complete analytic model (sAc) where $f_{\mathrm{peak}}(t)$ is constant and equal to $f_{\mathrm{peak}}=\langle f_{\mathrm{peak}}^{t\in[0, t_*]}\rangle$, we perform the fits for the models Ac, sAc and compare the fitting factors $FFs$. 

\begin{figure}[h]
	\includegraphics[scale=0.32]{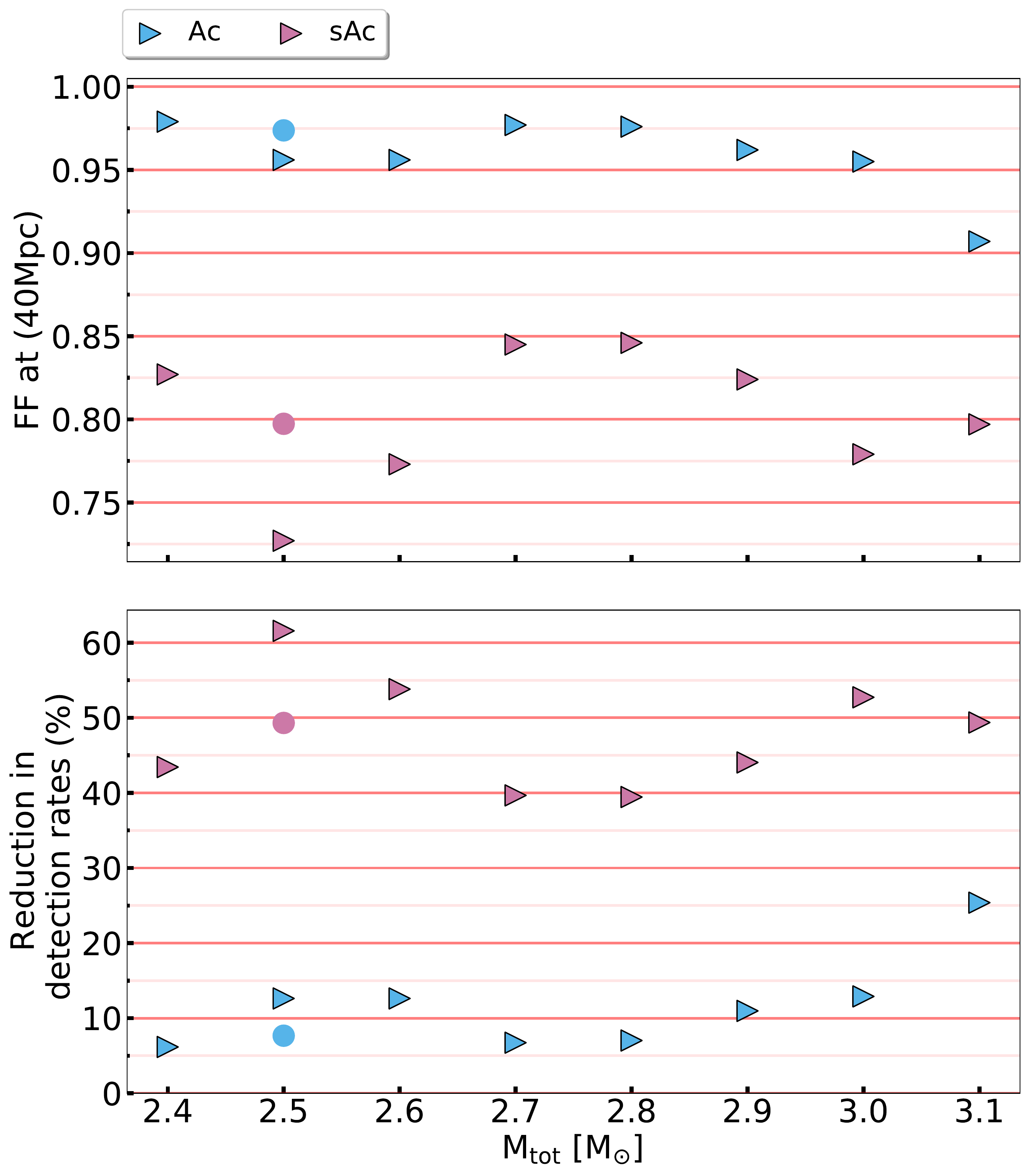} 
	\caption{\label{fig:sequence-Ac-sAc-match-ET.pdf}{\it Top panel}: Fitting factors $FFs$ for the Ac and sAc analytic fits. {\it Bottom panel}: Reduction in detection rates for the Ac and sAc analytic fits. The circles indicate the $FF$ (blue) and reduction in detection rates (pink) for the sAc model fit for the HR simulation (see Appendix~\ref{Resolution study}).}
\end{figure}

Table \ref{comparison of TMs} and Fig.~\ref{fig:sequence-Ac-sAc-match-ET.pdf} (top panel) show the fitting factors along the mass sequence. The model sAc leads to small fitting factors $FFs$, ranging from 0.727 to 0.846. In contrast, the Ac model performs significantly better. In terms of the reduction in detection rates, the sAc model is significantly worse than the Ac model. We note that for the two highest mass models ($M_\mathrm{tot}=3.0,3.1~M_\odot$) substituting $\langle f_\mathrm{peak}^{t\in[0,t_*]}\rangle$ with $f_\mathrm{peak}$ leads to $FFs$ close to the ones obtained with Ac.

\begin{table} 
	\begin{ruledtabular}
		\begin{tabular}{c|cc|cc}
		\ &\multicolumn{2}{l|}{\makecell{Fitting factors ($FFs$)\\\ }}&\multicolumn{2}{l}{\makecell{Reduction in detection \\rates (\%)}} \vspace{0.2cm}  \\ \hline 
			\textrm{$ M_\mathrm{tot}[M_\odot]$}&
			\textrm{Ac}&
			\textrm{sAc}&
			\textrm{Ac}&
			\textrm{sAc}\\
			\colrule

2.4& 0.979& 0.827& 6.17& 43.44\\
2.5& 0.956& 0.727& 12.63& 61.58\\
2.6& 0.956& 0.773& 12.63& 53.81\\
2.7& 0.977& 0.845& 6.74& 39.66\\
2.8& 0.976& 0.846& 7.03& 39.45\\
2.9& 0.962& 0.824& 10.97& 44.05\\
3.0& 0.955& 0.779& 12.90& 52.73\\
3.1& 0.907& 0.797& 25.39& 49.37\\

		\end{tabular}
	\end{ruledtabular}
	\caption{Fitting factors $FFs$ and reduction in detection rates (\%) for the Ac and sAc analytic models for the post-merger GW emission (see Tab.~\ref{Table with definitions} for definitions).}
	\label{comparison of TMs}
\end{table}

\section{Parameters of the analytic model}\label{Parameters of the analytic model}
In this section we discuss the parameters of the analytic model and their dependence on the total binary mass $M_\mathrm{tot}$. We find a systematic dependence on $M_\mathrm{tot}$ for all the parameters of the model and employ polynomial fits to obtain analytic descriptions of the respective dependencies. We first focus on the analytic description of $f_\mathrm{peak}(t)$ and the parameters which determine the 2-segment piecewise function,  Eq.~(\ref{fpeak_piecewise}). We then discuss the amplitudes $A_\mathrm{i}$, timescales $\tau_\mathrm{i}$ and normalization factor $\mathcal{N}$. We address the initial phases $\phi_\mathrm{i}$ where we find additional correlations between these parameters. Finally, we employ empirical relations for all the parameters of the analytic model (Ac) and discuss a purely analytic model which uses exclusively analytic functions. 

\subsection{$\mathbf{f_\mathrm{peak}(t)}$ parametrization}\label{fpeak parameters}

\begin{figure}
    \includegraphics[scale=0.312]{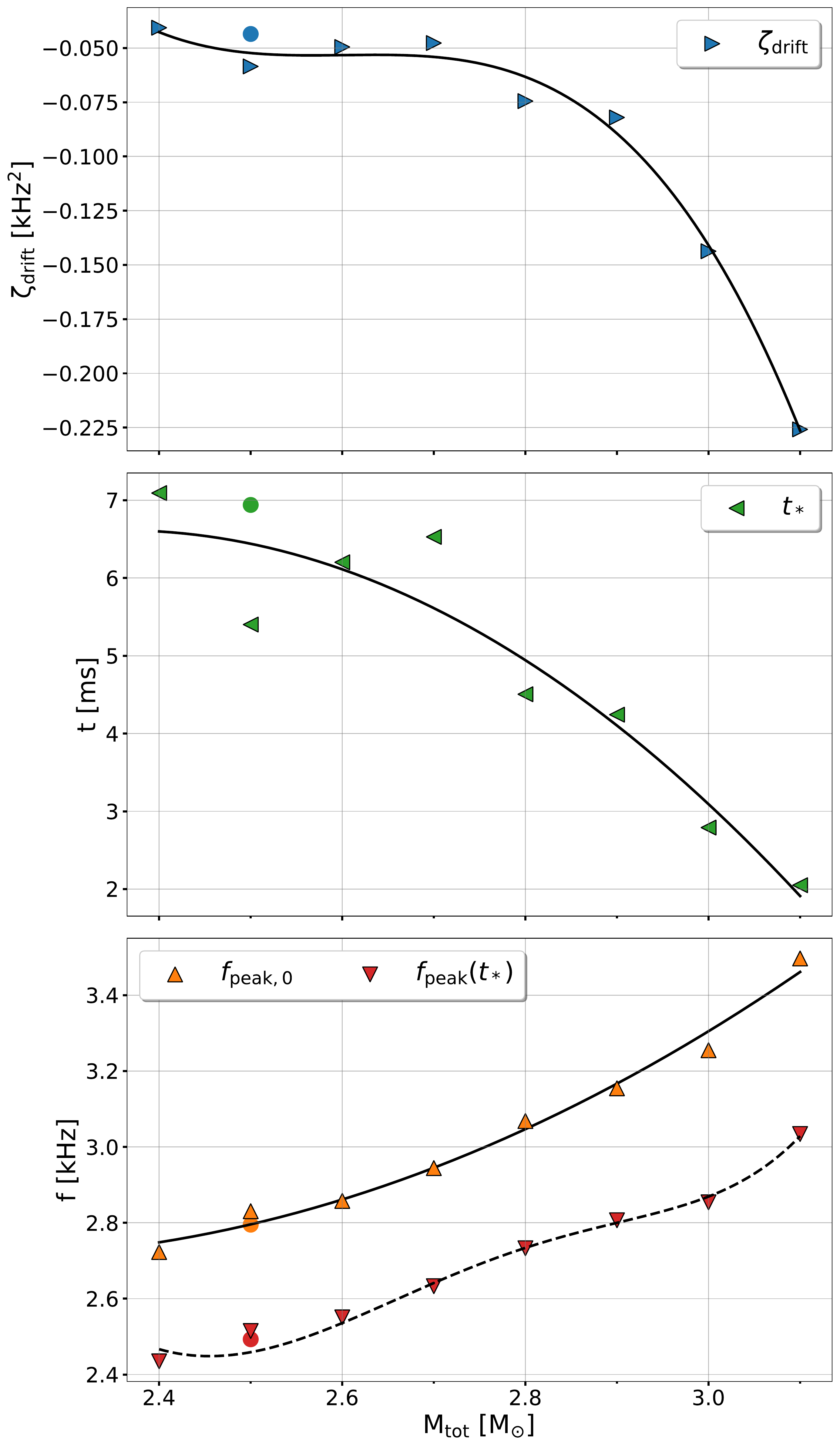}
    \caption{\label{fig:fpeak-relations.pdf} {\it Top panel}: $\zeta_{\mathrm{drift}}$ parameter along the mass sequence. The black curve shows a third order polynomial fit. {\it Middle panel}: $t_{*}$ parameter along the mass sequence. The black curve shows a second order polynomial fit. {\it Bottom panel}: $f_{\mathrm{peak,0}}$ parameter (orange) along the mass sequence. The black solid curve shows a second order polynomial fit. In addition, data points (red) along the mass sequence are shown for $f_{\mathrm{peak}}(t_*)$, which is determined by $\zeta_\mathrm{drift}$, $f_{\mathrm{peak,0}}$, $t_*$. The black dashed curve is determined by polynomial fits to $\zeta_\mathrm{drift}$, $f_{\mathrm{peak,0}}$, $t_*$. Cyan circles indicate the $f_\mathrm{peak}$ extracted from the GW spectra (see Fig.~\ref{fig:sequence-FFT.pdf}). The circles indicate the respective parameters (for each figure) for the HR simulation (see Appendix~\ref{Resolution study}).}
\end{figure}

Figure~\ref{fig:fpeak-relations.pdf} shows the extracted parameters $\zeta_\mathrm{drift}, t_*, f_\mathrm{peak,0}$ as functions of total mass $M_\mathrm{tot}$ for our sequence of simulations. We find that these parameters follow specific dependencies. The dependence of the parameters $t_*, f_\mathrm{peak,0}$ and $\zeta_\mathrm{drift}$ can be modelled by second and third order polynomials (black lines) respectively, given by
\begin{eqnarray}\label{equations-fpeak(t)-fits_1}
\zeta_\mathrm{drift} &=& -1.420 \cdot M_\mathrm{tot}^3+11.085\cdot  M_\mathrm{tot}^2\nonumber\\
&\ &- 28.834\cdot M_\mathrm{tot}+24.943,\\
\label{equations-fpeak(t)-fits_2}f_\mathrm{peak,0} &=& +0.908\cdot M_\mathrm{tot}^2-3.974\cdot M_\mathrm{tot}+7.058,\\
\label{equations-fpeak(t)-fits_3}t_{*} &=& -8.523\cdot M_\mathrm{tot}^2+40.179\cdot M_\mathrm{tot}-40.741.
\end{eqnarray}

Figure \ref{fig:fpeak-relations.pdf} (bottom) shows the final frequency $f_\mathrm{peak}(t=t_*)$, which by definition is determined by the parameters $\zeta_\mathrm{drift}, f_\mathrm{peak,0}, t_*$ (the black dashed curve is determined by Eq.~\eqref{equations-fpeak(t)-fits_1}-\eqref{equations-fpeak(t)-fits_3}).

The parameter $f_\mathrm{peak}(t_*)$ (see Fig.~\ref{fig:fpeak-relations.pdf}) approximately coincides with $f_\mathrm{peak}$ (maximum of the peak in the GW spectra) and increases with the total mass $M_\mathrm{tot}$ (since the remnant becomes more compact). $f_\mathrm{peak,0}$ exhibits a similar dependence on the total mass $M_\mathrm{tot}$. As previously mentioned, $f_\mathrm{peak}(t)$ evolves faster and more significantly for high-mass configurations. This is  confirmed by the difference $\Delta f_{\mathrm{peak}}=f_\mathrm{peak,0}-f_\mathrm{peak}(t_*)$, which increases with total mass $M_\mathrm{tot}$ from $0.288$ kHz for the model with the lowest mass to $0.462$ kHz for the configuration with $M_\mathrm{tot}=3.1M_\odot$.

The duration of the frequency drift, $t_*$, is a decreasing function of the total binary mass $M_\mathrm{tot}$. We note that in particular $t_*$ possibly exhibits a dependence on the numerical scheme, resolution and physics of the simulation tool, which can affect the angular momentum redistribution of the remnant and possibly prolong or shorten the drift.

The slope parameter $\zeta_\mathrm{drift}$ is approximately constant ($\approx -0.060 \mathrm{kHz}^2$) for $M_\mathrm{tot}\leq2.8\ M_\odot$. However, a rapid decrease occurs as the total mass $M_\mathrm{tot}$ approaches $M_\mathrm{thres}$ (see Fig.~\ref{fig:fpeak-relations.pdf}). Such a trend may not be unexpected as a result of an accelerated evolution of the remnant (in the early post-merger phase) due to the strong gravity. 

If it is possible to extract $\zeta_\mathrm{drift}$, $t_*$ and $\Delta f_\mathrm{peak}$, one may use this information to estimate the proximity to a prompt collapse. To this end, the occurrence of a faster frequency evolution for high-mass binaries should be confirmed for other EoS models, possibly considering $\zeta_\mathrm{drift}$, $t_*$, $\Delta f_\mathrm{peak}$ relative to $f_\mathrm{peak}$, instead of absolute values.

\subsection{Amplitudes, timescales, normalization factor} \label{Amplitudes and timescales}
In this subsection we discuss the properties of the dominant component's parameters $A_\mathrm{peak}$, $\tau_\mathrm{peak}$, the parameters of the secondary components $A_\mathrm{spiral}$, $A_\mathrm{2\pm0}$, $\tau_\mathrm{spiral}$, $\tau_\mathrm{2\pm0}$ and the normalization factor $\mathcal{N}$. We employ the parameters determined for the complete analytic model (Ac).

\subsubsection{${A_\mathrm{peak}}$, ${\tau_\mathrm{peak}}$   }

\begin{figure*}\label{TM_fitted_parameters-2}
    \centering
    \subfloat[\label{fig:TM_fitted_parameters_2_1}]{%
    \includegraphics[scale=0.33]{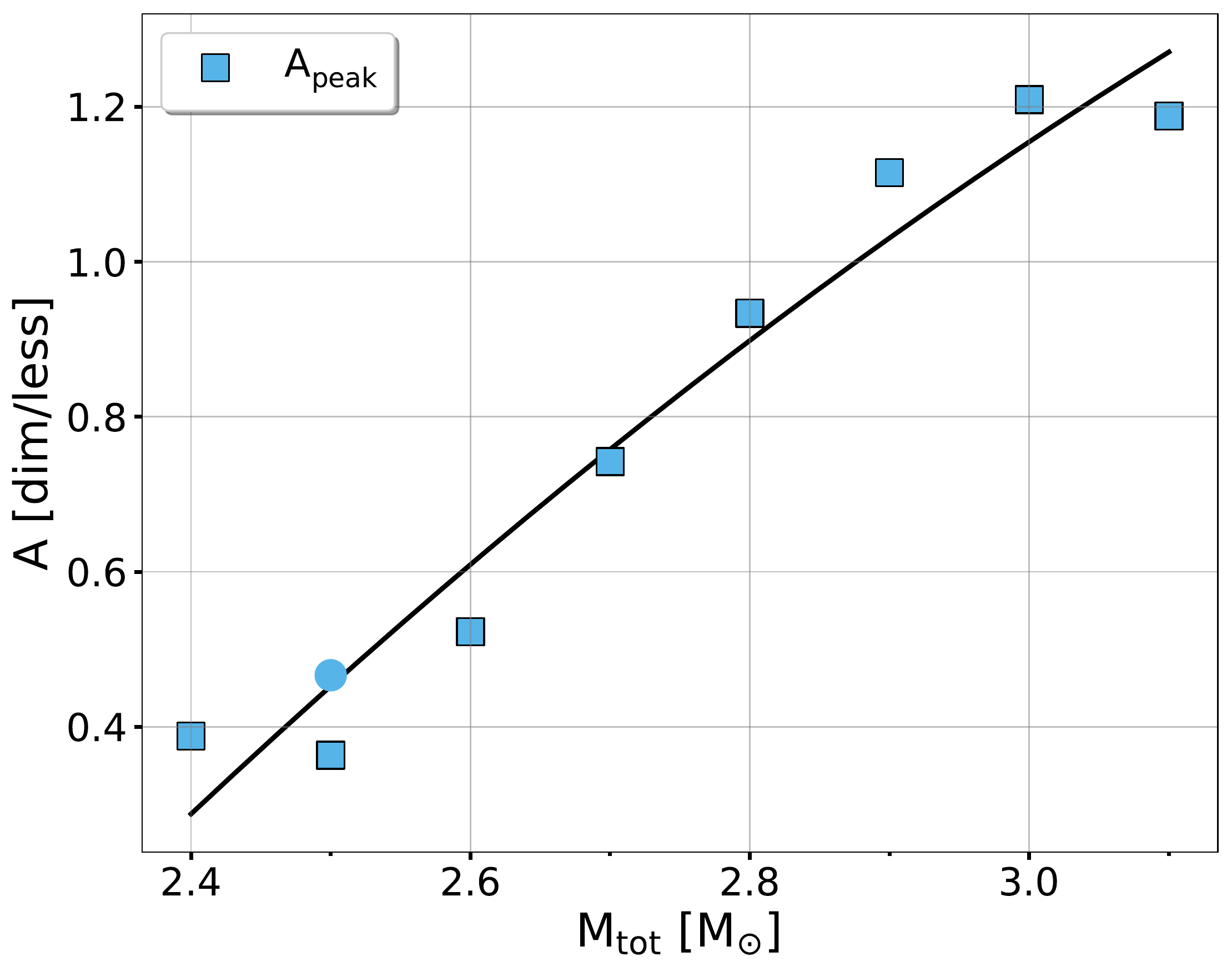}}
    \hspace*{12pt}
    \subfloat[\label{fig:TM_fitted_parameters_2_2}]{
    \includegraphics[scale=0.33]{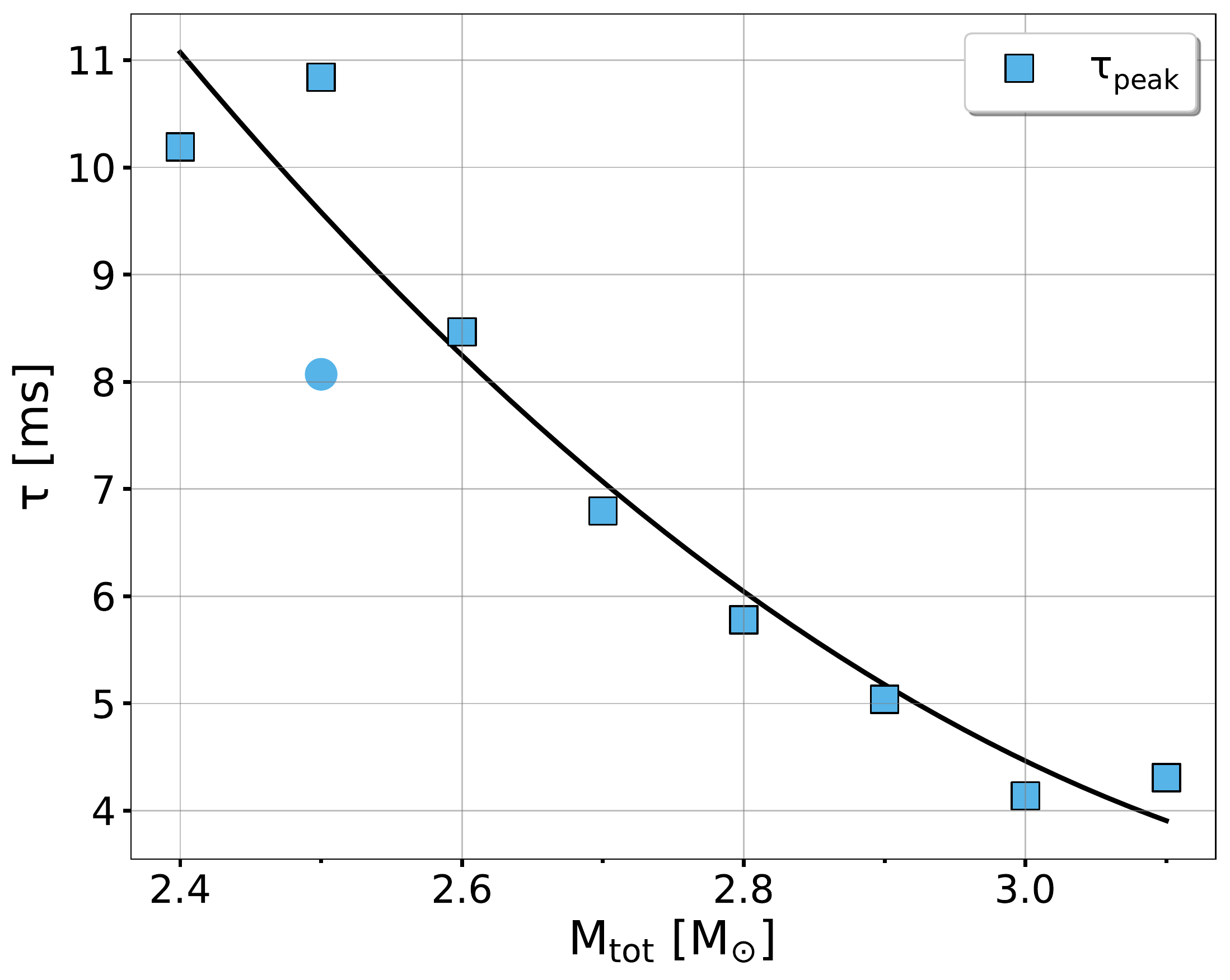}}
    
    \subfloat[\label{fig:TM_fitted_parameters_2_3}]{%
    \includegraphics[scale=0.33]{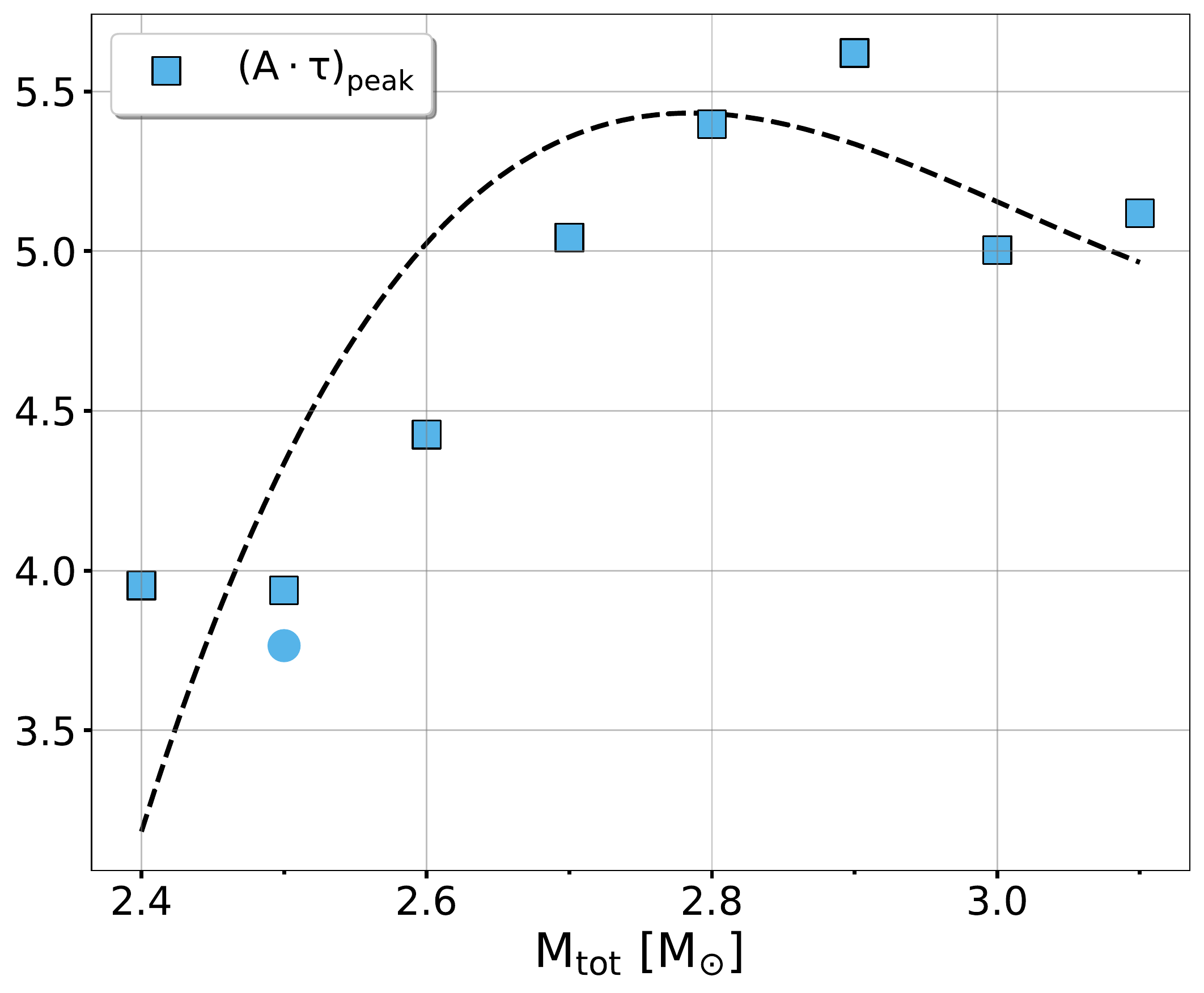}}
    \caption{{\it Top left panel}: Analytic model dimensionless amplitude $A_{\mathrm{peak}}$ for $r\cdot h_+(t)$ for the analytic model which employs the $f_{\mathrm{peak}}^\mathrm{{analytic}}(t)$ description. Black curve corresponds to second order polynomial fit. {\it Top right panel}: Analytic model timescale $\tau_{\mathrm{peak}}$ for the analytic model which employs the $f_{\mathrm{peak}}^\mathrm{{analytic}}(t)$ description. Black curve corresponds to second order polynomial fit. {\it Bottom panel}: Analytic model products $(A\cdot\tau)_{\mathrm{peak}}$. Black dashed curves determined by fits to $A_{\mathrm{peak}},\tau_{\mathrm{peak}}$. The blue circles indicate the respective parameters (for each figure) for the HR simulation (see Appendix~\ref{Resolution study}).}
    
\end{figure*}
Figures \ref{fig:TM_fitted_parameters_2_1},\ref{fig:TM_fitted_parameters_2_2} show the parameters $A_\mathrm{peak}$, $\tau_\mathrm{peak}$. These parameters follow dependencies, which can be modelled by second-order polynomial fits given by
\begin{eqnarray}
 \label{Apeak of mtot}A_\mathrm{peak} &=& -0.409\cdot M_\mathrm{tot}^2+3.657\cdot M_\mathrm{tot}-6.130, \\
 \label{Tpeak of mtot}\tau_\mathrm{peak} &=& +7.782\cdot M_\mathrm{tot}^2-53.040\cdot M_\mathrm{tot}+93.542.
\end{eqnarray}
$A_\mathrm{peak}$ increases with $M_\mathrm{tot}$, which may be expected, since the involved masses are higher and also the initial excitation is more pronounced. $\tau_\mathrm{peak}$ decreases as the total binary mass $M_\mathrm{tot}$ increases, indicating a stronger damping.

\subsubsection{${A_\mathrm{spiral}, A_{2\pm0}},{\tau_\mathrm{spiral}, \tau_{2\pm0}}$ }
\begin{figure*}\label{TM_fitted_parameters}
    \centering
    \subfloat[\label{fig:TM_fitted_parameters_1}]{%
    \includegraphics[scale=0.33]{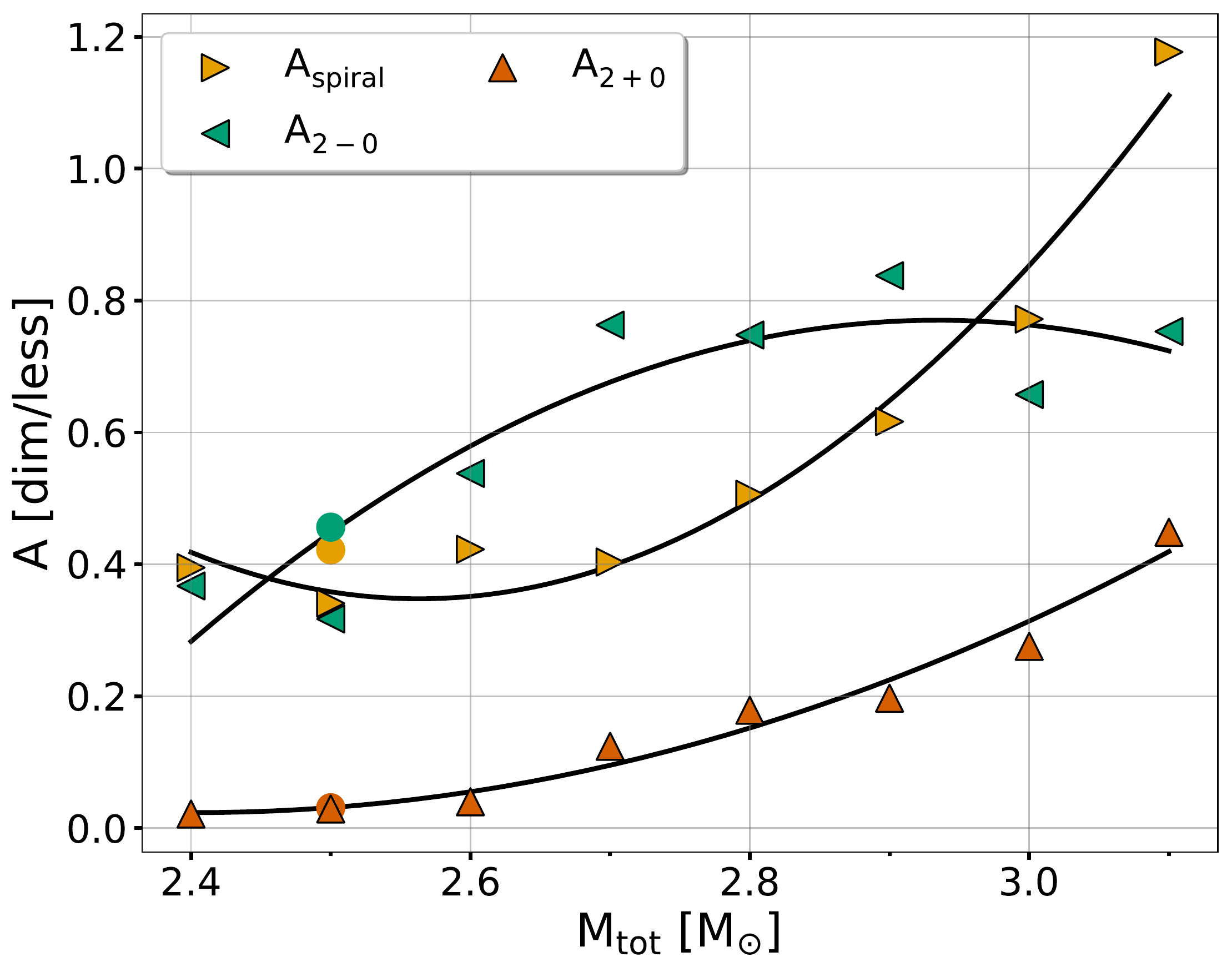}}
    \hspace*{12pt}
    \subfloat[\label{fig:TM_fitted_parameters_2}]{
    \includegraphics[scale=0.33]{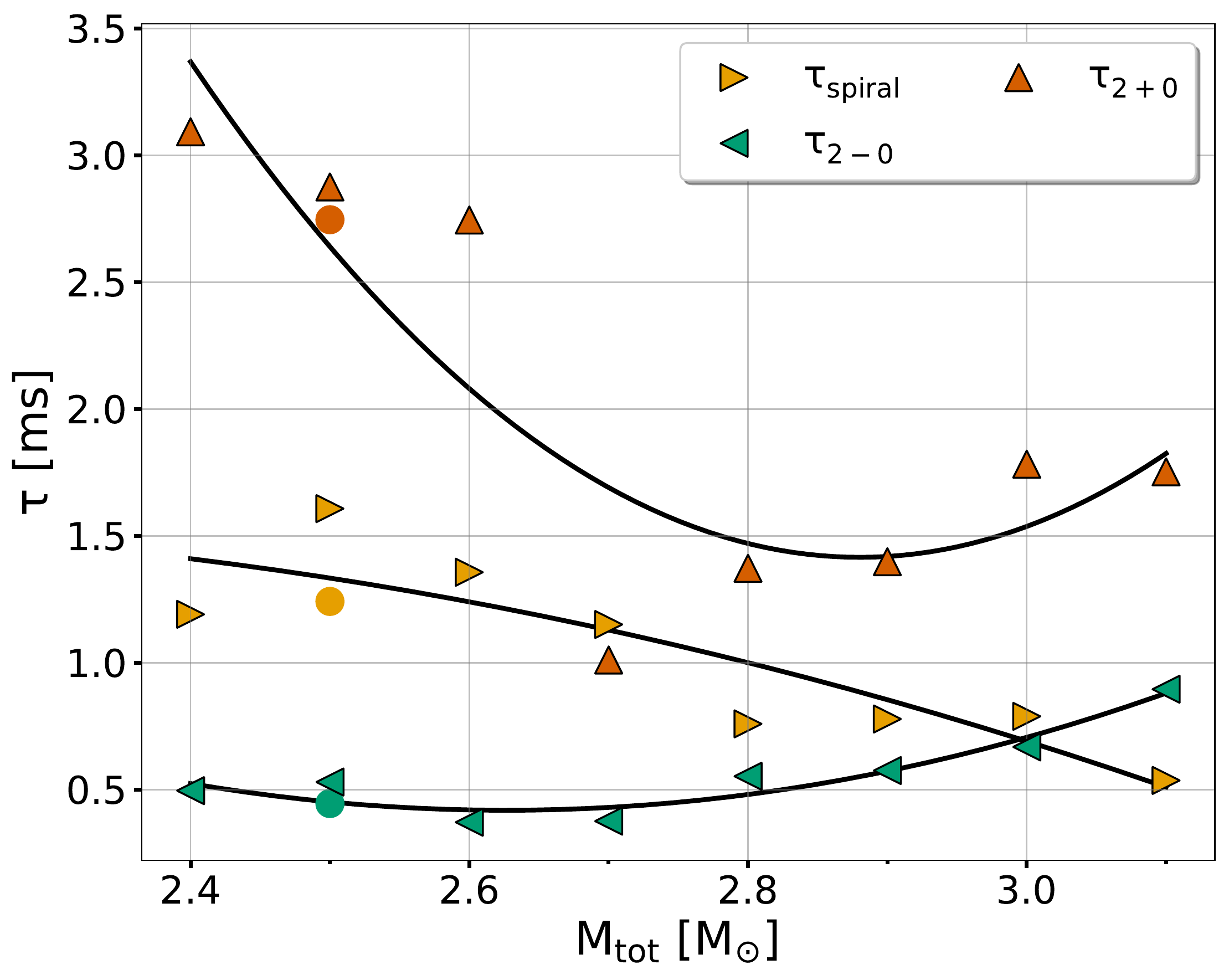}}
    
    \subfloat[\label{fig:TM_fitted_parameters_3}]{%
    \includegraphics[scale=0.33]{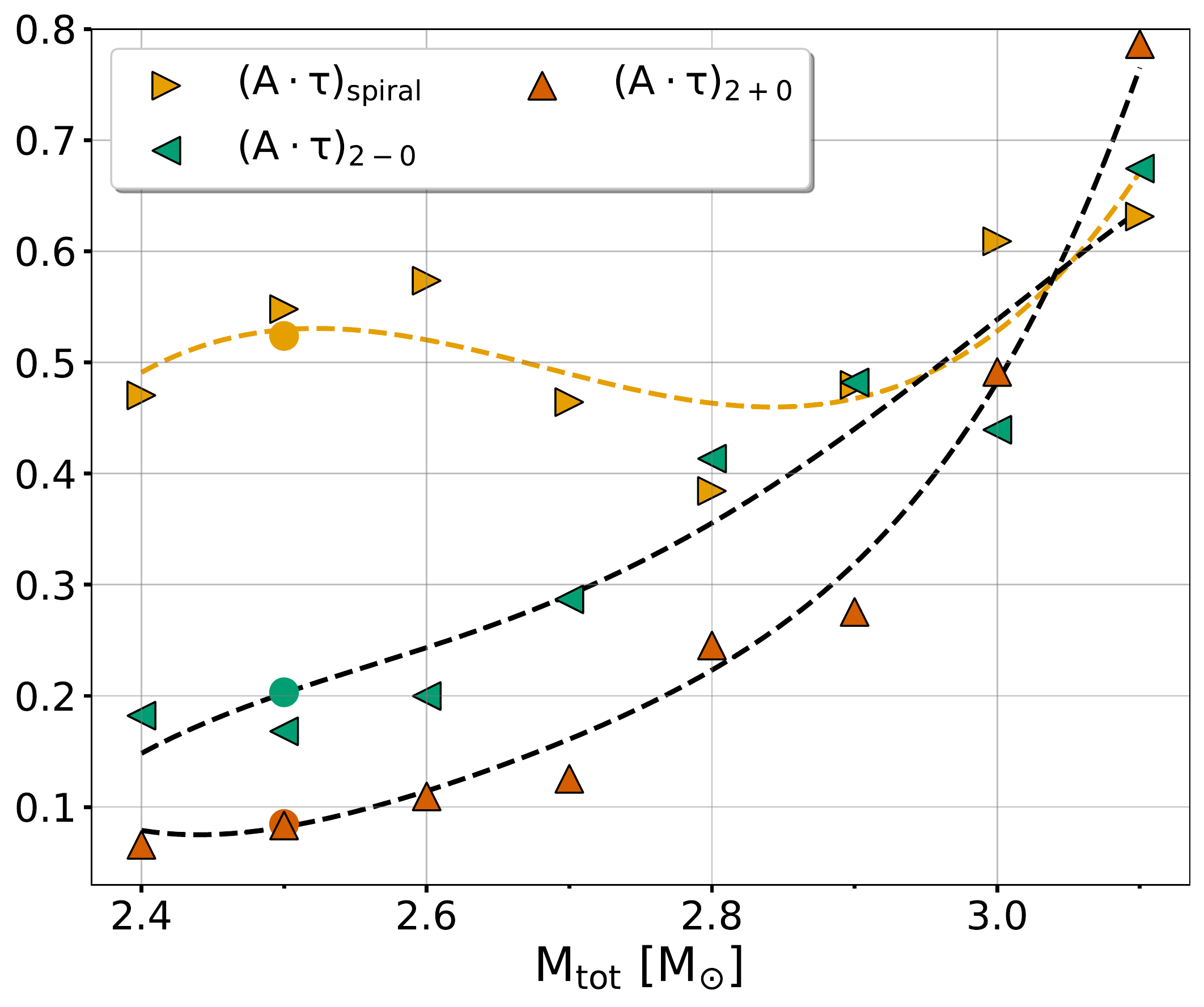}}
    \hspace*{12pt}        
    \subfloat[\label{fig:TM_fitted_parameters_4}]{%
    \includegraphics[scale=0.33]{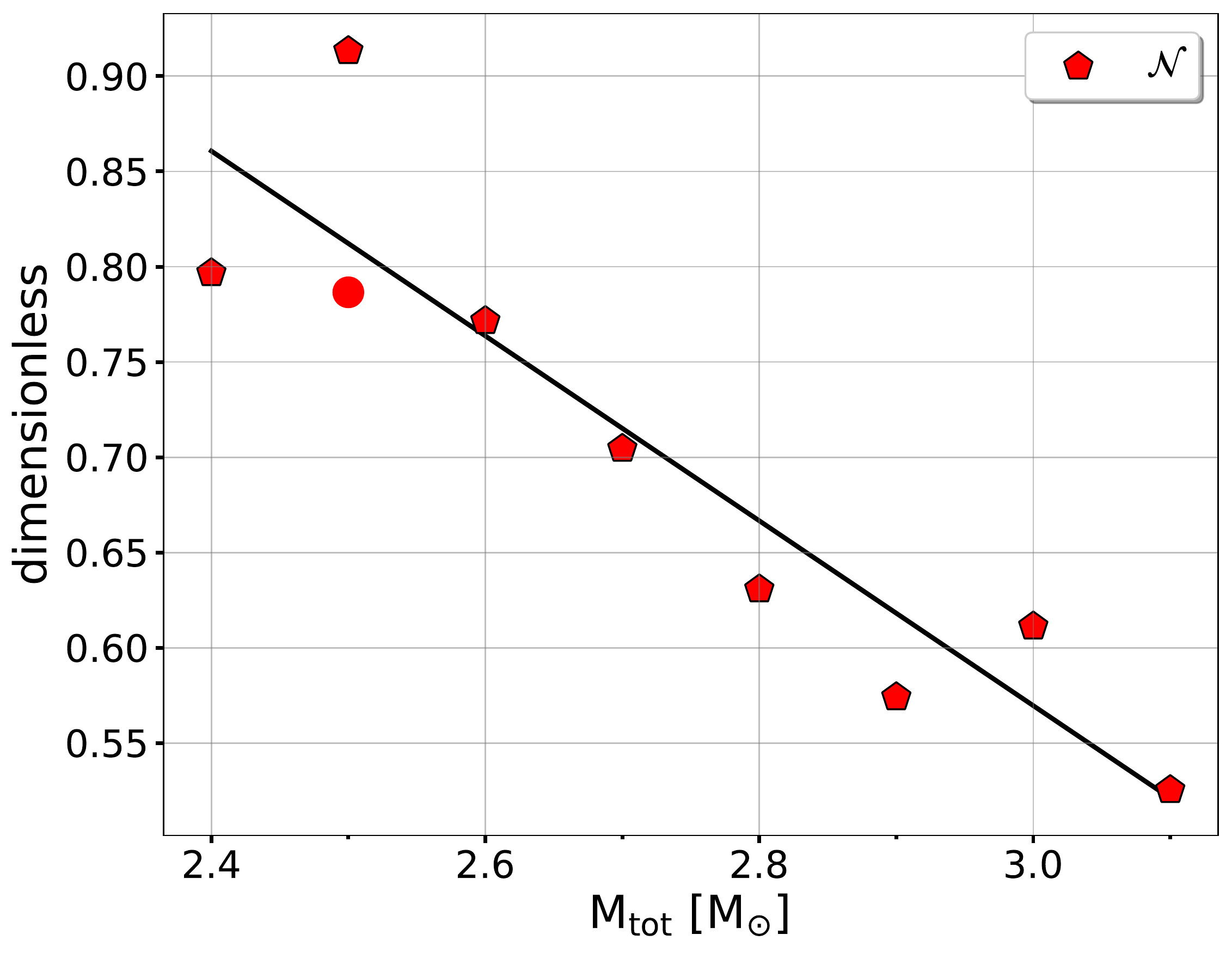}}

    \caption{{\it Top left panel}: Analytic model dimensionless amplitudes $A_{\mathrm{spiral}}$, $A_{2-0}$, $A_{2+0}$ for $r\cdot h_+(t)$ extracted from spectrograms. Black curves correspond to second-order polynomial fits. {\it Top right panel}: Analytic model timescales $\tau_{\mathrm{spiral}}$, $\tau_{2-0}$, $\tau_{2+0}$ extracted from spectrograms. Black curves correspond to second-order polynomial fits. {\it Bottom left panel}: Analytic model products $(A\cdot\tau)_{\mathrm{spiral}}$, $(A\cdot\tau)_{2-0}$, $(A\cdot\tau)_{2+0}$. Black dashed curves determined by polynomial fits to $A_i,\tau_i$ for i=spiral,$2\pm0$. Yellow dashed curve corresponds to the fourth order polynomial fit to $(A\cdot\tau)_{\mathrm{spiral}}$. {\it Bottom right panel}: Analytic model correction factor $\mathcal{N}$ for the analytic model which employs the $f_{\mathrm{peak}}^{\tiny \mathrm{analytic}}(t)$ description. Black curve corresponds to a linear fit. The colored circles indicate the respective parameters (for each quantity and figure) for the HR simulation (see Appendix~\ref{Resolution study}).}
    
\end{figure*}

Figures \ref{fig:TM_fitted_parameters_1},\ref{fig:TM_fitted_parameters_2} display the parameters $A_i$, $\tau_i$ (for $\mathrm{i=spiral,2\pm0}$). The amplitudes $A_\mathrm{i}$ and timescales $\tau_\mathrm{i}$ correlate with the total mass $M_\mathrm{tot}$ and follow specific trends. We quantify these dependencies by performing second-order polynomial fits resulting in
\begin{eqnarray} 
\label{Tspiral of mtot}\tau_\mathrm{spiral} &=& -0.874\cdot M_\mathrm{tot}^2+3.521\cdot M_\mathrm{tot}-2.005, \\
\label{T2-0 of mtot}\tau_{2-0} &=& +2.057\cdot M_\mathrm{tot}^2-10.804\cdot M_\mathrm{tot}+14.606,\\
\label{T2+0 of mtot}\tau_{2+0} &=& +8.469\cdot M_\mathrm{tot}^2-48.785\cdot M_\mathrm{tot}+71.671,
\end{eqnarray}
\begin{eqnarray}
\label{Aspiral of mtot}A_\mathrm{spiral} &=& +2.649\cdot M_\mathrm{tot}^2-13.580\cdot M_\mathrm{tot}+17.752,\\
\label{A2-0 of mtot}A_{2-0} &=& -1.704\cdot M_\mathrm{tot}^2+10.004\cdot M_\mathrm{tot}-13.909,\\
\label{A2+0 of mtot}A_{2+0} &=& +0.816\cdot M_\mathrm{tot}^2-3.920\cdot M_\mathrm{tot}+4.734.
\end{eqnarray}
These relations are not particularly tight, especially for $A_{2-0}$ and $\tau_\mathrm{spiral}$, which is likely caused by the difficulty to precisely extract secondary features from the complex signal. However, the amplitudes of all secondary features clearly increase with mass. 

As the total mass $M_\mathrm{tot}$ increases, the components $f_\mathrm{2\pm0}$ become more prominent and this is seen in $A_{2\pm0}$ too (see Fig.~\ref{fig:TM_fitted_parameters_1}). This is understandable, because the radial oscillation mode is more strongly excited for high-mass models. For low-mass configurations, the coupling to the radial oscillation is significantly suppressed (see Fig.~\ref{fig:lapses_all}), and consequently the amplitudes of the couplings $f_{2-0}$ and $f_{2+0}$ should be small, which is only the case for the $f_{2+0}$ component. We suspect that the relatively high amplitude $A_{2-0}$ for small $M_\mathrm{tot}$ is an artifact of the fit and is compensated by a very small decay timescale. The weakness of the radial oscillation implies that the $\tau_{2\pm0}$ are not very meaningful measures for low-mass systems. For higher total binary masses one can see a mild increase of $\tau_{2\pm0}$, which is in line with the behavior in Fig.~\ref{fig:lapses_all}. The timescales of the spiral component exhibit a mild decrease, corresponding to a faster dissipation of the tidal bulges. The amplitude of the $f_\mathrm{spiral}$ component similarly increases with $M_\mathrm{tot}$. 

Furthermore, we consider the product $(A\cdot \tau)_\mathrm{i}$ as a quantitative measure for the strength of a secondary feature. Figure~\ref{fig:TM_fitted_parameters_3} shows the products $(A\cdot \tau)_\mathrm{i}$ for each frequency component. We use Eqs.~\eqref{Tspiral of mtot}-\eqref{A2+0 of mtot} to derive analytic expressions displayed by dashed curves. The products $(A\cdot \tau)_\mathrm{2\pm0}$ increase systematically with $M_\mathrm{tot}$ as expected and closely follow the analytic expressions. The product $(A\cdot \tau)_\mathrm{spiral}$ is roughly constant \footnote{$(A\cdot \tau)_\mathrm{spiral}$ shows a large scatter from the derived analytic expression (using Eq.~\eqref{Aspiral of mtot},~\eqref{Tspiral of mtot}), however, we find that a fourth order polynomial fit describes well the trend.}.

The strength of the secondary components quantified as in Fig.~\ref{fig:TM_fitted_parameters_3} resembles the behavior which was anticipated in \cite{AB2015},  and reproduces different types of post-merger GW emission: for low-mass binaries the $f_\mathrm{spiral}$ component is dominant (Type III in the notation of~\cite{AB2015}), for intermediate masses the strength of $f_\mathrm{spiral}$ and $f_{2-0}$ is roughly comparable (Type II), and for models with very high $M_\mathrm{tot}$ the couplings with the radial oscillation are dominant over $f_\mathrm{spiral}$ (Type I). The products $(A\cdot \tau)_\mathrm{i}$ may thus serve as a quantitative measure to classify different types of post-merger dynamics and GW emission including the morphology of the spectrum.

We note that the method we use for the derivation of $A_\mathrm{i}$, $\tau_\mathrm{i}$ introduces a bias whenever $f_{2-0}$ and $f_{\mathrm{spiral}}$ are close (see \ref{Amplitudes and timescales}). The latter is possibly one of the reasons for the scattering of $A_\mathrm{i}$, $\tau_\mathrm{i}$ from the analytic fits. 

\subsubsection{$\mathcal{N}$}

Figure \ref{fig:TM_fitted_parameters_4} shows the normalization factor $\mathcal{N}$ as a function of total mass $M_\mathrm{tot}$ for fits with the complete analytic model (Ac). We find a linear dependence on $M_\mathrm{tot}$ modelled by 
\begin{eqnarray}
\label{N of Mtot}\mathcal{N} &=& -0.485\cdot M_\mathrm{tot}+2.025
\end{eqnarray}
$\mathcal{N}$ becomes less important (close to 1) for low-mass configurations and more significant (close to 0.50) for high-mass configurations. 

One reason for this trend may be that for estimating $A_\mathrm{i}$, $\tau_\mathrm{i}$ we treat each component separately. In low-mass configurations the components $f_{\mathrm{spiral}}$  and $f_{2-0}$ are well separated and therefore the parameters $A_\mathrm{i}$, $\tau_\mathrm{i}$ are accurately derived. However, this is not the case for high-mass configurations, where the peaks overlap, and thus the parameters may be overestimated and the correction becomes necessary. 

Another reason may be the fact that the $f_{2\pm0}$ components are significantly weaker than the $f_\mathrm{spiral}$ components for low-mass systems (see e.g. the products $(A\cdot \tau)_\mathrm{i}$). Hence, their contribution to the total signal is minor and a single secondary feature does not require significant corrections by the normalization factor.

\subsection{Initial phases ${\phi_\mathrm{peak}}$, $\phi_\mathrm{spiral}$, $\phi_\mathrm{2\pm0}$ }

\begin{figure}[h]
	\includegraphics[scale=0.33]{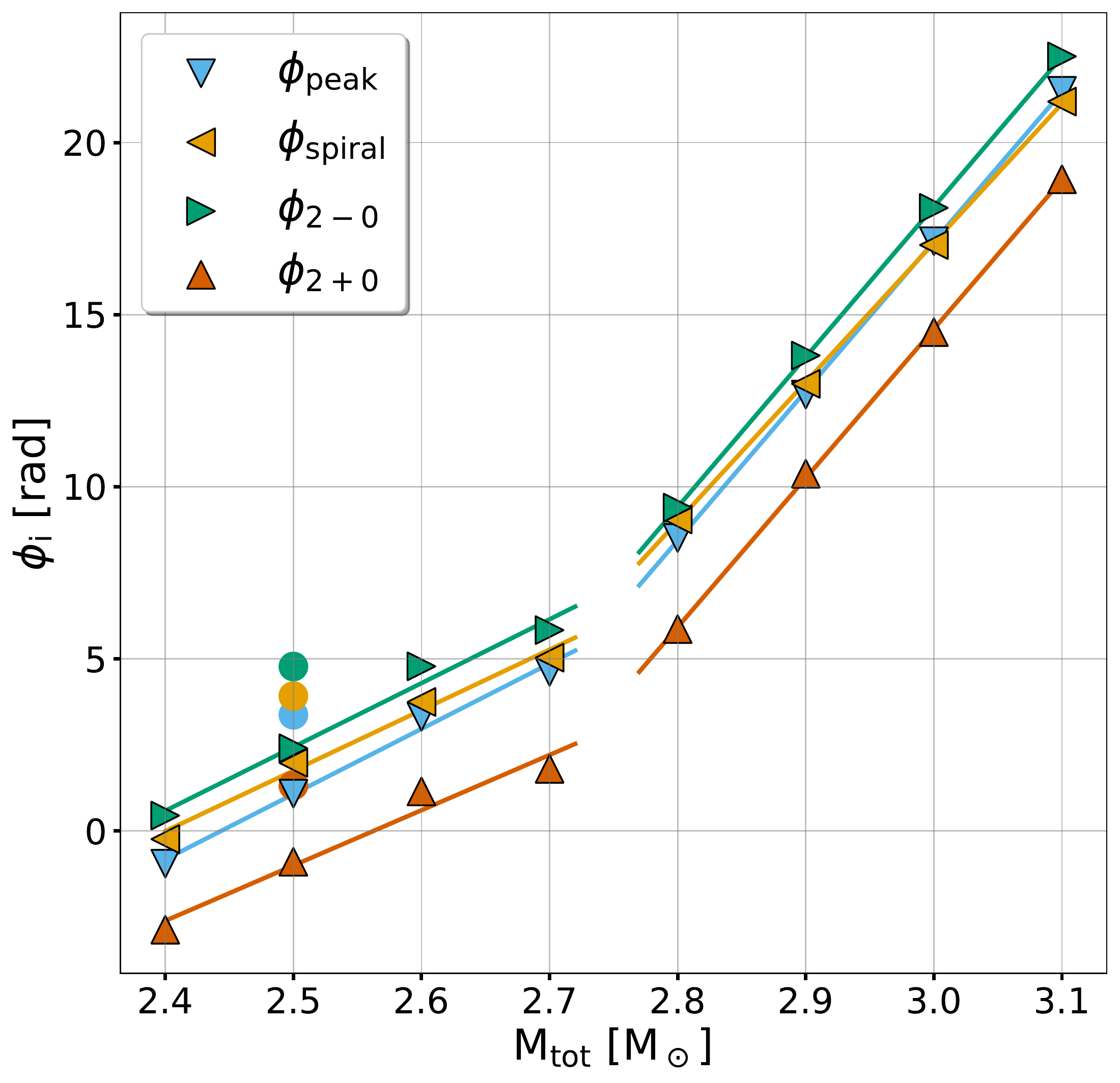} 
	\caption{\label{fig:sequence_phi_of_Mtot.pdf} Initial phases $\phi_\mathrm{i}$ (for i=peak, spiral, $2\pm0$) for the analytic model which employs the $f_{\mathrm{peak}}^\mathrm{analytic}(t)$ description as a function of total binary mass. Colored curves correspond to piecewise linear fits. The colored circles indicate the respective parameters for the HR simulation (see Appendix~\ref{Resolution study}).}
\end{figure}

In this subsection we discuss the properties of the initial phases $\phi_\mathrm{i}$ (for $\mathrm{i=peak,spiral,2\pm0}$) for all the models in the sequence of simulations. For the analysis we add multiples of $2\pi$ to the initial phases $\phi_\mathrm{i}$ such that $\phi_\mathrm{i}(M_\mathrm{tot})$ becomes an increasing function (see Fig.~\ref{fig:sequence_phi_of_Mtot.pdf}).

\begin{figure}[h]
	\includegraphics[scale=0.33]{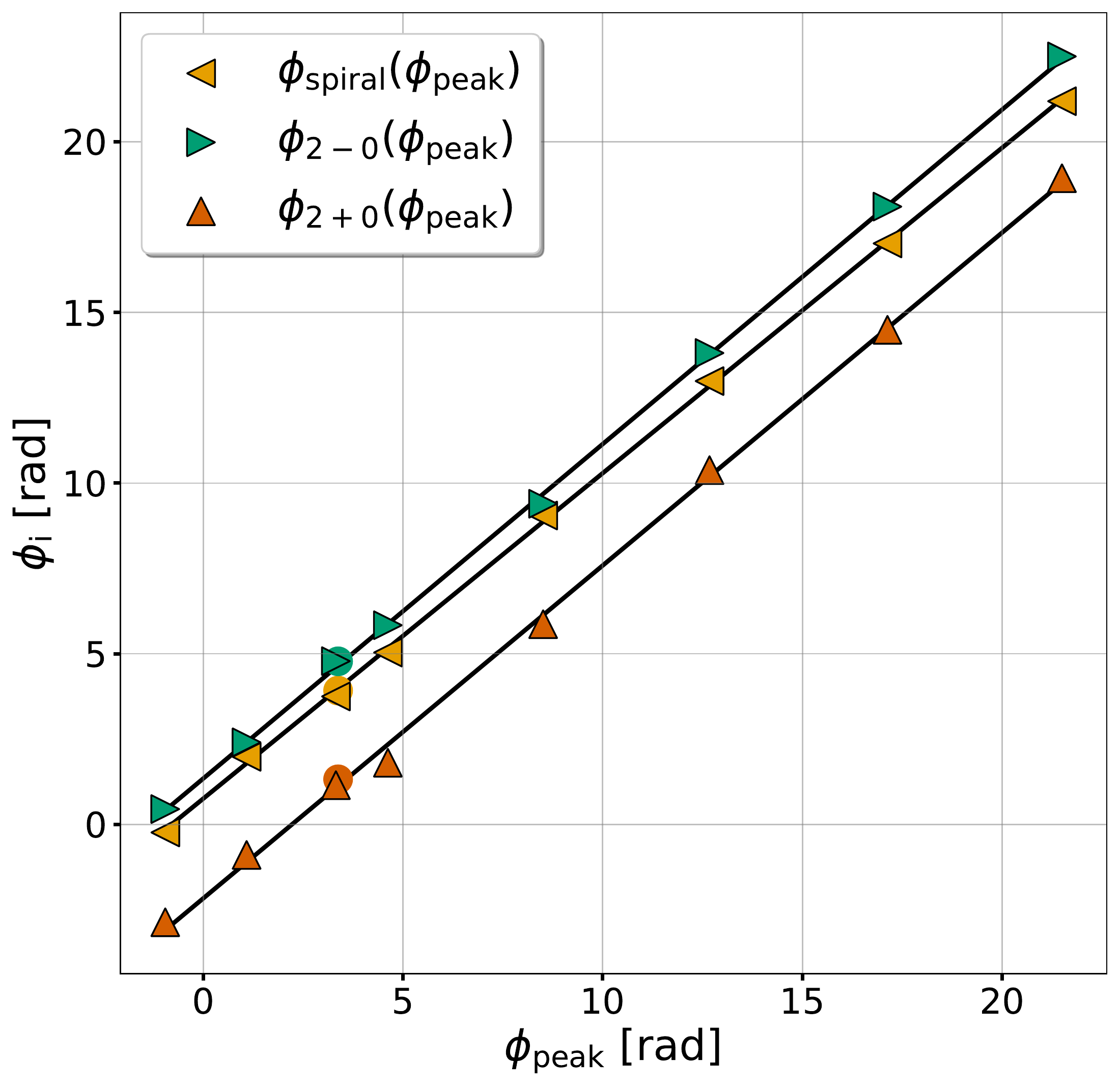} 
	\caption{\label{fig:sequence_phi_of_phipeak.pdf} Initial phases $\phi_\mathrm{i}$ (for i=spiral, $2\pm0$) with respect to $\phi_{\mathrm{peak}}$ for the analytic model which employs the$f_{\mathrm{peak}}^\mathrm{{analytic}}(t)$ description. Black curves correspond to linear fits. The colored circles indicate the respective parameters for the HR simulation (see Appendix~\ref{Resolution study}).}
\end{figure}
We find a tight correlation between $\phi_\mathrm{i}$ and the total mass $M_\mathrm{tot}$. We model this dependence with a 2-segment piecewise function consisting of two linear fits which intersect at total mass of $M_\mathrm{tot}=2.7~M_\odot$ (see Appendix~\ref{A:Analytic model}). 

These remarkably tight correlations imply that the properties of the gravitational phase $\phi(t)$ (see Eq.~\eqref{gravitaional phase}) in the early post-merger phase depend systematically on the total mass $M_\mathrm{tot}$.

Furthermore, we find tight correlations between the initial phases $\phi_\mathrm{spiral}$, $\phi_\mathrm{2\pm0}$ and $\phi_\mathrm{peak}$ as shown in Fig.~\ref{fig:sequence_phi_of_phipeak.pdf}.  We model these correlations with linear fits given by
\begin{eqnarray}
\label{phispiral-phipeak}\phi_\mathrm{spiral} &=& +0.953\cdot \phi_\mathrm{peak}+0.756\\
\label{phi2-0-phipeak}\phi_{2-0} &=& +0.980\cdot \phi_\mathrm{peak}+1.345\\
\label{phi2+0-phipeak}\phi_{2+0} &=& +0.975\cdot \phi_\mathrm{peak}-2.166 .
\end{eqnarray}

The slope parameters in Eq.~\eqref{phispiral-phipeak}-\eqref{phi2+0-phipeak} are approximately equal and differ at most by 3\%. We note that the slopes are also close to 1, which would imply a constant difference in phase between the $f_\mathrm{peak}$ component and the secondary components. It may well be that these relations and the ones shown in Fig.~\ref{fig:sequence_phi_of_Mtot.pdf} are in reality even tighter and the small but finite scatter results from finite resolution in the simulations or the fitting procedure. Such tight relations can be employed to reduce the complexity of the analytic fit by reducing the parameter space.

We find that using the $\phi_\mathrm{peak}(M_\mathrm{tot})$ and $\phi_\mathrm{spiral}(\phi_\mathrm{peak})$, $\phi_\mathrm{2\pm0}(\phi_\mathrm{peak})$
relations (see Eq.~\eqref{phispiral-phipeak} to Eq.~\eqref{phi2+0-phipeak})  one may reduce the number of the analytic model's parameters (and thus the complexity of the fitting procedure) and obtain good fits to the data. We test this by substituting the initial phases $\phi_\mathrm{i}$ with the predictions made by Eq.~\eqref{phipeak-Mtot}, Eq.~\eqref{phispiral-phipeak}-\eqref{phi2+0-phipeak} and find that the $FFs$ only differ by a few percent (0.5-3\%) compared to fits to the analytic model. When we perform a phase alignment in the waveforms the $FFs$ differ by at most by $\approx$1\% .

We overplot the initial phases  $\phi_\mathrm{i}$ (for i=peak, spiral, $2\pm0$) for the HR simulation in Figs.~\ref{fig:sequence_phi_of_Mtot.pdf} and~\ref{fig:sequence_phi_of_phipeak.pdf} (see Appendix~\ref{Resolution study}). These appear to be slightly larger than the ones from the mass sequence simulations, however, their relative difference is similar to the mass sequence simulations as corroborated by Fig.~\ref{fig:sequence_phi_of_phipeak.pdf}.

Furthermore, we also find that these tight correlations between the initial phases (Eq~\eqref{phispiral-phipeak}-\eqref{phi2+0-phipeak}) are unaffected by residual eccentricities in the ID (see Fig.~\ref{fig:sequence_ecc_phi_of_phipeak.pdf} and Appendix~\ref{Effect of the eccentricity in the ID}).

\subsection{Purely analytic model}
We consider a \textit{purely analytic model} $\mathcal{P}(M_\mathrm{tot},t)$ which uses the analytic functions Eq.~\eqref{equations-fpeak(t)-fits_1}-\eqref{N of Mtot},\eqref{phipeak-Mtot}-\eqref{phi2+0-Mtot} and thus depends only on $M_\mathrm{tot}$. We evaluate its performance by computing the respective $FFs$. Table~\ref{Fitting Factors purely analytic model} shows the $FFs$ for the analytic model $\mathcal{P}(M_\mathrm{tot},t)$ compared to the Ac analytic fits. The $FFs$ drop significantly as expected, however, the majority of the fits still result in $FFs\gtrsim 0.80$. The $FFs$ can be further improved by considering an analytic model where $\phi_\mathrm{peak}$ is treated as a free parameter, denoted by $\mathcal{P}(M_\mathrm{tot},t;\phi_\mathrm{peak})$. In this case, almost all configurations lead $FFs\gtrsim 0.85$ (see Tab.~\ref{Fitting Factors purely analytic model}).

These considerations show that it may be possible to determine the different analytic functions Eq.~\eqref{equations-fpeak(t)-fits_1}-\eqref{N of Mtot},\eqref{phipeak-Mtot}-\eqref{phi2+0-Mtot} (or only piecewise linear segments of these functions) by several simulations and anticipated observations and then use those functions to interpolate the model in $M_\mathrm{tot}$.

\begin{table}[h]
	\begin{ruledtabular}
		\begin{tabular}{c|ccc}
		\multicolumn{4}{l}{{Fitting Factors ($FFs$) }}\vspace{0.2cm}  \\ \hline 
			\textrm{$ M_\mathrm{tot}[M_\odot]$}&
			\textrm{Ac}&
			\textrm{$\mathcal{P}(M_\mathrm{tot},t)$}&
            \textrm{$\mathcal{P}(M_\mathrm{tot},t;\phi_\mathrm{peak})$}\\
			\colrule
			
2.4& 0.979& 0.653& 0.801\\
2.5& 0.956& 0.795& 0.847\\
2.6& 0.956& 0.912& 0.913\\
2.7& 0.977& 0.878& 0.922\\
2.8& 0.976& 0.878& 0.899\\
2.9& 0.962& 0.848& 0.905\\
3.0& 0.955& 0.595& 0.864\\
3.1& 0.907& 0.887& 0.898\\
	
		\end{tabular}
	\end{ruledtabular}
	\caption{Fitting factors $FFs$ for the analytic model Ac fits, the purely analytic model $\mathcal{P}(M_\mathrm{tot},t)$, and the analytic model with one free parameter $\mathcal{P}(M_\mathrm{tot},t;\phi_\mathrm{peak})$.}
	\label{Fitting Factors purely analytic model}
\end{table}

\section{Models close to prompt collapse} \label{Clode to prompt collapse}
In this section we analyze the spectral properties of configurations with a total mass $M_\mathrm{tot}$ close to threshold mass for prompt collapse $M_\mathrm{thres}$. Fitting factors decrease for these high-mass models, which possibly points to an incompleteness of our analytic model. We separately consider two modifications to the analytic model in order to increase the $FF$. First, we include a dynamical evolution of $f_{2\pm0}(t)$. Second, we incorporate the $f_\mathrm{spiral-0}$ component, i.e. an additional coupling between $f_\mathrm{spiral}$ and $f_0$ (see Subsect.~\ref{SubSec. spiral-0 coupling}). Table~\ref{Table with definitions for modified models} summarizes information for the extended analytic models.

\begin{table}[h]
	\begin{ruledtabular}
		\begin{tabular}{p{3.cm}|c| p{3.cm}}
			\textrm{Model description}& \textrm{Name} &
            \textrm{Components}\\
			\colrule
\texttt{Extended analytic model 1} & M1 & $f_\mathrm{peak}^\mathrm{analytic}(t)$, $f_\mathrm{spiral}$, $f_{2-0}(t)$, $f_{2+0}(t)$ \\\hline
\texttt{Extended analytic model 2} & M2 & $f_\mathrm{peak}^\mathrm{analytic}(t)$, $f_\mathrm{spiral}$, $f_{2-0}$, $f_{2+0}$, $f_\mathrm{spiral-0}$ \\\hline
		\end{tabular}
	\end{ruledtabular}
	\caption{Definitions for the two extended analytic models. When the time argument is explicitly written, a time-dependent description is employed for that particular component. }
	\label{Table with definitions for modified models} 
\end{table}

\subsection{Extended analytic models and GW fits} \label{Extended analytic model}

To assess the importance of the time evolution of $f_{2\pm0}(t)$, we extracted $f_{2\pm0}(t)$ from spectrograms (see \ref{Sec. mass sequence}) and inserted the numerically extracted values into the analytic model (see Subsect.~\ref{semi-analytic model}). We do not further discuss a parametrization of $f_{2\pm0}(t)$ because we find below that even the complete numerical description of $f_{2\pm0}(t)$ yields only a minor improvement. 

\begin{figure}[h]
	\includegraphics[scale=0.33]{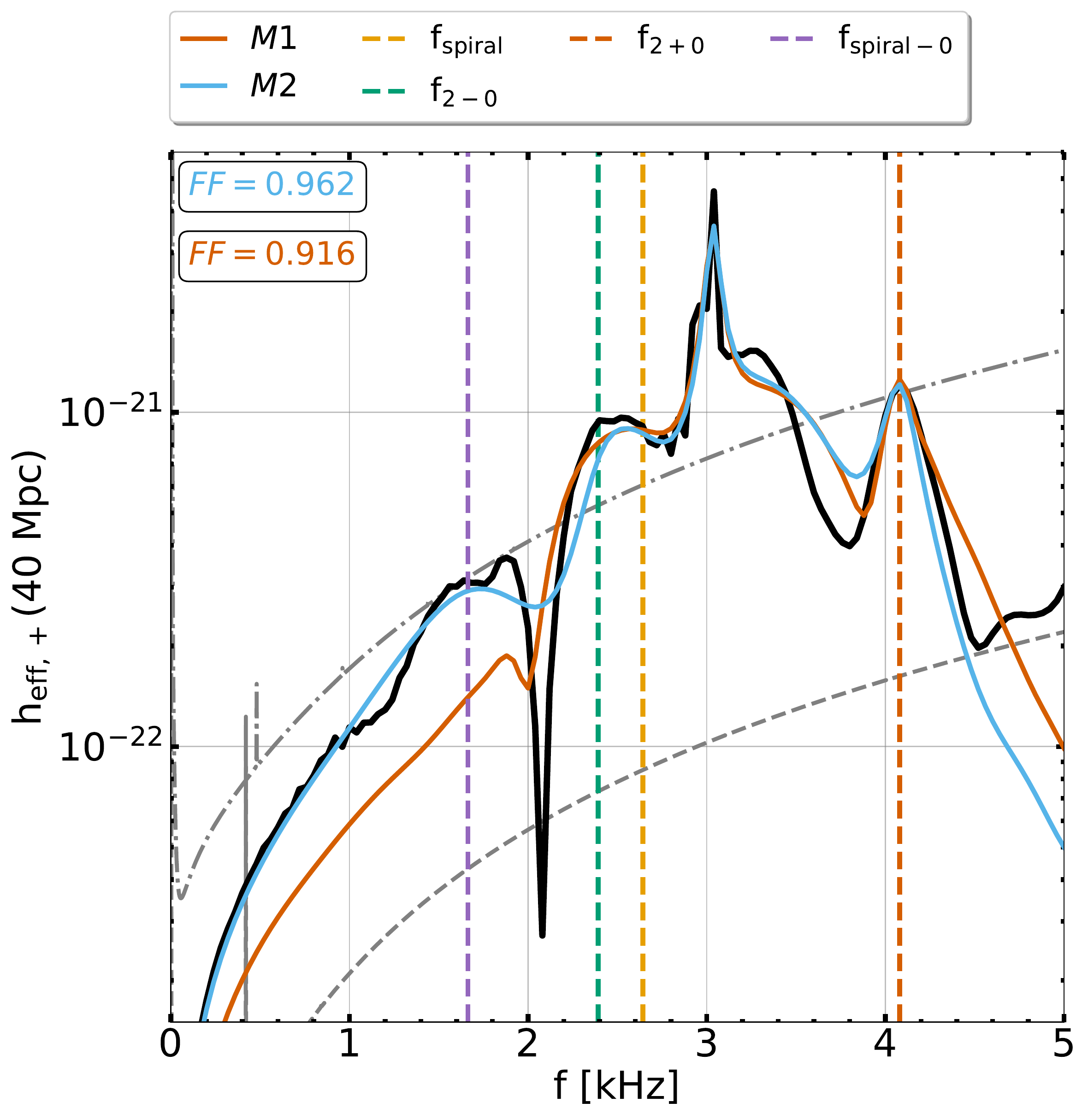} 
	\caption{\label{fig:dumpFFT_dynamictoymodel_components_mpa1-m1.55.pdf} Effective GW spectra $h_\mathrm{eff,+}(f)$ for simulation, and analytic models for $\mathrm{M_\mathrm{tot}=3.1M_\odot}$ model. Black line corresponds to the simulation. Colored curves illustrate the analytic model fits as described by the labels. Colored boxes show the corresponding fitting factors $FFs$. Dashed vertical lines indicate secondary frequencies.}
\end{figure}

Figure \ref{fig:dumpFFT_dynamictoymodel_components_mpa1-m1.55.pdf} shows the fits to the simulation with total binary mass $M_\mathrm{tot}=3.1~M_\odot$ for the extended analytic models. The introduction of the time-evolving components $f_{2\pm 0}(t)$ leads to a mild increase of the fitting factor: $FF_\mathrm{new,1}=0.916$ compared to the original of $FF_\mathrm{old}=0.907$ (see \ref{GW fits}). This increase in $FF$ slightly improves the reduction in detection rates from $25.39\%$ to $23.14\%$. The model with the dynamical $f_{2\pm 0}(t)$ qualitatively reproduces a small peak at approximately 1.9~kHz (orange curve in Fig.~\ref{fig:dumpFFT_dynamictoymodel_components_mpa1-m1.55.pdf}), but still does not yield a good description of the simulation below 2~kHz.

The blue curve in Fig.~\ref{fig:dumpFFT_dynamictoymodel_components_mpa1-m1.55.pdf} includes the $f_\mathrm{spiral-0}$ component, whose frequency we assume to be constant and we also adopt constant values for $f_{2\pm 0}$ as in the original model. We do not incorporate an $f_\mathrm{spiral+0}$ component in our modified analytic model since we do not observe a distinct peak in the GW spectrum at the respective frequency. The parameters $A_\mathrm{spiral-0}$, $\tau_\mathrm{spiral-0}$ are derived from the spectrograms as described in \ref{ATs for secondary components}.

The inclusion of the $f_{\mathrm{spiral-0}}$ component substantially increases the fitting factor $FF_\mathrm{new,2}=0.962$. This leads to a significant improvement regarding the reduction in the detection rates of $11\%$. The importance of the $f_{\mathrm{spiral-0}}$ component is also apparent in the GW spectrum (compare orange and blue curve below 2~kHz). As previously mentioned, the strength of $f_\mathrm{spiral-0}$ relative to frequency dependent sensitivity curve is similar to $f_{2+0}$ (for this mass configuration) and thus has a large impact on $FF$.

We remark that the first modification (time-evolving $f_{2\pm 0 }(t)$) only slightly improves the analytic fits but increases the complexity of the model  since a parametrization of $f_{2\pm 0 }(t)$ would require a number of additional parameters. The second modification (inclusion of $f_\mathrm{spiral-0}$) improves significantly the analytic fits ($FFs$) and only introduces a minimum of new parameters ($A_\mathrm{spiral-0}$, $\tau_\mathrm{spiral-0}$, $\phi_\mathrm{spiral-0}$, while the frequency is already given by the other components). 

\subsection{Additional spectral features}

\begin{figure}[h]
	\includegraphics[scale=0.33]{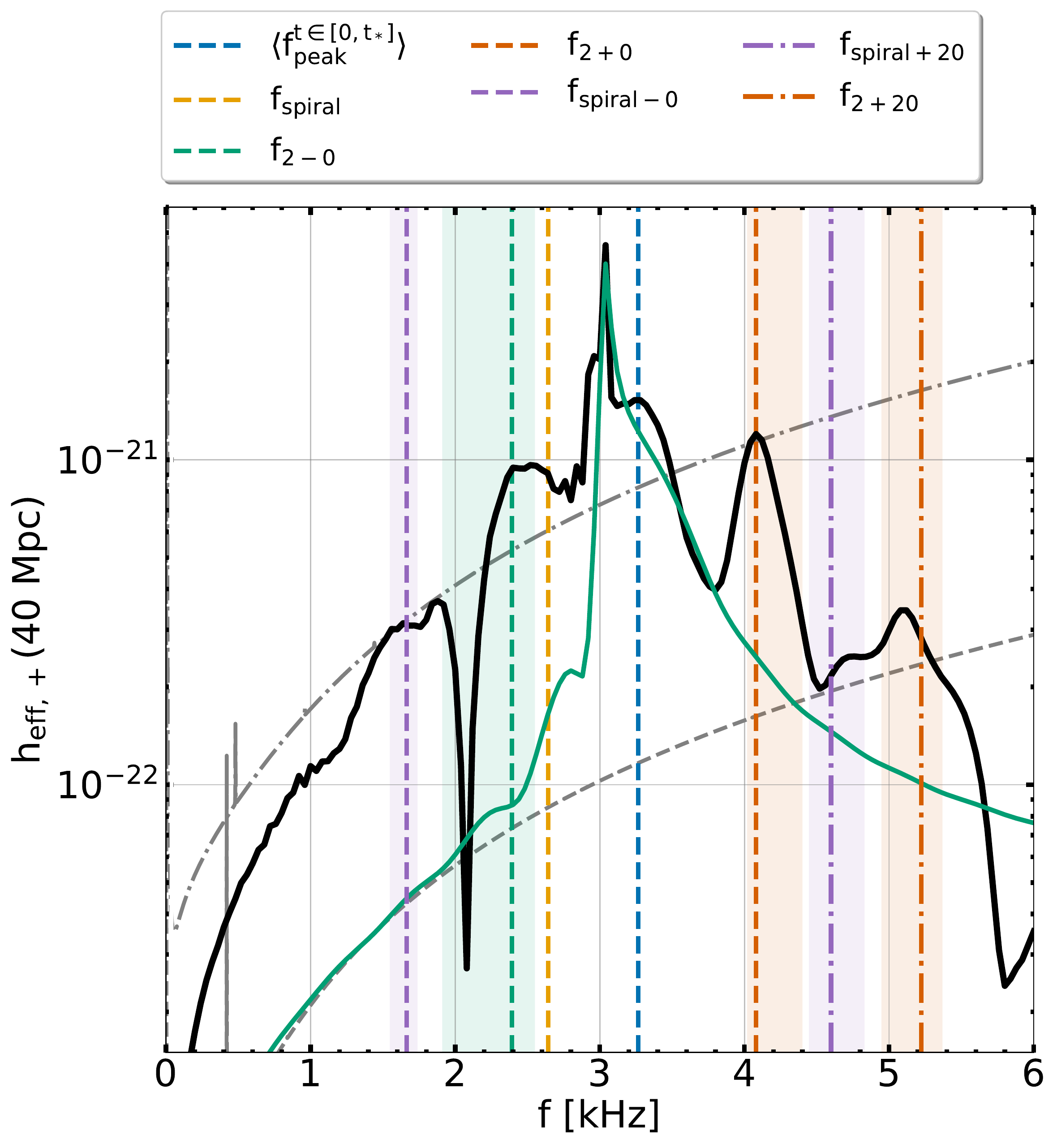} 
	\caption{\label{fig:dumpFFT_allfreq_components_mpa1-m1.55.pdf} Effective GW spectra $h_\mathrm{eff,+}(f)$ for simulation, and simplified analytic model for $\mathrm{M_\mathrm{tot}=3.1M_\odot}$ model. Black line corresponds to the simulation. Green line displays the simplified 1-component analytic model (A1). Dashed, dash-dotted vertical lines indicate secondary frequencies $\langle f_{\mathrm{peak}}^{t \in [0,t_*]}\rangle$, $f_{\mathrm{spiral}}$, $f_{\mathrm{2\pm0}}$, $f_{\mathrm{spiral-0}}$, $f_{\mathrm{2}+20}$, $f_{\mathrm{spiral+20}}$. Shaded areas visualize their respective spread due to the time evolving frequencies.}
\end{figure}

We finally note that the different components and their couplings provide explanations for basically every feature in the GW spectrum up to about 6~kHz if one additionally considers higher order combination tones. This is shown in Fig.~\ref{fig:dumpFFT_allfreq_components_mpa1-m1.55.pdf}, where we in addition draw the fit for the simplified analytic model (green curve). We estimate those additional frequencies employing the dominant frequency at early times $\langle f_{\mathrm{peak}}^{t\in[0, t_*] }\rangle$ and using expressions $f_{2+20}\approx \langle f_{\mathrm{peak}}^{t\in[0, t_*] }\rangle + 2\cdot f_0$ and $f_\mathrm{spiral+20}\approx f_\mathrm{spiral} + 2\cdot f_0$. We derive the respective frequency ranges inserting the time evolution of $f_0(t)$ and $f_\mathrm{peak}^\mathrm{analytic}(t)$. The estimated frequency ranges for $f_{2+20}$ and $f_\mathrm{spiral+20}$ match relatively well with peaks in the GW spectrum. We note that the frequencies $f_{2+20}$ and $f_\mathrm{spiral+20}$ are also expected to follow empirical relations, which can be exploited in more sophisticated analytic models. 

Finally, we remark that the frequency component $f_{2-20}$ is most likely less important than $f_\mathrm{spiral-0}$. Our estimation using the expression $f_{2-20}\approx \langle f_{\mathrm{peak}}^{t\in[0, t_*] }\rangle - 2\cdot f_0$ leads to $f_{2-20}=1.309$~kHz, which is significantly lower than the peak in the GW spectrum ($f_\mathrm{spiral-0}=1.664$~kHz). 

Another  relevant feature is the re-excitation of the quadrupolar mode $f_{\rm peak}$, which occurs roughly 10~ms after merging in high-mass models, see Appendix \ref{Appendix-sequence}, possibly due to the excitation of a low-$|T/W|$ rotational instability (see, e.g. \citep{Passamonti_Andersson_2020,Xie_etal_2020,PhysRevD.101.064052} and references therein). This feature is not captured by the adopted single exponential decay of the amplitude.

\section{Conclusions}

In this work, we investigate the spectral properties of the GW emission for a mass sequence of binary neutron star mergers and introduce an analytic model for the post-merger GW emission, which employs exponentially decaying sinusoids. We discuss the features of the GW spectra and their dependence on total binary mass. We also study the time evolution of certain frequency components using spectrograms. Notably, we find that the $f_\mathrm{peak}$ mode exhibits a time evolution which can be split in two phases: a) a rapid initial drift b) an approximately constant frequency at late times. We identify a new mechanism which explains a low frequency peak (in the GW spectra) occurring in high mass configurations. It is caused by a coupling between the antipodal bulges ($f_\mathrm{spiral}$) and the quasi-radial mode ($f_0$). Our analytic model incorporates a time-dependent $f_\mathrm{peak}(t)$ and three secondary components ($f_\mathrm{spiral}$, $f_{2\pm0}$). We evaluate the model's performance using the noise-weighted fitting factor $FF$ and find good agreement with the simulations with $FF>0.95$ for the majority of the models. We explore the dependencies of the analytic model's parameters, and correlations among them, on the total mass $M_\mathrm{tot}$. Finally, we include potential modifications to the analytic model for the configurations with $M_\mathrm{tot}$ close to $M_\mathrm{thres}$. 

All models of the mass sequence exhibit a time-dependent $f_\mathrm{peak}(t)$. In their GW spectra, $f_\mathrm{peak}$ is one-sided towards to high frequencies. We model this evolution with a 2-segment piecewise function Eq.~\eqref{fpeak_piecewise} and quantify the drift using spectrograms of the simulation signals. The analytic model reproduces remarkably well the one-sided $f_\mathrm{peak}$ structure and thus it confirms that our choice is sufficient for the description of such time-dependent $f_\mathrm{peak}(t)$. We note that $\langle f_\mathrm{peak}^{t \in [0, t_*}\rangle$ is a good measure of $f_\mathrm{peak}$ in the early post-merger phase. We find that the parameters characterizing the time evolution of $f_\mathrm{peak}(t)$ ($\zeta_\mathrm{drift}$, $t_*$, $f_\mathrm{peak,0}$) show a dependence on total mass $M_\mathrm{tot}$. The frequency evolution becomes faster for high-mass configurations. We note that such trends may provide information on the proximity to prompt collapse. 

We confirm that the post-merger GW spectra follow the classification scheme introduced in \cite{AB2015}. As the total mass $M_\mathrm{tot}$ increases and the quasi-radial mode is stronger excited, the secondary components $f_{2\pm0}$ become more pronounced and there is an overlap between $f_{2-0}$ and $f_\mathrm{spiral}$. In low mass configurations a low frequency modulation $f_\mathrm{peak}-f_\mathrm{spiral}$ is seen in $\alpha_\mathrm{min}(t)$ with comparable strength to that of the quasi-radial mode.

Using the analytic model's parameters, amplitudes and decay timescales $A_\mathrm{i}$, $\tau_\mathrm{i}$, we find that the products $(A\cdot \tau)_i$ for the mass sequence can be used to quantitatively define the strength of secondary components (for $\mathrm{i=spiral,2\pm0}$), and allow a {\it quantitative classification } of the different types of spectra as in~\cite{AB2015}.

Furthermore, we identify a new mechanism generating a potentially relevant secondary GW feature: in high mass configurations the coupling between $f_\mathrm{spiral}$ and $f_0$ leads to frequencies at approximately $f_\mathrm{spiral\pm 0}\approx f_\mathrm{spiral}\pm f_0$. We note that relative to the sensitivity curve, $f_\mathrm{spiral-0}$ is comparable to $f_{2+0}$. 

We hypothesize couplings to the overtones of the quasi-radial mode, such as $f_{2+20}$ and $f_\mathrm{spiral+20}$, and identify frequency peaks in the GW spectrum near their vicinity such  that we explain nearly all visible frequency peaks.

The analytic model leads to fitting factors $FFs$ (assuming the sensitivity curve of ET) in the range of $[0.907-0.979]$ where the majority of the models has $FF>0.95$ and only the $M_\mathrm{tot}=3.1 M_\odot$ configuration (close to $M_\mathrm{thres}$) has $FF=0.907$. We find that for this configuration the inclusion of the $f_\mathrm{spiral-0}$ component significantly increases the fitting factor to $FF=0.962$. We further assessed our analytic model by considering simplified analytic models with fewer frequency components. We find that post-merger GW templates should incorporate at least two secondary components such as $f_\mathrm{spiral}$ and $f_\mathrm{2-0}$. The simplified model with one secondary component leads to a large reduction in detection rates. Using an additional simplified analytic model which incorporates a constant $f_\mathrm{peak}$ and three secondary components, we find that an accurate description of $f_\mathrm{peak}(t)$ is crucial for obtaining higher $FFs$, at least for the particular EoS studied here. 

We find systematic dependencies for all the analytic model's parameters with respect to the total binary mass $M_\mathrm{tot}$. $A_\mathrm{i}$ and $\tau_\mathrm{i}$ (for $\mathrm{i=peak,spiral,2\pm0}$) correlate with $M_\mathrm{tot}$ and follow trends which we model using second-order polynomials. Some of these trends, such as $A_{2-0}$ and $\tau_\mathrm{spiral}$, are not particularly tight but the dependence on mass is clear. We also find tight correlations between the initial phases $\phi_\mathrm{i}$ (for $\mathrm{i=peak,spiral,2\pm0}$) of each component with the total mass $M_\mathrm{tot}$, and between the secondary component initial phases $\phi_\mathrm{spiral}$, $\phi_\mathrm{2\pm0}$ and $\phi_\mathrm{peak}$ which may suggest a constant phase difference between the $f_\mathrm{peak}$ and secondary components.

One possible limitation of our analytic model is that it includes a relatively large number of parameters, which results from the complexity of the problem. However, all of them show a clear dependence on the total binary mass $M_\mathrm{tot}$, and can be modelled by analytic relations. These can potentially be used to decrease the parameter space in data analysis techniques. Furthermore, we note that the tight relations between the initial phases $\phi_\mathrm{peak}$ and $\phi_\mathrm{spiral}$, $\phi_\mathrm{2\pm0}$ should be further explored.

In our analysis, we applied a hierarchical procedure to initially estimate a subset of the analytic model's parameters and to then determined the remaining parameters using a curve fitting procedure (trust-region-reflective algorithm). Ideally, one would employ more sophisticated parameter estimation techniques, which provide distributions in the parameter space.

In future work, we plan to evaluate the model's performance for a large sample of EoSs and for un-equal mass mergers.

\section*{Acknowledgements}
T.S thanks Georgios Lioutas for the useful discussions. The work of T.S. is supported by the Klaus Tschira Foundation. T.S. is Fellow of the International Max Planck Research School for Astronomy and Cosmic Physics at the University of Heidelberg (IMPRS-HD) and acknowledges financial support from IMPRS-HD. T.S. acknowledges support by the High Performance and Cloud Computing Group at the Zentrum f{\"u}r Datenverarbeitung of the University of T{\"u}bingen, the state of Baden-W{\"u}rttemberg through bwHPC and the German Research Foundation (DFG) through grant no INST 37/935-1 FUGG. A.B. acknowledges support by the European Research Council (ERC) under the European Union’s Horizon 2020 research and innovation programme under grant agreement No.\ 759253, and support by Deutsche Forschungsgemeinschaft (DFG, German Research Foundation) - Project-ID 279384907 - SFB 1245 and DFG - Project-ID  138713538 - SFB  881  (“The  Milky  Way  System”, subproject  A10) and support by the State of Hesse within the Cluster Project ELEMENTS. N.S. gratefully acknowledges the Italian Istituto Nazionale di Fisica Nucleare (INFN), the French Centre National de la Recherche Scientifique (CNRS) and the Netherlands Organization for Scientific Research, for the construction and operation of the Virgo detector and the creation and support of the EGO consortium. Computing time was provided, in part, by allocations on the ARIS supercomputing facility of GRNET in Athens (SIMGRAV, SIMDIFF and BNSMERGE allocations) and by the “Aristoteles Cluster” at AUTh.

\appendix

\section{Numerical setup}\label{Numerical setup}
In this section we further discuss our numerical setup. We address the impact of residual eccentricity in the ID, of the initial orbital separation, of the numerical resolution, and of pi-symmetry on the spectral features.

\subsection{Effect of residual eccentricity in ID and of initial orbital separation}\label{Effect of the eccentricity in the ID}

In this section we address the effect of the residual eccentricity in the ID to the spectral features. To minimize the initial eccentricity in the ID we implement the prescription introduced in \cite{PhysRevD.90.064006} and adapt it to the field equations solved within LORENE \cite{Lorene:web}. 

We carry out two additional simulations with total binary mass $M_\mathrm{tot}=2.5~M_\odot$ (as in the reference simulation) and initial separation distance of $d=50~$km. The reduction of eccentricity is achieved with an iterative procedure (described in \cite{PhysRevD.90.064006}), which uses a few revolutions during the inspiral. It performs better at large initial separation, e.g. $d=50~$km, which is why we choose a larger $d$ for these tests. Otherwise the numerical setup is the same as for the mass sequence simulations (see Sect.~\ref{Sec:Methods}). Considering these two additional simulations we can assess the impact of eccentricity on the spectral features since this is the only parameter, which differs between those two calculations. We refer to the simulation with the quasi-circular ID and the simulation with reduced eccentricity as QC and RE, respectively.

In order to compute the separation distance between the two NSs, we assume that the center of mass of the star coincides with the location of the maximum rest-mass density $\rho_\mathrm{max}$. We use these coordinates ($x_\mathrm{max},y_\mathrm{max}$) in the orbital plane and define the separation distance by

\begin{eqnarray}
d(t) = 2\cdot \sqrt{x_\mathrm{max}^2 +y_\mathrm{max}^2}.
\end{eqnarray}
where the factor 2 reflects the pi-symmetry of the system. 
\begin{figure}[h!]
	\includegraphics[scale=0.33]{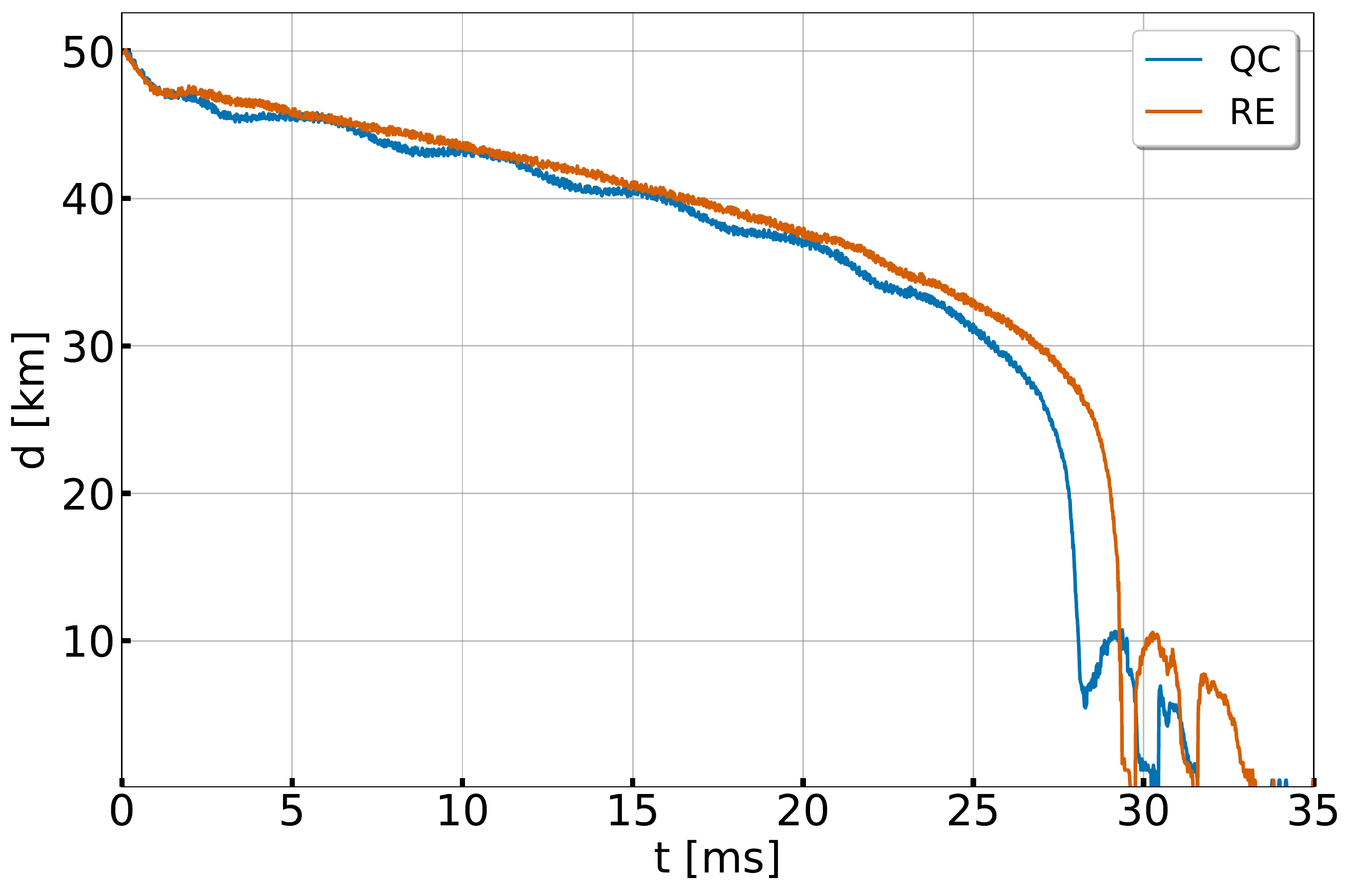} 
	\caption{\label{fig:dumpGW_sep_mpa1-m1_25.pdf}Time evolution of the coordinate separation distance $d(t)$ for the simulations QC (blue) and RE (orange).}
\end{figure}

Figure~\ref{fig:dumpGW_sep_mpa1-m1_25.pdf} shows the time evolution of the coordinate separation distance $d(t)$ for both simulations. We estimate the eccentricity using the method described in \cite{PhysRevD.90.064006}. The residual eccentricity in the QC simulation is $e\approx0.0088$ and the reduced residual eccentricity of the RE simulation is $e\approx0.00089$. The QC simulation exhibits small modulations in the separation distance $d(t)$, while in the RE simulation these oscillations disappear, as expected due to the reduced eccentricity. 

\begin{figure}[h!]
	\includegraphics[scale=0.33]{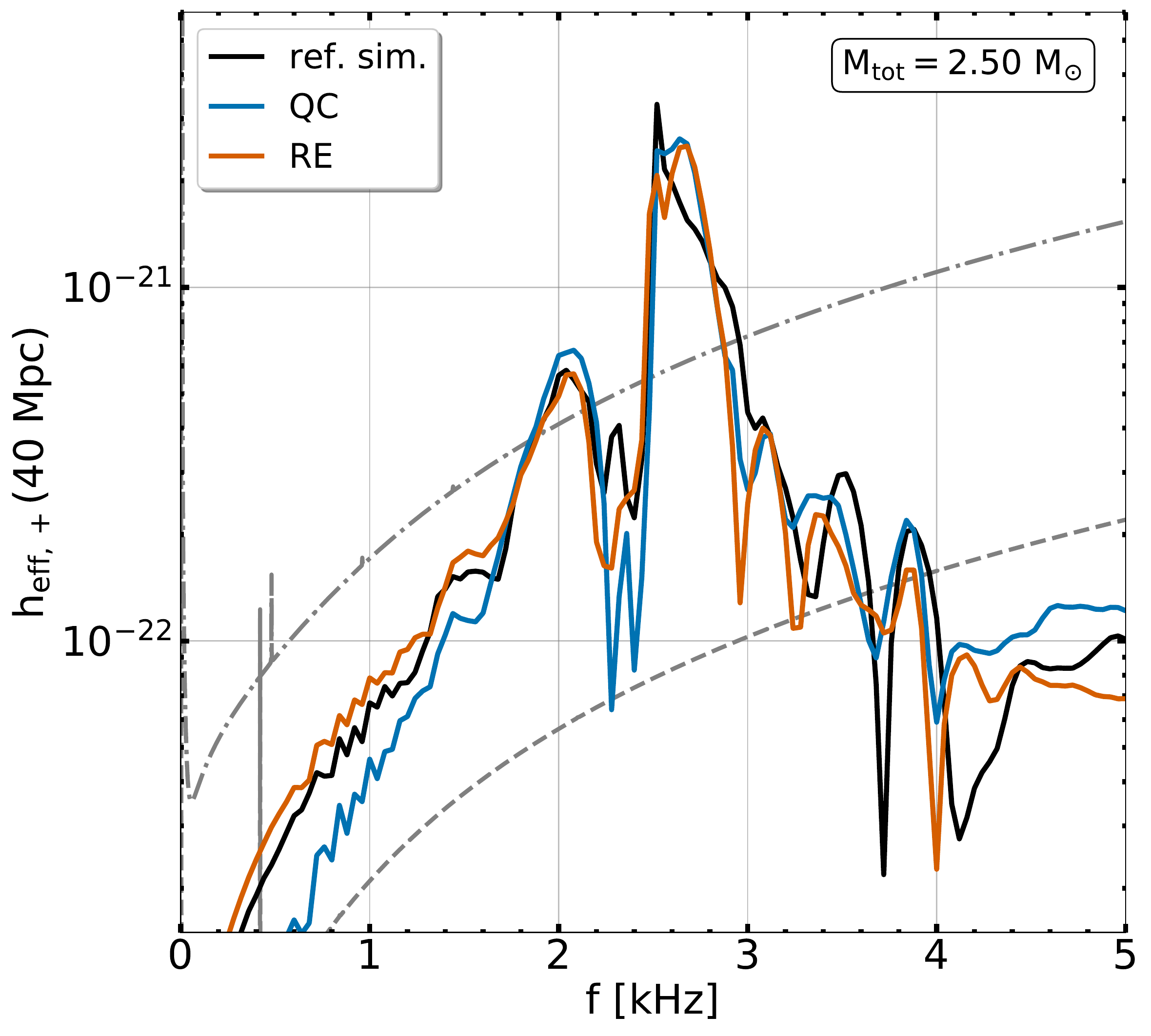} 
	\caption{\label{fig:dumpFFT_eccruns_comparison.pdf} Effective GW spectra  $h_\mathrm{eff,+}(f)$ for the reference simulation (black) and simulations QC (blue) and RE (orange). The dash dotted curves denote the design sensitivity Advanced LIGO \cite{LIGOScientific:2014pky} and of the Einstein Telescope \cite{EinsteinTelescope2010}, respectively.}
\end{figure}

\begin{figure}[h!]
	\includegraphics[scale=0.33]{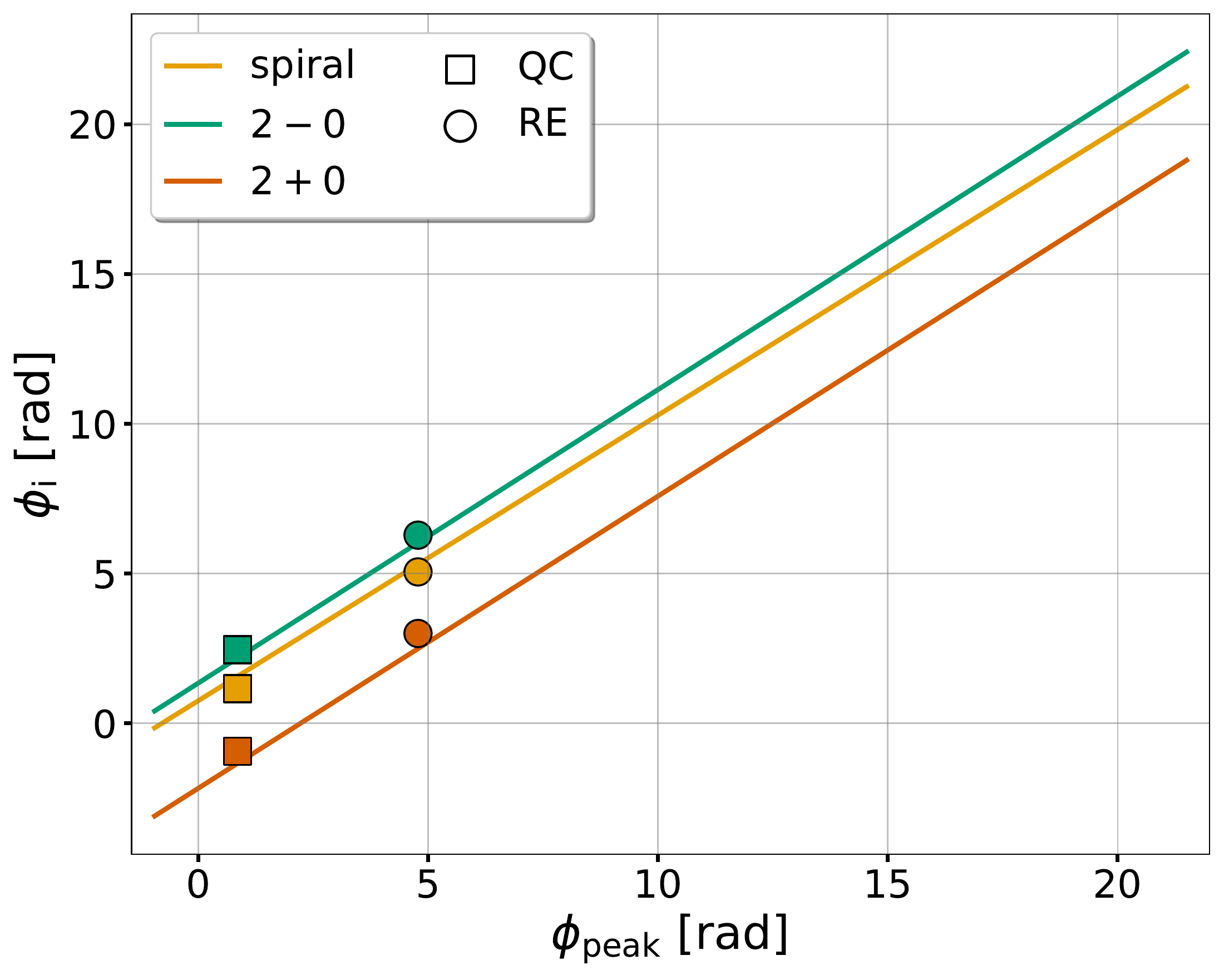} 
	\caption{\label{fig:sequence_ecc_phi_of_phipeak.pdf}Initial phases $\phi_\mathrm{i}$ (for i=spiral, $2\pm0$) with respect to $\phi_{\mathrm{peak}}$ for the analytic model (Ac) fits for the simulations QC (square) and RE (circle). Colored lines correspond to Eq.~\eqref{phispiral-phipeak}-\eqref{phi2+0-phipeak}. }
\end{figure}

Figure~\ref{fig:dumpFFT_eccruns_comparison.pdf} displays the GW spectra for the reference simulation and the simulations QC and RE. The spectra agree in the general features and their morphology, and the frequency peaks coincide. They exhibit an $f_\mathrm{peak}$ which is one-sided, and a dominant secondary peak, $f_\mathrm{spiral}$, with comparable strength. Therefore it is unlikely that the residual eccentricity in the ID affects the mechanisms for the frequency of evolution of $f_\mathrm{peak}$ or the formation of the antipodal bulges and thus $f_\mathrm{spiral}$. There are small differences in the amplitudes of the frequency peaks, which might be explained by differences in the impact velocities during the collision. However, these differences can also be seen between the reference simulation and the simulation QC, which only differ in initial separation distance. In particular, the morphology of the main peak is to some extent affected by the initial orbital separation.

We compute the analytic (Ac) and semi-analytic (Sc) model fits for the two simulations and find large fittings factors of $FF\gtrsim 0.970$ (see Tab.~\ref{Fitting factors for QC and RE}). We overplot the initial phases $\phi_\mathrm{i}$ (for $\mathrm{i=peak,spiral,2\pm0}$) together with Eq.~\eqref{phispiral-phipeak}-\eqref{phi2+0-phipeak} in Fig.~\ref{fig:sequence_ecc_phi_of_phipeak.pdf}. We find that the tight correlations between the phases still hold and the impact of residual eccentricities is negligible. We note that the initial phases $\phi_\mathrm{i}$ (for $i=\mathrm{peak,spiral,2\pm0}$) are shifted with additions or subtractions of multiples of $2\pi$.

\begin{table}[h]
	\begin{ruledtabular}
		\begin{tabular}{c|cc}
		\multicolumn{3}{l}{{Fitting Factors ($FFs$) }}\vspace{0.2cm}  \\ \hline 
			\textrm{Simulation}&
			\textrm{Sc}&
			\textrm{Ac}\\
			\colrule
		
QC& 0.969& 0.978\\
RE& 0.981& 0.979\\
		\end{tabular}
	\end{ruledtabular}
	\caption{Fitting factors $FFs$ for the analytic (Ac) and semi-analytic (Sc) model fits for the simulations QC and RE.}
	\label{Fitting factors for QC and RE}
\end{table}

\subsection{Resolution study}\label{Resolution study}
In this section we discuss the impact of resolution on the spectral properties. We consider an additional high resolution simulation with total binary mass $M_\mathrm{tot}=2.5~M_\odot$ (as in the reference model) and finest grid spacing of $dx=185~$m (keeping same number of refinement levels). Apart from the resolution, the numerical setup is identical to the one described in Sec.~\ref{Sec:Methods}. We refer to the high resolution simulation as HR. 

\begin{figure}[h!]
	\includegraphics[scale=0.33]{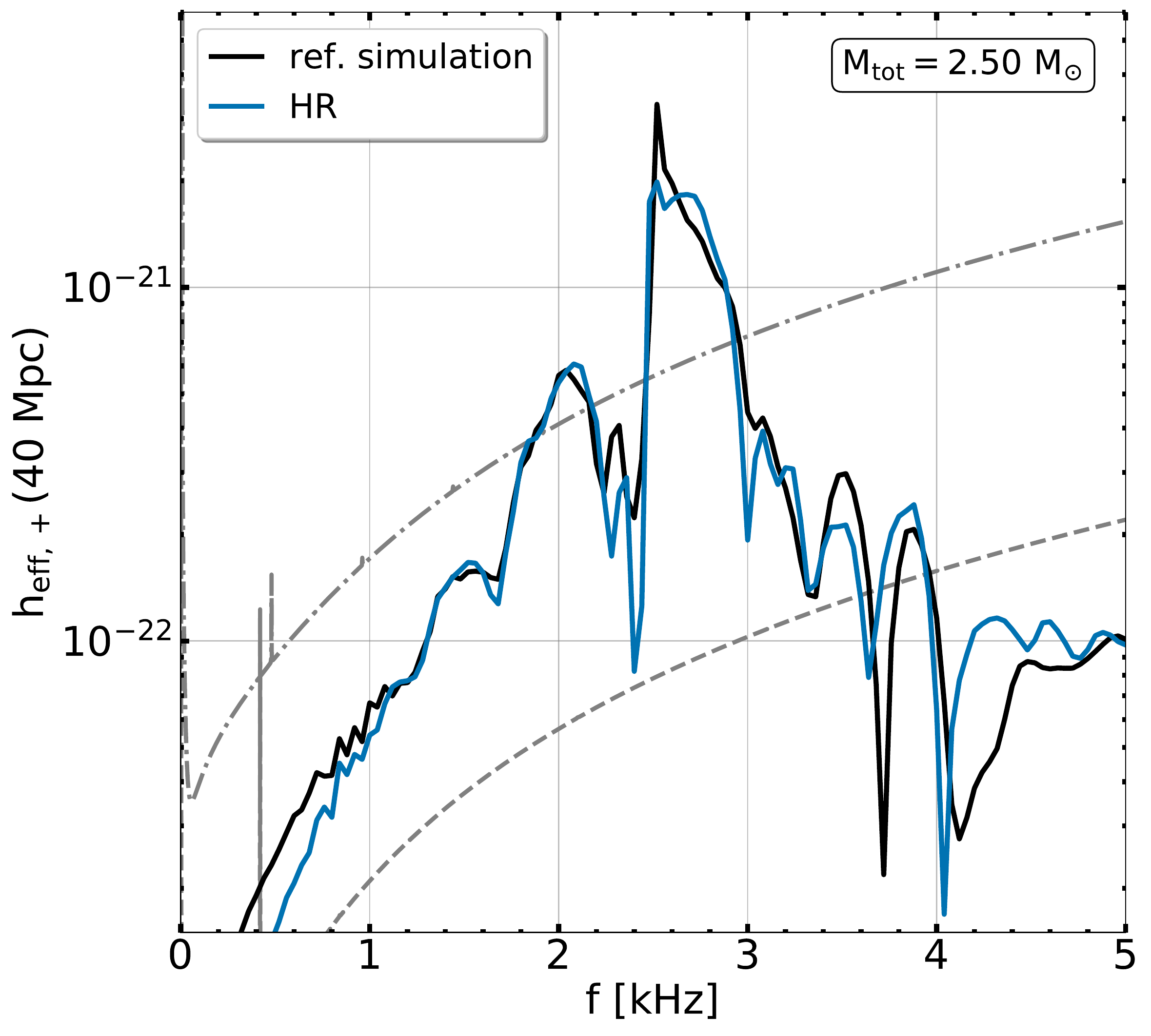} 
	\caption{\label{fig:dumpFFT_HR_mpa1-m1_25.pdf}Effective GW spectra  $h_\mathrm{eff,+}(f)$ for the reference simulation (black) and the high resolution simulation HR (blue). The dash dotted curves denote the design sensitivity Advanced LIGO \cite{LIGOScientific:2014pky} and of the Einstein Telescope \cite{EinsteinTelescope2010}, respectively.}
\end{figure}

\begin{figure}[h!]
	\includegraphics[scale=0.33]{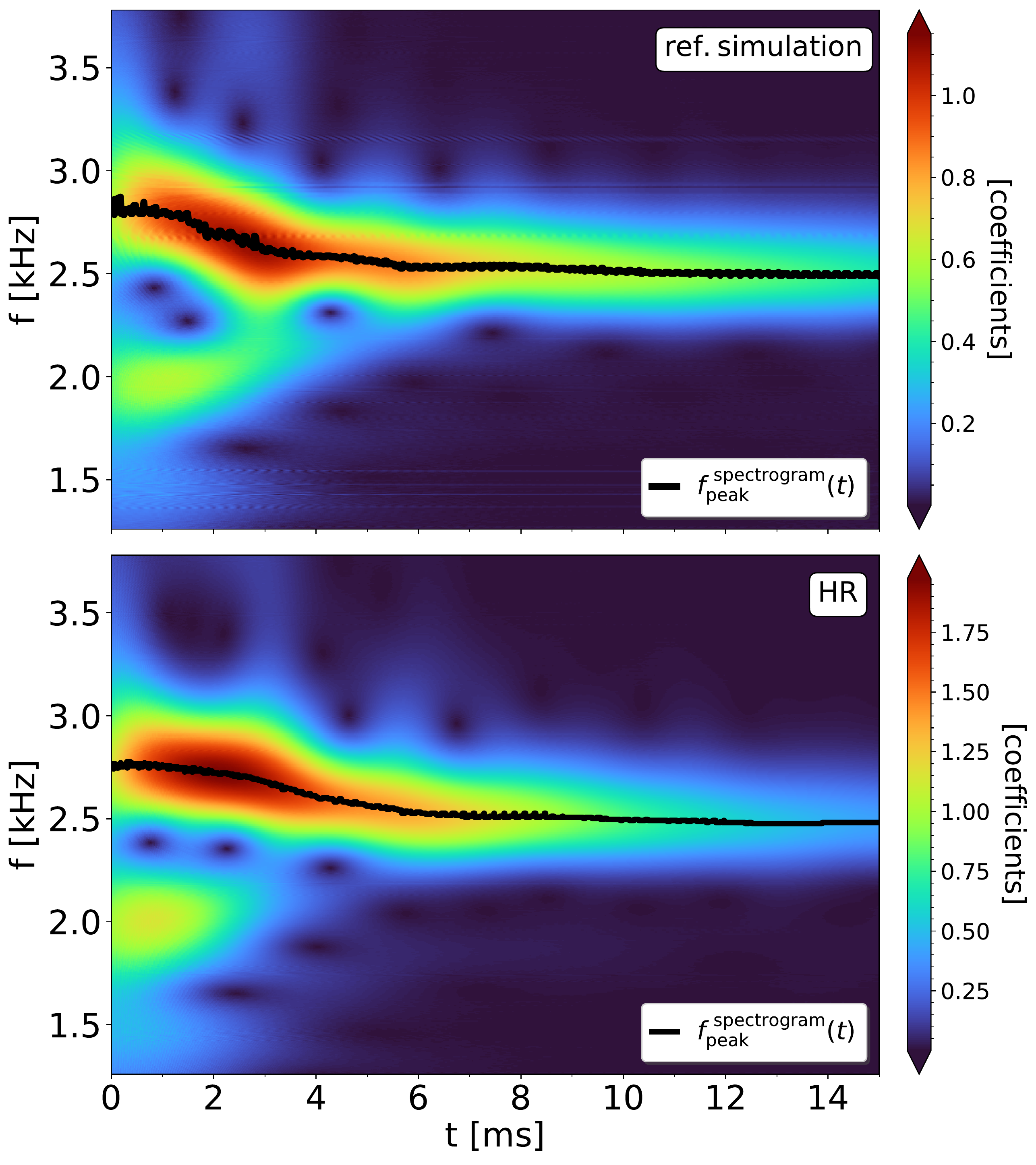} 
	\caption{\label{fig:dump_specComp_HR_m1_25.pdf} {\it Top panel}: spectrogram of $h_+(t)$ for the reference simulation. {\it Bottom panel}: spectrogram of  $h_+(t)$ for the high resolution simulation HR. The black curves correspond to the numerically extracted $f_{\mathrm{peak}}^{\rm spectrogram}(t)$ for the reference simulation and HR, respectively. }
\end{figure}

Figure~\ref{fig:dumpFFT_HR_mpa1-m1_25.pdf} displays the GW spectra for the reference and HR simulation. The agreement between the frequency peaks is remarkable, although there are small differences in the morphology of the main peak. The time evolution of $f_\mathrm{peak}(t)$ agrees well in both simulations, and we observe a very good agreement between the secondary frequencies, especially for $f_\mathrm{spiral}$, $f_{2-0}$. Figure~\ref{fig:dump_specComp_HR_m1_25.pdf} shows the spectrograms for the two simulations and confirms that, in spite of the differences in the structure of the main peak shown in the spectra of Figure~\ref{fig:dumpFFT_HR_mpa1-m1_25.pdf}, the time-evolution of $f_{\rm peak}(t)$ is qualitatively similar in both cases and it can thus be described by the same analytic model that we describe in the main text. 

We also compute the analytic (Ac) and semi-analytic (Sc) model fits for the HR simulation and find large fitting factors of $FF\gtrsim0.970$ (see Tab.~\ref{Fitting factors for HR}). We note that for this particular configuration the $FFs$ are even larger than the ones obtained for the reference simulation.

\begin{table}[h]
	\begin{ruledtabular}
		\begin{tabular}{c|cc}
		\multicolumn{3}{l}{{Fitting Factors ($FFs$) }}\vspace{0.2cm}  \\ \hline 
			\textrm{Simulation}&
			\textrm{Sc}&
			\textrm{Ac}\\
			\colrule
		
ref. & 0.969& 0.956\\
HR& 0.978& 0.974\\
		\end{tabular}
	\end{ruledtabular}
	\caption{Fitting factors $FFs$ for the analytic (Ac) and semi-analytic (Sc) model fits for the HR simulation.}
	\label{Fitting factors for HR}
\end{table}

\subsection{Effect of pi-symmetry}\label{Effect of pi-symmetry}

\begin{figure}[h!]
	\includegraphics[scale=0.33]{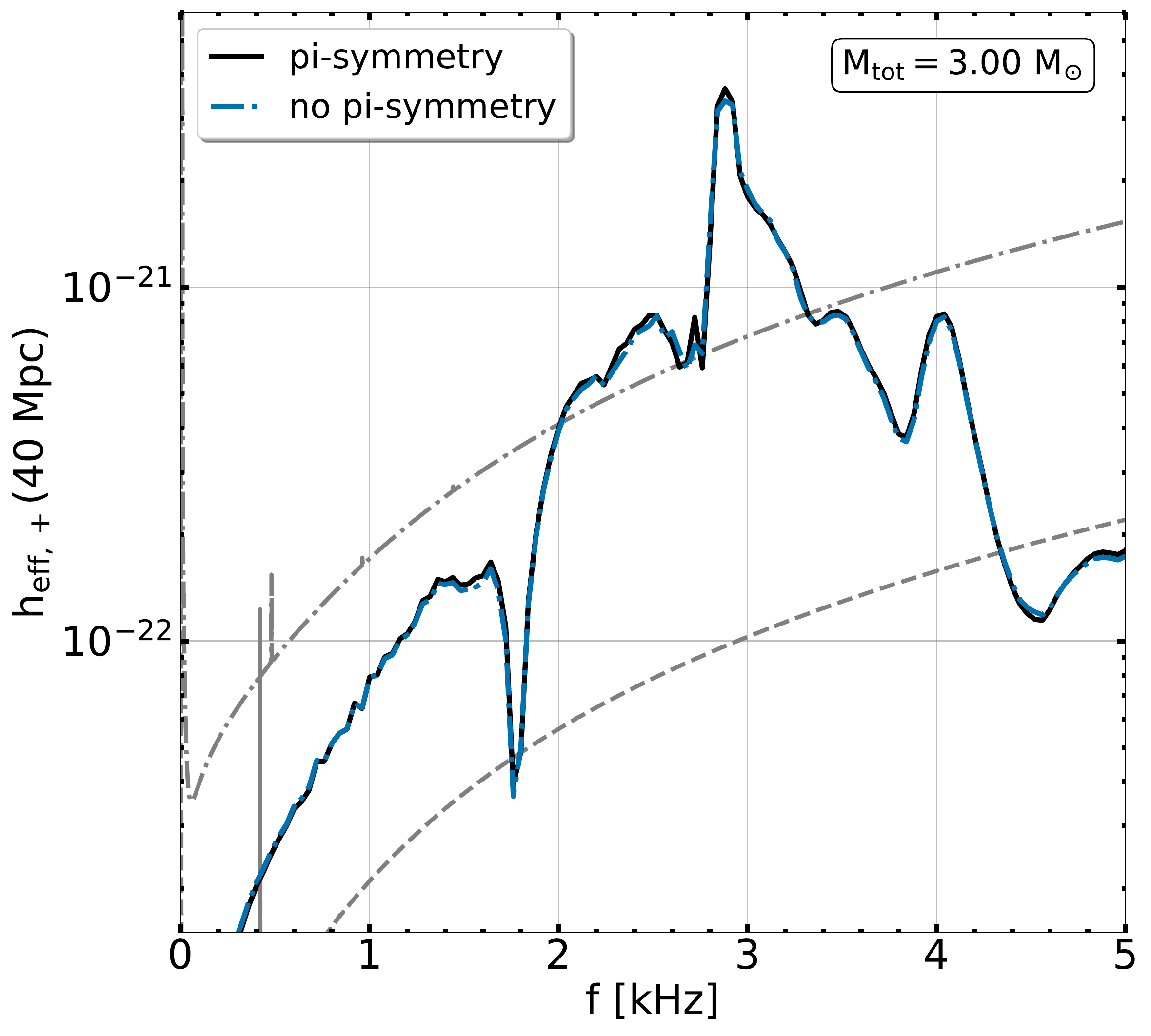} 
	\caption{\label{fig:dumpFFT_nopi_mpa1-m1_50.pdf} Effective GW spectra  $h_\mathrm{eff,+}(f)$ for the $M_\mathrm{tot}=3.0~M_\odot$  simulation with pi-symmetry (black) and without pi-symmetry (blue). The dash dotted curves denote the design sensitivity Advanced LIGO \cite{LIGOScientific:2014pky} and of the Einstein Telescope \cite{EinsteinTelescope2010}, respectively.}
\end{figure}

In this section we discuss the impact of imposing pi-symmetry during the simulations. We carry out additional simulations using the same numerical setup as described in Sect.~\ref{Sec:Methods} but without pi-symmetry. We run additional models for  $M_\mathrm{tot}=2.5~M_\odot,~2.7~M_\odot,~2.9~M_\odot$ and $3.0~M_\odot$. The respective spectra agree very well with the simulations using pi-symmetry. As example, we show the calculation for $M_\mathrm{tot}=3.0~M_\odot$ in Fig.~\ref{fig:dumpFFT_nopi_mpa1-m1_50.pdf}.
This particular configuration is discussed in Subsect.~\ref{SubSec. spiral-0 coupling}. We conclude that imposing pi-symmetry does not impact the spectral features and in particular the $f_\mathrm{spiral\pm0}$ coupling is unaffected by the pi-symmetry.

\section{Spectral properties of the mass sequence models}\label{A:Spectral properties of the mass sequence models}

\begin{figure*}[h]
	\includegraphics[ scale=0.36]{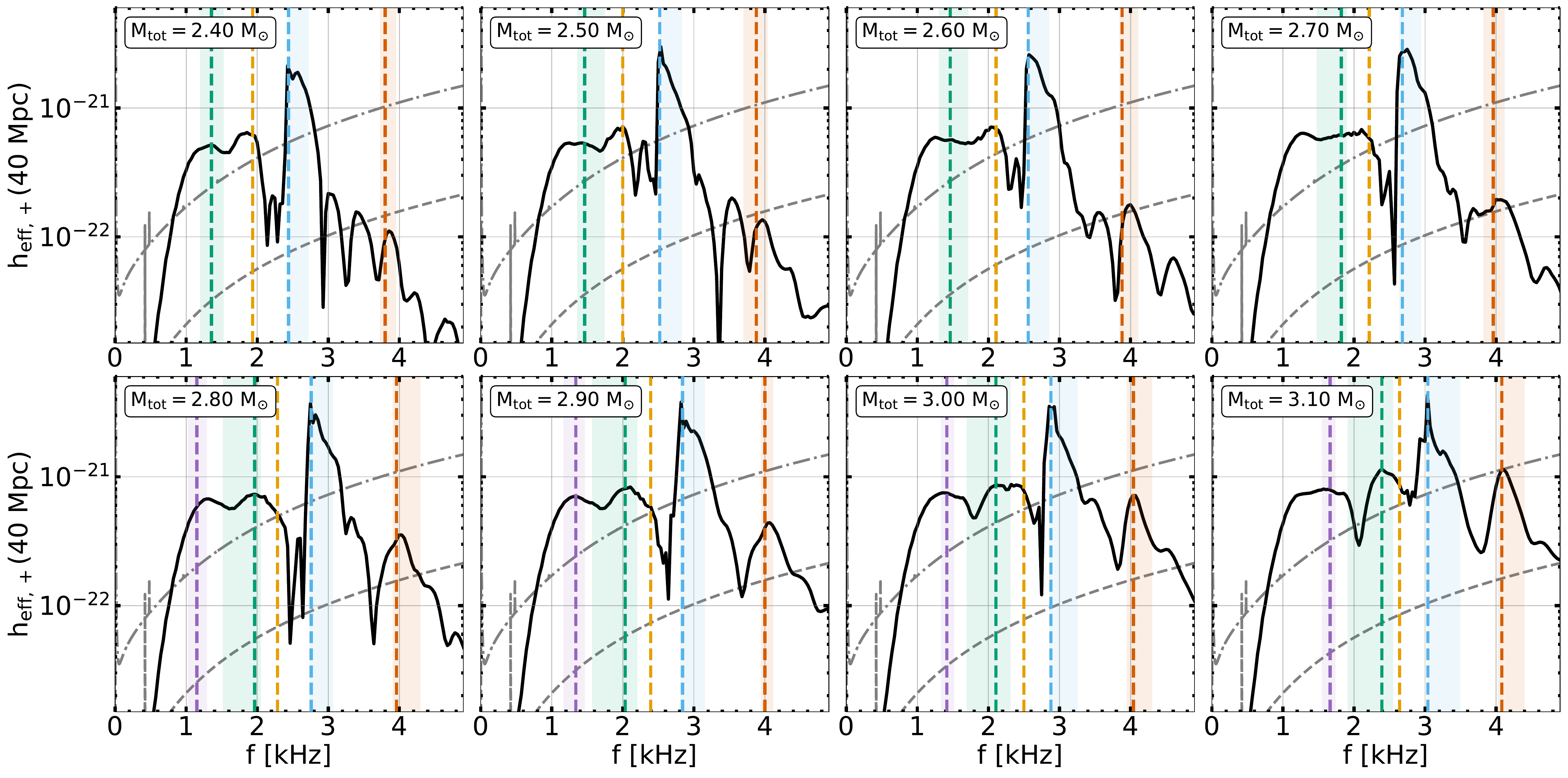} 
	\caption{\label{fig:sequence-FFT-fullGW.pdf}As Fig.~\ref{fig:dumpFFT_evolution_mpa1-m1.25.pdf} but including the inspiral signal.}
\end{figure*}

\begin{figure*}[h]
	\includegraphics[scale=0.35]{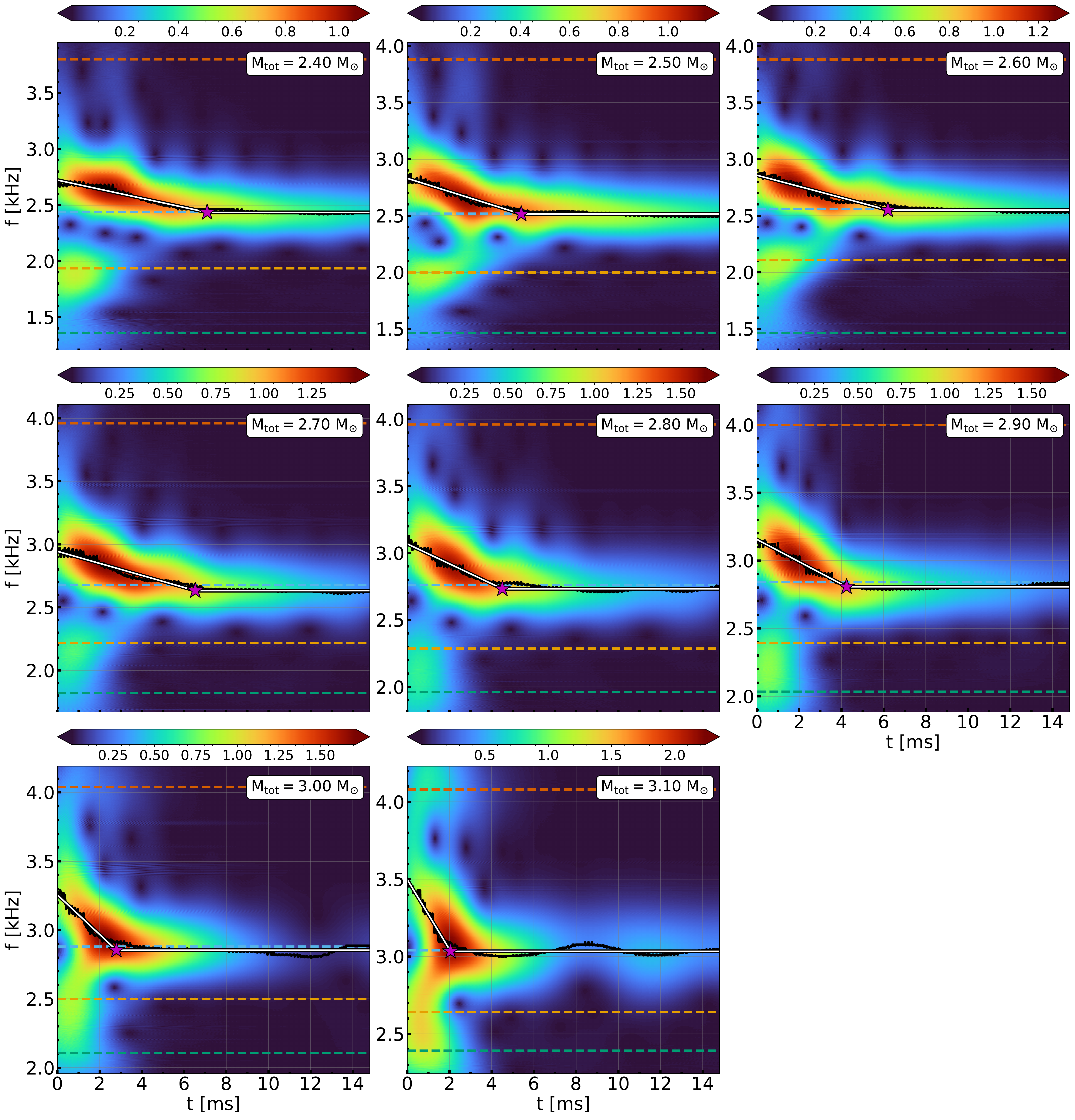} 
	\caption{\label{fig:sequence-SPECs.pdf}As Fig.~\ref{fig:dumpSPEC_mpa1-m1.25} but for all models in our mass sequence.}
\end{figure*}

\begin{figure*}[h]
	\includegraphics[scale=0.35]{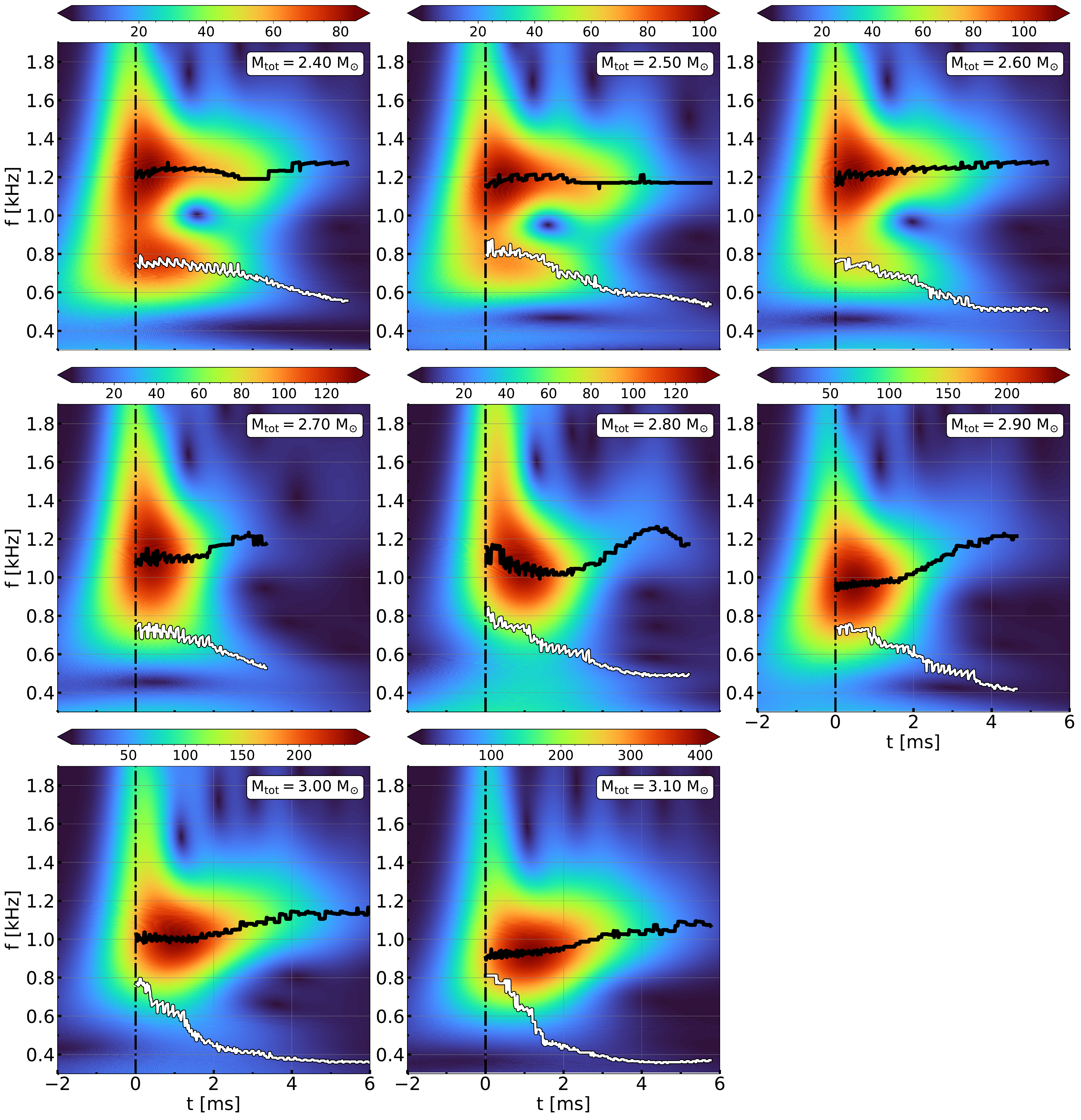} 
	\caption{\label{fig:sequence-SPECs_f0.pdf}As Fig.~\ref{fig:dumpSPEC_f0_from_ac_mpa1-m1.25.pdf} but for all models of our mass sequence.}
\end{figure*}

We present supplementary figures for the mass sequence models. Figure~\ref{fig:sequence-FFT-fullGW.pdf} shows the GW spectra including the inspiral signal. Figure~\ref{fig:sequence-SPECs.pdf} displays the spectrograms for the post-merger GW signal $h_+(t)$. Figure~\ref{fig:sequence-SPECs_f0.pdf} provides the spectrograms for the minimum lapse function $\alpha_\mathrm{min}(t)$ starting at a few milliseconds before the merging phase.

\section{Analytic model}\label{A:Analytic model}

\subsection{Spectrogram analysis}
\begin{figure*}
    \centering
    \subfloat[\label{fig:dump_Amp_of_t_m1.25_1.pdf}]{%
    \includegraphics[scale=0.33]{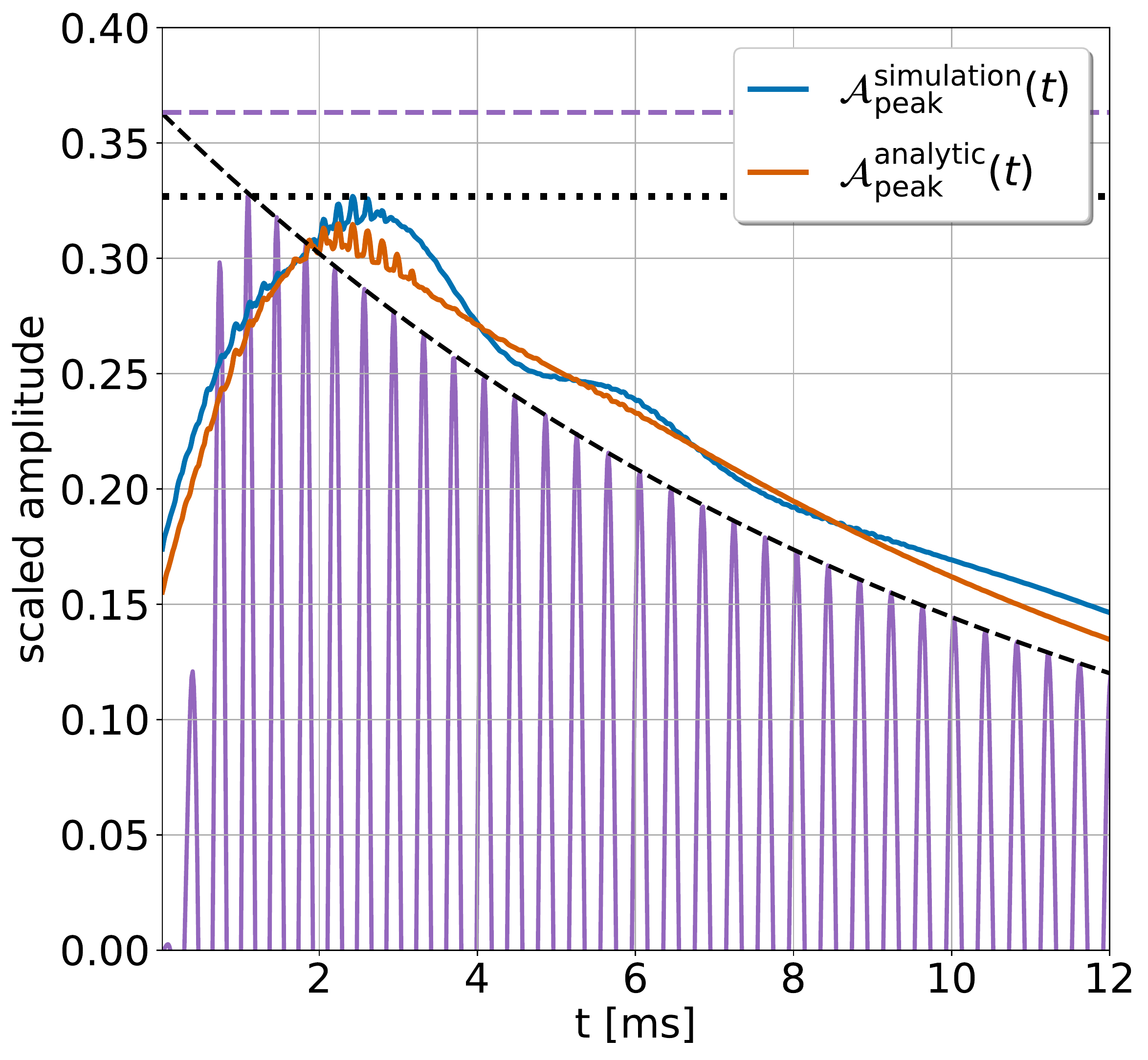}}
    \hspace*{12pt}
    \subfloat[\label{fig:dump_Amp_of_t_m1.25_2.pdf}]{
    \includegraphics[scale=0.33]{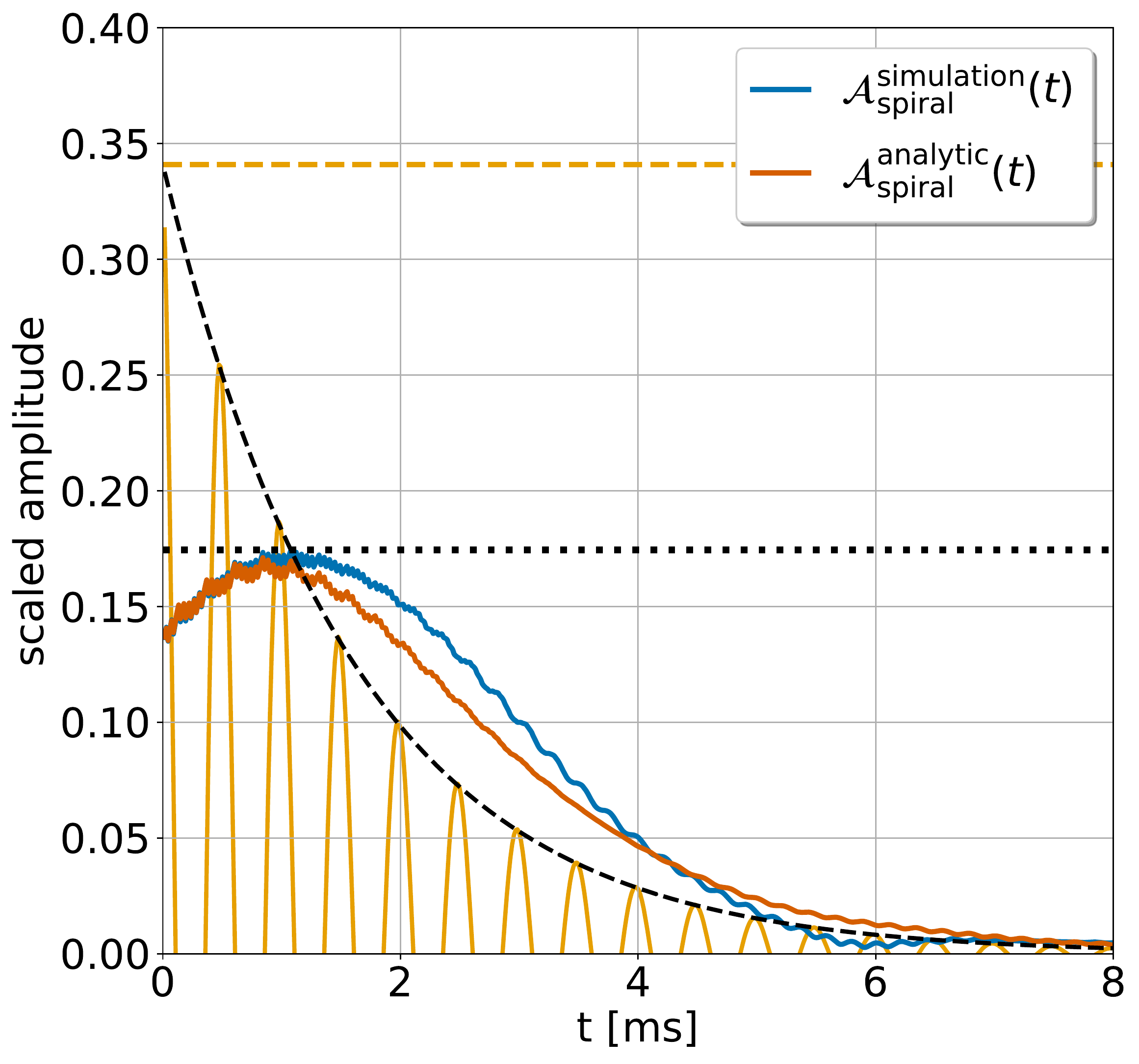}}
    
    \subfloat[\label{fig:dump_Amp_of_t_m1.25_3.pdf}]{%
    \includegraphics[scale=0.33]{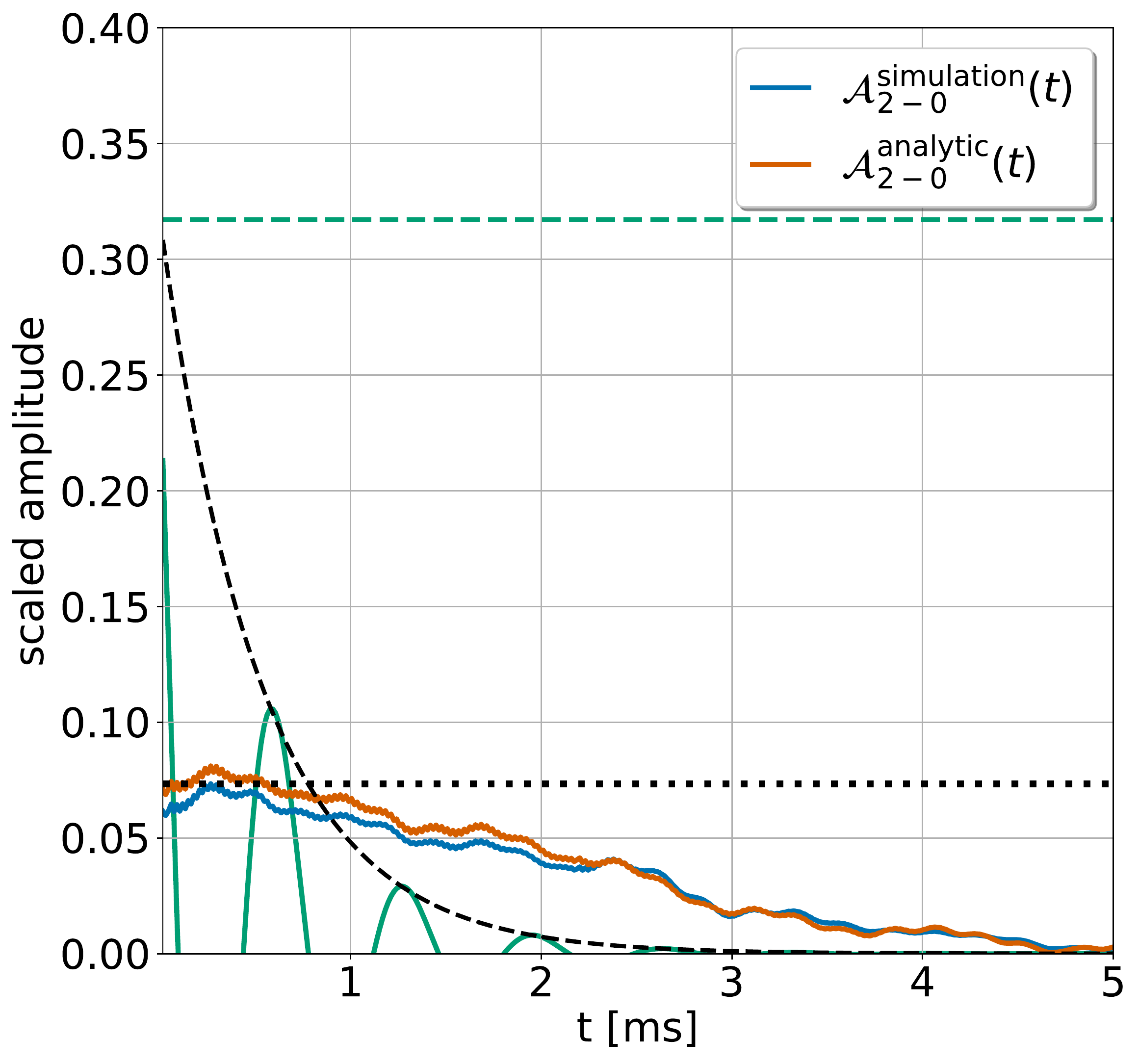}}
    \caption{{\it Top left panel}: Coefficient curves $\mathcal{A}_{\mathrm{peak}}(t)$ for the $f_{\mathrm{peak}}$ component extracted from the spectrograms of simulation and analytic model. Purple dashed line indicates amplitude $A_{\mathrm{peak}}$. Black dotted horizontal line shows maximum of $\mathcal{A}_{\mathrm{peak}}(t)$ for simulation. Purple sinusoidal function shows $f_{\mathrm{peak}}$ component as used in the analytic model. Dashed black curve shows its exponential decay. {\it Top right panel}: Coefficient curves $\mathcal{A}_{\mathrm{spiral}}(t)$ for the $f_{\mathrm{spiral}}$ component extracted from spectrograms of simulation and analytic model. Yellow dashed line shows the amplitude $A_{\mathrm{spiral}}$. Black dotted horizontal line indicates the maximum of $\mathcal{A_{\mathrm{spiral}}}(t)$ for simulation. Yellow sinusoidal function shows $f_{\mathrm{spiral}}$ component as used in the analytic model. Dashed black curve shows its exponential decay. {\it Bottom panel}: Coefficient curves $\mathcal{A}_{\mathrm{2-0}}(t)$ for the $f_{\mathrm{2-0}}$ component extracted from spectrograms of simulation and analytic model. Yellow dashed line shows amplitude $A_{\mathrm{2-0}}$. Black dotted horizontal line indicates maximum of $\mathcal{A_{\mathrm{2-0}}}(t)$ for simulation. Yellow sinusoidal function shows $f_{\mathrm{2-0}}$ component as used in the analytic model. Dashed black curve shows its exponential decay.}
        
    \label{fig:my_label}
\end{figure*}

To further evaluate the analytic model, we quantitatively analyze the spectrograms of the numerical simulation and of the analytic fit. Figures Fig.~\ref{fig:dump_Amp_of_t_m1.25_1.pdf} to Fig.~\ref{fig:dump_Amp_of_t_m1.25_3.pdf} depict the wavelet coefficients $\mathcal{A}_\mathrm{peak}(t)$, $\mathcal{A}_\mathrm{spiral}(t)$, $\mathcal{A}_\mathrm{2-0}(t)$, which we extract at the corresponding frequencies from the spectrograms as function of time, for the simulation (blue line) and for the analytic fit (orange line). The analytic model performs well in reproducing the three coefficient of the simulation data. For illustration purposes, we omit the normalization factor $\mathcal{N}$ and overlay the corresponding exponentially decaying sinusoid functions for each frequency component while we rescale the coefficient curves by a constant factor which ensures that the maxima of $\Big(h_\mathrm{peak}(t)\mathcal{W}(t;s)\Big)$ and $\mathcal{A}_\mathrm{peak}(t)$ coincide.

We note that the three components exhibit different magnitudes of the coefficient curves $\mathcal{A}_\mathrm{i}(t)$, whereas the amplitudes $A_\mathrm{i}$ ($\mathrm{i=peak,spiral,2-0}$) of the analytic model are roughly comparable. For our reference simulation with a total binary mass $M_\mathrm{tot}=2.5 M_\odot$, we actually expect that the $f_\mathrm{spiral}$ component is the strongest secondary feature, which is also suggested by the GW spectrum, and in fact the maxima of the coefficients show this hierarchy. We thus remark that the amplitudes $A_\mathrm{i}$ of the analytic model may have only a limited physical meaning, while other quantities, such as the surface area under $\mathcal{A}_\mathrm{i}(t)$, the maxima of $\mathcal{A}_\mathrm{i}(t)$ or the product $A_\mathrm{i}\cdot \tau_i$ may turn out to be more representative for the merger dynamics and GW emission.

\subsection{Sequence of simulations with different masses}
\label{Appendix-sequence}
\begin{figure*}[h]
	\includegraphics[scale=0.35]{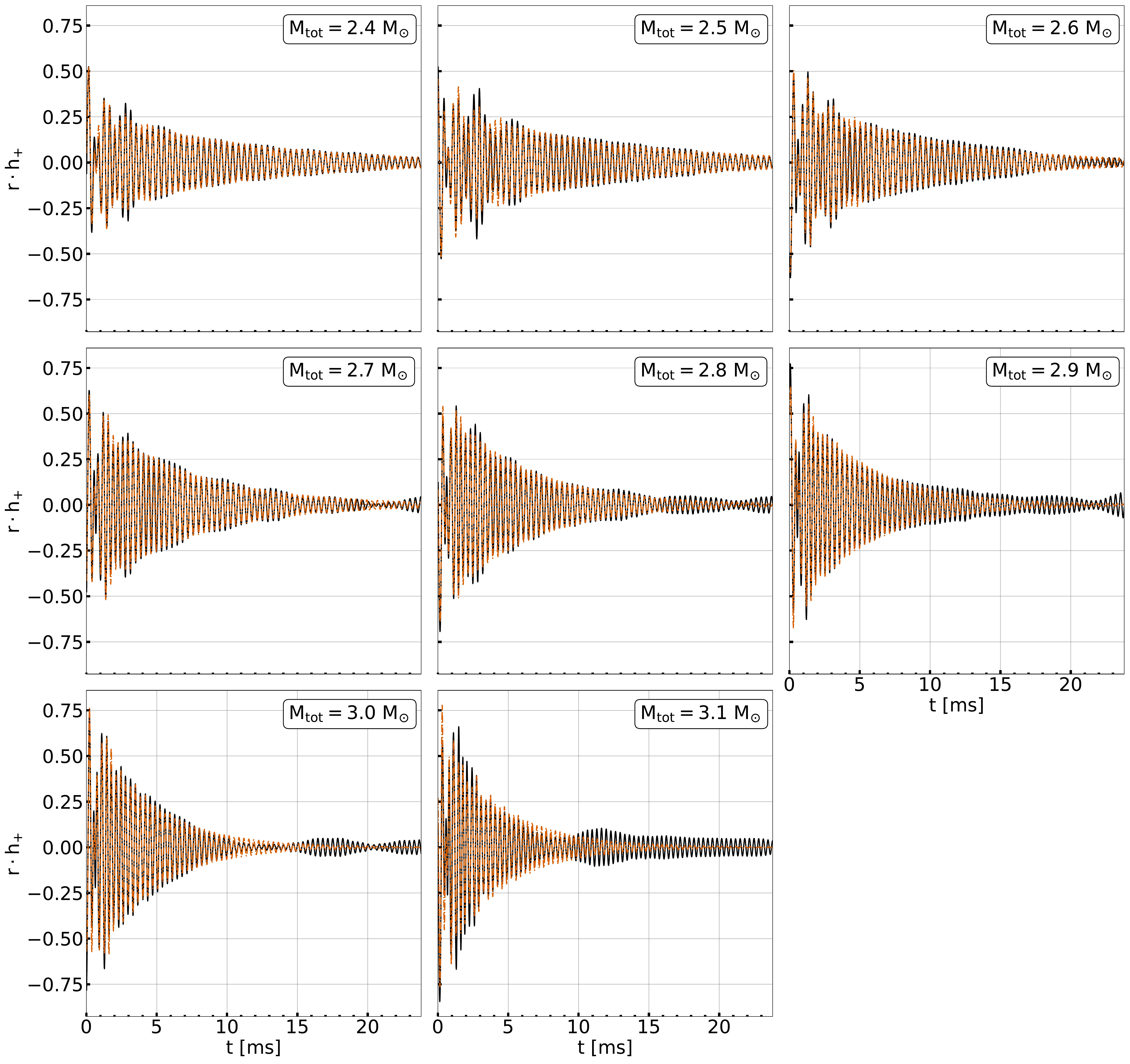} 
	\caption{\label{fig:sequence-GW-model.pdf} As Fig.~\ref{fig:dumpGW_model_mpa1-m1.25.pdf} but for all models of our mass sequence.}
\end{figure*}

Figure~\ref{fig:sequence-GW-model.pdf} shows the time-domain signals for the simulation and the analytic model along the sequence of simulations with different masses. Notice the possible excitation of a low-$|T/W|$ rotational instability in the highest-mass model, after $\sim 10$ms from the onset of merger \citep{Passamonti_Andersson_2020,Xie_etal_2020,PhysRevD.101.064052}.

\subsection{Initial phases}
We find that the initial phases $\phi_\mathrm{peak}$, $\phi_\mathrm{spiral}$, $\phi_\mathrm{2\pm0}$ correlate with the total binary mass $M_\mathrm{tot}$. We model this dependence with a 2-segment piecewise function consisting of two linear fits which intersect at $M_\mathrm{tot}=2.7~M_\odot$. These are given by

\begin{eqnarray}\label{phis_of Mtot}
\label{phipeak-Mtot}\phi_{\mathrm{peak}} =    \left\{
\begin{array}{ll}
+18.957\cdot M_\mathrm{tot}-46.321  & \mbox{for } M_\mathrm{tot}\leq 2.7~M_\odot \\
+43.425\cdot M_\mathrm{tot}-113.152 &\mbox{for }  M_\mathrm{tot} >  2.7~M_\odot \nonumber \\
\end{array} 
\right. 
\\
\\
\label{phispiral-Mtot}\phi_{\mathrm{spiral}} =    \left\{
\begin{array}{ll}
+17.580\cdot M_\mathrm{tot}-42.199 & \mbox{for } M_\mathrm{tot}\leq 2.7~M_\odot \\
+40.448\cdot M_\mathrm{tot}-104.258 &\mbox{for }  M_\mathrm{tot} >  2.7~M_\odot \nonumber \\
\end{array} 
\right. 
\\
\\
\label{phi2-0-Mtot}\phi_{\mathrm{2-0}} =    \left\{
\begin{array}{ll}
+18.541\cdot M_\mathrm{tot}-43.911 & \mbox{for } M_\mathrm{tot}\leq 2.7~M_\odot \\
+43.613\cdot M_\mathrm{tot}-112.705 & \mbox{for } M_\mathrm{tot} >  2.7~M_\odot \nonumber \\
\end{array} 
\right. 
\\
\\
\label{phi2+0-Mtot}\phi_{\mathrm{2+0}} =    \left\{
\begin{array}{ll}
+16.064\cdot M_\mathrm{tot}-41.163 & \mbox{for } M_\mathrm{tot}\leq 2.7~M_\odot \\
+43.309\cdot M_\mathrm{tot}-115.341 & \mbox{for } M_\mathrm{tot} >  2.7~M_\odot \nonumber \\
\end{array} 
\right. 
\\
\end{eqnarray}

\subsection{Empirical relations}\label{Empirical relations}
The fits are carried out for the signals $r\cdot h_s(t)$ (for $s=+,\times$). Table~\ref{Units} provides information about the analytic model's parameters. 

\begin{table}
	\begin{ruledtabular}
		\begin{tabular}{l|cc}
		\multicolumn{3}{l}{{Analytic model's parameters}}\vspace{0.2cm}  \\ \hline 
			\textrm{Symbol}&
			\textrm{Unit}&
			\textrm{Equation}\\
			\colrule
        $M$& $M_\odot$& -\\
        $h_\mathrm{eff,+}(f)=f\cdot \widetilde{h}_+(f)$& dimensionless& -\\
        Sensitivity curve $\equiv \sqrt{S_h(f)\cdot f}$& dimensionless& -\\
        $f_\mathrm{peak}$& kHz& -\\
        $f_\mathrm{spiral}$& kHz& -\\
        $f_\mathrm{2\pm0}$& kHz& -\\
        $f_\mathrm{spiral\pm0}$& kHz& -\\
        $\zeta_\mathrm{drift}$& $\mathrm{kHz^2}$& \eqref{equations-fpeak(t)-fits_1}\\
        $f_\mathrm{peak,0}$& $\mathrm{kHz}$& \eqref{equations-fpeak(t)-fits_2}\\
        $t_*$& $\mathrm{ms}$& \eqref{equations-fpeak(t)-fits_3}\\
        $A_\mathrm{peak}$& dimensionless&\eqref{Apeak of mtot}\\
        $\tau_\mathrm{peak}$& ms&\eqref{Tpeak of mtot}\\
        $\tau_\mathrm{spiral}$& ms&\eqref{Tspiral of mtot}\\
        $\tau_\mathrm{2-0}$& ms&\eqref{T2-0 of mtot}\\
        $\tau_\mathrm{2+0}$& ms&\eqref{T2+0 of mtot}\\
        $A_\mathrm{spiral}$& dimensionless&\eqref{Aspiral of mtot}\\
        $A_\mathrm{2-0}$& dimensionless&\eqref{A2-0 of mtot}\\
        $A_\mathrm{2+0}$& dimensionless&\eqref{A2+0 of mtot}\\
        $\mathcal{N}$& dimensionless&\eqref{N of Mtot}\\
        $\phi_\mathrm{peak}$& rad&\eqref{phipeak-Mtot}\\
        $\phi_\mathrm{spiral}$& rad&\eqref{phispiral-phipeak},\eqref{phispiral-Mtot}\\
        $\phi_\mathrm{2-0}$& rad&\eqref{phi2-0-phipeak},\eqref{phi2-0-Mtot}\\
        $\phi_\mathrm{2+0}$& rad&\eqref{phi2+0-phipeak},\eqref{phi2+0-Mtot}
		\end{tabular}
	\end{ruledtabular}
	\caption{The analytic model's parameters.}
	\label{Units}
\end{table}


\begin{thebibliography}{120}%
\makeatletter
\providecommand \@ifxundefined [1]{%
 \@ifx{#1\undefined}
}%
\providecommand \@ifnum [1]{%
 \ifnum #1\expandafter \@firstoftwo
 \else \expandafter \@secondoftwo
 \fi
}%
\providecommand \@ifx [1]{%
 \ifx #1\expandafter \@firstoftwo
 \else \expandafter \@secondoftwo
 \fi
}%
\providecommand \natexlab [1]{#1}%
\providecommand \enquote  [1]{``#1''}%
\providecommand \bibnamefont  [1]{#1}%
\providecommand \bibfnamefont [1]{#1}%
\providecommand \citenamefont [1]{#1}%
\providecommand \href@noop [0]{\@secondoftwo}%
\providecommand \href [0]{\begingroup \@sanitize@url \@href}%
\providecommand \@href[1]{\@@startlink{#1}\@@href}%
\providecommand \@@href[1]{\endgroup#1\@@endlink}%
\providecommand \@sanitize@url [0]{\catcode `\\12\catcode `\$12\catcode
  `\&12\catcode `\#12\catcode `\^12\catcode `\_12\catcode `\%12\relax}%
\providecommand \@@startlink[1]{}%
\providecommand \@@endlink[0]{}%
\providecommand \url  [0]{\begingroup\@sanitize@url \@url }%
\providecommand \@url [1]{\endgroup\@href {#1}{\urlprefix }}%
\providecommand \urlprefix  [0]{URL }%
\providecommand \Eprint [0]{\href }%
\providecommand \doibase [0]{https://doi.org/}%
\providecommand \selectlanguage [0]{\@gobble}%
\providecommand \bibinfo  [0]{\@secondoftwo}%
\providecommand \bibfield  [0]{\@secondoftwo}%
\providecommand \translation [1]{[#1]}%
\providecommand \BibitemOpen [0]{}%
\providecommand \bibitemStop [0]{}%
\providecommand \bibitemNoStop [0]{.\EOS\space}%
\providecommand \EOS [0]{\spacefactor3000\relax}%
\providecommand \BibitemShut  [1]{\csname bibitem#1\endcsname}%
\let\auto@bib@innerbib\@empty
\bibitem [{\citenamefont {{Abbott}}\ \emph
  {et~al.}(2017{\natexlab{a}})\citenamefont {{Abbott}}, \citenamefont
  {{Abbott}}, \citenamefont {{Abbott}}, \citenamefont {{Acernese}},
  \citenamefont {{Ackley}}, \citenamefont {{Adams}}, \citenamefont {{Adams}},
  \citenamefont {{Addesso}}, \citenamefont {{Adhikari}}, \citenamefont {{Adya}}
  \emph {et~al.}}]{2017PhRvL.119p1101A}%
  \BibitemOpen
  \bibfield  {author} {\bibinfo {author} {\bibfnamefont {B.~P.}\ \bibnamefont
  {{Abbott}}}, \bibinfo {author} {\bibfnamefont {R.}~\bibnamefont {{Abbott}}},
  \bibinfo {author} {\bibfnamefont {T.~D.}\ \bibnamefont {{Abbott}}}, \bibinfo
  {author} {\bibfnamefont {F.}~\bibnamefont {{Acernese}}}, \bibinfo {author}
  {\bibfnamefont {K.}~\bibnamefont {{Ackley}}}, \bibinfo {author}
  {\bibfnamefont {C.}~\bibnamefont {{Adams}}}, \bibinfo {author} {\bibfnamefont
  {T.}~\bibnamefont {{Adams}}}, \bibinfo {author} {\bibfnamefont
  {P.}~\bibnamefont {{Addesso}}}, \bibinfo {author} {\bibfnamefont {R.~X.}\
  \bibnamefont {{Adhikari}}}, \bibinfo {author} {\bibfnamefont {V.~B.}\
  \bibnamefont {{Adya}}}, \emph {et~al.} (\bibinfo {collaboration} {LIGO
  Scientific Collaboration and Virgo Collaboration}),\ }\bibfield  {title}
  {\bibinfo {title} {{GW170817: Observation of Gravitational Waves from a
  Binary Neutron Star Inspiral}},\ }\href
  {https://doi.org/10.1103/PhysRevLett.119.161101} {\bibfield  {journal}
  {\bibinfo  {journal} {Phys. Rev. Lett.}\ }\textbf {\bibinfo {volume} {119}},\
  \bibinfo {eid} {161101} (\bibinfo {year} {2017}{\natexlab{a}})},\ \Eprint
  {https://arxiv.org/abs/1710.05832} {arXiv:1710.05832 [gr-qc]} \BibitemShut
  {NoStop}%
\bibitem [{\citenamefont {{Abbott}}\ \emph {et~al.}(2020)\citenamefont
  {{Abbott}}, \citenamefont {{Abbott}}, \citenamefont {{Abbott}}, \citenamefont
  {{Abraham}}, \citenamefont {{Acernese}}, \citenamefont {{Ackley}},
  \citenamefont {{Adams}}, \citenamefont {{Adhikari}}, \citenamefont {{Adya}},
  \citenamefont {{Affeldt}} \emph {et~al.}}]{2020ApJ...892L...3A}%
  \BibitemOpen
  \bibfield  {author} {\bibinfo {author} {\bibfnamefont {B.~P.}\ \bibnamefont
  {{Abbott}}}, \bibinfo {author} {\bibfnamefont {R.}~\bibnamefont {{Abbott}}},
  \bibinfo {author} {\bibfnamefont {T.~D.}\ \bibnamefont {{Abbott}}}, \bibinfo
  {author} {\bibfnamefont {S.}~\bibnamefont {{Abraham}}}, \bibinfo {author}
  {\bibfnamefont {F.}~\bibnamefont {{Acernese}}}, \bibinfo {author}
  {\bibfnamefont {K.}~\bibnamefont {{Ackley}}}, \bibinfo {author}
  {\bibfnamefont {C.}~\bibnamefont {{Adams}}}, \bibinfo {author} {\bibfnamefont
  {R.~X.}\ \bibnamefont {{Adhikari}}}, \bibinfo {author} {\bibfnamefont
  {V.~B.}\ \bibnamefont {{Adya}}}, \bibinfo {author} {\bibfnamefont
  {C.}~\bibnamefont {{Affeldt}}}, \emph {et~al.},\ }\bibfield  {title}
  {\bibinfo {title} {{GW190425: Observation of a Compact Binary Coalescence
  with Total Mass {\ensuremath{\sim}} 3.4 M$_{\odot}$}},\ }\href
  {https://doi.org/10.3847/2041-8213/ab75f5} {\bibfield  {journal} {\bibinfo
  {journal} {Astrophys. J. Lett.}\ }\textbf {\bibinfo {volume} {892}},\
  \bibinfo {eid} {L3} (\bibinfo {year} {2020})},\ \Eprint
  {https://arxiv.org/abs/2001.01761} {arXiv:2001.01761 [astro-ph.HE]}
  \BibitemShut {NoStop}%
\bibitem [{\citenamefont {Abbott}\ \emph
  {et~al.}(2018{\natexlab{a}})\citenamefont {Abbott} \emph
  {et~al.}}]{Aasi:2013wya}%
  \BibitemOpen
  \bibfield  {author} {\bibinfo {author} {\bibfnamefont {B.~P.}\ \bibnamefont
  {Abbott}} \emph {et~al.} (\bibinfo {collaboration} {KAGRA, LIGO Scientific,
  VIRGO}),\ }\bibfield  {title} {\bibinfo {title} {{Prospects for Observing and
  Localizing Gravitational-Wave Transients with Advanced LIGO, Advanced Virgo
  and KAGRA}},\ }\href {https://doi.org/10.1007/s41114-018-0012-9} {\bibfield
  {journal} {\bibinfo  {journal} {Living Rev. Rel.}\ }\textbf {\bibinfo
  {volume} {21}},\ \bibinfo {pages} {3} (\bibinfo {year}
  {2018}{\natexlab{a}})},\ \Eprint {https://arxiv.org/abs/1304.0670}
  {arXiv:1304.0670 [gr-qc]} \BibitemShut {NoStop}%
\bibitem [{\citenamefont {Abbott}\ \emph {et~al.}(2019)\citenamefont {Abbott}
  \emph {et~al.}}]{LIGOScientific:2018hze}%
  \BibitemOpen
  \bibfield  {author} {\bibinfo {author} {\bibfnamefont {B.~P.}\ \bibnamefont
  {Abbott}} \emph {et~al.} (\bibinfo {collaboration} {LIGO Scientific,
  Virgo}),\ }\bibfield  {title} {\bibinfo {title} {{Properties of the binary
  neutron star merger GW170817}},\ }\href
  {https://doi.org/10.1103/PhysRevX.9.011001} {\bibfield  {journal} {\bibinfo
  {journal} {Phys. Rev. X}\ }\textbf {\bibinfo {volume} {9}},\ \bibinfo {pages}
  {011001} (\bibinfo {year} {2019})},\ \Eprint
  {https://arxiv.org/abs/1805.11579} {arXiv:1805.11579 [gr-qc]} \BibitemShut
  {NoStop}%
\bibitem [{\citenamefont {{Chatziioannou}}(2020)}]{2020GReGr..52..109C}%
  \BibitemOpen
  \bibfield  {author} {\bibinfo {author} {\bibfnamefont {K.}~\bibnamefont
  {{Chatziioannou}}},\ }\bibfield  {title} {\bibinfo {title} {{Neutron-star
  tidal deformability and equation-of-state constraints}},\ }\href
  {https://doi.org/10.1007/s10714-020-02754-3} {\bibfield  {journal} {\bibinfo
  {journal} {General Relativity and Gravitation}\ }\textbf {\bibinfo {volume}
  {52}},\ \bibinfo {eid} {109} (\bibinfo {year} {2020})},\ \Eprint
  {https://arxiv.org/abs/2006.03168} {arXiv:2006.03168 [gr-qc]} \BibitemShut
  {NoStop}%
\bibitem [{\citenamefont {{Dietrich}}\ \emph
  {et~al.}(2021{\natexlab{a}})\citenamefont {{Dietrich}}, \citenamefont
  {{Hinderer}},\ and\ \citenamefont {{Samajdar}}}]{2021GReGr..53...27D}%
  \BibitemOpen
  \bibfield  {author} {\bibinfo {author} {\bibfnamefont {T.}~\bibnamefont
  {{Dietrich}}}, \bibinfo {author} {\bibfnamefont {T.}~\bibnamefont
  {{Hinderer}}},\ and\ \bibinfo {author} {\bibfnamefont {A.}~\bibnamefont
  {{Samajdar}}},\ }\bibfield  {title} {\bibinfo {title} {{Interpreting binary
  neutron star mergers: describing the binary neutron star dynamics, modelling
  gravitational waveforms, and analyzing detections}},\ }\href
  {https://doi.org/10.1007/s10714-020-02751-6} {\bibfield  {journal} {\bibinfo
  {journal} {General Relativity and Gravitation}\ }\textbf {\bibinfo {volume}
  {53}},\ \bibinfo {eid} {27} (\bibinfo {year} {2021}{\natexlab{a}})},\ \Eprint
  {https://arxiv.org/abs/2004.02527} {arXiv:2004.02527 [gr-qc]} \BibitemShut
  {NoStop}%
\bibitem [{\citenamefont {{Bauswein}}\ \emph {et~al.}(2017)\citenamefont
  {{Bauswein}}, \citenamefont {{Just}}, \citenamefont {{Janka}},\ and\
  \citenamefont {{Stergioulas}}}]{Bauswein_etal_2017}%
  \BibitemOpen
  \bibfield  {author} {\bibinfo {author} {\bibfnamefont {A.}~\bibnamefont
  {{Bauswein}}}, \bibinfo {author} {\bibfnamefont {O.}~\bibnamefont {{Just}}},
  \bibinfo {author} {\bibfnamefont {H.-T.}\ \bibnamefont {{Janka}}},\ and\
  \bibinfo {author} {\bibfnamefont {N.}~\bibnamefont {{Stergioulas}}},\
  }\bibfield  {title} {\bibinfo {title} {{Neutron-star Radius Constraints from
  GW170817 and Future Detections}},\ }\href
  {https://doi.org/10.3847/2041-8213/aa9994} {\bibfield  {journal} {\bibinfo
  {journal} {Astrophys. J. Lett.}\ }\textbf {\bibinfo {volume} {850}},\
  \bibinfo {eid} {L34} (\bibinfo {year} {2017})},\ \Eprint
  {https://arxiv.org/abs/1710.06843} {arXiv:1710.06843 [astro-ph.HE]}
  \BibitemShut {NoStop}%
\bibitem [{\citenamefont {Abbott}\ \emph
  {et~al.}(2018{\natexlab{b}})\citenamefont {Abbott} \emph
  {et~al.}}]{Abbott:2018exr}%
  \BibitemOpen
  \bibfield  {author} {\bibinfo {author} {\bibfnamefont {B.~P.}\ \bibnamefont
  {Abbott}} \emph {et~al.} (\bibinfo {collaboration} {LIGO Scientific,
  Virgo}),\ }\bibfield  {title} {\bibinfo {title} {{GW170817: Measurements of
  neutron star radii and equation of state}},\ }\href
  {https://doi.org/10.1103/PhysRevLett.121.161101} {\bibfield  {journal}
  {\bibinfo  {journal} {Phys. Rev. Lett.}\ }\textbf {\bibinfo {volume} {121}},\
  \bibinfo {pages} {161101} (\bibinfo {year} {2018}{\natexlab{b}})},\ \Eprint
  {https://arxiv.org/abs/1805.11581} {arXiv:1805.11581 [gr-qc]} \BibitemShut
  {NoStop}%
\bibitem [{\citenamefont {Radice}\ and\ \citenamefont
  {Dai}(2019)}]{Radice2019Apr}%
  \BibitemOpen
  \bibfield  {author} {\bibinfo {author} {\bibfnamefont {D.}~\bibnamefont
  {Radice}}\ and\ \bibinfo {author} {\bibfnamefont {L.}~\bibnamefont {Dai}},\
  }\bibfield  {title} {\bibinfo {title} {Multimessenger parameter estimation of
  gw170817},\ }\href {https://doi.org/10.1140/epja/i2019-12716-4} {\bibfield
  {journal} {\bibinfo  {journal} {The European Physical Journal A}\ }\textbf
  {\bibinfo {volume} {55}},\ \bibinfo {pages} {50} (\bibinfo {year}
  {2019})}\BibitemShut {NoStop}%
\bibitem [{\citenamefont {{Capano}}\ \emph {et~al.}(2020)\citenamefont
  {{Capano}}, \citenamefont {{Tews}}, \citenamefont {{Brown}}, \citenamefont
  {{Margalit}}, \citenamefont {{De}}, \citenamefont {{Kumar}}, \citenamefont
  {{Brown}}, \citenamefont {{Krishnan}},\ and\ \citenamefont
  {{Reddy}}}]{2020NatAs...4..625C}%
  \BibitemOpen
  \bibfield  {author} {\bibinfo {author} {\bibfnamefont {C.~D.}\ \bibnamefont
  {{Capano}}}, \bibinfo {author} {\bibfnamefont {I.}~\bibnamefont {{Tews}}},
  \bibinfo {author} {\bibfnamefont {S.~M.}\ \bibnamefont {{Brown}}}, \bibinfo
  {author} {\bibfnamefont {B.}~\bibnamefont {{Margalit}}}, \bibinfo {author}
  {\bibfnamefont {S.}~\bibnamefont {{De}}}, \bibinfo {author} {\bibfnamefont
  {S.}~\bibnamefont {{Kumar}}}, \bibinfo {author} {\bibfnamefont {D.~A.}\
  \bibnamefont {{Brown}}}, \bibinfo {author} {\bibfnamefont {B.}~\bibnamefont
  {{Krishnan}}},\ and\ \bibinfo {author} {\bibfnamefont {S.}~\bibnamefont
  {{Reddy}}},\ }\bibfield  {title} {\bibinfo {title} {{Stringent constraints on
  neutron-star radii from multimessenger observations and nuclear theory}},\
  }\href {https://doi.org/10.1038/s41550-020-1014-6} {\bibfield  {journal}
  {\bibinfo  {journal} {Nature Astronomy}\ }\textbf {\bibinfo {volume} {4}},\
  \bibinfo {pages} {625} (\bibinfo {year} {2020})},\ \Eprint
  {https://arxiv.org/abs/1908.10352} {arXiv:1908.10352 [astro-ph.HE]}
  \BibitemShut {NoStop}%
\bibitem [{\citenamefont {{Dietrich}}\ \emph {et~al.}(2020)\citenamefont
  {{Dietrich}}, \citenamefont {{Coughlin}}, \citenamefont {{Pang}},
  \citenamefont {{Bulla}}, \citenamefont {{Heinzel}}, \citenamefont {{Issa}},
  \citenamefont {{Tews}},\ and\ \citenamefont
  {{Antier}}}]{2020Sci...370.1450D}%
  \BibitemOpen
  \bibfield  {author} {\bibinfo {author} {\bibfnamefont {T.}~\bibnamefont
  {{Dietrich}}}, \bibinfo {author} {\bibfnamefont {M.~W.}\ \bibnamefont
  {{Coughlin}}}, \bibinfo {author} {\bibfnamefont {P.~T.~H.}\ \bibnamefont
  {{Pang}}}, \bibinfo {author} {\bibfnamefont {M.}~\bibnamefont {{Bulla}}},
  \bibinfo {author} {\bibfnamefont {J.}~\bibnamefont {{Heinzel}}}, \bibinfo
  {author} {\bibfnamefont {L.}~\bibnamefont {{Issa}}}, \bibinfo {author}
  {\bibfnamefont {I.}~\bibnamefont {{Tews}}},\ and\ \bibinfo {author}
  {\bibfnamefont {S.}~\bibnamefont {{Antier}}},\ }\bibfield  {title} {\bibinfo
  {title} {{Multimessenger constraints on the neutron-star equation of state
  and the Hubble constant}},\ }\href {https://doi.org/10.1126/science.abb4317}
  {\bibfield  {journal} {\bibinfo  {journal} {Science}\ }\textbf {\bibinfo
  {volume} {370}},\ \bibinfo {pages} {1450} (\bibinfo {year} {2020})},\ \Eprint
  {https://arxiv.org/abs/2002.11355} {arXiv:2002.11355 [astro-ph.HE]}
  \BibitemShut {NoStop}%
\bibitem [{\citenamefont {{Landry}}\ \emph {et~al.}(2020)\citenamefont
  {{Landry}}, \citenamefont {{Essick}},\ and\ \citenamefont
  {{Chatziioannou}}}]{2020PhRvD.101l3007L}%
  \BibitemOpen
  \bibfield  {author} {\bibinfo {author} {\bibfnamefont {P.}~\bibnamefont
  {{Landry}}}, \bibinfo {author} {\bibfnamefont {R.}~\bibnamefont {{Essick}}},\
  and\ \bibinfo {author} {\bibfnamefont {K.}~\bibnamefont {{Chatziioannou}}},\
  }\bibfield  {title} {\bibinfo {title} {{Nonparametric constraints on neutron
  star matter with existing and upcoming gravitational wave and pulsar
  observations}},\ }\href {https://doi.org/10.1103/PhysRevD.101.123007}
  {\bibfield  {journal} {\bibinfo  {journal} {Phys. Rev. D}\ }\textbf {\bibinfo
  {volume} {101}},\ \bibinfo {eid} {123007} (\bibinfo {year} {2020})},\ \Eprint
  {https://arxiv.org/abs/2003.04880} {arXiv:2003.04880 [astro-ph.HE]}
  \BibitemShut {NoStop}%
\bibitem [{\citenamefont {{Breschi}}\ \emph
  {et~al.}(2021{\natexlab{a}})\citenamefont {{Breschi}}, \citenamefont
  {{Perego}}, \citenamefont {{Bernuzzi}}, \citenamefont {{Del Pozzo}},
  \citenamefont {{Nedora}}, \citenamefont {{Radice}},\ and\ \citenamefont
  {{Vescovi}}}]{2021arXiv210101201B}%
  \BibitemOpen
  \bibfield  {author} {\bibinfo {author} {\bibfnamefont {M.}~\bibnamefont
  {{Breschi}}}, \bibinfo {author} {\bibfnamefont {A.}~\bibnamefont {{Perego}}},
  \bibinfo {author} {\bibfnamefont {S.}~\bibnamefont {{Bernuzzi}}}, \bibinfo
  {author} {\bibfnamefont {W.}~\bibnamefont {{Del Pozzo}}}, \bibinfo {author}
  {\bibfnamefont {V.}~\bibnamefont {{Nedora}}}, \bibinfo {author}
  {\bibfnamefont {D.}~\bibnamefont {{Radice}}},\ and\ \bibinfo {author}
  {\bibfnamefont {D.}~\bibnamefont {{Vescovi}}},\ }\bibfield  {title} {\bibinfo
  {title} {{AT2017gfo: Bayesian inference and model selection of
  multi-component kilonovae and constraints on the neutron star equation of
  state}},\ }\href@noop {} {\bibfield  {journal} {\bibinfo  {journal} {arXiv
  e-prints}\ ,\ \bibinfo {eid} {arXiv:2101.01201}} (\bibinfo {year}
  {2021}{\natexlab{a}})},\ \Eprint {https://arxiv.org/abs/2101.01201}
  {arXiv:2101.01201 [astro-ph.HE]} \BibitemShut {NoStop}%
\bibitem [{\citenamefont {{Raaijmakers}}\ \emph {et~al.}(2021)\citenamefont
  {{Raaijmakers}}, \citenamefont {{Greif}}, \citenamefont {{Hebeler}},
  \citenamefont {{Hinderer}}, \citenamefont {{Nissanke}}, \citenamefont
  {{Schwenk}}, \citenamefont {{Riley}}, \citenamefont {{Watts}}, \citenamefont
  {{Lattimer}},\ and\ \citenamefont {{Ho}}}]{Raaijmakers2021sep}%
  \BibitemOpen
  \bibfield  {author} {\bibinfo {author} {\bibfnamefont {G.}~\bibnamefont
  {{Raaijmakers}}}, \bibinfo {author} {\bibfnamefont {S.~K.}\ \bibnamefont
  {{Greif}}}, \bibinfo {author} {\bibfnamefont {K.}~\bibnamefont {{Hebeler}}},
  \bibinfo {author} {\bibfnamefont {T.}~\bibnamefont {{Hinderer}}}, \bibinfo
  {author} {\bibfnamefont {S.}~\bibnamefont {{Nissanke}}}, \bibinfo {author}
  {\bibfnamefont {A.}~\bibnamefont {{Schwenk}}}, \bibinfo {author}
  {\bibfnamefont {T.~E.}\ \bibnamefont {{Riley}}}, \bibinfo {author}
  {\bibfnamefont {A.~L.}\ \bibnamefont {{Watts}}}, \bibinfo {author}
  {\bibfnamefont {J.~M.}\ \bibnamefont {{Lattimer}}},\ and\ \bibinfo {author}
  {\bibfnamefont {W.~C.~G.}\ \bibnamefont {{Ho}}},\ }\bibfield  {title}
  {\bibinfo {title} {{Constraints on the Dense Matter Equation of State and
  Neutron Star Properties from NICER's Mass-Radius Estimate of PSR J0740+6620
  and Multimessenger Observations}},\ }\href
  {https://doi.org/10.3847/2041-8213/ac089a} {\bibfield  {journal} {\bibinfo
  {journal} {\apjl}\ }\textbf {\bibinfo {volume} {918}},\ \bibinfo {eid} {L29}
  (\bibinfo {year} {2021})},\ \Eprint {https://arxiv.org/abs/2105.06981}
  {arXiv:2105.06981 [astro-ph.HE]} \BibitemShut {NoStop}%
\bibitem [{\citenamefont {{Legred}}\ \emph {et~al.}(2021)\citenamefont
  {{Legred}}, \citenamefont {{Chatziioannou}}, \citenamefont {{Essick}},
  \citenamefont {{Han}},\ and\ \citenamefont {{Landry}}}]{2021PhRvD.104f3003L}%
  \BibitemOpen
  \bibfield  {author} {\bibinfo {author} {\bibfnamefont {I.}~\bibnamefont
  {{Legred}}}, \bibinfo {author} {\bibfnamefont {K.}~\bibnamefont
  {{Chatziioannou}}}, \bibinfo {author} {\bibfnamefont {R.}~\bibnamefont
  {{Essick}}}, \bibinfo {author} {\bibfnamefont {S.}~\bibnamefont {{Han}}},\
  and\ \bibinfo {author} {\bibfnamefont {P.}~\bibnamefont {{Landry}}},\
  }\bibfield  {title} {\bibinfo {title} {{Impact of the PSR J 0740+6620 radius
  constraint on the properties of high-density matter}},\ }\href
  {https://doi.org/10.1103/PhysRevD.104.063003} {\bibfield  {journal} {\bibinfo
   {journal} {\prd}\ }\textbf {\bibinfo {volume} {104}},\ \bibinfo {eid}
  {063003} (\bibinfo {year} {2021})},\ \Eprint
  {https://arxiv.org/abs/2106.05313} {arXiv:2106.05313 [astro-ph.HE]}
  \BibitemShut {NoStop}%
\bibitem [{\citenamefont {{Pang}}\ \emph {et~al.}(2021)\citenamefont {{Pang}},
  \citenamefont {{Tews}}, \citenamefont {{Coughlin}}, \citenamefont {{Bulla}},
  \citenamefont {{Van Den Broeck}},\ and\ \citenamefont
  {{Dietrich}}}]{2021arXiv210508688P}%
  \BibitemOpen
  \bibfield  {author} {\bibinfo {author} {\bibfnamefont {P.~T.~H.}\
  \bibnamefont {{Pang}}}, \bibinfo {author} {\bibfnamefont {I.}~\bibnamefont
  {{Tews}}}, \bibinfo {author} {\bibfnamefont {M.~W.}\ \bibnamefont
  {{Coughlin}}}, \bibinfo {author} {\bibfnamefont {M.}~\bibnamefont {{Bulla}}},
  \bibinfo {author} {\bibfnamefont {C.}~\bibnamefont {{Van Den Broeck}}},\ and\
  \bibinfo {author} {\bibfnamefont {T.}~\bibnamefont {{Dietrich}}},\ }\bibfield
   {title} {\bibinfo {title} {{Nuclear-Physics Multi-Messenger Astrophysics
  Constraints on the Neutron-Star Equation of State: Adding NICER's PSR
  J0740+6620 Measurement}},\ }\href@noop {} {\bibfield  {journal} {\bibinfo
  {journal} {arXiv e-prints}\ ,\ \bibinfo {eid} {arXiv:2105.08688}} (\bibinfo
  {year} {2021})},\ \Eprint {https://arxiv.org/abs/2105.08688}
  {arXiv:2105.08688 [astro-ph.HE]} \BibitemShut {NoStop}%
\bibitem [{\citenamefont {{Nicholl}}\ \emph {et~al.}(2021)\citenamefont
  {{Nicholl}}, \citenamefont {{Margalit}}, \citenamefont {{Schmidt}},
  \citenamefont {{Smith}}, \citenamefont {{Ridley}},\ and\ \citenamefont
  {{Nuttall}}}]{2021MNRAS.505.3016N}%
  \BibitemOpen
  \bibfield  {author} {\bibinfo {author} {\bibfnamefont {M.}~\bibnamefont
  {{Nicholl}}}, \bibinfo {author} {\bibfnamefont {B.}~\bibnamefont
  {{Margalit}}}, \bibinfo {author} {\bibfnamefont {P.}~\bibnamefont
  {{Schmidt}}}, \bibinfo {author} {\bibfnamefont {G.~P.}\ \bibnamefont
  {{Smith}}}, \bibinfo {author} {\bibfnamefont {E.~J.}\ \bibnamefont
  {{Ridley}}},\ and\ \bibinfo {author} {\bibfnamefont {J.}~\bibnamefont
  {{Nuttall}}},\ }\bibfield  {title} {\bibinfo {title} {{Tight multimessenger
  constraints on the neutron star equation of state from GW170817 and a forward
  model for kilonova light-curve synthesis}},\ }\href
  {https://doi.org/10.1093/mnras/stab1523} {\bibfield  {journal} {\bibinfo
  {journal} {\mnras}\ }\textbf {\bibinfo {volume} {505}},\ \bibinfo {pages}
  {3016} (\bibinfo {year} {2021})},\ \Eprint {https://arxiv.org/abs/2102.02229}
  {arXiv:2102.02229 [astro-ph.HE]} \BibitemShut {NoStop}%
\bibitem [{\citenamefont {{Del Pozzo}}\ \emph {et~al.}(2013)\citenamefont {{Del
  Pozzo}}, \citenamefont {Li}, \citenamefont {Agathos}, \citenamefont {{Van Den
  Broeck}},\ and\ \citenamefont {Vitale}}]{DelPozzo2013}%
  \BibitemOpen
  \bibfield  {author} {\bibinfo {author} {\bibfnamefont {W.}~\bibnamefont {{Del
  Pozzo}}}, \bibinfo {author} {\bibfnamefont {T.~G.~F.}\ \bibnamefont {Li}},
  \bibinfo {author} {\bibfnamefont {M.}~\bibnamefont {Agathos}}, \bibinfo
  {author} {\bibfnamefont {C.}~\bibnamefont {{Van Den Broeck}}},\ and\ \bibinfo
  {author} {\bibfnamefont {S.}~\bibnamefont {Vitale}},\ }\bibfield  {title}
  {\bibinfo {title} {{Demonstrating the Feasibility of Probing the Neutron-Star
  Equation of State with Second-Generation Gravitational-Wave Detectors}},\
  }\href {https://doi.org/10.1103/PhysRevLett.111.071101} {\bibfield  {journal}
  {\bibinfo  {journal} {Phys. Rev. Lett.}\ }\textbf {\bibinfo {volume} {111}},\
  \bibinfo {pages} {071101} (\bibinfo {year} {2013})}\BibitemShut {NoStop}%
\bibitem [{\citenamefont {{Chatziioannou}}\ \emph {et~al.}(2015)\citenamefont
  {{Chatziioannou}}, \citenamefont {{Yagi}}, \citenamefont {{Klein}},
  \citenamefont {{Cornish}},\ and\ \citenamefont
  {{Yunes}}}]{2015PhRvD..92j4008C}%
  \BibitemOpen
  \bibfield  {author} {\bibinfo {author} {\bibfnamefont {K.}~\bibnamefont
  {{Chatziioannou}}}, \bibinfo {author} {\bibfnamefont {K.}~\bibnamefont
  {{Yagi}}}, \bibinfo {author} {\bibfnamefont {A.}~\bibnamefont {{Klein}}},
  \bibinfo {author} {\bibfnamefont {N.}~\bibnamefont {{Cornish}}},\ and\
  \bibinfo {author} {\bibfnamefont {N.}~\bibnamefont {{Yunes}}},\ }\bibfield
  {title} {\bibinfo {title} {{Probing the internal composition of neutron stars
  with gravitational waves}},\ }\href
  {https://doi.org/10.1103/PhysRevD.92.104008} {\bibfield  {journal} {\bibinfo
  {journal} {Phys. Rev. D}\ }\textbf {\bibinfo {volume} {92}},\ \bibinfo {eid}
  {104008} (\bibinfo {year} {2015})},\ \Eprint
  {https://arxiv.org/abs/1508.02062} {arXiv:1508.02062 [gr-qc]} \BibitemShut
  {NoStop}%
\bibitem [{\citenamefont {{Lackey}}\ and\ \citenamefont
  {{Wade}}(2015)}]{2015PhRvD..91d3002L}%
  \BibitemOpen
  \bibfield  {author} {\bibinfo {author} {\bibfnamefont {B.~D.}\ \bibnamefont
  {{Lackey}}}\ and\ \bibinfo {author} {\bibfnamefont {L.}~\bibnamefont
  {{Wade}}},\ }\bibfield  {title} {\bibinfo {title} {{Reconstructing the
  neutron-star equation of state with gravitational-wave detectors from a
  realistic population of inspiralling binary neutron stars}},\ }\href
  {https://doi.org/10.1103/PhysRevD.91.043002} {\bibfield  {journal} {\bibinfo
  {journal} {Phys. Rev. D}\ }\textbf {\bibinfo {volume} {91}},\ \bibinfo {eid}
  {043002} (\bibinfo {year} {2015})},\ \Eprint
  {https://arxiv.org/abs/1410.8866} {arXiv:1410.8866 [gr-qc]} \BibitemShut
  {NoStop}%
\bibitem [{\citenamefont {{Hernandez Vivanco}}\ \emph
  {et~al.}(2019)\citenamefont {{Hernandez Vivanco}}, \citenamefont {{Smith}},
  \citenamefont {{Thrane}}, \citenamefont {{Lasky}}, \citenamefont {{Talbot}},\
  and\ \citenamefont {{Raymond}}}]{2019PhRvD.100j3009H}%
  \BibitemOpen
  \bibfield  {author} {\bibinfo {author} {\bibfnamefont {F.}~\bibnamefont
  {{Hernandez Vivanco}}}, \bibinfo {author} {\bibfnamefont {R.}~\bibnamefont
  {{Smith}}}, \bibinfo {author} {\bibfnamefont {E.}~\bibnamefont {{Thrane}}},
  \bibinfo {author} {\bibfnamefont {P.~D.}\ \bibnamefont {{Lasky}}}, \bibinfo
  {author} {\bibfnamefont {C.}~\bibnamefont {{Talbot}}},\ and\ \bibinfo
  {author} {\bibfnamefont {V.}~\bibnamefont {{Raymond}}},\ }\bibfield  {title}
  {\bibinfo {title} {{Measuring the neutron star equation of state with
  gravitational waves: The first forty binary neutron star merger
  observations}},\ }\href {https://doi.org/10.1103/PhysRevD.100.103009}
  {\bibfield  {journal} {\bibinfo  {journal} {Phys. Rev. D}\ }\textbf {\bibinfo
  {volume} {100}},\ \bibinfo {eid} {103009} (\bibinfo {year} {2019})},\ \Eprint
  {https://arxiv.org/abs/1909.02698} {arXiv:1909.02698 [gr-qc]} \BibitemShut
  {NoStop}%
\bibitem [{\citenamefont {{Chatziioannou}}\ and\ \citenamefont
  {{Han}}(2020)}]{2020PhRvD.101d4019C}%
  \BibitemOpen
  \bibfield  {author} {\bibinfo {author} {\bibfnamefont {K.}~\bibnamefont
  {{Chatziioannou}}}\ and\ \bibinfo {author} {\bibfnamefont {S.}~\bibnamefont
  {{Han}}},\ }\bibfield  {title} {\bibinfo {title} {{Studying strong phase
  transitions in neutron stars with gravitational waves}},\ }\href
  {https://doi.org/10.1103/PhysRevD.101.044019} {\bibfield  {journal} {\bibinfo
   {journal} {Phys. Rev. D}\ }\textbf {\bibinfo {volume} {101}},\ \bibinfo
  {eid} {044019} (\bibinfo {year} {2020})},\ \Eprint
  {https://arxiv.org/abs/1911.07091} {arXiv:1911.07091 [gr-qc]} \BibitemShut
  {NoStop}%
\bibitem [{\citenamefont {{Abbott}}\ \emph
  {et~al.}(2017{\natexlab{b}})\citenamefont {{Abbott}}, \citenamefont
  {{Abbott}}, \citenamefont {{Abbott}}, \citenamefont {{Acernese}},
  \citenamefont {{Ackley}}, \citenamefont {{Adams}}, \citenamefont {{Adams}},
  \citenamefont {{Addesso}}, \citenamefont {{Adhikari}}, \citenamefont {{Adya}}
  \emph {et~al.}}]{2017ApJ...851L..16A}%
  \BibitemOpen
  \bibfield  {author} {\bibinfo {author} {\bibfnamefont {B.~P.}\ \bibnamefont
  {{Abbott}}}, \bibinfo {author} {\bibfnamefont {R.}~\bibnamefont {{Abbott}}},
  \bibinfo {author} {\bibfnamefont {T.~D.}\ \bibnamefont {{Abbott}}}, \bibinfo
  {author} {\bibfnamefont {F.}~\bibnamefont {{Acernese}}}, \bibinfo {author}
  {\bibfnamefont {K.}~\bibnamefont {{Ackley}}}, \bibinfo {author}
  {\bibfnamefont {C.}~\bibnamefont {{Adams}}}, \bibinfo {author} {\bibfnamefont
  {T.}~\bibnamefont {{Adams}}}, \bibinfo {author} {\bibfnamefont
  {P.}~\bibnamefont {{Addesso}}}, \bibinfo {author} {\bibfnamefont {R.~X.}\
  \bibnamefont {{Adhikari}}}, \bibinfo {author} {\bibfnamefont {V.~B.}\
  \bibnamefont {{Adya}}}, \emph {et~al.} (\bibinfo {collaboration} {LIGO
  Scientific Collaboration and Virgo Collaboration}),\ }\bibfield  {title}
  {\bibinfo {title} {{Search for Post-merger Gravitational Waves from the
  Remnant of the Binary Neutron Star Merger GW170817}},\ }\href
  {https://doi.org/10.3847/2041-8213/aa9a35} {\bibfield  {journal} {\bibinfo
  {journal} {Astrophys. J. Lett.}\ }\textbf {\bibinfo {volume} {851}},\
  \bibinfo {eid} {L16} (\bibinfo {year} {2017}{\natexlab{b}})},\ \Eprint
  {https://arxiv.org/abs/1710.09320} {arXiv:1710.09320 [astro-ph.HE]}
  \BibitemShut {NoStop}%
\bibitem [{\citenamefont {Abbott}\ \emph {et~al.}(2020)\citenamefont {Abbott}
  \emph {et~al.}}]{KAGRA:2020npa}%
  \BibitemOpen
  \bibfield  {author} {\bibinfo {author} {\bibfnamefont {B.~P.}\ \bibnamefont
  {Abbott}} \emph {et~al.} (\bibinfo {collaboration} {KAGRA, LIGO Scientific,
  Virgo}),\ }\bibfield  {title} {\bibinfo {title} {{Prospects for observing and
  localizing gravitational-wave transients with Advanced LIGO, Advanced Virgo
  and KAGRA}},\ }\href {https://doi.org/10.1007/s41114-020-00026-9} {\bibfield
  {journal} {\bibinfo  {journal} {Living Rev. Rel.}\ }\textbf {\bibinfo
  {volume} {23}},\ \bibinfo {pages} {3} (\bibinfo {year} {2020})}\BibitemShut
  {NoStop}%
\bibitem [{\citenamefont {{Martynov}}\ \emph {et~al.}(2019)\citenamefont
  {{Martynov}}, \citenamefont {{Miao}}, \citenamefont {{Yang}}, \citenamefont
  {{Vivanco}}, \citenamefont {{Thrane}}, \citenamefont {{Smith}}, \citenamefont
  {{Lasky}}, \citenamefont {{East}}, \citenamefont {{Adhikari}}, \citenamefont
  {{Bauswein}}, \citenamefont {{Brooks}}, \citenamefont {{Chen}}, \citenamefont
  {{Corbitt}}, \citenamefont {{Freise}}, \citenamefont {{Grote}}, \citenamefont
  {{Levin}}, \citenamefont {{Zhao}},\ and\ \citenamefont
  {{Vecchio}}}]{2019PhRvD..99j2004M}%
  \BibitemOpen
  \bibfield  {author} {\bibinfo {author} {\bibfnamefont {D.}~\bibnamefont
  {{Martynov}}}, \bibinfo {author} {\bibfnamefont {H.}~\bibnamefont {{Miao}}},
  \bibinfo {author} {\bibfnamefont {H.}~\bibnamefont {{Yang}}}, \bibinfo
  {author} {\bibfnamefont {F.~H.}\ \bibnamefont {{Vivanco}}}, \bibinfo {author}
  {\bibfnamefont {E.}~\bibnamefont {{Thrane}}}, \bibinfo {author}
  {\bibfnamefont {R.}~\bibnamefont {{Smith}}}, \bibinfo {author} {\bibfnamefont
  {P.}~\bibnamefont {{Lasky}}}, \bibinfo {author} {\bibfnamefont {W.~E.}\
  \bibnamefont {{East}}}, \bibinfo {author} {\bibfnamefont {R.}~\bibnamefont
  {{Adhikari}}}, \bibinfo {author} {\bibfnamefont {A.}~\bibnamefont
  {{Bauswein}}}, \bibinfo {author} {\bibfnamefont {A.}~\bibnamefont
  {{Brooks}}}, \bibinfo {author} {\bibfnamefont {Y.}~\bibnamefont {{Chen}}},
  \bibinfo {author} {\bibfnamefont {T.}~\bibnamefont {{Corbitt}}}, \bibinfo
  {author} {\bibfnamefont {A.}~\bibnamefont {{Freise}}}, \bibinfo {author}
  {\bibfnamefont {H.}~\bibnamefont {{Grote}}}, \bibinfo {author} {\bibfnamefont
  {Y.}~\bibnamefont {{Levin}}}, \bibinfo {author} {\bibfnamefont
  {C.}~\bibnamefont {{Zhao}}},\ and\ \bibinfo {author} {\bibfnamefont
  {A.}~\bibnamefont {{Vecchio}}},\ }\bibfield  {title} {\bibinfo {title}
  {{Exploring the sensitivity of gravitational wave detectors to neutron star
  physics}},\ }\href {https://doi.org/10.1103/PhysRevD.99.102004} {\bibfield
  {journal} {\bibinfo  {journal} {\prd}\ }\textbf {\bibinfo {volume} {99}},\
  \bibinfo {eid} {102004} (\bibinfo {year} {2019})},\ \Eprint
  {https://arxiv.org/abs/1901.03885} {arXiv:1901.03885 [astro-ph.IM]}
  \BibitemShut {NoStop}%
\bibitem [{\citenamefont {Ackley}\ \emph {et~al.}(2020)\citenamefont {Ackley}
  \emph {et~al.}}]{Ackley:2020atn}%
  \BibitemOpen
  \bibfield  {author} {\bibinfo {author} {\bibfnamefont {K.}~\bibnamefont
  {Ackley}} \emph {et~al.},\ }\bibfield  {title} {\bibinfo {title} {{Neutron
  Star Extreme Matter Observatory: A kilohertz-band gravitational-wave detector
  in the global network}},\ }\href {https://doi.org/10.1017/pasa.2020.39}
  {\bibfield  {journal} {\bibinfo  {journal} {Publ. Astron. Soc. Austral.}\
  }\textbf {\bibinfo {volume} {37}},\ \bibinfo {pages} {e047} (\bibinfo {year}
  {2020})},\ \Eprint {https://arxiv.org/abs/2007.03128} {arXiv:2007.03128
  [astro-ph.HE]} \BibitemShut {NoStop}%
\bibitem [{\citenamefont {{Ganapathy}}\ \emph {et~al.}(2021)\citenamefont
  {{Ganapathy}}, \citenamefont {{McCuller}}, \citenamefont {{Rollins}},
  \citenamefont {{Hall}}, \citenamefont {{Barsotti}},\ and\ \citenamefont
  {{Evans}}}]{2021PhRvD.103b2002G}%
  \BibitemOpen
  \bibfield  {author} {\bibinfo {author} {\bibfnamefont {D.}~\bibnamefont
  {{Ganapathy}}}, \bibinfo {author} {\bibfnamefont {L.}~\bibnamefont
  {{McCuller}}}, \bibinfo {author} {\bibfnamefont {J.~G.}\ \bibnamefont
  {{Rollins}}}, \bibinfo {author} {\bibfnamefont {E.~D.}\ \bibnamefont
  {{Hall}}}, \bibinfo {author} {\bibfnamefont {L.}~\bibnamefont {{Barsotti}}},\
  and\ \bibinfo {author} {\bibfnamefont {M.}~\bibnamefont {{Evans}}},\
  }\bibfield  {title} {\bibinfo {title} {{Tuning Advanced LIGO to kilohertz
  signals from neutron-star collisions}},\ }\href
  {https://doi.org/10.1103/PhysRevD.103.022002} {\bibfield  {journal} {\bibinfo
   {journal} {Phys. Rev. D}\ }\textbf {\bibinfo {volume} {103}},\ \bibinfo
  {eid} {022002} (\bibinfo {year} {2021})},\ \Eprint
  {https://arxiv.org/abs/2010.15735} {arXiv:2010.15735 [astro-ph.IM]}
  \BibitemShut {NoStop}%
\bibitem [{\citenamefont {{Page}}\ \emph {et~al.}(2021)\citenamefont {{Page}},
  \citenamefont {{Goryachev}}, \citenamefont {{Miao}}, \citenamefont {{Chen}},
  \citenamefont {{Ma}}, \citenamefont {{Mason}}, \citenamefont {{Rossi}},
  \citenamefont {{Blair}}, \citenamefont {{Ju}}, \citenamefont {{Blair}},
  \citenamefont {{Schliesser}}, \citenamefont {{Tobar}},\ and\ \citenamefont
  {{Zhao}}}]{2021CmPhy...4...27P}%
  \BibitemOpen
  \bibfield  {author} {\bibinfo {author} {\bibfnamefont {M.~A.}\ \bibnamefont
  {{Page}}}, \bibinfo {author} {\bibfnamefont {M.}~\bibnamefont {{Goryachev}}},
  \bibinfo {author} {\bibfnamefont {H.}~\bibnamefont {{Miao}}}, \bibinfo
  {author} {\bibfnamefont {Y.}~\bibnamefont {{Chen}}}, \bibinfo {author}
  {\bibfnamefont {Y.}~\bibnamefont {{Ma}}}, \bibinfo {author} {\bibfnamefont
  {D.}~\bibnamefont {{Mason}}}, \bibinfo {author} {\bibfnamefont
  {M.}~\bibnamefont {{Rossi}}}, \bibinfo {author} {\bibfnamefont {C.~D.}\
  \bibnamefont {{Blair}}}, \bibinfo {author} {\bibfnamefont {L.}~\bibnamefont
  {{Ju}}}, \bibinfo {author} {\bibfnamefont {D.~G.}\ \bibnamefont {{Blair}}},
  \bibinfo {author} {\bibfnamefont {A.}~\bibnamefont {{Schliesser}}}, \bibinfo
  {author} {\bibfnamefont {M.~E.}\ \bibnamefont {{Tobar}}},\ and\ \bibinfo
  {author} {\bibfnamefont {C.}~\bibnamefont {{Zhao}}},\ }\bibfield  {title}
  {\bibinfo {title} {{Gravitational wave detectors with broadband high
  frequency sensitivity}},\ }\href {https://doi.org/10.1038/s42005-021-00526-2}
  {\bibfield  {journal} {\bibinfo  {journal} {Communications Physics}\ }\textbf
  {\bibinfo {volume} {4}},\ \bibinfo {eid} {27} (\bibinfo {year} {2021})},\
  \Eprint {https://arxiv.org/abs/2007.08766} {arXiv:2007.08766
  [physics.optics]} \BibitemShut {NoStop}%
\bibitem [{\citenamefont {{Sarin}}\ and\ \citenamefont
  {{Lasky}}(2021{\natexlab{a}})}]{2021arXiv211010892S}%
  \BibitemOpen
  \bibfield  {author} {\bibinfo {author} {\bibfnamefont {N.}~\bibnamefont
  {{Sarin}}}\ and\ \bibinfo {author} {\bibfnamefont {P.~D.}\ \bibnamefont
  {{Lasky}}},\ }\bibfield  {title} {\bibinfo {title} {{Multimessenger astronomy
  with a kHz-band gravitational-wave observatory}},\ }\href@noop {} {\bibfield
  {journal} {\bibinfo  {journal} {arXiv e-prints}\ ,\ \bibinfo {eid}
  {arXiv:2110.10892}} (\bibinfo {year} {2021}{\natexlab{a}})},\ \Eprint
  {https://arxiv.org/abs/2110.10892} {arXiv:2110.10892 [astro-ph.HE]}
  \BibitemShut {NoStop}%
\bibitem [{\citenamefont {Abbott}\ \emph
  {et~al.}(2017{\natexlab{a}})\citenamefont {Abbott} \emph
  {et~al.}}]{LIGOScientific:2016wof}%
  \BibitemOpen
  \bibfield  {author} {\bibinfo {author} {\bibfnamefont {B.~P.}\ \bibnamefont
  {Abbott}} \emph {et~al.} (\bibinfo {collaboration} {LIGO Scientific}),\
  }\bibfield  {title} {\bibinfo {title} {{Exploring the Sensitivity of Next
  Generation Gravitational Wave Detectors}},\ }\href
  {https://doi.org/10.1088/1361-6382/aa51f4} {\bibfield  {journal} {\bibinfo
  {journal} {Class. Quant. Grav.}\ }\textbf {\bibinfo {volume} {34}},\ \bibinfo
  {pages} {044001} (\bibinfo {year} {2017}{\natexlab{a}})},\ \Eprint
  {https://arxiv.org/abs/1607.08697} {arXiv:1607.08697 [astro-ph.IM]}
  \BibitemShut {NoStop}%
\bibitem [{\citenamefont {Maggiore}\ \emph {et~al.}(2020)\citenamefont
  {Maggiore} \emph {et~al.}}]{Maggiore:2019uih}%
  \BibitemOpen
  \bibfield  {author} {\bibinfo {author} {\bibfnamefont {M.}~\bibnamefont
  {Maggiore}} \emph {et~al.},\ }\bibfield  {title} {\bibinfo {title} {{Science
  Case for the Einstein Telescope}},\ }\href
  {https://doi.org/10.1088/1475-7516/2020/03/050} {\bibfield  {journal}
  {\bibinfo  {journal} {JCAP}\ }\textbf {\bibinfo {volume} {03}},\ \bibinfo
  {pages} {050}},\ \Eprint {https://arxiv.org/abs/1912.02622} {arXiv:1912.02622
  [astro-ph.CO]} \BibitemShut {NoStop}%
\bibitem [{\citenamefont {{Rasio}}\ and\ \citenamefont
  {{Shapiro}}(1992)}]{1992ApJ...401..226R}%
  \BibitemOpen
  \bibfield  {author} {\bibinfo {author} {\bibfnamefont {F.~A.}\ \bibnamefont
  {{Rasio}}}\ and\ \bibinfo {author} {\bibfnamefont {S.~L.}\ \bibnamefont
  {{Shapiro}}},\ }\bibfield  {title} {\bibinfo {title} {{Hydrodynamical
  Evolution of Coalescing Binary Neutron Stars}},\ }\href
  {https://doi.org/10.1086/172055} {\bibfield  {journal} {\bibinfo  {journal}
  {Astrophys. J.}\ }\textbf {\bibinfo {volume} {401}},\ \bibinfo {pages} {226}
  (\bibinfo {year} {1992})}\BibitemShut {NoStop}%
\bibitem [{\citenamefont {{Shibata}}(2005)}]{2005PhRvL..94t1101S}%
  \BibitemOpen
  \bibfield  {author} {\bibinfo {author} {\bibfnamefont {M.}~\bibnamefont
  {{Shibata}}},\ }\bibfield  {title} {\bibinfo {title} {{Constraining Nuclear
  Equations of State Using Gravitational Waves from Hypermassive Neutron
  Stars}},\ }\href {https://doi.org/10.1103/PhysRevLett.94.201101} {\bibfield
  {journal} {\bibinfo  {journal} {Phys. Rev. Lett.}\ }\textbf {\bibinfo
  {volume} {94}},\ \bibinfo {eid} {201101} (\bibinfo {year} {2005})},\ \Eprint
  {https://arxiv.org/abs/gr-qc/0504082} {arXiv:gr-qc/0504082 [gr-qc]}
  \BibitemShut {NoStop}%
\bibitem [{\citenamefont {Bauswein}\ and\ \citenamefont
  {Janka}(2012{\natexlab{a}})}]{PhysRevLett.108.011101}%
  \BibitemOpen
  \bibfield  {author} {\bibinfo {author} {\bibfnamefont {A.}~\bibnamefont
  {Bauswein}}\ and\ \bibinfo {author} {\bibfnamefont {H.-T.}\ \bibnamefont
  {Janka}},\ }\bibfield  {title} {\bibinfo {title} {Measuring neutron-star
  properties via gravitational waves from neutron-star mergers},\ }\href
  {https://doi.org/10.1103/PhysRevLett.108.011101} {\bibfield  {journal}
  {\bibinfo  {journal} {Phys. Rev. Lett.}\ }\textbf {\bibinfo {volume} {108}},\
  \bibinfo {pages} {011101} (\bibinfo {year} {2012}{\natexlab{a}})}\BibitemShut
  {NoStop}%
\bibitem [{\citenamefont {Hotokezaka}\ \emph {et~al.}(2013)\citenamefont
  {Hotokezaka}, \citenamefont {Kiuchi}, \citenamefont {Kyutoku}, \citenamefont
  {Muranushi}, \citenamefont {Sekiguchi}, \citenamefont {Shibata},\ and\
  \citenamefont {Taniguchi}}]{Hotokezaka2013}%
  \BibitemOpen
  \bibfield  {author} {\bibinfo {author} {\bibfnamefont {K.}~\bibnamefont
  {Hotokezaka}}, \bibinfo {author} {\bibfnamefont {K.}~\bibnamefont {Kiuchi}},
  \bibinfo {author} {\bibfnamefont {K.}~\bibnamefont {Kyutoku}}, \bibinfo
  {author} {\bibfnamefont {T.}~\bibnamefont {Muranushi}}, \bibinfo {author}
  {\bibfnamefont {Y.-i.}\ \bibnamefont {Sekiguchi}}, \bibinfo {author}
  {\bibfnamefont {M.}~\bibnamefont {Shibata}},\ and\ \bibinfo {author}
  {\bibfnamefont {K.}~\bibnamefont {Taniguchi}},\ }\bibfield  {title} {\bibinfo
  {title} {Remnant massive neutron stars of binary neutron star mergers:
  Evolution process and gravitational waveform},\ }\href
  {https://doi.org/10.1103/PhysRevD.88.044026} {\bibfield  {journal} {\bibinfo
  {journal} {Phys. Rev. D}\ }\textbf {\bibinfo {volume} {88}},\ \bibinfo
  {pages} {044026} (\bibinfo {year} {2013})}\BibitemShut {NoStop}%
\bibitem [{\citenamefont {{Clark}}\ \emph {et~al.}(2014)\citenamefont
  {{Clark}}, \citenamefont {{Bauswein}}, \citenamefont {{Cadonati}},
  \citenamefont {{Janka}}, \citenamefont {{Pankow}},\ and\ \citenamefont
  {{Stergioulas}}}]{2014PhRvD..90f2004C}%
  \BibitemOpen
  \bibfield  {author} {\bibinfo {author} {\bibfnamefont {J.}~\bibnamefont
  {{Clark}}}, \bibinfo {author} {\bibfnamefont {A.}~\bibnamefont {{Bauswein}}},
  \bibinfo {author} {\bibfnamefont {L.}~\bibnamefont {{Cadonati}}}, \bibinfo
  {author} {\bibfnamefont {H.~T.}\ \bibnamefont {{Janka}}}, \bibinfo {author}
  {\bibfnamefont {C.}~\bibnamefont {{Pankow}}},\ and\ \bibinfo {author}
  {\bibfnamefont {N.}~\bibnamefont {{Stergioulas}}},\ }\bibfield  {title}
  {\bibinfo {title} {{Prospects for high frequency burst searches following
  binary neutron star coalescence with advanced gravitational wave
  detectors}},\ }\href {https://doi.org/10.1103/PhysRevD.90.062004} {\bibfield
  {journal} {\bibinfo  {journal} {Phys. Rev. D}\ }\textbf {\bibinfo {volume}
  {90}},\ \bibinfo {eid} {062004} (\bibinfo {year} {2014})},\ \Eprint
  {https://arxiv.org/abs/1406.5444} {arXiv:1406.5444 [astro-ph.HE]}
  \BibitemShut {NoStop}%
\bibitem [{\citenamefont {Bauswein}\ and\ \citenamefont
  {Stergioulas}(2015)}]{AB2015}%
  \BibitemOpen
  \bibfield  {author} {\bibinfo {author} {\bibfnamefont {A.}~\bibnamefont
  {Bauswein}}\ and\ \bibinfo {author} {\bibfnamefont {N.}~\bibnamefont
  {Stergioulas}},\ }\bibfield  {title} {\bibinfo {title} {Unified picture of
  the post-merger dynamics and gravitational wave emission in neutron star
  mergers},\ }\href {https://doi.org/10.1103/PhysRevD.91.124056} {\bibfield
  {journal} {\bibinfo  {journal} {Phys. Rev. D}\ }\textbf {\bibinfo {volume}
  {91}},\ \bibinfo {pages} {124056} (\bibinfo {year} {2015})}\BibitemShut
  {NoStop}%
\bibitem [{\citenamefont {Bernuzzi}\ \emph {et~al.}(2015)\citenamefont
  {Bernuzzi}, \citenamefont {Dietrich},\ and\ \citenamefont
  {Nagar}}]{Bernuzzi2015}%
  \BibitemOpen
  \bibfield  {author} {\bibinfo {author} {\bibfnamefont {S.}~\bibnamefont
  {Bernuzzi}}, \bibinfo {author} {\bibfnamefont {T.}~\bibnamefont {Dietrich}},\
  and\ \bibinfo {author} {\bibfnamefont {A.}~\bibnamefont {Nagar}},\ }\bibfield
   {title} {\bibinfo {title} {Modeling the complete gravitational wave spectrum
  of neutron star mergers},\ }\href
  {https://doi.org/10.1103/PhysRevLett.115.091101} {\bibfield  {journal}
  {\bibinfo  {journal} {Phys. Rev. Lett.}\ }\textbf {\bibinfo {volume} {115}},\
  \bibinfo {pages} {091101} (\bibinfo {year} {2015})}\BibitemShut {NoStop}%
\bibitem [{\citenamefont {{Clark}}\ \emph {et~al.}(2016)\citenamefont
  {{Clark}}, \citenamefont {{Bauswein}}, \citenamefont {{Stergioulas}},\ and\
  \citenamefont {{Shoemaker}}}]{Clark2016}%
  \BibitemOpen
  \bibfield  {author} {\bibinfo {author} {\bibfnamefont {J.~A.}\ \bibnamefont
  {{Clark}}}, \bibinfo {author} {\bibfnamefont {A.}~\bibnamefont {{Bauswein}}},
  \bibinfo {author} {\bibfnamefont {N.}~\bibnamefont {{Stergioulas}}},\ and\
  \bibinfo {author} {\bibfnamefont {D.}~\bibnamefont {{Shoemaker}}},\
  }\bibfield  {title} {\bibinfo {title} {{Observing gravitational waves from
  the post-merger phase of binary neutron star coalescence}},\ }\href
  {https://doi.org/10.1088/0264-9381/33/8/085003} {\bibfield  {journal}
  {\bibinfo  {journal} {Classical and Quantum Gravity}\ }\textbf {\bibinfo
  {volume} {33}},\ \bibinfo {eid} {085003} (\bibinfo {year} {2016})},\ \Eprint
  {https://arxiv.org/abs/1509.08522} {arXiv:1509.08522 [astro-ph.HE]}
  \BibitemShut {NoStop}%
\bibitem [{\citenamefont {{Bauswein}}\ \emph {et~al.}(2016)\citenamefont
  {{Bauswein}}, \citenamefont {{Stergioulas}},\ and\ \citenamefont
  {{Janka}}}]{AB2016}%
  \BibitemOpen
  \bibfield  {author} {\bibinfo {author} {\bibfnamefont {A.}~\bibnamefont
  {{Bauswein}}}, \bibinfo {author} {\bibfnamefont {N.}~\bibnamefont
  {{Stergioulas}}},\ and\ \bibinfo {author} {\bibfnamefont {H.-T.}\
  \bibnamefont {{Janka}}},\ }\bibfield  {title} {\bibinfo {title} {{Exploring
  properties of high-density matter through remnants of neutron-star
  mergers}},\ }\href {https://doi.org/10.1140/epja/i2016-16056-7} {\bibfield
  {journal} {\bibinfo  {journal} {European Physical Journal A}\ }\textbf
  {\bibinfo {volume} {52}},\ \bibinfo {eid} {56} (\bibinfo {year} {2016})},\
  \Eprint {https://arxiv.org/abs/1508.05493} {arXiv:1508.05493 [astro-ph.HE]}
  \BibitemShut {NoStop}%
\bibitem [{\citenamefont {Rezzolla}\ and\ \citenamefont
  {Takami}(2016)}]{PhysRevD.93.124051}%
  \BibitemOpen
  \bibfield  {author} {\bibinfo {author} {\bibfnamefont {L.}~\bibnamefont
  {Rezzolla}}\ and\ \bibinfo {author} {\bibfnamefont {K.}~\bibnamefont
  {Takami}},\ }\bibfield  {title} {\bibinfo {title} {Gravitational-wave signal
  from binary neutron stars: A systematic analysis of the spectral
  properties},\ }\href {https://doi.org/10.1103/PhysRevD.93.124051} {\bibfield
  {journal} {\bibinfo  {journal} {Phys. Rev. D}\ }\textbf {\bibinfo {volume}
  {93}},\ \bibinfo {pages} {124051} (\bibinfo {year} {2016})}\BibitemShut
  {NoStop}%
\bibitem [{\citenamefont {Foucart}\ \emph {et~al.}(2016)\citenamefont
  {Foucart}, \citenamefont {Haas}, \citenamefont {Duez}, \citenamefont
  {O'Connor}, \citenamefont {Ott}, \citenamefont {Roberts}, \citenamefont
  {Kidder}, \citenamefont {Lippuner}, \citenamefont {Pfeiffer},\ and\
  \citenamefont {Scheel}}]{Foucart2016}%
  \BibitemOpen
  \bibfield  {author} {\bibinfo {author} {\bibfnamefont {F.}~\bibnamefont
  {Foucart}}, \bibinfo {author} {\bibfnamefont {R.}~\bibnamefont {Haas}},
  \bibinfo {author} {\bibfnamefont {M.~D.}\ \bibnamefont {Duez}}, \bibinfo
  {author} {\bibfnamefont {E.}~\bibnamefont {O'Connor}}, \bibinfo {author}
  {\bibfnamefont {C.~D.}\ \bibnamefont {Ott}}, \bibinfo {author} {\bibfnamefont
  {L.}~\bibnamefont {Roberts}}, \bibinfo {author} {\bibfnamefont {L.~E.}\
  \bibnamefont {Kidder}}, \bibinfo {author} {\bibfnamefont {J.}~\bibnamefont
  {Lippuner}}, \bibinfo {author} {\bibfnamefont {H.~P.}\ \bibnamefont
  {Pfeiffer}},\ and\ \bibinfo {author} {\bibfnamefont {M.~A.}\ \bibnamefont
  {Scheel}},\ }\bibfield  {title} {\bibinfo {title} {Low mass binary neutron
  star mergers: Gravitational waves and neutrino emission},\ }\href
  {https://doi.org/10.1103/PhysRevD.93.044019} {\bibfield  {journal} {\bibinfo
  {journal} {Phys. Rev. D}\ }\textbf {\bibinfo {volume} {93}},\ \bibinfo
  {pages} {044019} (\bibinfo {year} {2016})}\BibitemShut {NoStop}%
\bibitem [{\citenamefont {{Lehner}}\ \emph {et~al.}(2016)\citenamefont
  {{Lehner}}, \citenamefont {{Liebling}}, \citenamefont {{Palenzuela}},
  \citenamefont {{Caballero}}, \citenamefont {{O'Connor}}, \citenamefont
  {{Anderson}},\ and\ \citenamefont {{Neilsen}}}]{Lehner2016}%
  \BibitemOpen
  \bibfield  {author} {\bibinfo {author} {\bibfnamefont {L.}~\bibnamefont
  {{Lehner}}}, \bibinfo {author} {\bibfnamefont {S.~L.}\ \bibnamefont
  {{Liebling}}}, \bibinfo {author} {\bibfnamefont {C.}~\bibnamefont
  {{Palenzuela}}}, \bibinfo {author} {\bibfnamefont {O.~L.}\ \bibnamefont
  {{Caballero}}}, \bibinfo {author} {\bibfnamefont {E.}~\bibnamefont
  {{O'Connor}}}, \bibinfo {author} {\bibfnamefont {M.}~\bibnamefont
  {{Anderson}}},\ and\ \bibinfo {author} {\bibfnamefont {D.}~\bibnamefont
  {{Neilsen}}},\ }\bibfield  {title} {\bibinfo {title} {{Unequal mass binary
  neutron star mergers and multimessenger signals}},\ }\href
  {https://doi.org/10.1088/0264-9381/33/18/184002} {\bibfield  {journal}
  {\bibinfo  {journal} {Classical and Quantum Gravity}\ }\textbf {\bibinfo
  {volume} {33}},\ \bibinfo {eid} {184002} (\bibinfo {year} {2016})},\ \Eprint
  {https://arxiv.org/abs/1603.00501} {arXiv:1603.00501 [gr-qc]} \BibitemShut
  {NoStop}%
\bibitem [{\citenamefont {Chatziioannou}\ \emph {et~al.}(2017)\citenamefont
  {Chatziioannou}, \citenamefont {Clark}, \citenamefont {Bauswein},
  \citenamefont {Millhouse}, \citenamefont {Littenberg},\ and\ \citenamefont
  {Cornish}}]{PhysRevD.96.124035}%
  \BibitemOpen
  \bibfield  {author} {\bibinfo {author} {\bibfnamefont {K.}~\bibnamefont
  {Chatziioannou}}, \bibinfo {author} {\bibfnamefont {J.~A.}\ \bibnamefont
  {Clark}}, \bibinfo {author} {\bibfnamefont {A.}~\bibnamefont {Bauswein}},
  \bibinfo {author} {\bibfnamefont {M.}~\bibnamefont {Millhouse}}, \bibinfo
  {author} {\bibfnamefont {T.~B.}\ \bibnamefont {Littenberg}},\ and\ \bibinfo
  {author} {\bibfnamefont {N.}~\bibnamefont {Cornish}},\ }\bibfield  {title}
  {\bibinfo {title} {Inferring the post-merger gravitational wave emission from
  binary neutron star coalescences},\ }\href
  {https://doi.org/10.1103/PhysRevD.96.124035} {\bibfield  {journal} {\bibinfo
  {journal} {Phys. Rev. D}\ }\textbf {\bibinfo {volume} {96}},\ \bibinfo
  {pages} {124035} (\bibinfo {year} {2017})}\BibitemShut {NoStop}%
\bibitem [{\citenamefont {Maione}\ \emph {et~al.}(2017)\citenamefont {Maione},
  \citenamefont {De~Pietri}, \citenamefont {Feo},\ and\ \citenamefont
  {L\"offler}}]{PhysRevD.96.063011}%
  \BibitemOpen
  \bibfield  {author} {\bibinfo {author} {\bibfnamefont {F.}~\bibnamefont
  {Maione}}, \bibinfo {author} {\bibfnamefont {R.}~\bibnamefont {De~Pietri}},
  \bibinfo {author} {\bibfnamefont {A.}~\bibnamefont {Feo}},\ and\ \bibinfo
  {author} {\bibfnamefont {F.}~\bibnamefont {L\"offler}},\ }\bibfield  {title}
  {\bibinfo {title} {Spectral analysis of gravitational waves from binary
  neutron star merger remnants},\ }\href
  {https://doi.org/10.1103/PhysRevD.96.063011} {\bibfield  {journal} {\bibinfo
  {journal} {Phys. Rev. D}\ }\textbf {\bibinfo {volume} {96}},\ \bibinfo
  {pages} {063011} (\bibinfo {year} {2017})}\BibitemShut {NoStop}%
\bibitem [{\citenamefont {{Endrizzi}}\ \emph {et~al.}(2018)\citenamefont
  {{Endrizzi}}, \citenamefont {{Logoteta}}, \citenamefont {{Giacomazzo}},
  \citenamefont {{Bombaci}}, \citenamefont {{Kastaun}},\ and\ \citenamefont
  {{Ciolfi}}}]{2018PhRvD..98d3015E}%
  \BibitemOpen
  \bibfield  {author} {\bibinfo {author} {\bibfnamefont {A.}~\bibnamefont
  {{Endrizzi}}}, \bibinfo {author} {\bibfnamefont {D.}~\bibnamefont
  {{Logoteta}}}, \bibinfo {author} {\bibfnamefont {B.}~\bibnamefont
  {{Giacomazzo}}}, \bibinfo {author} {\bibfnamefont {I.}~\bibnamefont
  {{Bombaci}}}, \bibinfo {author} {\bibfnamefont {W.}~\bibnamefont
  {{Kastaun}}},\ and\ \bibinfo {author} {\bibfnamefont {R.}~\bibnamefont
  {{Ciolfi}}},\ }\bibfield  {title} {\bibinfo {title} {{Effects of chiral
  effective field theory equation of state on binary neutron star mergers}},\
  }\href {https://doi.org/10.1103/PhysRevD.98.043015} {\bibfield  {journal}
  {\bibinfo  {journal} {\prd}\ }\textbf {\bibinfo {volume} {98}},\ \bibinfo
  {eid} {043015} (\bibinfo {year} {2018})},\ \Eprint
  {https://arxiv.org/abs/1806.09832} {arXiv:1806.09832 [astro-ph.HE]}
  \BibitemShut {NoStop}%
\bibitem [{\citenamefont {{Bauswein}}\ \emph {et~al.}(2019)\citenamefont
  {{Bauswein}}, \citenamefont {{Bastian}}, \citenamefont {{Blaschke}},
  \citenamefont {{Chatziioannou}}, \citenamefont {{Clark}}, \citenamefont
  {{Fischer}},\ and\ \citenamefont {{Oertel}}}]{2019PhRvL.122f1102B}%
  \BibitemOpen
  \bibfield  {author} {\bibinfo {author} {\bibfnamefont {A.}~\bibnamefont
  {{Bauswein}}}, \bibinfo {author} {\bibfnamefont {N.-U.~F.}\ \bibnamefont
  {{Bastian}}}, \bibinfo {author} {\bibfnamefont {D.~B.}\ \bibnamefont
  {{Blaschke}}}, \bibinfo {author} {\bibfnamefont {K.}~\bibnamefont
  {{Chatziioannou}}}, \bibinfo {author} {\bibfnamefont {J.~A.}\ \bibnamefont
  {{Clark}}}, \bibinfo {author} {\bibfnamefont {T.}~\bibnamefont {{Fischer}}},\
  and\ \bibinfo {author} {\bibfnamefont {M.}~\bibnamefont {{Oertel}}},\
  }\bibfield  {title} {\bibinfo {title} {{Identifying a First-Order Phase
  Transition in Neutron-Star Mergers through Gravitational Waves}},\ }\href
  {https://doi.org/10.1103/PhysRevLett.122.061102} {\bibfield  {journal}
  {\bibinfo  {journal} {\prl}\ }\textbf {\bibinfo {volume} {122}},\ \bibinfo
  {eid} {061102} (\bibinfo {year} {2019})},\ \Eprint
  {https://arxiv.org/abs/1809.01116} {arXiv:1809.01116 [astro-ph.HE]}
  \BibitemShut {NoStop}%
\bibitem [{\citenamefont {{Bauswein}}\ and\ \citenamefont
  {{Stergioulas}}(2019)}]{Bauswein2019sep}%
  \BibitemOpen
  \bibfield  {author} {\bibinfo {author} {\bibfnamefont {A.}~\bibnamefont
  {{Bauswein}}}\ and\ \bibinfo {author} {\bibfnamefont {N.}~\bibnamefont
  {{Stergioulas}}},\ }\bibfield  {title} {\bibinfo {title} {{Spectral
  classification of gravitational-wave emission and equation of state
  constraints in binary neutron star mergers}},\ }\href
  {https://doi.org/10.1088/1361-6471/ab2b90} {\bibfield  {journal} {\bibinfo
  {journal} {Journal of Physics G Nuclear Physics}\ }\textbf {\bibinfo {volume}
  {46}},\ \bibinfo {eid} {113002} (\bibinfo {year} {2019})},\ \Eprint
  {https://arxiv.org/abs/1901.06969} {arXiv:1901.06969 [gr-qc]} \BibitemShut
  {NoStop}%
\bibitem [{\citenamefont {Breschi}\ \emph {et~al.}(2019)\citenamefont
  {Breschi}, \citenamefont {Bernuzzi}, \citenamefont {Zappa}, \citenamefont
  {Agathos}, \citenamefont {Perego}, \citenamefont {Radice},\ and\
  \citenamefont {Nagar}}]{PhysRevD.100.104029}%
  \BibitemOpen
  \bibfield  {author} {\bibinfo {author} {\bibfnamefont {M.}~\bibnamefont
  {Breschi}}, \bibinfo {author} {\bibfnamefont {S.}~\bibnamefont {Bernuzzi}},
  \bibinfo {author} {\bibfnamefont {F.}~\bibnamefont {Zappa}}, \bibinfo
  {author} {\bibfnamefont {M.}~\bibnamefont {Agathos}}, \bibinfo {author}
  {\bibfnamefont {A.}~\bibnamefont {Perego}}, \bibinfo {author} {\bibfnamefont
  {D.}~\bibnamefont {Radice}},\ and\ \bibinfo {author} {\bibfnamefont
  {A.}~\bibnamefont {Nagar}},\ }\bibfield  {title} {\bibinfo {title} {Kilohertz
  gravitational waves from binary neutron star remnants: Time-domain model and
  constraints on extreme matter},\ }\href
  {https://doi.org/10.1103/PhysRevD.100.104029} {\bibfield  {journal} {\bibinfo
   {journal} {Phys. Rev. D}\ }\textbf {\bibinfo {volume} {100}},\ \bibinfo
  {pages} {104029} (\bibinfo {year} {2019})}\BibitemShut {NoStop}%
\bibitem [{\citenamefont {Easter}\ \emph {et~al.}(2019)\citenamefont {Easter},
  \citenamefont {Lasky}, \citenamefont {Casey}, \citenamefont {Rezzolla},\ and\
  \citenamefont {Takami}}]{Easter2019}%
  \BibitemOpen
  \bibfield  {author} {\bibinfo {author} {\bibfnamefont {P.~J.}\ \bibnamefont
  {Easter}}, \bibinfo {author} {\bibfnamefont {P.~D.}\ \bibnamefont {Lasky}},
  \bibinfo {author} {\bibfnamefont {A.~R.}\ \bibnamefont {Casey}}, \bibinfo
  {author} {\bibfnamefont {L.}~\bibnamefont {Rezzolla}},\ and\ \bibinfo
  {author} {\bibfnamefont {K.}~\bibnamefont {Takami}},\ }\bibfield  {title}
  {\bibinfo {title} {Computing fast and reliable gravitational waveforms of
  binary neutron star merger remnants},\ }\href
  {https://doi.org/10.1103/PhysRevD.100.043005} {\bibfield  {journal} {\bibinfo
   {journal} {Phys. Rev. D}\ }\textbf {\bibinfo {volume} {100}},\ \bibinfo
  {pages} {043005} (\bibinfo {year} {2019})}\BibitemShut {NoStop}%
\bibitem [{\citenamefont {{Weih}}\ \emph {et~al.}(2020)\citenamefont {{Weih}},
  \citenamefont {{Hanauske}},\ and\ \citenamefont
  {{Rezzolla}}}]{2020PhRvL.124q1103W}%
  \BibitemOpen
  \bibfield  {author} {\bibinfo {author} {\bibfnamefont {L.~R.}\ \bibnamefont
  {{Weih}}}, \bibinfo {author} {\bibfnamefont {M.}~\bibnamefont {{Hanauske}}},\
  and\ \bibinfo {author} {\bibfnamefont {L.}~\bibnamefont {{Rezzolla}}},\
  }\bibfield  {title} {\bibinfo {title} {{Postmerger Gravitational-Wave
  Signatures of Phase Transitions in Binary Mergers}},\ }\href
  {https://doi.org/10.1103/PhysRevLett.124.171103} {\bibfield  {journal}
  {\bibinfo  {journal} {\prl}\ }\textbf {\bibinfo {volume} {124}},\ \bibinfo
  {eid} {171103} (\bibinfo {year} {2020})},\ \Eprint
  {https://arxiv.org/abs/1912.09340} {arXiv:1912.09340 [gr-qc]} \BibitemShut
  {NoStop}%
\bibitem [{\citenamefont {Torres-Rivas}\ \emph {et~al.}(2019)\citenamefont
  {Torres-Rivas}, \citenamefont {Chatziioannou}, \citenamefont {Bauswein},\
  and\ \citenamefont {Clark}}]{PhysRevD.99.044014}%
  \BibitemOpen
  \bibfield  {author} {\bibinfo {author} {\bibfnamefont {A.}~\bibnamefont
  {Torres-Rivas}}, \bibinfo {author} {\bibfnamefont {K.}~\bibnamefont
  {Chatziioannou}}, \bibinfo {author} {\bibfnamefont {A.}~\bibnamefont
  {Bauswein}},\ and\ \bibinfo {author} {\bibfnamefont {J.~A.}\ \bibnamefont
  {Clark}},\ }\bibfield  {title} {\bibinfo {title} {Observing the post-merger
  signal of gw170817-like events with improved gravitational-wave detectors},\
  }\href {https://doi.org/10.1103/PhysRevD.99.044014} {\bibfield  {journal}
  {\bibinfo  {journal} {Phys. Rev. D}\ }\textbf {\bibinfo {volume} {99}},\
  \bibinfo {pages} {044014} (\bibinfo {year} {2019})}\BibitemShut {NoStop}%
\bibitem [{\citenamefont {Tsang}\ \emph {et~al.}(2019)\citenamefont {Tsang},
  \citenamefont {Dietrich},\ and\ \citenamefont {Van
  Den~Broeck}}]{PhysRevD.100.044047}%
  \BibitemOpen
  \bibfield  {author} {\bibinfo {author} {\bibfnamefont {K.~W.}\ \bibnamefont
  {Tsang}}, \bibinfo {author} {\bibfnamefont {T.}~\bibnamefont {Dietrich}},\
  and\ \bibinfo {author} {\bibfnamefont {C.}~\bibnamefont {Van Den~Broeck}},\
  }\bibfield  {title} {\bibinfo {title} {Modeling the postmerger gravitational
  wave signal and extracting binary properties from future binary neutron star
  detections},\ }\href {https://doi.org/10.1103/PhysRevD.100.044047} {\bibfield
   {journal} {\bibinfo  {journal} {Phys. Rev. D}\ }\textbf {\bibinfo {volume}
  {100}},\ \bibinfo {pages} {044047} (\bibinfo {year} {2019})}\BibitemShut
  {NoStop}%
\bibitem [{\citenamefont {{Blacker}}\ \emph {et~al.}(2020)\citenamefont
  {{Blacker}}, \citenamefont {{Bastian}}, \citenamefont {{Bauswein}},
  \citenamefont {{Blaschke}}, \citenamefont {{Fischer}}, \citenamefont
  {{Oertel}}, \citenamefont {{Soultanis}},\ and\ \citenamefont
  {{Typel}}}]{2020PhRvD.102l3023B}%
  \BibitemOpen
  \bibfield  {author} {\bibinfo {author} {\bibfnamefont {S.}~\bibnamefont
  {{Blacker}}}, \bibinfo {author} {\bibfnamefont {N.-U.~F.}\ \bibnamefont
  {{Bastian}}}, \bibinfo {author} {\bibfnamefont {A.}~\bibnamefont
  {{Bauswein}}}, \bibinfo {author} {\bibfnamefont {D.~B.}\ \bibnamefont
  {{Blaschke}}}, \bibinfo {author} {\bibfnamefont {T.}~\bibnamefont
  {{Fischer}}}, \bibinfo {author} {\bibfnamefont {M.}~\bibnamefont {{Oertel}}},
  \bibinfo {author} {\bibfnamefont {T.}~\bibnamefont {{Soultanis}}},\ and\
  \bibinfo {author} {\bibfnamefont {S.}~\bibnamefont {{Typel}}},\ }\bibfield
  {title} {\bibinfo {title} {{Constraining the onset density of the
  hadron-quark phase transition with gravitational-wave observations}},\ }\href
  {https://doi.org/10.1103/PhysRevD.102.123023} {\bibfield  {journal} {\bibinfo
   {journal} {\prd}\ }\textbf {\bibinfo {volume} {102}},\ \bibinfo {eid}
  {123023} (\bibinfo {year} {2020})},\ \Eprint
  {https://arxiv.org/abs/2006.03789} {arXiv:2006.03789 [astro-ph.HE]}
  \BibitemShut {NoStop}%
\bibitem [{\citenamefont {{Bauswein}}\ and\ \citenamefont
  {{Blacker}}(2020)}]{2020EPJST.229.3595B}%
  \BibitemOpen
  \bibfield  {author} {\bibinfo {author} {\bibfnamefont {A.}~\bibnamefont
  {{Bauswein}}}\ and\ \bibinfo {author} {\bibfnamefont {S.}~\bibnamefont
  {{Blacker}}},\ }\bibfield  {title} {\bibinfo {title} {{Impact of quark
  deconfinement in neutron star mergers and hybrid star mergers}},\ }\href
  {https://doi.org/10.1140/epjst/e2020-000138-7} {\bibfield  {journal}
  {\bibinfo  {journal} {European Physical Journal Special Topics}\ }\textbf
  {\bibinfo {volume} {229}},\ \bibinfo {pages} {3595} (\bibinfo {year}
  {2020})},\ \Eprint {https://arxiv.org/abs/2006.16183} {arXiv:2006.16183
  [astro-ph.HE]} \BibitemShut {NoStop}%
\bibitem [{\citenamefont {{Vretinaris}}\ \emph {et~al.}(2020)\citenamefont
  {{Vretinaris}}, \citenamefont {{Stergioulas}},\ and\ \citenamefont
  {{Bauswein}}}]{2020PhRvD.101h4039V}%
  \BibitemOpen
  \bibfield  {author} {\bibinfo {author} {\bibfnamefont {S.}~\bibnamefont
  {{Vretinaris}}}, \bibinfo {author} {\bibfnamefont {N.}~\bibnamefont
  {{Stergioulas}}},\ and\ \bibinfo {author} {\bibfnamefont {A.}~\bibnamefont
  {{Bauswein}}},\ }\bibfield  {title} {\bibinfo {title} {{Empirical relations
  for gravitational-wave asteroseismology of binary neutron star mergers}},\
  }\href {https://doi.org/10.1103/PhysRevD.101.084039} {\bibfield  {journal}
  {\bibinfo  {journal} {Phys. Rev. D}\ }\textbf {\bibinfo {volume} {101}},\
  \bibinfo {eid} {084039} (\bibinfo {year} {2020})},\ \Eprint
  {https://arxiv.org/abs/1910.10856} {arXiv:1910.10856 [gr-qc]} \BibitemShut
  {NoStop}%
\bibitem [{\citenamefont {Easter}\ \emph {et~al.}(2020)\citenamefont {Easter},
  \citenamefont {Ghonge}, \citenamefont {Lasky}, \citenamefont {Casey},
  \citenamefont {Clark}, \citenamefont {Hernandez~Vivanco},\ and\ \citenamefont
  {Chatziioannou}}]{Easter2020}%
  \BibitemOpen
  \bibfield  {author} {\bibinfo {author} {\bibfnamefont {P.~J.}\ \bibnamefont
  {Easter}}, \bibinfo {author} {\bibfnamefont {S.}~\bibnamefont {Ghonge}},
  \bibinfo {author} {\bibfnamefont {P.~D.}\ \bibnamefont {Lasky}}, \bibinfo
  {author} {\bibfnamefont {A.~R.}\ \bibnamefont {Casey}}, \bibinfo {author}
  {\bibfnamefont {J.~A.}\ \bibnamefont {Clark}}, \bibinfo {author}
  {\bibfnamefont {F.}~\bibnamefont {Hernandez~Vivanco}},\ and\ \bibinfo
  {author} {\bibfnamefont {K.}~\bibnamefont {Chatziioannou}},\ }\bibfield
  {title} {\bibinfo {title} {Detection and parameter estimation of binary
  neutron star merger remnants},\ }\href
  {https://doi.org/10.1103/PhysRevD.102.043011} {\bibfield  {journal} {\bibinfo
   {journal} {Phys. Rev. D}\ }\textbf {\bibinfo {volume} {102}},\ \bibinfo
  {pages} {043011} (\bibinfo {year} {2020})}\BibitemShut {NoStop}%
\bibitem [{\citenamefont {{Haster}}\ \emph {et~al.}(2020)\citenamefont
  {{Haster}}, \citenamefont {{Chatziioannou}}, \citenamefont {{Bauswein}},\
  and\ \citenamefont {{Clark}}}]{2020PhRvL.125z1101H}%
  \BibitemOpen
  \bibfield  {author} {\bibinfo {author} {\bibfnamefont {C.-J.}\ \bibnamefont
  {{Haster}}}, \bibinfo {author} {\bibfnamefont {K.}~\bibnamefont
  {{Chatziioannou}}}, \bibinfo {author} {\bibfnamefont {A.}~\bibnamefont
  {{Bauswein}}},\ and\ \bibinfo {author} {\bibfnamefont {J.~A.}\ \bibnamefont
  {{Clark}}},\ }\bibfield  {title} {\bibinfo {title} {{Inference of the Neutron
  Star Equation of State from Cosmological Distances}},\ }\href
  {https://doi.org/10.1103/PhysRevLett.125.261101} {\bibfield  {journal}
  {\bibinfo  {journal} {\prl}\ }\textbf {\bibinfo {volume} {125}},\ \bibinfo
  {eid} {261101} (\bibinfo {year} {2020})},\ \Eprint
  {https://arxiv.org/abs/2004.11334} {arXiv:2004.11334 [gr-qc]} \BibitemShut
  {NoStop}%
\bibitem [{\citenamefont {{Friedman}}\ and\ \citenamefont
  {{Stergioulas}}(2020)}]{2020IJMPD..2941015F}%
  \BibitemOpen
  \bibfield  {author} {\bibinfo {author} {\bibfnamefont {J.~L.}\ \bibnamefont
  {{Friedman}}}\ and\ \bibinfo {author} {\bibfnamefont {N.}~\bibnamefont
  {{Stergioulas}}},\ }\bibfield  {title} {\bibinfo {title} {{Astrophysical
  implications of neutron star inspiral and coalescence}},\ }\href
  {https://doi.org/10.1142/S0218271820410151} {\bibfield  {journal} {\bibinfo
  {journal} {International Journal of Modern Physics D}\ }\textbf {\bibinfo
  {volume} {29}},\ \bibinfo {eid} {2041015-632} (\bibinfo {year} {2020})},\
  \Eprint {https://arxiv.org/abs/2005.14135} {arXiv:2005.14135 [astro-ph.HE]}
  \BibitemShut {NoStop}%
\bibitem [{\citenamefont {{Kiuchi}}\ \emph {et~al.}(2020)\citenamefont
  {{Kiuchi}}, \citenamefont {{Kawaguchi}}, \citenamefont {{Kyutoku}},
  \citenamefont {{Sekiguchi}},\ and\ \citenamefont
  {{Shibata}}}]{2020PhRvD.101h4006K}%
  \BibitemOpen
  \bibfield  {author} {\bibinfo {author} {\bibfnamefont {K.}~\bibnamefont
  {{Kiuchi}}}, \bibinfo {author} {\bibfnamefont {K.}~\bibnamefont
  {{Kawaguchi}}}, \bibinfo {author} {\bibfnamefont {K.}~\bibnamefont
  {{Kyutoku}}}, \bibinfo {author} {\bibfnamefont {Y.}~\bibnamefont
  {{Sekiguchi}}},\ and\ \bibinfo {author} {\bibfnamefont {M.}~\bibnamefont
  {{Shibata}}},\ }\bibfield  {title} {\bibinfo {title} {{Sub-radian-accuracy
  gravitational waves from coalescing binary neutron stars in numerical
  relativity. II. Systematic study on the equation of state, binary mass, and
  mass ratio}},\ }\href {https://doi.org/10.1103/PhysRevD.101.084006}
  {\bibfield  {journal} {\bibinfo  {journal} {\prd}\ }\textbf {\bibinfo
  {volume} {101}},\ \bibinfo {eid} {084006} (\bibinfo {year} {2020})},\ \Eprint
  {https://arxiv.org/abs/1907.03790} {arXiv:1907.03790 [astro-ph.HE]}
  \BibitemShut {NoStop}%
\bibitem [{\citenamefont {{Prakash}}\ \emph {et~al.}(2021)\citenamefont
  {{Prakash}}, \citenamefont {{Radice}}, \citenamefont {{Logoteta}},
  \citenamefont {{Perego}}, \citenamefont {{Nedora}}, \citenamefont
  {{Bombaci}}, \citenamefont {{Kashyap}}, \citenamefont {{Bernuzzi}},\ and\
  \citenamefont {{Endrizzi}}}]{2021PhRvD.104h3029P}%
  \BibitemOpen
  \bibfield  {author} {\bibinfo {author} {\bibfnamefont {A.}~\bibnamefont
  {{Prakash}}}, \bibinfo {author} {\bibfnamefont {D.}~\bibnamefont {{Radice}}},
  \bibinfo {author} {\bibfnamefont {D.}~\bibnamefont {{Logoteta}}}, \bibinfo
  {author} {\bibfnamefont {A.}~\bibnamefont {{Perego}}}, \bibinfo {author}
  {\bibfnamefont {V.}~\bibnamefont {{Nedora}}}, \bibinfo {author}
  {\bibfnamefont {I.}~\bibnamefont {{Bombaci}}}, \bibinfo {author}
  {\bibfnamefont {R.}~\bibnamefont {{Kashyap}}}, \bibinfo {author}
  {\bibfnamefont {S.}~\bibnamefont {{Bernuzzi}}},\ and\ \bibinfo {author}
  {\bibfnamefont {A.}~\bibnamefont {{Endrizzi}}},\ }\bibfield  {title}
  {\bibinfo {title} {{Signatures of deconfined quark phases in binary neutron
  star mergers}},\ }\href {https://doi.org/10.1103/PhysRevD.104.083029}
  {\bibfield  {journal} {\bibinfo  {journal} {\prd}\ }\textbf {\bibinfo
  {volume} {104}},\ \bibinfo {eid} {083029} (\bibinfo {year} {2021})},\ \Eprint
  {https://arxiv.org/abs/2106.07885} {arXiv:2106.07885 [astro-ph.HE]}
  \BibitemShut {NoStop}%
\bibitem [{\citenamefont {{Most}}\ and\ \citenamefont
  {{Raithel}}(2021)}]{2021arXiv210706804M}%
  \BibitemOpen
  \bibfield  {author} {\bibinfo {author} {\bibfnamefont {E.~R.}\ \bibnamefont
  {{Most}}}\ and\ \bibinfo {author} {\bibfnamefont {C.~A.}\ \bibnamefont
  {{Raithel}}},\ }\bibfield  {title} {\bibinfo {title} {{Impact of the nuclear
  symmetry energy on the post-merger phase of a binary neutron star
  coalescence}},\ }\href@noop {} {\bibfield  {journal} {\bibinfo  {journal}
  {arXiv e-prints}\ ,\ \bibinfo {eid} {arXiv:2107.06804}} (\bibinfo {year}
  {2021})},\ \Eprint {https://arxiv.org/abs/2107.06804} {arXiv:2107.06804
  [astro-ph.HE]} \BibitemShut {NoStop}%
\bibitem [{\citenamefont {{Breschi}}\ \emph
  {et~al.}(2021{\natexlab{b}})\citenamefont {{Breschi}}, \citenamefont
  {{Bernuzzi}}, \citenamefont {{Godzieba}}, \citenamefont {{Perego}},\ and\
  \citenamefont {{Radice}}}]{2021arXiv211006957B}%
  \BibitemOpen
  \bibfield  {author} {\bibinfo {author} {\bibfnamefont {M.}~\bibnamefont
  {{Breschi}}}, \bibinfo {author} {\bibfnamefont {S.}~\bibnamefont
  {{Bernuzzi}}}, \bibinfo {author} {\bibfnamefont {D.}~\bibnamefont
  {{Godzieba}}}, \bibinfo {author} {\bibfnamefont {A.}~\bibnamefont
  {{Perego}}},\ and\ \bibinfo {author} {\bibfnamefont {D.}~\bibnamefont
  {{Radice}}},\ }\bibfield  {title} {\bibinfo {title} {{Constraints on the
  neutron star's maximum densities from postmerger gravitational-waves with
  third-generation observations}},\ }\href@noop {} {\bibfield  {journal}
  {\bibinfo  {journal} {arXiv e-prints}\ ,\ \bibinfo {eid} {arXiv:2110.06957}}
  (\bibinfo {year} {2021}{\natexlab{b}})},\ \Eprint
  {https://arxiv.org/abs/2110.06957} {arXiv:2110.06957 [gr-qc]} \BibitemShut
  {NoStop}%
\bibitem [{\citenamefont {{Liebling}}\ \emph {et~al.}(2021)\citenamefont
  {{Liebling}}, \citenamefont {{Palenzuela}},\ and\ \citenamefont
  {{Lehner}}}]{Liebling2021}%
  \BibitemOpen
  \bibfield  {author} {\bibinfo {author} {\bibfnamefont {S.~L.}\ \bibnamefont
  {{Liebling}}}, \bibinfo {author} {\bibfnamefont {C.}~\bibnamefont
  {{Palenzuela}}},\ and\ \bibinfo {author} {\bibfnamefont {L.}~\bibnamefont
  {{Lehner}}},\ }\bibfield  {title} {\bibinfo {title} {{Effects of high density
  phase transitions on neutron star dynamics}},\ }\href
  {https://doi.org/10.1088/1361-6382/abf898} {\bibfield  {journal} {\bibinfo
  {journal} {Classical and Quantum Gravity}\ }\textbf {\bibinfo {volume}
  {38}},\ \bibinfo {eid} {115007} (\bibinfo {year} {2021})},\ \Eprint
  {https://arxiv.org/abs/2010.12567} {arXiv:2010.12567 [gr-qc]} \BibitemShut
  {NoStop}%
\bibitem [{\citenamefont {Lioutas}\ \emph {et~al.}(2021)\citenamefont
  {Lioutas}, \citenamefont {Bauswein},\ and\ \citenamefont
  {Stergioulas}}]{Lioutas2021}%
  \BibitemOpen
  \bibfield  {author} {\bibinfo {author} {\bibfnamefont {G.}~\bibnamefont
  {Lioutas}}, \bibinfo {author} {\bibfnamefont {A.}~\bibnamefont {Bauswein}},\
  and\ \bibinfo {author} {\bibfnamefont {N.}~\bibnamefont {Stergioulas}},\
  }\bibfield  {title} {\bibinfo {title} {Frequency deviations in universal
  relations of isolated neutron stars and postmerger remnants},\ }\href
  {https://doi.org/10.1103/PhysRevD.104.043011} {\bibfield  {journal} {\bibinfo
   {journal} {Phys. Rev. D}\ }\textbf {\bibinfo {volume} {104}},\ \bibinfo
  {pages} {043011} (\bibinfo {year} {2021})}\BibitemShut {NoStop}%
\bibitem [{\citenamefont {{Ruiz}}\ \emph {et~al.}(2021)\citenamefont {{Ruiz}},
  \citenamefont {{Tsokaros}},\ and\ \citenamefont
  {{Shapiro}}}]{2021arXiv211011968R}%
  \BibitemOpen
  \bibfield  {author} {\bibinfo {author} {\bibfnamefont {M.}~\bibnamefont
  {{Ruiz}}}, \bibinfo {author} {\bibfnamefont {A.}~\bibnamefont {{Tsokaros}}},\
  and\ \bibinfo {author} {\bibfnamefont {S.~L.}\ \bibnamefont {{Shapiro}}},\
  }\bibfield  {title} {\bibinfo {title} {{Jet Launching from Merging Magnetized
  Binary Neutron Stars with Realistic Equations of State}},\ }\href@noop {}
  {\bibfield  {journal} {\bibinfo  {journal} {arXiv e-prints}\ ,\ \bibinfo
  {eid} {arXiv:2110.11968}} (\bibinfo {year} {2021})},\ \Eprint
  {https://arxiv.org/abs/2110.11968} {arXiv:2110.11968 [astro-ph.HE]}
  \BibitemShut {NoStop}%
\bibitem [{\citenamefont {Zhu}\ and\ \citenamefont
  {Rezzolla}(2021)}]{PhysRevD.104.083004}%
  \BibitemOpen
  \bibfield  {author} {\bibinfo {author} {\bibfnamefont {Z.}~\bibnamefont
  {Zhu}}\ and\ \bibinfo {author} {\bibfnamefont {L.}~\bibnamefont {Rezzolla}},\
  }\bibfield  {title} {\bibinfo {title} {Fully general-relativistic simulations
  of isolated and binary strange quark stars},\ }\href
  {https://doi.org/10.1103/PhysRevD.104.083004} {\bibfield  {journal} {\bibinfo
   {journal} {Phys. Rev. D}\ }\textbf {\bibinfo {volume} {104}},\ \bibinfo
  {pages} {083004} (\bibinfo {year} {2021})}\BibitemShut {NoStop}%
\bibitem [{\citenamefont {Messenger}\ \emph {et~al.}(2014)\citenamefont
  {Messenger}, \citenamefont {Takami}, \citenamefont {Gossan}, \citenamefont
  {Rezzolla},\ and\ \citenamefont {Sathyaprakash}}]{Messenger2014}%
  \BibitemOpen
  \bibfield  {author} {\bibinfo {author} {\bibfnamefont {C.}~\bibnamefont
  {Messenger}}, \bibinfo {author} {\bibfnamefont {K.}~\bibnamefont {Takami}},
  \bibinfo {author} {\bibfnamefont {S.}~\bibnamefont {Gossan}}, \bibinfo
  {author} {\bibfnamefont {L.}~\bibnamefont {Rezzolla}},\ and\ \bibinfo
  {author} {\bibfnamefont {B.~S.}\ \bibnamefont {Sathyaprakash}},\ }\bibfield
  {title} {\bibinfo {title} {Source redshifts from gravitational-wave
  observations of binary neutron star mergers},\ }\href
  {https://doi.org/10.1103/PhysRevX.4.041004} {\bibfield  {journal} {\bibinfo
  {journal} {Phys. Rev. X}\ }\textbf {\bibinfo {volume} {4}},\ \bibinfo {pages}
  {041004} (\bibinfo {year} {2014})}\BibitemShut {NoStop}%
\bibitem [{\citenamefont {Bose}\ \emph {et~al.}(2018)\citenamefont {Bose},
  \citenamefont {Chakravarti}, \citenamefont {Rezzolla}, \citenamefont
  {Sathyaprakash},\ and\ \citenamefont {Takami}}]{Bose2018}%
  \BibitemOpen
  \bibfield  {author} {\bibinfo {author} {\bibfnamefont {S.}~\bibnamefont
  {Bose}}, \bibinfo {author} {\bibfnamefont {K.}~\bibnamefont {Chakravarti}},
  \bibinfo {author} {\bibfnamefont {L.}~\bibnamefont {Rezzolla}}, \bibinfo
  {author} {\bibfnamefont {B.~S.}\ \bibnamefont {Sathyaprakash}},\ and\
  \bibinfo {author} {\bibfnamefont {K.}~\bibnamefont {Takami}},\ }\bibfield
  {title} {\bibinfo {title} {Neutron-star radius from a population of binary
  neutron star mergers},\ }\href
  {https://doi.org/10.1103/PhysRevLett.120.031102} {\bibfield  {journal}
  {\bibinfo  {journal} {Phys. Rev. Lett.}\ }\textbf {\bibinfo {volume} {120}},\
  \bibinfo {pages} {031102} (\bibinfo {year} {2018})}\BibitemShut {NoStop}%
\bibitem [{\citenamefont {Yang}\ \emph {et~al.}(2018)\citenamefont {Yang},
  \citenamefont {Paschalidis}, \citenamefont {Yagi}, \citenamefont {Lehner},
  \citenamefont {Pretorius},\ and\ \citenamefont {Yunes}}]{Yang2018}%
  \BibitemOpen
  \bibfield  {author} {\bibinfo {author} {\bibfnamefont {H.}~\bibnamefont
  {Yang}}, \bibinfo {author} {\bibfnamefont {V.}~\bibnamefont {Paschalidis}},
  \bibinfo {author} {\bibfnamefont {K.}~\bibnamefont {Yagi}}, \bibinfo {author}
  {\bibfnamefont {L.}~\bibnamefont {Lehner}}, \bibinfo {author} {\bibfnamefont
  {F.}~\bibnamefont {Pretorius}},\ and\ \bibinfo {author} {\bibfnamefont
  {N.}~\bibnamefont {Yunes}},\ }\bibfield  {title} {\bibinfo {title}
  {Gravitational wave spectroscopy of binary neutron star merger remnants with
  mode stacking},\ }\href {https://doi.org/10.1103/PhysRevD.97.024049}
  {\bibfield  {journal} {\bibinfo  {journal} {Phys. Rev. D}\ }\textbf {\bibinfo
  {volume} {97}},\ \bibinfo {pages} {024049} (\bibinfo {year}
  {2018})}\BibitemShut {NoStop}%
\bibitem [{\citenamefont {{Whittaker}}\ \emph {et~al.}(2022)\citenamefont
  {{Whittaker}}, \citenamefont {{East}}, \citenamefont {{Green}}, \citenamefont
  {{Lehner}},\ and\ \citenamefont {{Yang}}}]{2022arXiv220106461W}%
  \BibitemOpen
  \bibfield  {author} {\bibinfo {author} {\bibfnamefont {T.}~\bibnamefont
  {{Whittaker}}}, \bibinfo {author} {\bibfnamefont {W.~E.}\ \bibnamefont
  {{East}}}, \bibinfo {author} {\bibfnamefont {S.~R.}\ \bibnamefont {{Green}}},
  \bibinfo {author} {\bibfnamefont {L.}~\bibnamefont {{Lehner}}},\ and\
  \bibinfo {author} {\bibfnamefont {H.}~\bibnamefont {{Yang}}},\ }\bibfield
  {title} {\bibinfo {title} {{Using machine learning to parametrize postmerger
  signals from binary neutron stars}},\ }\href@noop {} {\bibfield  {journal}
  {\bibinfo  {journal} {arXiv e-prints}\ ,\ \bibinfo {eid} {arXiv:2201.06461}}
  (\bibinfo {year} {2022})},\ \Eprint {https://arxiv.org/abs/2201.06461}
  {arXiv:2201.06461 [gr-qc]} \BibitemShut {NoStop}%
\bibitem [{\citenamefont {{Punturo}}\ \emph
  {et~al.}(2010{\natexlab{a}})\citenamefont {{Punturo}}, \citenamefont
  {{Abernathy}}, \citenamefont {{Acernese}}, \citenamefont {{Allen}},
  \citenamefont {{Andersson}}, \citenamefont {{Arun}}, \citenamefont
  {{Barone}}, \citenamefont {{Barr}}, \citenamefont {{Barsuglia}},
  \citenamefont {{Beker}} \emph {et~al.}}]{2010CQGra..27s4002P}%
  \BibitemOpen
  \bibfield  {author} {\bibinfo {author} {\bibfnamefont {M.}~\bibnamefont
  {{Punturo}}}, \bibinfo {author} {\bibfnamefont {M.}~\bibnamefont
  {{Abernathy}}}, \bibinfo {author} {\bibfnamefont {F.}~\bibnamefont
  {{Acernese}}}, \bibinfo {author} {\bibfnamefont {B.}~\bibnamefont {{Allen}}},
  \bibinfo {author} {\bibfnamefont {N.}~\bibnamefont {{Andersson}}}, \bibinfo
  {author} {\bibfnamefont {K.}~\bibnamefont {{Arun}}}, \bibinfo {author}
  {\bibfnamefont {F.}~\bibnamefont {{Barone}}}, \bibinfo {author}
  {\bibfnamefont {B.}~\bibnamefont {{Barr}}}, \bibinfo {author} {\bibfnamefont
  {M.}~\bibnamefont {{Barsuglia}}}, \bibinfo {author} {\bibnamefont {{Beker}}},
  \emph {et~al.},\ }\bibfield  {title} {\bibinfo {title} {{The Einstein
  Telescope: a third-generation gravitational wave observatory}},\ }\href
  {https://doi.org/10.1088/0264-9381/27/19/194002} {\bibfield  {journal}
  {\bibinfo  {journal} {Classical and Quantum Gravity}\ }\textbf {\bibinfo
  {volume} {27}},\ \bibinfo {eid} {194002} (\bibinfo {year}
  {2010}{\natexlab{a}})}\BibitemShut {NoStop}%
\bibitem [{\citenamefont {Müther}\ \emph {et~al.}(1987)\citenamefont
  {Müther}, \citenamefont {Prakash},\ and\ \citenamefont
  {Ainsworth}}]{MUTHER1987469}%
  \BibitemOpen
  \bibfield  {author} {\bibinfo {author} {\bibfnamefont {H.}~\bibnamefont
  {Müther}}, \bibinfo {author} {\bibfnamefont {M.}~\bibnamefont {Prakash}},\
  and\ \bibinfo {author} {\bibfnamefont {T.}~\bibnamefont {Ainsworth}},\
  }\bibfield  {title} {\bibinfo {title} {The nuclear symmetry energy in
  relativistic brueckner-hartree-fock calculations},\ }\href
  {https://doi.org/https://doi.org/10.1016/0370-2693(87)91611-X} {\bibfield
  {journal} {\bibinfo  {journal} {Physics Letters B}\ }\textbf {\bibinfo
  {volume} {199}},\ \bibinfo {pages} {469 } (\bibinfo {year}
  {1987})}\BibitemShut {NoStop}%
\bibitem [{\citenamefont {Abbott}\ \emph
  {et~al.}(2017{\natexlab{b}})\citenamefont {Abbott}, \citenamefont {Abbott},
  \citenamefont {Abbott}, \citenamefont {Acernese}, \citenamefont {Ackley},
  \citenamefont {Adams}, \citenamefont {Adams}, \citenamefont {Addesso},
  \citenamefont {Adhikari}, \citenamefont {Adya} \emph {et~al.}}]{GW170817}%
  \BibitemOpen
  \bibfield  {author} {\bibinfo {author} {\bibfnamefont {B.~P.}\ \bibnamefont
  {Abbott}}, \bibinfo {author} {\bibfnamefont {R.}~\bibnamefont {Abbott}},
  \bibinfo {author} {\bibfnamefont {T.~D.}\ \bibnamefont {Abbott}}, \bibinfo
  {author} {\bibfnamefont {F.}~\bibnamefont {Acernese}}, \bibinfo {author}
  {\bibfnamefont {K.}~\bibnamefont {Ackley}}, \bibinfo {author} {\bibfnamefont
  {C.}~\bibnamefont {Adams}}, \bibinfo {author} {\bibfnamefont
  {T.}~\bibnamefont {Adams}}, \bibinfo {author} {\bibfnamefont
  {P.}~\bibnamefont {Addesso}}, \bibinfo {author} {\bibfnamefont {R.~X.}\
  \bibnamefont {Adhikari}}, \bibinfo {author} {\bibfnamefont {V.~B.}\
  \bibnamefont {Adya}}, \emph {et~al.} (\bibinfo {collaboration} {LIGO
  Scientific Collaboration and Virgo Collaboration}),\ }\bibfield  {title}
  {\bibinfo {title} {Gw170817: Observation of gravitational waves from a binary
  neutron star inspiral},\ }\href
  {https://doi.org/10.1103/PhysRevLett.119.161101} {\bibfield  {journal}
  {\bibinfo  {journal} {Phys. Rev. Lett.}\ }\textbf {\bibinfo {volume} {119}},\
  \bibinfo {pages} {161101} (\bibinfo {year} {2017}{\natexlab{b}})}\BibitemShut
  {NoStop}%
\bibitem [{\citenamefont {Antoniadis}\ \emph {et~al.}(2013)\citenamefont
  {Antoniadis} \emph {et~al.}}]{Antoniadis:2013pzd}%
  \BibitemOpen
  \bibfield  {author} {\bibinfo {author} {\bibfnamefont {J.}~\bibnamefont
  {Antoniadis}} \emph {et~al.},\ }\bibfield  {title} {\bibinfo {title} {{A
  Massive Pulsar in a Compact Relativistic Binary}},\ }\href
  {https://doi.org/10.1126/science.1233232} {\bibfield  {journal} {\bibinfo
  {journal} {Science}\ }\textbf {\bibinfo {volume} {340}},\ \bibinfo {pages}
  {6131} (\bibinfo {year} {2013})},\ \Eprint {https://arxiv.org/abs/1304.6875}
  {arXiv:1304.6875 [astro-ph.HE]} \BibitemShut {NoStop}%
\bibitem [{\citenamefont {Bauswein}\ \emph {et~al.}(2021)\citenamefont
  {Bauswein}, \citenamefont {Blacker}, \citenamefont {Lioutas}, \citenamefont
  {Soultanis}, \citenamefont {Vijayan},\ and\ \citenamefont
  {Stergioulas}}]{AB2021june}%
  \BibitemOpen
  \bibfield  {author} {\bibinfo {author} {\bibfnamefont {A.}~\bibnamefont
  {Bauswein}}, \bibinfo {author} {\bibfnamefont {S.}~\bibnamefont {Blacker}},
  \bibinfo {author} {\bibfnamefont {G.}~\bibnamefont {Lioutas}}, \bibinfo
  {author} {\bibfnamefont {T.}~\bibnamefont {Soultanis}}, \bibinfo {author}
  {\bibfnamefont {V.}~\bibnamefont {Vijayan}},\ and\ \bibinfo {author}
  {\bibfnamefont {N.}~\bibnamefont {Stergioulas}},\ }\bibfield  {title}
  {\bibinfo {title} {Systematics of prompt black-hole formation in neutron star
  mergers},\ }\href {https://doi.org/10.1103/PhysRevD.103.123004} {\bibfield
  {journal} {\bibinfo  {journal} {Phys. Rev. D}\ }\textbf {\bibinfo {volume}
  {103}},\ \bibinfo {pages} {123004} (\bibinfo {year} {2021})}\BibitemShut
  {NoStop}%
\bibitem [{LORENE()}]{Lorene:web}%
  \BibitemOpen
  LORENE,\ \href@noop {} {\bibinfo {title} {{LORENE}: {L}angage {O}bjet pour la
  {RE}lativit\'e {N}um\'eriqu{E}}},\ \bibinfo {note} {webpage:
  http://www.lorene.obspm.fr/}\BibitemShut {NoStop}%
\bibitem [{\citenamefont {Gourgoulhon}\ \emph {et~al.}(2001)\citenamefont
  {Gourgoulhon}, \citenamefont {Grandcl\'ement}, \citenamefont {Taniguchi},
  \citenamefont {Marck},\ and\ \citenamefont {Bonazzola}}]{Gourgoulhon2001}%
  \BibitemOpen
  \bibfield  {author} {\bibinfo {author} {\bibfnamefont {E.}~\bibnamefont
  {Gourgoulhon}}, \bibinfo {author} {\bibfnamefont {P.}~\bibnamefont
  {Grandcl\'ement}}, \bibinfo {author} {\bibfnamefont {K.}~\bibnamefont
  {Taniguchi}}, \bibinfo {author} {\bibfnamefont {J.-A.}\ \bibnamefont
  {Marck}},\ and\ \bibinfo {author} {\bibfnamefont {S.}~\bibnamefont
  {Bonazzola}},\ }\bibfield  {title} {\bibinfo {title} {Quasiequilibrium
  sequences of synchronized and irrotational binary neutron stars in general
  relativity: Method and tests},\ }\href
  {https://doi.org/10.1103/PhysRevD.63.064029} {\bibfield  {journal} {\bibinfo
  {journal} {Phys. Rev. D}\ }\textbf {\bibinfo {volume} {63}},\ \bibinfo
  {pages} {064029} (\bibinfo {year} {2001})}\BibitemShut {NoStop}%
\bibitem [{\citenamefont {Etienne}\ \emph {et~al.}(2021)\citenamefont
  {Etienne}, \citenamefont {Brandt}, \citenamefont {Diener}, \citenamefont
  {Gabella}, \citenamefont {Gracia-Linares}, \citenamefont {Haas},
  \citenamefont {Kedia}, \citenamefont {Alcubierre}, \citenamefont {Alic},
  \citenamefont {Allen} \emph {et~al.}}]{EinsteinToolkit:2021_05}%
  \BibitemOpen
  \bibfield  {author} {\bibinfo {author} {\bibfnamefont {Z.}~\bibnamefont
  {Etienne}}, \bibinfo {author} {\bibfnamefont {S.~R.}\ \bibnamefont {Brandt}},
  \bibinfo {author} {\bibfnamefont {P.}~\bibnamefont {Diener}}, \bibinfo
  {author} {\bibfnamefont {W.~E.}\ \bibnamefont {Gabella}}, \bibinfo {author}
  {\bibfnamefont {M.}~\bibnamefont {Gracia-Linares}}, \bibinfo {author}
  {\bibfnamefont {R.}~\bibnamefont {Haas}}, \bibinfo {author} {\bibfnamefont
  {A.}~\bibnamefont {Kedia}}, \bibinfo {author} {\bibfnamefont
  {M.}~\bibnamefont {Alcubierre}}, \bibinfo {author} {\bibfnamefont
  {D.}~\bibnamefont {Alic}}, \bibinfo {author} {\bibfnamefont {G.}~\bibnamefont
  {Allen}}, \emph {et~al.},\ }\href {https://doi.org/10.5281/zenodo.4884780}
  {\bibinfo {title} {The einstein toolkit}} (\bibinfo {year} {2021}),\ \bibinfo
  {note} {to find out more, visit http://einsteintoolkit.org}\BibitemShut
  {NoStop}%
\bibitem [{\citenamefont {Baiotti}\ \emph {et~al.}(2005)\citenamefont
  {Baiotti}, \citenamefont {Hawke}, \citenamefont {Montero}, \citenamefont
  {L{\"o}ffler}, \citenamefont {Rezzolla}, \citenamefont {Stergioulas},
  \citenamefont {Font},\ and\ \citenamefont {Seidel}}]{Baiotti:2004wn}%
  \BibitemOpen
  \bibfield  {author} {\bibinfo {author} {\bibfnamefont {L.}~\bibnamefont
  {Baiotti}}, \bibinfo {author} {\bibfnamefont {I.}~\bibnamefont {Hawke}},
  \bibinfo {author} {\bibfnamefont {P.~J.}\ \bibnamefont {Montero}}, \bibinfo
  {author} {\bibfnamefont {F.}~\bibnamefont {L{\"o}ffler}}, \bibinfo {author}
  {\bibfnamefont {L.}~\bibnamefont {Rezzolla}}, \bibinfo {author}
  {\bibfnamefont {N.}~\bibnamefont {Stergioulas}}, \bibinfo {author}
  {\bibfnamefont {J.~A.}\ \bibnamefont {Font}},\ and\ \bibinfo {author}
  {\bibfnamefont {E.}~\bibnamefont {Seidel}},\ }\bibfield  {title} {\bibinfo
  {title} {{Three-dimensional relativistic simulations of rotating neutron star
  collapse to a Kerr black hole}},\ }\href
  {https://doi.org/10.1103/PhysRevD.71.024035} {\bibfield  {journal} {\bibinfo
  {journal} {Phys. Rev. D}\ }\textbf {\bibinfo {volume} {71}},\ \bibinfo
  {pages} {024035} (\bibinfo {year} {2005})},\ \Eprint
  {https://arxiv.org/abs/arXiv:gr-qc/0403029} {arXiv:gr-qc/0403029}
  \BibitemShut {NoStop}%
\bibitem [{\citenamefont {M{\"o}sta}\ \emph {et~al.}(2014)\citenamefont
  {M{\"o}sta}, \citenamefont {Mundim}, \citenamefont {Faber}, \citenamefont
  {Haas}, \citenamefont {Noble}, \citenamefont {Bode}, \citenamefont
  {L{\"o}ffler}, \citenamefont {Ott}, \citenamefont {Reisswig},\ and\
  \citenamefont {Schnetter}}]{Moesta:2013dna}%
  \BibitemOpen
  \bibfield  {author} {\bibinfo {author} {\bibfnamefont {P.}~\bibnamefont
  {M{\"o}sta}}, \bibinfo {author} {\bibfnamefont {B.~C.}\ \bibnamefont
  {Mundim}}, \bibinfo {author} {\bibfnamefont {J.~A.}\ \bibnamefont {Faber}},
  \bibinfo {author} {\bibfnamefont {R.}~\bibnamefont {Haas}}, \bibinfo {author}
  {\bibfnamefont {S.~C.}\ \bibnamefont {Noble}}, \bibinfo {author}
  {\bibfnamefont {T.}~\bibnamefont {Bode}}, \bibinfo {author} {\bibfnamefont
  {F.}~\bibnamefont {L{\"o}ffler}}, \bibinfo {author} {\bibfnamefont {C.~D.}\
  \bibnamefont {Ott}}, \bibinfo {author} {\bibfnamefont {C.}~\bibnamefont
  {Reisswig}},\ and\ \bibinfo {author} {\bibfnamefont {E.}~\bibnamefont
  {Schnetter}},\ }\bibfield  {title} {\bibinfo {title} {{{GRHydro}: {A} new
  open source general-relativistic magnetohydrodynamics code for the {E}instein
  {T}oolkit}},\ }\href {https://doi.org/10.1088/0264-9381/31/1/015005}
  {\bibfield  {journal} {\bibinfo  {journal} {Classical and Quantum Gravity}\
  }\textbf {\bibinfo {volume} {31}},\ \bibinfo {pages} {015005} (\bibinfo
  {year} {2014})},\ \Eprint {https://arxiv.org/abs/arXiv:1304.5544 [gr-qc]}
  {arXiv:1304.5544 [gr-qc]} \BibitemShut {NoStop}%
\bibitem [{\citenamefont {Banyuls}\ \emph {et~al.}(1997)\citenamefont
  {Banyuls}, \citenamefont {Font}, \citenamefont {Ibanez}, \citenamefont
  {Marti},\ and\ \citenamefont {Miralles}}]{Banyuls:1997}%
  \BibitemOpen
  \bibfield  {author} {\bibinfo {author} {\bibfnamefont {F.}~\bibnamefont
  {Banyuls}}, \bibinfo {author} {\bibfnamefont {J.~A.}\ \bibnamefont {Font}},
  \bibinfo {author} {\bibfnamefont {J.~M.}\ \bibnamefont {Ibanez}}, \bibinfo
  {author} {\bibfnamefont {J.~M.}\ \bibnamefont {Marti}},\ and\ \bibinfo
  {author} {\bibfnamefont {J.~A.}\ \bibnamefont {Miralles}},\ }\bibfield
  {title} {\bibinfo {title} {Numerical $\lbrace$3 + 1$\rbrace$ general
  relativistic hydrodynamics: A local characteristic approach},\ }\href
  {https://doi.org/10.1086/303604} {\bibfield  {journal} {\bibinfo  {journal}
  {The Astrophysical Journal}\ }\textbf {\bibinfo {volume} {476}},\ \bibinfo
  {pages} {221} (\bibinfo {year} {1997})}\BibitemShut {NoStop}%
\bibitem [{\citenamefont {Font}(2008)}]{Font:2007zz}%
  \BibitemOpen
  \bibfield  {author} {\bibinfo {author} {\bibfnamefont {J.~A.}\ \bibnamefont
  {Font}},\ }\bibfield  {title} {\bibinfo {title} {{{N}umerical {H}ydrodynamics
  and {M}agnetohydrodynamics in {G}eneral {R}elativity}},\ }\href
  {http://www.livingreviews.org/lrr-2008-7} {\bibfield  {journal} {\bibinfo
  {journal} {Living Rev. Relativity}\ }\textbf {\bibinfo {volume} {11}}
  (\bibinfo {year} {2008})}\BibitemShut {NoStop}%
\bibitem [{\citenamefont {Harten}\ \emph {et~al.}(1983)\citenamefont {Harten},
  \citenamefont {Lax},\ and\ \citenamefont {van Leer}}]{Harten:1983on}%
  \BibitemOpen
  \bibfield  {author} {\bibinfo {author} {\bibfnamefont {A.}~\bibnamefont
  {Harten}}, \bibinfo {author} {\bibfnamefont {P.~D.}\ \bibnamefont {Lax}},\
  and\ \bibinfo {author} {\bibfnamefont {B.}~\bibnamefont {van Leer}},\
  }\bibfield  {title} {\bibinfo {title} {On upstream differencing and
  {G}odunov-type schemes for hyperbolic conservation laws},\ }\href@noop {}
  {\bibfield  {journal} {\bibinfo  {journal} {SIAM review}\ }\textbf {\bibinfo
  {volume} {25}},\ \bibinfo {pages} {35} (\bibinfo {year} {1983})}\BibitemShut
  {NoStop}%
\bibitem [{\citenamefont {Liu}\ \emph {et~al.}(1994)\citenamefont {Liu},
  \citenamefont {Osher},\ and\ \citenamefont {Chan}}]{WENO:1994}%
  \BibitemOpen
  \bibfield  {author} {\bibinfo {author} {\bibfnamefont {X.-D.}\ \bibnamefont
  {Liu}}, \bibinfo {author} {\bibfnamefont {S.}~\bibnamefont {Osher}},\ and\
  \bibinfo {author} {\bibfnamefont {T.}~\bibnamefont {Chan}},\ }\bibfield
  {title} {\bibinfo {title} {Weighted essentially non-oscillatory schemes},\
  }\href {https://doi.org/https://doi.org/10.1006/jcph.1994.1187} {\bibfield
  {journal} {\bibinfo  {journal} {Journal of Computational Physics}\ }\textbf
  {\bibinfo {volume} {115}},\ \bibinfo {pages} {200 } (\bibinfo {year}
  {1994})}\BibitemShut {NoStop}%
\bibitem [{\citenamefont {Jiang}\ and\ \citenamefont {Shu}(1996)}]{WENO:1996}%
  \BibitemOpen
  \bibfield  {author} {\bibinfo {author} {\bibfnamefont {G.-S.}\ \bibnamefont
  {Jiang}}\ and\ \bibinfo {author} {\bibfnamefont {C.-W.}\ \bibnamefont
  {Shu}},\ }\bibfield  {title} {\bibinfo {title} {Efficient implementation of
  weighted eno schemes},\ }\href
  {https://doi.org/https://doi.org/10.1006/jcph.1996.0130} {\bibfield
  {journal} {\bibinfo  {journal} {Journal of Computational Physics}\ }\textbf
  {\bibinfo {volume} {126}},\ \bibinfo {pages} {202 } (\bibinfo {year}
  {1996})}\BibitemShut {NoStop}%
\bibitem [{\citenamefont {Bernuzzi}\ and\ \citenamefont
  {Hilditch}(2010)}]{PhysRevD.81.084003}%
  \BibitemOpen
  \bibfield  {author} {\bibinfo {author} {\bibfnamefont {S.}~\bibnamefont
  {Bernuzzi}}\ and\ \bibinfo {author} {\bibfnamefont {D.}~\bibnamefont
  {Hilditch}},\ }\bibfield  {title} {\bibinfo {title} {Constraint violation in
  free evolution schemes: Comparing the bssnok formulation with a conformal
  decomposition of the z4 formulation},\ }\href
  {https://doi.org/10.1103/PhysRevD.81.084003} {\bibfield  {journal} {\bibinfo
  {journal} {Phys. Rev. D}\ }\textbf {\bibinfo {volume} {81}},\ \bibinfo
  {pages} {084003} (\bibinfo {year} {2010})}\BibitemShut {NoStop}%
\bibitem [{\citenamefont {Hilditch}\ \emph {et~al.}(2013)\citenamefont
  {Hilditch}, \citenamefont {Bernuzzi}, \citenamefont {Thierfelder},
  \citenamefont {Cao}, \citenamefont {Tichy},\ and\ \citenamefont
  {Br\"ugmann}}]{PhysRevD.88.084057}%
  \BibitemOpen
  \bibfield  {author} {\bibinfo {author} {\bibfnamefont {D.}~\bibnamefont
  {Hilditch}}, \bibinfo {author} {\bibfnamefont {S.}~\bibnamefont {Bernuzzi}},
  \bibinfo {author} {\bibfnamefont {M.}~\bibnamefont {Thierfelder}}, \bibinfo
  {author} {\bibfnamefont {Z.}~\bibnamefont {Cao}}, \bibinfo {author}
  {\bibfnamefont {W.}~\bibnamefont {Tichy}},\ and\ \bibinfo {author}
  {\bibfnamefont {B.}~\bibnamefont {Br\"ugmann}},\ }\bibfield  {title}
  {\bibinfo {title} {Compact binary evolutions with the z4c formulation},\
  }\href {https://doi.org/10.1103/PhysRevD.88.084057} {\bibfield  {journal}
  {\bibinfo  {journal} {Phys. Rev. D}\ }\textbf {\bibinfo {volume} {88}},\
  \bibinfo {pages} {084057} (\bibinfo {year} {2013})}\BibitemShut {NoStop}%
\bibitem [{\citenamefont {Pollney}\ \emph {et~al.}(2011)\citenamefont
  {Pollney}, \citenamefont {Reisswig}, \citenamefont {Schnetter}, \citenamefont
  {Dorband},\ and\ \citenamefont {Diener}}]{PhysRevD.83.044045}%
  \BibitemOpen
  \bibfield  {author} {\bibinfo {author} {\bibfnamefont {D.}~\bibnamefont
  {Pollney}}, \bibinfo {author} {\bibfnamefont {C.}~\bibnamefont {Reisswig}},
  \bibinfo {author} {\bibfnamefont {E.}~\bibnamefont {Schnetter}}, \bibinfo
  {author} {\bibfnamefont {N.}~\bibnamefont {Dorband}},\ and\ \bibinfo {author}
  {\bibfnamefont {P.}~\bibnamefont {Diener}},\ }\bibfield  {title} {\bibinfo
  {title} {High accuracy binary black hole simulations with an extended wave
  zone},\ }\href {https://doi.org/10.1103/PhysRevD.83.044045} {\bibfield
  {journal} {\bibinfo  {journal} {Phys. Rev. D}\ }\textbf {\bibinfo {volume}
  {83}},\ \bibinfo {pages} {044045} (\bibinfo {year} {2011})}\BibitemShut
  {NoStop}%
\bibitem [{\citenamefont {Reisswig}\ \emph {et~al.}(2013)\citenamefont
  {Reisswig}, \citenamefont {Ott}, \citenamefont {Abdikamalov}, \citenamefont
  {Haas}, \citenamefont {M\"osta},\ and\ \citenamefont
  {Schnetter}}]{PhysRevLett.111.151101}%
  \BibitemOpen
  \bibfield  {author} {\bibinfo {author} {\bibfnamefont {C.}~\bibnamefont
  {Reisswig}}, \bibinfo {author} {\bibfnamefont {C.~D.}\ \bibnamefont {Ott}},
  \bibinfo {author} {\bibfnamefont {E.}~\bibnamefont {Abdikamalov}}, \bibinfo
  {author} {\bibfnamefont {R.}~\bibnamefont {Haas}}, \bibinfo {author}
  {\bibfnamefont {P.}~\bibnamefont {M\"osta}},\ and\ \bibinfo {author}
  {\bibfnamefont {E.}~\bibnamefont {Schnetter}},\ }\bibfield  {title} {\bibinfo
  {title} {Formation and coalescence of cosmological supermassive-black-hole
  binaries in supermassive-star collapse},\ }\href
  {https://doi.org/10.1103/PhysRevLett.111.151101} {\bibfield  {journal}
  {\bibinfo  {journal} {Phys. Rev. Lett.}\ }\textbf {\bibinfo {volume} {111}},\
  \bibinfo {pages} {151101} (\bibinfo {year} {2013})}\BibitemShut {NoStop}%
\bibitem [{\citenamefont {Read}\ \emph {et~al.}(2009)\citenamefont {Read},
  \citenamefont {Lackey}, \citenamefont {Owen},\ and\ \citenamefont
  {Friedman}}]{PhysRevD.79.124032}%
  \BibitemOpen
  \bibfield  {author} {\bibinfo {author} {\bibfnamefont {J.~S.}\ \bibnamefont
  {Read}}, \bibinfo {author} {\bibfnamefont {B.~D.}\ \bibnamefont {Lackey}},
  \bibinfo {author} {\bibfnamefont {B.~J.}\ \bibnamefont {Owen}},\ and\
  \bibinfo {author} {\bibfnamefont {J.~L.}\ \bibnamefont {Friedman}},\
  }\bibfield  {title} {\bibinfo {title} {Constraints on a phenomenologically
  parametrized neutron-star equation of state},\ }\href
  {https://doi.org/10.1103/PhysRevD.79.124032} {\bibfield  {journal} {\bibinfo
  {journal} {Phys. Rev. D}\ }\textbf {\bibinfo {volume} {79}},\ \bibinfo
  {pages} {124032} (\bibinfo {year} {2009})}\BibitemShut {NoStop}%
\bibitem [{\citenamefont {Bauswein}\ \emph {et~al.}(2010)\citenamefont
  {Bauswein}, \citenamefont {Janka},\ and\ \citenamefont {Oechslin}}]{AB2010}%
  \BibitemOpen
  \bibfield  {author} {\bibinfo {author} {\bibfnamefont {A.}~\bibnamefont
  {Bauswein}}, \bibinfo {author} {\bibfnamefont {H.-T.}\ \bibnamefont
  {Janka}},\ and\ \bibinfo {author} {\bibfnamefont {R.}~\bibnamefont
  {Oechslin}},\ }\bibfield  {title} {\bibinfo {title} {Testing approximations
  of thermal effects in neutron star merger simulations},\ }\href
  {https://doi.org/10.1103/PhysRevD.82.084043} {\bibfield  {journal} {\bibinfo
  {journal} {Phys. Rev. D}\ }\textbf {\bibinfo {volume} {82}},\ \bibinfo
  {pages} {084043} (\bibinfo {year} {2010})}\BibitemShut {NoStop}%
\bibitem [{\citenamefont {Reisswig}\ and\ \citenamefont
  {Pollney}(2011)}]{FFI-Reisswig_2011}%
  \BibitemOpen
  \bibfield  {author} {\bibinfo {author} {\bibfnamefont {C.}~\bibnamefont
  {Reisswig}}\ and\ \bibinfo {author} {\bibfnamefont {D.}~\bibnamefont
  {Pollney}},\ }\bibfield  {title} {\bibinfo {title} {Notes on the integration
  of numerical relativity waveforms},\ }\href
  {https://doi.org/10.1088/0264-9381/28/19/195015} {\bibfield  {journal}
  {\bibinfo  {journal} {Classical and Quantum Gravity}\ }\textbf {\bibinfo
  {volume} {28}},\ \bibinfo {pages} {195015} (\bibinfo {year}
  {2011})}\BibitemShut {NoStop}%
\bibitem [{\citenamefont {{Zhuge}}\ \emph {et~al.}(1996)\citenamefont
  {{Zhuge}}, \citenamefont {{Centrella}},\ and\ \citenamefont
  {{McMillan}}}]{Zhuge1996}%
  \BibitemOpen
  \bibfield  {author} {\bibinfo {author} {\bibfnamefont {X.}~\bibnamefont
  {{Zhuge}}}, \bibinfo {author} {\bibfnamefont {J.~M.}\ \bibnamefont
  {{Centrella}}},\ and\ \bibinfo {author} {\bibfnamefont {S.~L.~W.}\
  \bibnamefont {{McMillan}}},\ }\bibfield  {title} {\bibinfo {title}
  {{Gravitational radiation from the coalescence of binary neutron stars:
  Effects due to the equation of state, spin, and mass ratio}},\ }\href
  {https://doi.org/10.1103/PhysRevD.54.7261} {\bibfield  {journal} {\bibinfo
  {journal} {\prd}\ }\textbf {\bibinfo {volume} {54}},\ \bibinfo {pages} {7261}
  (\bibinfo {year} {1996})},\ \Eprint {https://arxiv.org/abs/gr-qc/9610039}
  {arXiv:gr-qc/9610039 [gr-qc]} \BibitemShut {NoStop}%
\bibitem [{\citenamefont {Shibata}(2005)}]{Shibata2005a}%
  \BibitemOpen
  \bibfield  {author} {\bibinfo {author} {\bibfnamefont {M.}~\bibnamefont
  {Shibata}},\ }\bibfield  {title} {\bibinfo {title} {Constraining nuclear
  equations of state using gravitational waves from hypermassive neutron
  stars},\ }\href {https://doi.org/10.1103/PhysRevLett.94.201101} {\bibfield
  {journal} {\bibinfo  {journal} {Phys. Rev. Lett.}\ }\textbf {\bibinfo
  {volume} {94}},\ \bibinfo {pages} {201101} (\bibinfo {year}
  {2005})}\BibitemShut {NoStop}%
\bibitem [{\citenamefont {Shibata}\ \emph {et~al.}(2005)\citenamefont
  {Shibata}, \citenamefont {Taniguchi},\ and\ \citenamefont
  {Ury\ifmmode~\bar{u}\else \={u}\fi{}}}]{Shibata2005b}%
  \BibitemOpen
  \bibfield  {author} {\bibinfo {author} {\bibfnamefont {M.}~\bibnamefont
  {Shibata}}, \bibinfo {author} {\bibfnamefont {K.}~\bibnamefont {Taniguchi}},\
  and\ \bibinfo {author} {\bibfnamefont {K.~b.~o.}\ \bibnamefont
  {Ury\ifmmode~\bar{u}\else \={u}\fi{}}},\ }\bibfield  {title} {\bibinfo
  {title} {Merger of binary neutron stars with realistic equations of state in
  full general relativity},\ }\href
  {https://doi.org/10.1103/PhysRevD.71.084021} {\bibfield  {journal} {\bibinfo
  {journal} {Phys. Rev. D}\ }\textbf {\bibinfo {volume} {71}},\ \bibinfo
  {pages} {084021} (\bibinfo {year} {2005})}\BibitemShut {NoStop}%
\bibitem [{\citenamefont {Oechslin}\ and\ \citenamefont
  {Janka}(2007)}]{Oechslin2007}%
  \BibitemOpen
  \bibfield  {author} {\bibinfo {author} {\bibfnamefont {R.}~\bibnamefont
  {Oechslin}}\ and\ \bibinfo {author} {\bibfnamefont {H.-T.}\ \bibnamefont
  {Janka}},\ }\bibfield  {title} {\bibinfo {title} {Gravitational waves from
  relativistic neutron-star mergers with microphysical equations of state},\
  }\href {https://doi.org/10.1103/PhysRevLett.99.121102} {\bibfield  {journal}
  {\bibinfo  {journal} {Phys. Rev. Lett.}\ }\textbf {\bibinfo {volume} {99}},\
  \bibinfo {pages} {121102} (\bibinfo {year} {2007})}\BibitemShut {NoStop}%
\bibitem [{\citenamefont {Stergioulas}\ \emph {et~al.}(2011)\citenamefont
  {Stergioulas}, \citenamefont {Bauswein}, \citenamefont {Zagkouris},\ and\
  \citenamefont {Janka}}]{Stergioulas2011}%
  \BibitemOpen
  \bibfield  {author} {\bibinfo {author} {\bibfnamefont {N.}~\bibnamefont
  {Stergioulas}}, \bibinfo {author} {\bibfnamefont {A.}~\bibnamefont
  {Bauswein}}, \bibinfo {author} {\bibfnamefont {K.}~\bibnamefont
  {Zagkouris}},\ and\ \bibinfo {author} {\bibfnamefont {H.-T.}\ \bibnamefont
  {Janka}},\ }\bibfield  {title} {\bibinfo {title} {{Gravitational waves and
  non-axisymmetric oscillation modes in mergers of compact object binaries}},\
  }\href {https://doi.org/10.1111/j.1365-2966.2011.19493.x} {\bibfield
  {journal} {\bibinfo  {journal} {Monthly Notices of the Royal Astronomical
  Society}\ }\textbf {\bibinfo {volume} {418}},\ \bibinfo {pages} {427}
  (\bibinfo {year} {2011})},\ \Eprint
  {https://arxiv.org/abs/https://academic.oup.com/mnras/article-pdf/418/1/427/2849833/mnras0418-0427.pdf}
  {https://academic.oup.com/mnras/article-pdf/418/1/427/2849833/mnras0418-0427.pdf}
  \BibitemShut {NoStop}%
\bibitem [{\citenamefont {Bauswein}\ and\ \citenamefont
  {Janka}(2012{\natexlab{b}})}]{Bauswein2012Jan}%
  \BibitemOpen
  \bibfield  {author} {\bibinfo {author} {\bibfnamefont {A.}~\bibnamefont
  {Bauswein}}\ and\ \bibinfo {author} {\bibfnamefont {H.-T.}\ \bibnamefont
  {Janka}},\ }\bibfield  {title} {\bibinfo {title} {Measuring neutron-star
  properties via gravitational waves from neutron-star mergers},\ }\href
  {https://doi.org/10.1103/PhysRevLett.108.011101} {\bibfield  {journal}
  {\bibinfo  {journal} {Phys. Rev. Lett.}\ }\textbf {\bibinfo {volume} {108}},\
  \bibinfo {pages} {011101} (\bibinfo {year} {2012}{\natexlab{b}})}\BibitemShut
  {NoStop}%
\bibitem [{\citenamefont {Bauswein}\ \emph {et~al.}(2012)\citenamefont
  {Bauswein}, \citenamefont {Janka}, \citenamefont {Hebeler},\ and\
  \citenamefont {Schwenk}}]{Bauswein2012Sep}%
  \BibitemOpen
  \bibfield  {author} {\bibinfo {author} {\bibfnamefont {A.}~\bibnamefont
  {Bauswein}}, \bibinfo {author} {\bibfnamefont {H.-T.}\ \bibnamefont {Janka}},
  \bibinfo {author} {\bibfnamefont {K.}~\bibnamefont {Hebeler}},\ and\ \bibinfo
  {author} {\bibfnamefont {A.}~\bibnamefont {Schwenk}},\ }\bibfield  {title}
  {\bibinfo {title} {Equation-of-state dependence of the gravitational-wave
  signal from the ring-down phase of neutron-star mergers},\ }\href
  {https://doi.org/10.1103/PhysRevD.86.063001} {\bibfield  {journal} {\bibinfo
  {journal} {Phys. Rev. D}\ }\textbf {\bibinfo {volume} {86}},\ \bibinfo
  {pages} {063001} (\bibinfo {year} {2012})}\BibitemShut {NoStop}%
\bibitem [{\citenamefont {Takami}\ \emph {et~al.}(2015)\citenamefont {Takami},
  \citenamefont {Rezzolla},\ and\ \citenamefont {Baiotti}}]{Takami2015}%
  \BibitemOpen
  \bibfield  {author} {\bibinfo {author} {\bibfnamefont {K.}~\bibnamefont
  {Takami}}, \bibinfo {author} {\bibfnamefont {L.}~\bibnamefont {Rezzolla}},\
  and\ \bibinfo {author} {\bibfnamefont {L.}~\bibnamefont {Baiotti}},\
  }\bibfield  {title} {\bibinfo {title} {Spectral properties of the post-merger
  gravitational-wave signal from binary neutron stars},\ }\href
  {https://doi.org/10.1103/PhysRevD.91.064001} {\bibfield  {journal} {\bibinfo
  {journal} {Phys. Rev. D}\ }\textbf {\bibinfo {volume} {91}},\ \bibinfo
  {pages} {064001} (\bibinfo {year} {2015})}\BibitemShut {NoStop}%
\bibitem [{\citenamefont {{Baiotti}}(2019)}]{Baiotti2019}%
  \BibitemOpen
  \bibfield  {author} {\bibinfo {author} {\bibfnamefont {L.}~\bibnamefont
  {{Baiotti}}},\ }\bibfield  {title} {\bibinfo {title} {{Gravitational waves
  from neutron star mergers and their relation to the nuclear equation of
  state}},\ }\href {https://doi.org/10.1016/j.ppnp.2019.103714} {\bibfield
  {journal} {\bibinfo  {journal} {Progress in Particle and Nuclear Physics}\
  }\textbf {\bibinfo {volume} {109}},\ \bibinfo {eid} {103714} (\bibinfo {year}
  {2019})},\ \Eprint {https://arxiv.org/abs/1907.08534} {arXiv:1907.08534
  [astro-ph.HE]} \BibitemShut {NoStop}%
\bibitem [{\citenamefont {Friedman}\ and\ \citenamefont
  {Stergioulas}(2020)}]{Friedman2020}%
  \BibitemOpen
  \bibfield  {author} {\bibinfo {author} {\bibfnamefont {J.~L.}\ \bibnamefont
  {Friedman}}\ and\ \bibinfo {author} {\bibfnamefont {N.}~\bibnamefont
  {Stergioulas}},\ }\bibfield  {title} {\bibinfo {title} {Astrophysical
  implications of neutron star inspiral and coalescence},\ }\href
  {https://doi.org/10.1142/S0218271820410151} {\bibfield  {journal} {\bibinfo
  {journal} {International Journal of Modern Physics D}\ }\textbf {\bibinfo
  {volume} {29}},\ \bibinfo {pages} {2041015} (\bibinfo {year} {2020})},\
  \Eprint {https://arxiv.org/abs/https://doi.org/10.1142/S0218271820410151}
  {https://doi.org/10.1142/S0218271820410151} \BibitemShut {NoStop}%
\bibitem [{\citenamefont {{Bernuzzi}}(2020)}]{Bernuzzi2020}%
  \BibitemOpen
  \bibfield  {author} {\bibinfo {author} {\bibfnamefont {S.}~\bibnamefont
  {{Bernuzzi}}},\ }\bibfield  {title} {\bibinfo {title} {{Neutron star merger
  remnants}},\ }\href {https://doi.org/10.1007/s10714-020-02752-5} {\bibfield
  {journal} {\bibinfo  {journal} {General Relativity and Gravitation}\ }\textbf
  {\bibinfo {volume} {52}},\ \bibinfo {eid} {108} (\bibinfo {year} {2020})},\
  \Eprint {https://arxiv.org/abs/2004.06419} {arXiv:2004.06419 [astro-ph.HE]}
  \BibitemShut {NoStop}%
\bibitem [{\citenamefont {{Dietrich}}\ \emph
  {et~al.}(2021{\natexlab{b}})\citenamefont {{Dietrich}}, \citenamefont
  {{Hinderer}},\ and\ \citenamefont {{Samajdar}}}]{Diedrich2021mar}%
  \BibitemOpen
  \bibfield  {author} {\bibinfo {author} {\bibfnamefont {T.}~\bibnamefont
  {{Dietrich}}}, \bibinfo {author} {\bibfnamefont {T.}~\bibnamefont
  {{Hinderer}}},\ and\ \bibinfo {author} {\bibfnamefont {A.}~\bibnamefont
  {{Samajdar}}},\ }\bibfield  {title} {\bibinfo {title} {{Interpreting binary
  neutron star mergers: describing the binary neutron star dynamics, modelling
  gravitational waveforms, and analyzing detections}},\ }\href
  {https://doi.org/10.1007/s10714-020-02751-6} {\bibfield  {journal} {\bibinfo
  {journal} {General Relativity and Gravitation}\ }\textbf {\bibinfo {volume}
  {53}},\ \bibinfo {eid} {27} (\bibinfo {year} {2021}{\natexlab{b}})},\ \Eprint
  {https://arxiv.org/abs/2004.02527} {arXiv:2004.02527 [gr-qc]} \BibitemShut
  {NoStop}%
\bibitem [{\citenamefont {{Sarin}}\ and\ \citenamefont
  {{Lasky}}(2021{\natexlab{b}})}]{Sarin2021jun}%
  \BibitemOpen
  \bibfield  {author} {\bibinfo {author} {\bibfnamefont {N.}~\bibnamefont
  {{Sarin}}}\ and\ \bibinfo {author} {\bibfnamefont {P.~D.}\ \bibnamefont
  {{Lasky}}},\ }\bibfield  {title} {\bibinfo {title} {{The evolution of binary
  neutron star post-merger remnants: a review}},\ }\href
  {https://doi.org/10.1007/s10714-021-02831-1} {\bibfield  {journal} {\bibinfo
  {journal} {General Relativity and Gravitation}\ }\textbf {\bibinfo {volume}
  {53}},\ \bibinfo {eid} {59} (\bibinfo {year} {2021}{\natexlab{b}})},\ \Eprint
  {https://arxiv.org/abs/2012.08172} {arXiv:2012.08172 [astro-ph.HE]}
  \BibitemShut {NoStop}%
\bibitem [{\citenamefont {Lee}\ \emph {et~al.}(2019)\citenamefont {Lee},
  \citenamefont {Gommers}, \citenamefont {Waselewski}, \citenamefont
  {Wohlfahrt},\ and\ \citenamefont {O'Leary}}]{Lee2019}%
  \BibitemOpen
  \bibfield  {author} {\bibinfo {author} {\bibfnamefont {G.~R.}\ \bibnamefont
  {Lee}}, \bibinfo {author} {\bibfnamefont {R.}~\bibnamefont {Gommers}},
  \bibinfo {author} {\bibfnamefont {F.}~\bibnamefont {Waselewski}}, \bibinfo
  {author} {\bibfnamefont {K.}~\bibnamefont {Wohlfahrt}},\ and\ \bibinfo
  {author} {\bibfnamefont {A.}~\bibnamefont {O'Leary}},\ }\bibfield  {title}
  {\bibinfo {title} {Pywavelets: A python package for wavelet analysis},\
  }\href {https://doi.org/10.21105/joss.01237} {\bibfield  {journal} {\bibinfo
  {journal} {Journal of Open Source Software}\ }\textbf {\bibinfo {volume}
  {4}},\ \bibinfo {pages} {1237} (\bibinfo {year} {2019})}\BibitemShut
  {NoStop}%
\bibitem [{\citenamefont {Aasi}\ \emph {et~al.}(2015)\citenamefont {Aasi} \emph
  {et~al.}}]{LIGOScientific:2014pky}%
  \BibitemOpen
  \bibfield  {author} {\bibinfo {author} {\bibfnamefont {J.}~\bibnamefont
  {Aasi}} \emph {et~al.} (\bibinfo {collaboration} {LIGO Scientific}),\
  }\bibfield  {title} {\bibinfo {title} {{Advanced LIGO}},\ }\href
  {https://doi.org/10.1088/0264-9381/32/7/074001} {\bibfield  {journal}
  {\bibinfo  {journal} {Class. Quant. Grav.}\ }\textbf {\bibinfo {volume}
  {32}},\ \bibinfo {pages} {074001} (\bibinfo {year} {2015})},\ \Eprint
  {https://arxiv.org/abs/1411.4547} {arXiv:1411.4547 [gr-qc]} \BibitemShut
  {NoStop}%
\bibitem [{\citenamefont {{Punturo}}\ \emph
  {et~al.}(2010{\natexlab{b}})\citenamefont {{Punturo}}, \citenamefont
  {{Abernathy}}, \citenamefont {{Acernese}}, \citenamefont {{Allen}},
  \citenamefont {{Andersson}}, \citenamefont {{Arun}}, \citenamefont
  {{Barone}}, \citenamefont {{Barr}}, \citenamefont {{Barsuglia}},
  \citenamefont {{Beker}} \emph {et~al.}}]{EinsteinTelescope2010}%
  \BibitemOpen
  \bibfield  {author} {\bibinfo {author} {\bibfnamefont {M.}~\bibnamefont
  {{Punturo}}}, \bibinfo {author} {\bibfnamefont {M.}~\bibnamefont
  {{Abernathy}}}, \bibinfo {author} {\bibfnamefont {F.}~\bibnamefont
  {{Acernese}}}, \bibinfo {author} {\bibfnamefont {B.}~\bibnamefont {{Allen}}},
  \bibinfo {author} {\bibfnamefont {N.}~\bibnamefont {{Andersson}}}, \bibinfo
  {author} {\bibfnamefont {K.}~\bibnamefont {{Arun}}}, \bibinfo {author}
  {\bibfnamefont {F.}~\bibnamefont {{Barone}}}, \bibinfo {author}
  {\bibfnamefont {B.}~\bibnamefont {{Barr}}}, \bibinfo {author} {\bibfnamefont
  {M.}~\bibnamefont {{Barsuglia}}}, \bibinfo {author} {\bibfnamefont
  {M.}~\bibnamefont {{Beker}}}, \emph {et~al.},\ }\bibfield  {title} {\bibinfo
  {title} {{The Einstein Telescope: a third-generation gravitational wave
  observatory}},\ }\href {https://doi.org/10.1088/0264-9381/27/19/194002}
  {\bibfield  {journal} {\bibinfo  {journal} {Classical and Quantum Gravity}\
  }\textbf {\bibinfo {volume} {27}},\ \bibinfo {eid} {194002} (\bibinfo {year}
  {2010}{\natexlab{b}})}\BibitemShut {NoStop}%
\bibitem [{\citenamefont {Oechslin}\ \emph {et~al.}(2002)\citenamefont
  {Oechslin}, \citenamefont {Rosswog},\ and\ \citenamefont
  {Thielemann}}]{SPH1}%
  \BibitemOpen
  \bibfield  {author} {\bibinfo {author} {\bibfnamefont {R.}~\bibnamefont
  {Oechslin}}, \bibinfo {author} {\bibfnamefont {S.}~\bibnamefont {Rosswog}},\
  and\ \bibinfo {author} {\bibfnamefont {F.-K.}\ \bibnamefont {Thielemann}},\
  }\bibfield  {title} {\bibinfo {title} {Conformally flat smoothed particle
  hydrodynamics application to neutron star mergers},\ }\href
  {https://doi.org/10.1103/PhysRevD.65.103005} {\bibfield  {journal} {\bibinfo
  {journal} {Phys. Rev. D}\ }\textbf {\bibinfo {volume} {65}},\ \bibinfo
  {pages} {103005} (\bibinfo {year} {2002})}\BibitemShut {NoStop}%
\bibitem [{\citenamefont {{Oechslin, R.}}\ \emph {et~al.}(2007)\citenamefont
  {{Oechslin, R.}}, \citenamefont {{Janka, H.-T.}},\ and\ \citenamefont
  {{Marek, A.}}}]{SPH2}%
  \BibitemOpen
  \bibfield  {author} {\bibinfo {author} {\bibnamefont {{Oechslin, R.}}},
  \bibinfo {author} {\bibnamefont {{Janka, H.-T.}}},\ and\ \bibinfo {author}
  {\bibnamefont {{Marek, A.}}},\ }\bibfield  {title} {\bibinfo {title}
  {Relativistic neutron star merger simulations with non-zero temperature
  equations of state* - i. variation of binary parameters and equation of
  state},\ }\href {https://doi.org/10.1051/0004-6361:20066682} {\bibfield
  {journal} {\bibinfo  {journal} {A\&A}\ }\textbf {\bibinfo {volume} {467}},\
  \bibinfo {pages} {395} (\bibinfo {year} {2007})}\BibitemShut {NoStop}%
\bibitem [{\citenamefont {Vretinaris}\ \emph {et~al.}(2020)\citenamefont
  {Vretinaris}, \citenamefont {Stergioulas},\ and\ \citenamefont
  {Bauswein}}]{Vretinaris2020}%
  \BibitemOpen
  \bibfield  {author} {\bibinfo {author} {\bibfnamefont {S.}~\bibnamefont
  {Vretinaris}}, \bibinfo {author} {\bibfnamefont {N.}~\bibnamefont
  {Stergioulas}},\ and\ \bibinfo {author} {\bibfnamefont {A.}~\bibnamefont
  {Bauswein}},\ }\bibfield  {title} {\bibinfo {title} {Empirical relations for
  gravitational-wave asteroseismology of binary neutron star mergers},\ }\href
  {https://doi.org/10.1103/PhysRevD.101.084039} {\bibfield  {journal} {\bibinfo
   {journal} {Phys. Rev. D}\ }\textbf {\bibinfo {volume} {101}},\ \bibinfo
  {pages} {084039} (\bibinfo {year} {2020})}\BibitemShut {NoStop}%
\bibitem [{\citenamefont {Branch}\ \emph {et~al.}(1999)\citenamefont {Branch},
  \citenamefont {Coleman},\ and\ \citenamefont {Li}}]{TRF1}%
  \BibitemOpen
  \bibfield  {author} {\bibinfo {author} {\bibfnamefont {M.~A.}\ \bibnamefont
  {Branch}}, \bibinfo {author} {\bibfnamefont {T.~F.}\ \bibnamefont
  {Coleman}},\ and\ \bibinfo {author} {\bibfnamefont {Y.}~\bibnamefont {Li}},\
  }\bibfield  {title} {\bibinfo {title} {A subspace, interior, and conjugate
  gradient method for large-scale bound-constrained minimization problems},\
  }\href {https://doi.org/10.1137/S1064827595289108} {\bibfield  {journal}
  {\bibinfo  {journal} {SIAM Journal on Scientific Computing}\ }\textbf
  {\bibinfo {volume} {21}},\ \bibinfo {pages} {1} (\bibinfo {year} {1999})},\
  \Eprint {https://arxiv.org/abs/https://doi.org/10.1137/S1064827595289108}
  {https://doi.org/10.1137/S1064827595289108} \BibitemShut {NoStop}%
\bibitem [{\citenamefont {Byrd}\ \emph {et~al.}(1988)\citenamefont {Byrd},
  \citenamefont {Schnabel},\ and\ \citenamefont {Shultz}}]{TRF2}%
  \BibitemOpen
  \bibfield  {author} {\bibinfo {author} {\bibfnamefont {R.~H.}\ \bibnamefont
  {Byrd}}, \bibinfo {author} {\bibfnamefont {R.~B.}\ \bibnamefont {Schnabel}},\
  and\ \bibinfo {author} {\bibfnamefont {G.~A.}\ \bibnamefont {Shultz}},\
  }\bibfield  {title} {\bibinfo {title} {Approximate solution of the trust
  region problem by minimization over two-dimensional subspaces},\ }\href
  {https://doi.org/10.1007/BF01580735} {\bibfield  {journal} {\bibinfo
  {journal} {Mathematical Programming}\ }\textbf {\bibinfo {volume} {40}},\
  \bibinfo {pages} {247} (\bibinfo {year} {1988})}\BibitemShut {NoStop}%
\bibitem [{\citenamefont {Virtanen}\ \emph {et~al.}(2020)\citenamefont
  {Virtanen}, \citenamefont {Gommers}, \citenamefont {Oliphant}, \citenamefont
  {Haberland}, \citenamefont {Reddy}, \citenamefont {Cournapeau}, \citenamefont
  {Burovski}, \citenamefont {Peterson}, \citenamefont {Weckesser},
  \citenamefont {Bright} \emph {et~al.}}]{2020SciPy-NMeth}%
  \BibitemOpen
  \bibfield  {author} {\bibinfo {author} {\bibfnamefont {P.}~\bibnamefont
  {Virtanen}}, \bibinfo {author} {\bibfnamefont {R.}~\bibnamefont {Gommers}},
  \bibinfo {author} {\bibfnamefont {T.~E.}\ \bibnamefont {Oliphant}}, \bibinfo
  {author} {\bibfnamefont {M.}~\bibnamefont {Haberland}}, \bibinfo {author}
  {\bibfnamefont {T.}~\bibnamefont {Reddy}}, \bibinfo {author} {\bibfnamefont
  {D.}~\bibnamefont {Cournapeau}}, \bibinfo {author} {\bibfnamefont
  {E.}~\bibnamefont {Burovski}}, \bibinfo {author} {\bibfnamefont
  {P.}~\bibnamefont {Peterson}}, \bibinfo {author} {\bibfnamefont
  {W.}~\bibnamefont {Weckesser}}, \bibinfo {author} {\bibfnamefont
  {J.}~\bibnamefont {Bright}}, \emph {et~al.},\ }\bibfield  {title} {\bibinfo
  {title} {{{SciPy} 1.0: Fundamental Algorithms for Scientific Computing in
  Python}},\ }\href {https://doi.org/10.1038/s41592-019-0686-2} {\bibfield
  {journal} {\bibinfo  {journal} {Nature Methods}\ }\textbf {\bibinfo {volume}
  {17}},\ \bibinfo {pages} {261} (\bibinfo {year} {2020})}\BibitemShut
  {NoStop}%
\bibitem [{\citenamefont {Apostolatos}(1995)}]{Apostolatos1995}%
  \BibitemOpen
  \bibfield  {author} {\bibinfo {author} {\bibfnamefont {T.~A.}\ \bibnamefont
  {Apostolatos}},\ }\bibfield  {title} {\bibinfo {title} {Search templates for
  gravitational waves from precessing, inspiraling binaries},\ }\href
  {https://doi.org/10.1103/PhysRevD.52.605} {\bibfield  {journal} {\bibinfo
  {journal} {Phys. Rev. D}\ }\textbf {\bibinfo {volume} {52}},\ \bibinfo
  {pages} {605} (\bibinfo {year} {1995})}\BibitemShut {NoStop}%
\bibitem [{\citenamefont {Passamonti}\ and\ \citenamefont
  {Andersson}(2020)}]{Passamonti_Andersson_2020}%
  \BibitemOpen
  \bibfield  {author} {\bibinfo {author} {\bibfnamefont {A.}~\bibnamefont
  {Passamonti}}\ and\ \bibinfo {author} {\bibfnamefont {N.}~\bibnamefont
  {Andersson}},\ }\bibfield  {title} {\bibinfo {title} {{Merger-inspired
  rotation laws and the low-T/W instability in neutron stars}},\ }\href
  {https://doi.org/10.1093/mnras/staa2725} {\bibfield  {journal} {\bibinfo
  {journal} {\mnras}\ }\textbf {\bibinfo {volume} {498}},\ \bibinfo {pages}
  {5904} (\bibinfo {year} {2020})},\ \Eprint
  {https://arxiv.org/abs/https://academic.oup.com/mnras/article-pdf/498/4/5904/33838626/staa2725.pdf}
  {https://academic.oup.com/mnras/article-pdf/498/4/5904/33838626/staa2725.pdf}
  \BibitemShut {NoStop}%
\bibitem [{\citenamefont {{Xie}}\ \emph {et~al.}(2020)\citenamefont {{Xie}},
  \citenamefont {{Hawke}}, \citenamefont {{Passamonti}},\ and\ \citenamefont
  {{Andersson}}}]{Xie_etal_2020}%
  \BibitemOpen
  \bibfield  {author} {\bibinfo {author} {\bibfnamefont {X.}~\bibnamefont
  {{Xie}}}, \bibinfo {author} {\bibfnamefont {I.}~\bibnamefont {{Hawke}}},
  \bibinfo {author} {\bibfnamefont {A.}~\bibnamefont {{Passamonti}}},\ and\
  \bibinfo {author} {\bibfnamefont {N.}~\bibnamefont {{Andersson}}},\
  }\bibfield  {title} {\bibinfo {title} {{Instabilities in neutron-star
  postmerger remnants}},\ }\href {https://doi.org/10.1103/PhysRevD.102.044040}
  {\bibfield  {journal} {\bibinfo  {journal} {\prd}\ }\textbf {\bibinfo
  {volume} {102}},\ \bibinfo {eid} {044040} (\bibinfo {year} {2020})},\ \Eprint
  {https://arxiv.org/abs/2005.13696} {arXiv:2005.13696 [astro-ph.HE]}
  \BibitemShut {NoStop}%
\bibitem [{\citenamefont {De~Pietri}\ \emph {et~al.}(2020)\citenamefont
  {De~Pietri}, \citenamefont {Feo}, \citenamefont {Font}, \citenamefont
  {L\"offler}, \citenamefont {Pasquali},\ and\ \citenamefont
  {Stergioulas}}]{PhysRevD.101.064052}%
  \BibitemOpen
  \bibfield  {author} {\bibinfo {author} {\bibfnamefont {R.}~\bibnamefont
  {De~Pietri}}, \bibinfo {author} {\bibfnamefont {A.}~\bibnamefont {Feo}},
  \bibinfo {author} {\bibfnamefont {J.~A.}\ \bibnamefont {Font}}, \bibinfo
  {author} {\bibfnamefont {F.}~\bibnamefont {L\"offler}}, \bibinfo {author}
  {\bibfnamefont {M.}~\bibnamefont {Pasquali}},\ and\ \bibinfo {author}
  {\bibfnamefont {N.}~\bibnamefont {Stergioulas}},\ }\bibfield  {title}
  {\bibinfo {title} {Numerical-relativity simulations of long-lived remnants of
  binary neutron star mergers},\ }\href
  {https://doi.org/10.1103/PhysRevD.101.064052} {\bibfield  {journal} {\bibinfo
   {journal} {Phys. Rev. D}\ }\textbf {\bibinfo {volume} {101}},\ \bibinfo
  {pages} {064052} (\bibinfo {year} {2020})}\BibitemShut {NoStop}%
\bibitem [{\citenamefont {Kyutoku}\ \emph {et~al.}(2014)\citenamefont
  {Kyutoku}, \citenamefont {Shibata},\ and\ \citenamefont
  {Taniguchi}}]{PhysRevD.90.064006}%
  \BibitemOpen
  \bibfield  {author} {\bibinfo {author} {\bibfnamefont {K.}~\bibnamefont
  {Kyutoku}}, \bibinfo {author} {\bibfnamefont {M.}~\bibnamefont {Shibata}},\
  and\ \bibinfo {author} {\bibfnamefont {K.}~\bibnamefont {Taniguchi}},\
  }\bibfield  {title} {\bibinfo {title} {Reducing orbital eccentricity in
  initial data of binary neutron stars},\ }\href
  {https://doi.org/10.1103/PhysRevD.90.064006} {\bibfield  {journal} {\bibinfo
  {journal} {Phys. Rev. D}\ }\textbf {\bibinfo {volume} {90}},\ \bibinfo
  {pages} {064006} (\bibinfo {year} {2014})}\BibitemShut {NoStop}%
\end{thebibliography}
\end{document}